\documentclass[longauth]{aa} 
\usepackage{graphicx}
\usepackage{txfonts}
\usepackage{natbib}
\usepackage{xcolor}
\usepackage{supertabular}
\usepackage{subfig}
\usepackage{float}
\usepackage{hyperref}
\usepackage[section]{placeins}
\DeclareCaptionFormat{cont}{#1 (cont.)#2#3\par}

\def\as {\ifmmode {\rlap.}$\,$''$\,$\! \else ${\rlap.}$\,$''$\,$\!$\fi}

\begin{document} 

\title{The physical and chemical structure of high-mass star-forming regions}

\subtitle{Unraveling chemical complexity with the NOEMA large program ``CORE''}

\author{C.~Gieser
\inst{1}\fnmsep\thanks{Fellow of the International Max Planck Research School for Astronomy and Cosmic Physics at the University of Heidelberg (IMPRS-HD).}
\and
H.~Beuther\inst{1}
\and 
D.~Semenov\inst{1,2}
\and 
A.~Ahmadi\inst{3}
\and 
S.~Suri\inst{1}
\and
T.~M{\"o}ller\inst{4}
\and
M.T.~Beltr{\'a}n\inst{5}
\and
P.~Klaassen\inst{6}
\and
Q.~Zhang\inst{7}
\and
J.S.~Urquhart\inst{8}
\and
Th.~Henning\inst{1}
\and
S.~Feng\inst{9,10,11}
\and
R.~Galv{\'a}n-Madrid\inst{12}
\and
V.~de Souza Magalh\~{a}es\inst{13}
\and
L.~Moscadelli\inst{5}
\and
S.~Longmore\inst{14}
\and
S.~Leurini\inst{15}
\and
R.~Kuiper\inst{16}
\and
T.~Peters\inst{17}
\and
K.M.~Menten\inst{18}
\and
T.~Csengeri\inst{19}
\and
G.~Fuller\inst{20}
\and
F.~Wyrowski\inst{18}
\and
S.~Lumsden\inst{21}
\and
{\'A}.~S{\'a}nchez-Monge\inst{4}
\and
L.~Maud\inst{22}
\and
H.~Linz\inst{1}
\and
A.~Palau\inst{12}
\and
P.~Schilke\inst{4}
\and
J.~Pety\inst{13,23}
\and
R.~Pudritz\inst{24}
\and
J.M.~Winters\inst{13}
\and
V.~Pi{\'e}tu\inst{13}
}

\institute{Max Planck Institute for Astronomy, K{\"o}nigstuhl 17, 69117 Heidelberg, Germany,\\
\email{gieser@mpia.de}
\and 
Department of Chemistry, Ludwig Maximilian University, Butenandtstr. 5-13, 81377 Munich, Germany
\and
Leiden University, Niels Bohrweg 2, 2333 CA Leiden, Netherlands
\and
I. Physikalisches Institut, Universit{\"a}t zu K{\"o}ln, Z{\"u}lpicher Str. 77, D-50937, K{\"o}ln, Germany
\and
INAF, Osservatorio Astrofisico di Arcetri, Largo E. Fermi 5, I-50125 Firenze, Italy
\and
UK Astronomy Technology Centre, Royal Observatory Edinburgh, Blackford Hill, Edinburgh EH9 3HJ, UK
\and
Center for Astrophysics $|$ Harvard \& Smithsonian, 60 Garden Street, Cambridge, MA 02138, USA
\and
Centre for Astrophysics and Planetary Science, University of Kent, Canterbury, CT2 7NH, UK
\and
Academia Sinica Institute of Astronomy and Astrophysics, No.1, Sec. 4, Roosevelt Rd, Taipei 10617, Taiwan, Republic of China
\and
CAS Key Laboratory of FAST, National Astronomical Observatories, Chinese Academy of Sciences, Beijing 100101, People's Republic of China
\and
National Astronomical Observatory of Japan, National Institutes of Natural Sciences, 2-21-1 Osawa, Mitaka, Tokyo 181-8588, Japan
\and
Instituto de Radioastronom{\'i}a y Astrof{\'i}sica (IRyA), UNAM, Apdo. Postal 72-3 (Xangari), Morelia, Michoac{\'a}n 58089, Mexico
\and
Institut de Radioastronomie Millim\'{e}trique (IRAM), 300 Rue de la Piscine, F-38406 Saint Martin d'H\`{e}res, France
\and
Astrophysics Research Institute, Liverpool John Moores University, Liverpool, L3 5RF, UK
\and
INAF, Osservatorio Astronomico di Cagliari, Via della Scienza 5, I-09047, Selargius (CA), Italy
\and
Institute of Astronomy and Astrophysics, University of T{\"u}bingen, Auf der Morgenstelle 10, 72076, T{\"u}bingen, Germany
\and
Max-Planck-Institut f{\"u}r Astrophysik, Karl-Schwarzschild-Str. 1, 85748 Garching, Germany
\and
Max Planck Institut for Radioastronomie, Auf dem H{\"u}gel 69, 53121 Bonn, Germany
\and
Laboratoire d’astrophysique de Bordeaux, Univ. Bordeaux, CNRS, B18N, all{\`e}e Geoffroy Saint-Hilaire, 33615 Pessac, France
\and
Physics Department, UMIST, P.O. Box 88, Manchester M60 1QD, UK
\and
School of Physics and Astronomy, University of Leeds, Leeds LS2 9JT, United Kingdom
\and
European Southern Observatory, Karl-Schwarzschild-Str. 2, D-85748 Garching, Germany
\and
LERMA, Observatoire de Paris, PSL Research University, CNRS, Sorbonne Universit{\'e}s, 75014 Paris, France
\and
Department of Physics and Astronomy, McMaster University, 1280 Main St. W, Hamilton, ON L8S 4M1, Canada
}

\date{Received...; accepted...}
 
\abstract
{}
{Characterizing the physical and chemical properties of forming massive stars at the spatial resolution of individual high-mass cores lies at the heart of current star formation research.}
{We use sub-arcsecond resolution ($\sim$0\as4) observations with the NOrthern Extended Millimeter Array at 1.37\,mm to study the dust emission and molecular gas of 18 high-mass star-forming regions. With distances in the range of $0.7 - 5.5$\,kpc this corresponds to spatial scales down to $300 - 2\,300$\,au that are resolved by our observations. We combine the derived physical and chemical properties of individual cores in these regions to estimate their ages. The temperature structure of these regions are determined by fitting H$_{2}$CO and CH$_{3}$CN line emission. The density profiles are inferred from the 1.37\,mm continuum visibilities. The column densities of 11 different species are determined by fitting the emission lines with \texttt{XCLASS}.}
{Within the 18 observed regions, we identify 22 individual cores with associated 1.37\,mm continuum emission and with a radially decreasing temperature profile. We find an average temperature power-law index of $q = 0.4\pm0.1$ and an average density power-law index of $p = 2.0\pm0.2$ on scales on the order of several 1\,000\,au. Comparing these results with values of $p$ derived in the literature suggest that the density profiles remain unchanged from clump to core scales. The column densities relative to $N$(C$^{18}$O) between pairs of dense gas tracers show tight correlations. We apply the physical-chemical model MUlti Stage ChemicaL codE (\texttt{MUSCLE}) to the derived column densities of each core and find a mean chemical age of $\sim$60\,000\,yrs and an age spread of $20\,000 - 100\,000$\,yrs. With this paper we release all data products of the CORE project available at \url{https://www.mpia.de/core}.}
{The CORE sample reveals well constrained density and temperature power-law distributions. Furthermore, we characterize a large variety in molecular richness that can be explained by an age spread confirmed by our physical-chemical modeling. The hot molecular cores show the most emission lines, but we also find evolved cores at an evolutionary stage, in which most molecules are destroyed and thus the spectra appear line-poor again.}

\keywords{stars: formation -- interstellar medium: molecules -- astrochemistry}

\maketitle

\section{Introduction}\label{sec:introduction}

	The development of large telescopes and highly sensitive instruments allows us to investigate star-forming regions within and even outside of the Milky Way in a great detail \citep[for reviews, see, e.g.,][]{Larson1981,Shu1987,Kennicutt1998,McKee2007}. The study of high-mass star formation (HMSF) is challenging from an observational point of view \citep[for reviews, see, e.g.,][]{Beuther2007,Bonnell2007,Zinnecker2007,Smith2009,Tan2014,Krumholz2015,Schilke2015,Motte2018, Rosen2020}. High-mass stars are less common, located typically at large distances of several kiloparsecs and therefore difficult to observe at a high spatial resolution. The evolution of HMSF is fast, on the order of 10$^{5}$\,yrs as revealed by observations \citep[see, e.g., Table 2 in][]{Motte2018, Mottram2011} and theoretical models \citep{McKee2002,McKee2003,Kuiper2018}. Thus high-mass protostars are still deeply embedded within their parental molecular cloud when they reach the main sequence.
	
	In contrast to low-mass star formation, no clear consensus has yet been reached on the evolutionary stages of high-mass stars \citep[see, e.g., Fig. 8 in][]{Motte2018}. High-mass star-forming regions (HMSFRs) can be roughly categorized into several evolutionary stages based on the observed properties at infrared and radio wavelengths. Various classifications exist in the literature, e.g., based on infrared properties, \citet{Cooper2013} classifies massive young stellar objects (MYSOs) into three types: type I show strong H$_{2}$, but no ionized lines; type II are weaker in H$_{2}$, but show H{\sc i} emission in the Brackett series; type III have strong H{\sc i} and weak H$_{2}$ line emission. These near-infrared observations show that type III MYSOs are bluer than type I MYSOs indicating that they are more evolved. 
	
	In this study we follow the observationally driven nomenclature of four different evolutionary stages during HMSF by \citet{Gerner2014,Gerner2015}. When regions of a molecular cloud undergo gravitational collapse and fragmentation, protostars form in cold and dense clumps, often detected as infrared dark clouds \citep[IRDCs, e.g.,][]{Rathborne2006}. These clouds have typical temperatures of $\sim 10 - 20$\,K and are visible at (sub)mm wavelengths \citep[e.g.,][]{Pillai2006}, but show no or only weak emission in the infrared. As the collapse continues, due to the conservation of angular momentum, a disk-outflow system forms around the protostar and the temperature of the surrounding envelope increases due to central heating by the protostar \citep[e.g.,][]{SanchezMonge2013,Beltran2016}. Through accretion of the infalling gas, protostars increase their masses and luminosities and become visible at infrared wavelengths. Protostars during this stage are referred to as high-mass protostellar objects \citep[HMPOs, e.g.,][]{Sridharan2002,Williams2004,Beuther2010, DuarteCabral2013}. HMPOs are characterized by high bolometric luminosities, $L > 10^{4}$\,$L_{\odot}$, a strong thermal dust continuum, but only weak cm emission \citep{Sridharan2002,Beuther2002}. As the temperature of the surrounding envelope reaches $\gtrsim$100\,K, molecules that formed on the surfaces of dust grains and resided in ice layers, evaporate into the gas-phase revealing a rich molecular reservoir \citep[e.g.,][]{Osorio1999}. In this stage, the objects are referred to as hot molecular cores (HMCs) or ``hot cores'' \citep[e.g.,][]{Cesaroni1997}. HMPOs and HMCs have similar physical properties, however, the chemical composition of the gas around HMCs is richer due to higher temperatures in the envelope. Low-mass analogues are referred to as ``hot corinos'' \citep[e.g., IRAS\,16293-2422,][]{Bottinelli2004}. As the mass growth continues, the strong protostellar radiation dissociates and ionizes the surrounding envelope gas and an hyper/ultra-compact H{\sc ii} (HC/UCH{\sc ii}) region forms \citep[e.g.,][]{Wood1989,Hatchell1998,Churchwell2002,SanchezMonge2013b}. H{\sc ii} regions can be studied and classified through their strong free-free emission at cm wavelengths \citep[e.g.,][]{Peters2010b}. These four evolutionary stages are not sharp transitions and overlap, especially between the intermediate HMPO/HMC stages and there are HMCs which have already formed a HC/UCH{\sc ii} region. 
	
	Aside from the physical complexity of HMSFRs, observations reveal one of the most complex chemical compositions in the interstellar medium \citep[ISM, for a review, see, e.g.,][]{Herbst2009,Jorgensen2020}. To date about 200 molecules have been detected in the ISM \citep[an overview is found in][]{McGuire2018}, most of these can be observed toward the high-mass star-forming Galactic center molecular cloud Sagittarius B2 \citep[e.g.,][]{Belloche2013}. Spectroscopic studies at mm and sub-mm wavelengths have revealed the composition of the molecular gas in HMSFRs to be extremely diverse: nitrogen (N)-bearing species (such as HCN, CH$_{3}$CN); oxygen (O)-bearing species (such as H$_{2}$CO); sulfur (S)-bearing species (such as SO, SO$_{2}$). In shocked regions, enhanced abundances of S-bearing molecules such as SO$_{2}$ and silicon (Si)-bearing molecules such as SiO are observed \citep{Schilke1997}. During the HMC stage, a large variety of so-called complex organic molecules (COMs) evaporate into the gas-phase and produce line-rich spectra at mm wavelengths. We use the definition by \citet{Herbst2009} in which a COM contains six or more atoms. In the densest regions where protostars form, the chemical composition of the molecular gas has been studied thoroughly. An important finding is the chemical segregation of O- and N- bearing species \citep[e.g.,][]{Rodgers2001}. This has been observed and studied toward many bright HMCs such as Orion-KL \citep{Caselli1993,Feng2015}; W3\,H2O and W3\,OH \citep{Wyrowski1999,Qin2015}; AFGL\,2591 \citep{JimenezSerra2012, Gieser2019}; NGC7538\,IRS9, W3\,IRS5 and AFGL\,490 \citep{Fayolle2015}; G35.20-0.74N \citep{Allen2017}; SgrB2(N) \citep{Bonfand2017,Mills2018}; or AFGL\,4176 \citep{Johnston2020}.
	
	In this paper, we want to pin down the physical and chemical properties of HMSFRs from dust and molecular line emission and with this information investigate the chemical timescales. We analyze the physical structure and molecular content of 18 HMSFRs in combination with a physical-chemical model on spatial scales ranging from 10\,000\,au down to our resolution limit of 300\,au. We successfully tested the method presented in this paper on the well-studied hot core AFGL\,2591 VLA3 \citep{Gieser2019}. In this case-study, we used the molecular line and dust continuum data from the observations of the CORE project to derive the density and temperature structure and molecular column densities of this hot core and applied a physical-chemical model to estimate the chemical age. Here, we apply this method to all 18 CORE regions with the goal to compare and understand the chemical diversity we observe during HMSF.

	The paper is organized as followed: In Sect. \ref{sec:observations} we summarize the observations, data calibration and imaging techniques. In Sect. \ref{sec:physicalstructure} we derive the temperature and density structure of the regions. The molecular content is analyzed in Sect. \ref{sec:molecularcontent}. We apply a physical-chemical model to the observed molecular column densities in Sect. \ref{sec:chemicalmodeling}. The results of the physical and chemical properties of the regions are discussed in Sect. \ref{sec:discussion}. A summary and our conclusions are given in Sect. \ref{sec:conclusions}.

\section{Observations}\label{sec:observations}

\begin{table*}
\caption{Overview of the CORE sample. The isotopic ratios are taken from \citet{Wilson1994}, further explained in Sect. \ref{sec:temperaturestructure}, Sect. \ref{sec:XCLASSfitting}, and Sect. \ref{sec:MUSCLEsetup}.}
\label{tab:sample}
\centering
\begin{tabular}{lccccc|ccc}
\hline\hline
Region & \multicolumn{2}{c}{Coordinates} & Distance & Galactic Distance & Velocity & \multicolumn{3}{c}{Isotopic Ratios}\\
 & $\alpha$ & $\delta$ & $d$ & $d_\mathrm{gal}$ & $\varv_{\mathrm{LSR}}$ & $^{12}$C/$^{13}$C & $^{32}$S/$^{34}$S & $^{16}$O/$^{18}$O\\
 & J(2000) & J(2000) & (kpc) & (kpc) & (km s$^{-1}$) & & & \\
\hline
IRAS\,23033 & 23:05:25.00 & +60:08:15.5 & $4.3$ & $10.4$ & $-53.1$ & $86$ & $22$ & $649$\\ 
IRAS\,23151 & 23:17:21.01 & +59:28:47.5 & $3.3$ & $\,\,\, 9.8$ & $-54.4$ & $81$ & $22$ & $613$\\ 
IRAS\,23385 & 23:40:54.40 & +61:10:28.0 & $4.9$ & $11.1$ & $-50.2$ & $91$ & $22$ & $690$\\ 
AFGL\,2591 & 20:29:24.86 & +40:11:19.4 & $3.3$ & $\,\,\, 8.2$ & $-\,\,\, 5.5$ & $69$ & $22$ & $519$\\ 
CepA\,HW2 & 22:56:17.98 & +62:01:49.5 & $0.7$ & $\,\,\, 8.4$ & $-10.0$ & $71$ & $22$ & $531$\\ 
G084.9505 & 20:55:32.47 & +44:06:10.1 & $5.5$ & $\,\,\, 9.4$ & $-34.6$ & $78$ & $22$ & $590$\\ 
G094.6028 & 21:39:58.25 & +50:14:20.9 & $4.0$ & $\,\,\, 9.3$ & $+29.0$ & $77$ & $22$ & $584$\\ 
G100.38 & 22:16:10.35 & +52:21:34.7 & $3.5$ & $\,\,\, 9.4$ & $-37.6$ & $78$ & $22$ & $590$\\ 
G108.75 & 22:58:47.25 & +58:45:01.6 & $4.3$ & $10.3$ & $-51.5$ & $85$ & $22$ & $643$\\ 
G138.2957 & 03:01:31.32 & +60:29:13.2 & $2.9$ & $10.5$ & $-37.5$ & $86$ & $22$ & $654$\\ 
G139.9091 & 03:07:24.52 & +58:30:48.3 & $3.2$ & $10.8$ & $-40.5$ & $89$ & $22$ & $672$\\ 
G075.78 & 20:21:44.03 & +37:26:37.7 & $3.8$ & $\,\,\, 8.1$ & $-\,\,\, 8.0$ & $68$ & $22$ & $513$\\ 
IRAS\,21078 & 21:09:21.64 & +52:22:37.5 & $1.5$ & $\,\,\, 8.3$ & $-\,\,\, 6.1$ & $70$ & $22$ & $525$\\ 
NGC7538\,IRS9 & 23:14:01.68 & +61:27:19.1 & $2.7$ & $\,\,\, 9.5$ & $-57.0$ & $79$ & $22$ & $596$\\ 
S106 & 20:27:26.77 & +37:22:47.7 & $1.3$ & $\,\,\, 7.9$ & $-\,\,\, 1.0$ & $67$ & $22$ & $502$\\ 
S87\,IRS1 & 19:46:20.14 & +24:35:29.0 & $2.2$ & $\,\,\, 7.3$ & $+22.0$ & $62$ & $22$ & $466$\\ 
W3\,H2O & 02:27:04.60 & +61:52:24.7 & $2.0$ & $\,\,\, 9.6$ & $-48.5$ & $80$ & $22$ & $602$\\ 
W3\,IRS4 & 02:25:31.22 & +62:06:21.0 & $2.0$ & $\,\,\, 9.6$ & $-42.8$ & $80$ & $22$ & $602$\\ 
\hline 
\end{tabular}
\end{table*}

\begin{table*}
\caption{Overview of the CORE data products. The noise of the continuum data $\sigma_{\mathrm{cont}}$ is computed in a $3\as5\times3\as5$ rectangle in an emission-free region. $I_{\mathrm{peak}}$ is the continuum peak intensity. The average map noise of the merged (combined NOEMA + IRAM 30m) spectral line data $\sigma_{\mathrm{line,map}}$ is computed in a line-free range from a rest-frequency of 219.00\,GHz to 219.13\,GHz with a channel width of 3\,km\,s$^{-1}$ (2.2\,MHz) within the full primary beam.}
\label{tab:dataproducts}
\centering
\begin{tabular}{lcc|cc|cc|cc|ccc}
\hline\hline
Region & \multicolumn{4}{c}{Synthesized Beam} & \multicolumn{2}{|c}{\texttt{GILDAS}} & \multicolumn{2}{|c}{\texttt{CASA}} & \multicolumn{3}{|c}{\texttt{GILDAS}} \\
 & \multicolumn{2}{c}{Continuum Data} & \multicolumn{2}{|c}{Merged Line Data} & \multicolumn{2}{|c}{Standard Calibration} & \multicolumn{2}{|c}{Self-calibration} & \multicolumn{3}{|c}{{Self-calibration}} \\
 & $\theta_\mathrm{maj}\times \theta_\mathrm{min}$ & PA & $\theta_\mathrm{maj}\times \theta_\mathrm{min}$ & PA & $\sigma_{\mathrm{cont}}$ & $I_{\mathrm{peak}}$ & $\sigma_{\mathrm{cont}}$ & $I_{\mathrm{peak}}$ & $\sigma_{\mathrm{cont}}$ & $I_{\mathrm{peak}}$ & $\sigma_{\mathrm{line,map}}$ \\
 & ($''\times''$) & ($^{\circ}$) & ($''\times''$) & ($^{\circ}$) & \multicolumn{2}{|c}{(mJy beam$^{-1}$)} & \multicolumn{2}{|c}{(mJy beam$^{-1}$)}& \multicolumn{2}{|c}{(mJy beam$^{-1}$)} & (K) \\
\hline
IRAS\,23033 & 0.45$\times$0.37 & 47 & 0.43$\times$0.43 & 96 & 0.50 & $\,\,\, 28.21$ & 0.43 & $\,\,\, 38.76$ & 0.53 & $\,\,\, 38.13$ & 0.40\\ 
IRAS\,23151 & 0.45$\times$0.37 & 50 & 0.47$\times$0.38 & 54 & 0.26 & $\,\,\, 26.41$ & 0.26 & $\,\,\, 32.65$ & 0.25 & $\,\,\, 33.78$ & 0.48\\ 
IRAS\,23385 & 0.48$\times$0.43 & 58 & 0.49$\times$0.44 & 55 & 0.29 & $\,\,\, 14.61$ & 0.16 & $\,\,\, 18.03$ & 0.16 & $\,\,\, 17.84$ & 0.30\\ 
AFGL\,2591 & 0.47$\times$0.36 & 65 & 0.48$\times$0.37 & 65 & 0.53 & $\,\,\, 56.31$ & 0.50 & $\,\,\, 87.33$ & 0.51 & $\,\,\, 85.10$ & 0.65\\ 
CepA\,HW2 & 0.42$\times$0.41 & 41 & 0.44$\times$0.39 & 80 & 3.72 & $239.93$ & 2.39 & $441.07$ & 2.43 & $448.50$ & 0.68\\ 
G084.9505 & 0.43$\times$0.38 & 69 & 0.44$\times$0.39 & 72 & 0.10 & $\quad 4.37$ & 0.08 & $\quad 6.28$ & 0.08 & $\quad 5.19$ & 0.28\\ 
G094.6028 & 0.41$\times$0.39 & 77 & 0.41$\times$0.38 & 85 & 0.15 & $\,\,\, 10.24$ & 0.15 & $\,\,\, 13.65$ & 0.13 & $\,\,\, 12.84$ & 0.36\\ 
G100.38 & 0.49$\times$0.33 & 56 & 0.49$\times$0.34 & 55 & 0.07 & $\quad 6.35$ & 0.05 & $\quad 8.51$ & 0.06 & $\quad 7.28$ & 0.25\\ 
G108.75 & 0.50$\times$0.44 & 49 & 0.51$\times$0.44 & 49 & 0.17 & $\,\,\, 10.53$ & 0.12 & $\,\,\, 14.75$ & 0.13 & $\,\,\, 16.08$ & 0.27\\ 
G138.2957 & 0.50$\times$0.41 & 60 & 0.51$\times$0.41 & 59 & 0.16 & $\quad 6.28$ & 0.16 & $\quad 6.15$ & 0.10 & $\quad 6.28$ & 0.33\\ 
G139.9091 & 0.51$\times$0.40 & 56 & 0.52$\times$0.41 & 56 & 0.18 & $\quad 9.23$ & 0.20 & $\,\,\, 13.84$ & 0.15 & $\,\,\, 13.49$ & 0.34\\ 
G075.78 & 0.48$\times$0.37 & 60 & 0.49$\times$0.37 & 60 & 0.47 & $\,\,\, 46.31$ & 0.38 & $\,\,\, 64.50$ & 0.38 & $\,\,\, 71.67$ & 0.58\\ 
IRAS\,21078 & 0.48$\times$0.33 & 41 & 0.48$\times$0.33 & 41 & 0.41 & $\,\,\, 26.76$ & 0.44 & $\,\,\, 34.64$ & 0.44 & $\,\,\, 36.69$ & 0.41\\ 
NGC7538\,IRS9 & 0.44$\times$0.38 & 80 & 0.44$\times$0.38 & 80 & 0.32 & $\,\,\, 29.14$ & 0.19 & $\,\,\, 41.28$ & 0.21 & $\,\,\, 46.09$ & 0.42\\ 
S106 & 0.47$\times$0.34 & 47 & 0.48$\times$0.34 & 47 & 0.88 & $\,\,\, 77.41$ & 0.96 & $136.00$ & 1.20 & $150.47$ & 0.35\\ 
S87\,IRS1 & 0.54$\times$0.35 & 37 & 0.55$\times$0.36 & 37 & 0.12 & $\,\,\, 19.42$ & 0.17 & $\,\,\, 33.35$ & 0.18 & $\,\,\, 37.37$ & 0.44\\ 
W3\,H2O & 0.43$\times$0.32 & 86 & 0.45$\times$0.32 & 86 & 4.28 & $279.11$ & 2.54 & $449.84$ & 2.82 & $456.29$ & 0.74\\ 
W3\,IRS4 & 0.45$\times$0.32 & 82 & 0.46$\times$0.32 & 83 & 0.69 & $\,\,\, 21.73$ & 0.42 & $\,\,\, 39.34$ & 0.33 & $\,\,\, 51.50$ & 0.72\\ 
\hline 
\end{tabular}
\end{table*}

\begin{figure*}
\includegraphics[]{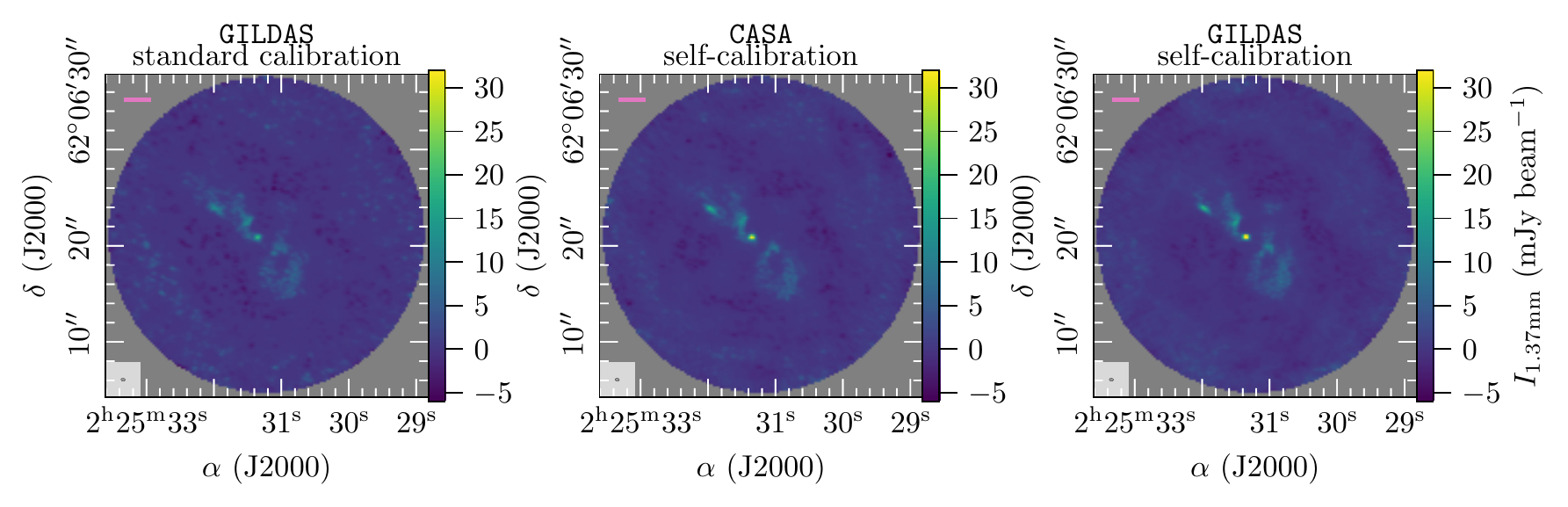}
\caption{Comparison of the \texttt{GILDAS} standard calibrated (left panel), \texttt{CASA} self-calibrated (middle panel) and \texttt{GILDAS} self-calibrated (right panel) W3\,IRS4 continuum data. In each panel, the beam size is shown in the bottom left corner and the pink bar in the top left corner indicates a linear spatial scale of 5\,000\,au.}
\label{fig:calibration_comparison_continuum}
\end{figure*}

\begin{figure}
\resizebox{\hsize}{!}{\includegraphics[]{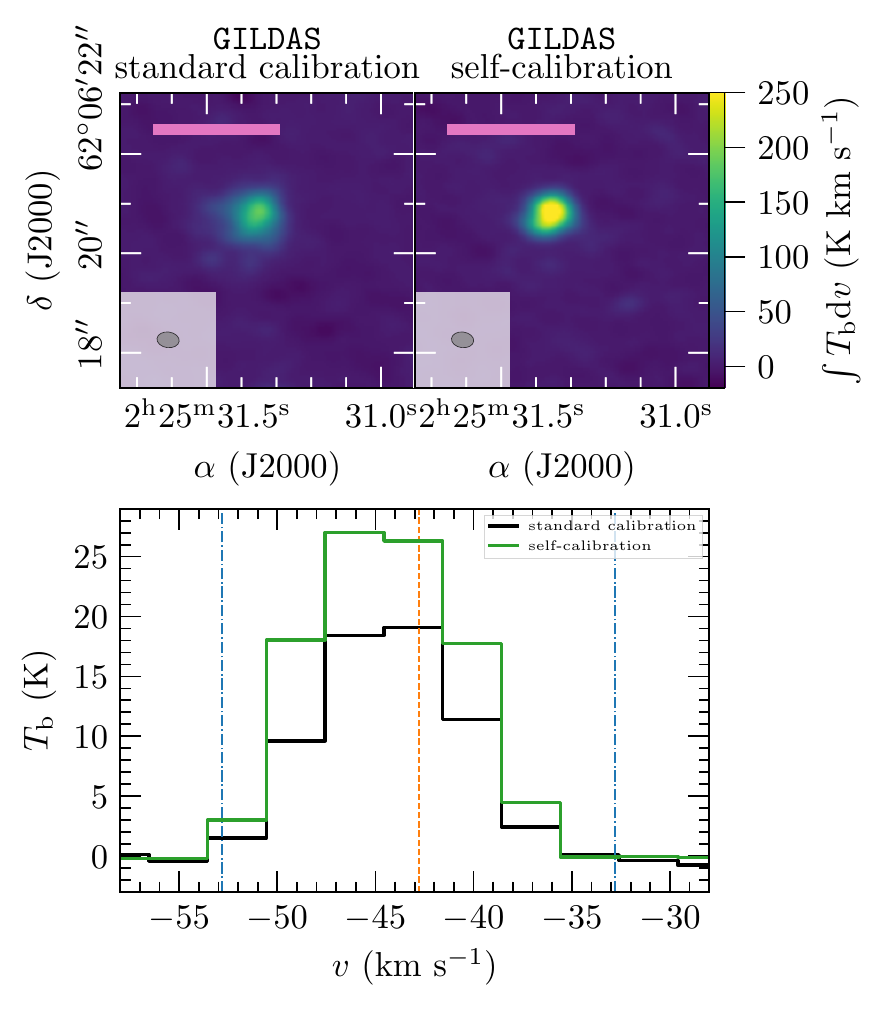}}
\caption{\textit{Top panels:} Comparison of the \texttt{GILDAS} standard calibration (left) and \texttt{GILDAS} self-calibration (right) of the W3\,IRS4 broad-band spectral line data zoomed in toward the position of the continuum peak. The integrated intensity of the CH$_{3}$CN $12_{3}-11_{3}$ transition is shown in color. In each panel, the beam size is shown in the bottom left corner and pink bar in the top left corner indicates a linear spatial scale of 5\,000\,au. \textit{Bottom panel:} CH$_{3}$CN $12_{3}-11_{3}$ spectrum at the position of the W3\,IRS4 continuum peak of the \texttt{GILDAS} standard calibrated data (black) and of the \texttt{GILDAS} self-calibrated data (green). The dashed orange line shows the systemic velocity of the region $\varv_{\mathrm{LSR}}$ and the dash-dotted blue lines show the lower and upper velocity limits ($\varv_{\mathrm{LSR}}\pm10$\,km\,s$^{-1}$) used for the integrated intensity map shown in the top panel.}
\label{fig:calibration_comparison_line}
\end{figure}

\begin{figure*}
\centering
\includegraphics[]{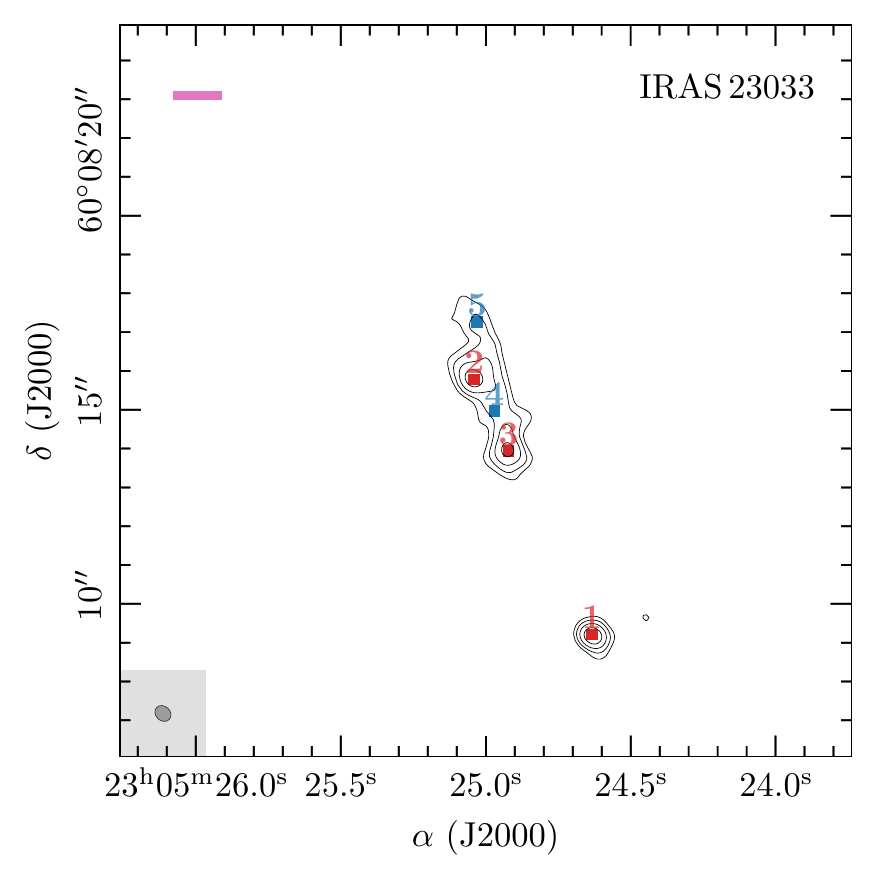}
\includegraphics[]{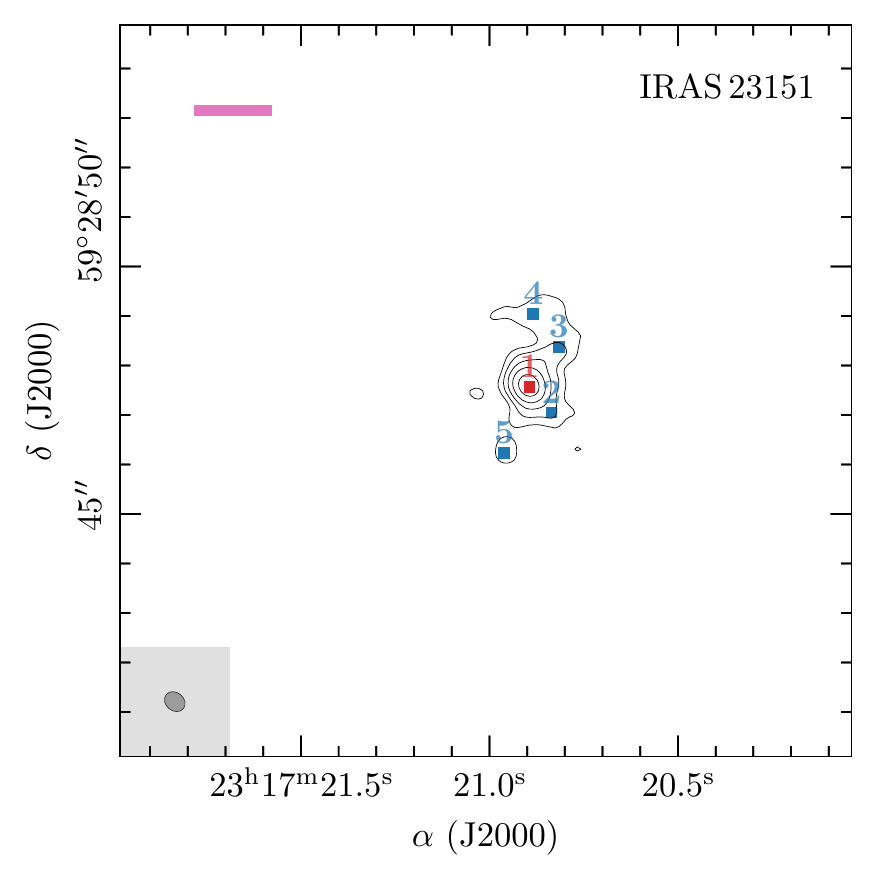}
\includegraphics[]{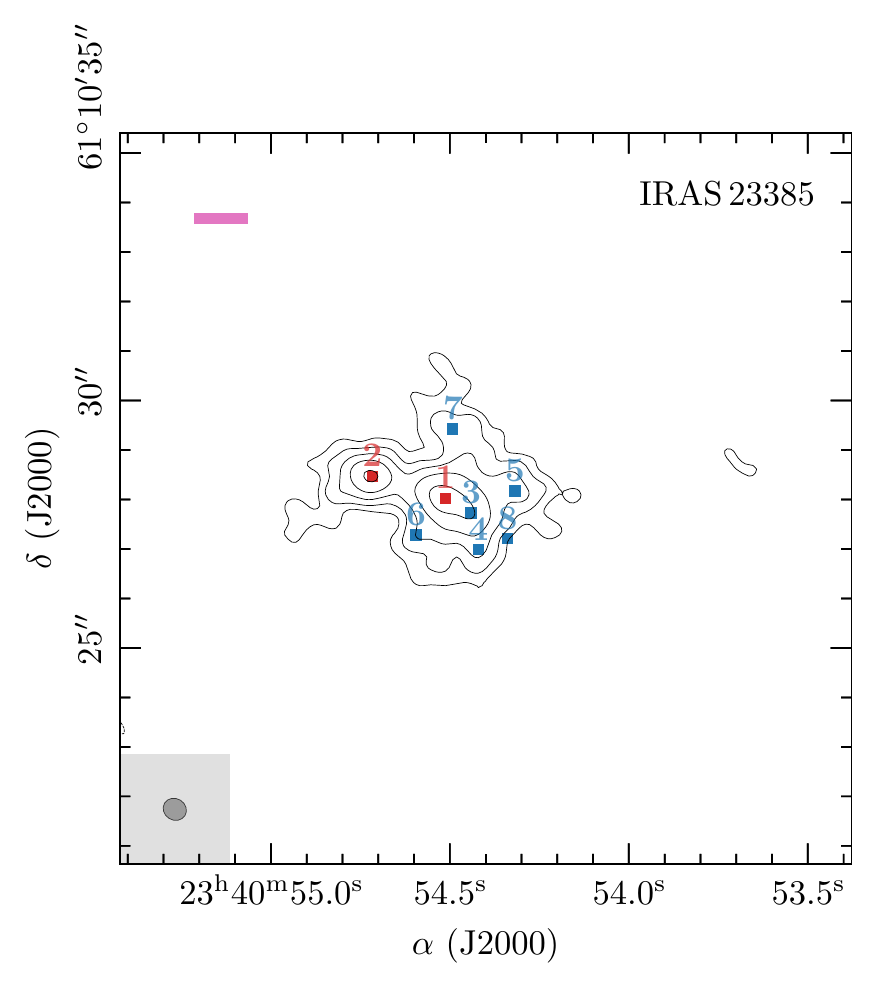}
\includegraphics[]{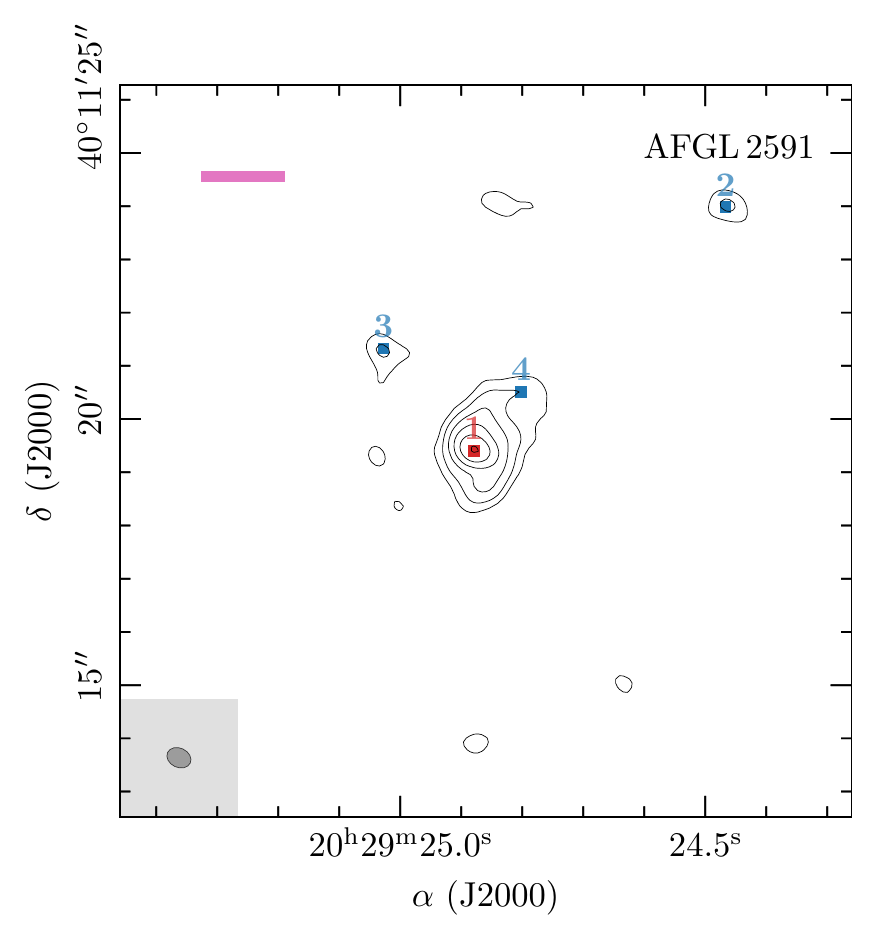}
\caption{1.37\,mm continuum emission. The selected positions analyzed in this paper (summarized in Table \ref{tab:positions}) are marked and labeled in red (core positions) and blue (all remaining positions). The dashed black contours show the $-5\sigma_\mathrm{cont}$ emission and the solid black contours start at 5$\sigma_\mathrm{cont}$ with contour steps increasing by a factor of 2 (e.g., $-$5, 5, 10, 20, 40, 80,...$\sigma_\mathrm{cont}$, see Table \ref{tab:dataproducts} for values of $\sigma_\mathrm{cont}$ for each region). In each panel, the beam size is shown in the bottom left corner and the pink bar in the top left corner indicates a linear spatial scale of 5\,000\,au. The field of view in each region is adjusted to only show the area with significant emission ($\geq 5\sigma_\mathrm{cont}$), and for regions with wide-spread emission, the primary beam is indicated by a black circle.}
\label{fig:continuum}
\end{figure*}

\begin{figure*}
\ContinuedFloat
\captionsetup{list=off,format=cont}
\centering
\includegraphics[]{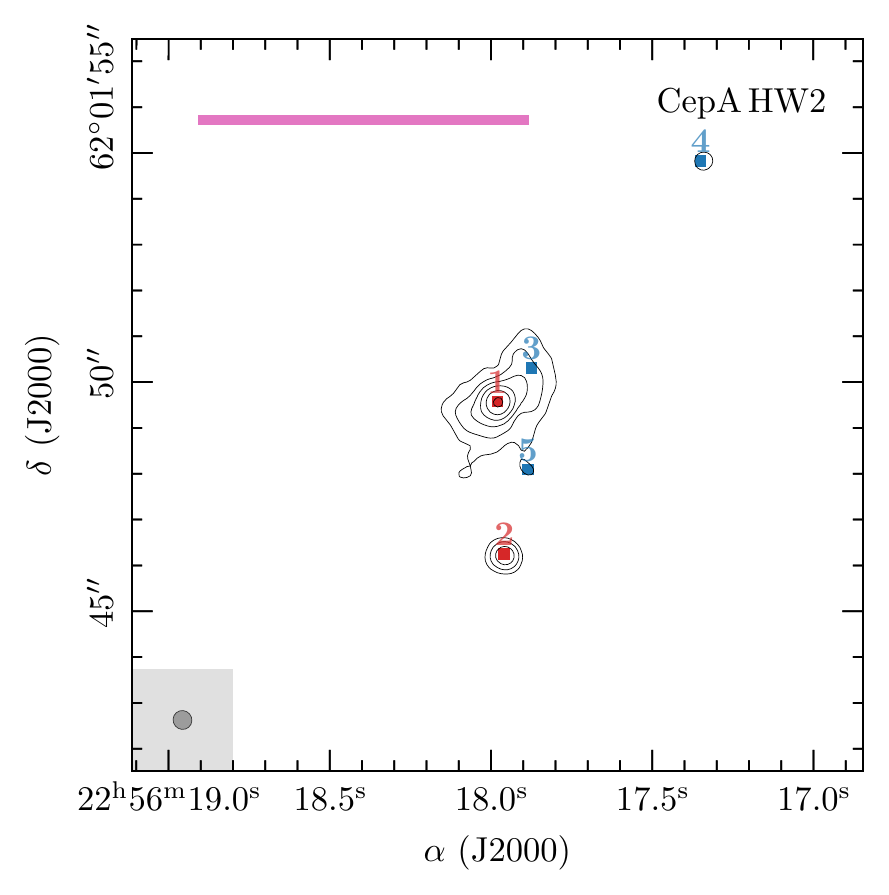}
\includegraphics[]{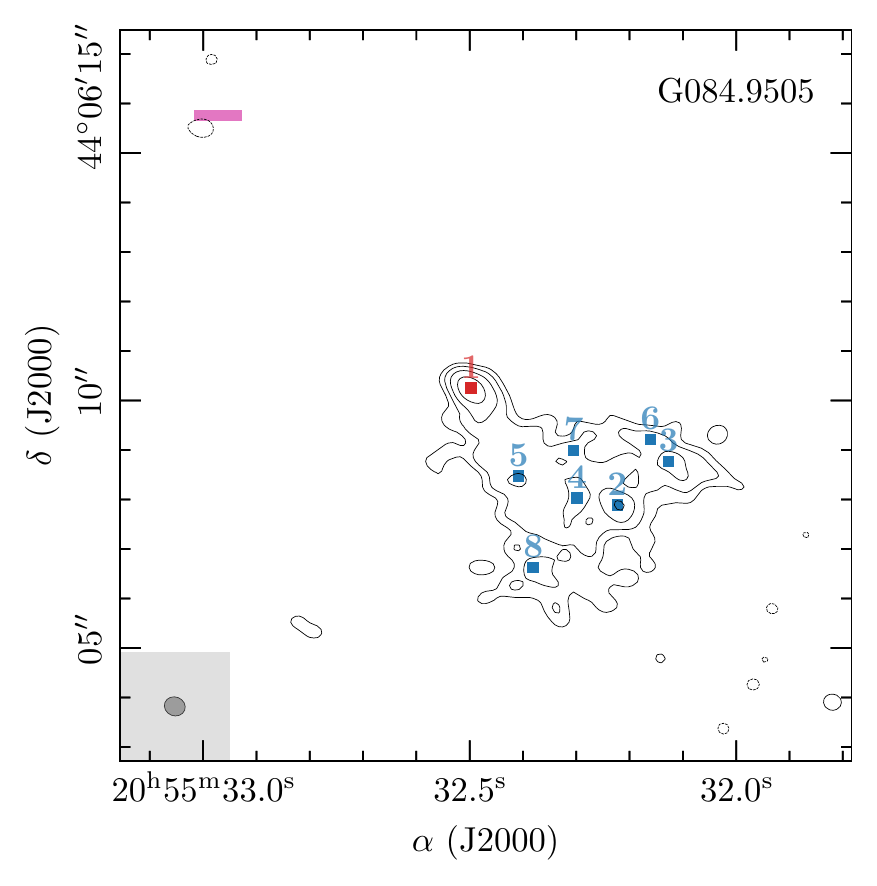}
\includegraphics[]{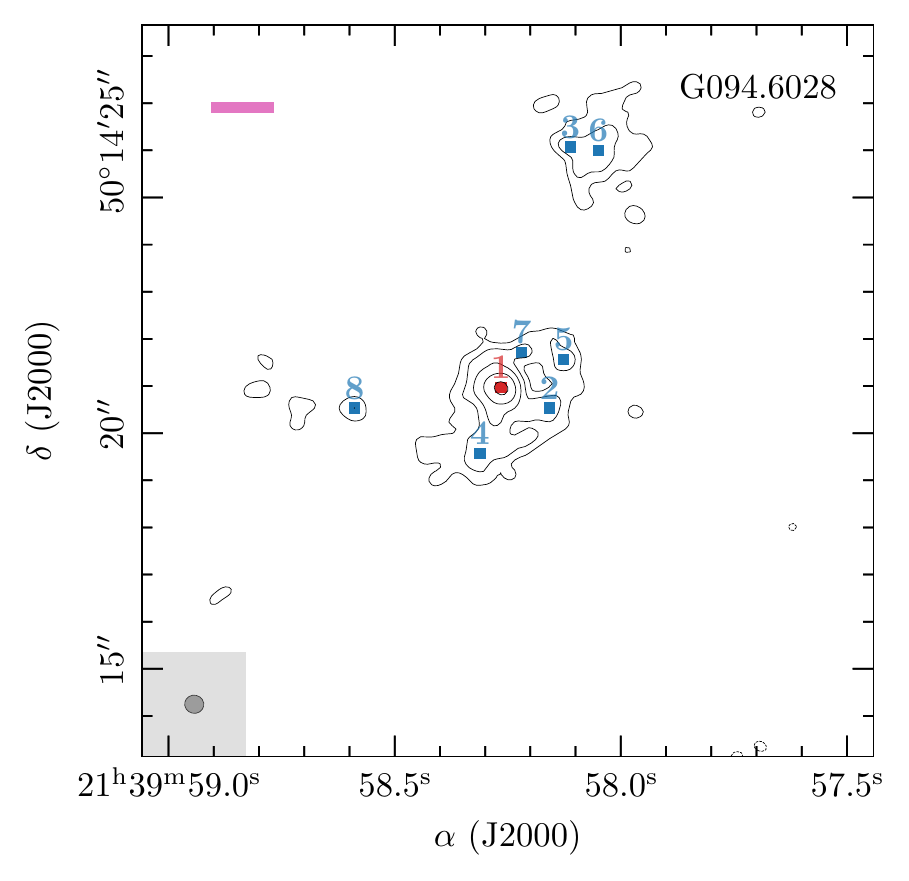}
\includegraphics[]{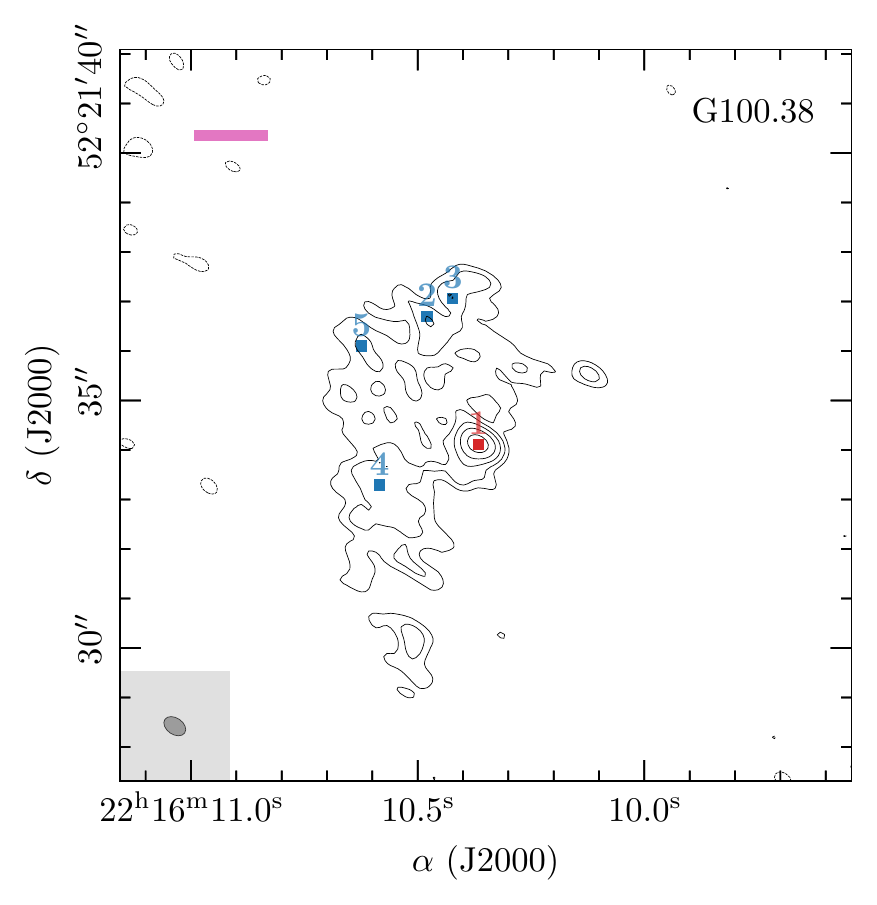}
\caption{1.37\,mm continuum emission. The selected positions analyzed in this paper (summarized in Table \ref{tab:positions}) are marked and labeled in red (core positions) and blue (all remaining positions). The dashed black contours show the $-5\sigma_\mathrm{cont}$ emission and the solid black contours start at 5$\sigma_\mathrm{cont}$ with contour steps increasing by a factor of 2 (e.g., $-$5, 5, 10, 20, 40, 80,...$\sigma_\mathrm{cont}$, see Table \ref{tab:dataproducts} for values of $\sigma_\mathrm{cont}$ for each region). In each panel, the beam size is shown in the bottom left corner and the pink bar in the top left corner indicates a linear spatial scale of 5\,000\,au. The field of view in each region is adjusted to only show the area with significant emission ($\geq 5\sigma_\mathrm{cont}$), and for regions with wide-spread emission, the primary beam is indicated by a black circle.}
\end{figure*}

\begin{figure*}
\ContinuedFloat
\captionsetup{list=off,format=cont}
\centering
\includegraphics[]{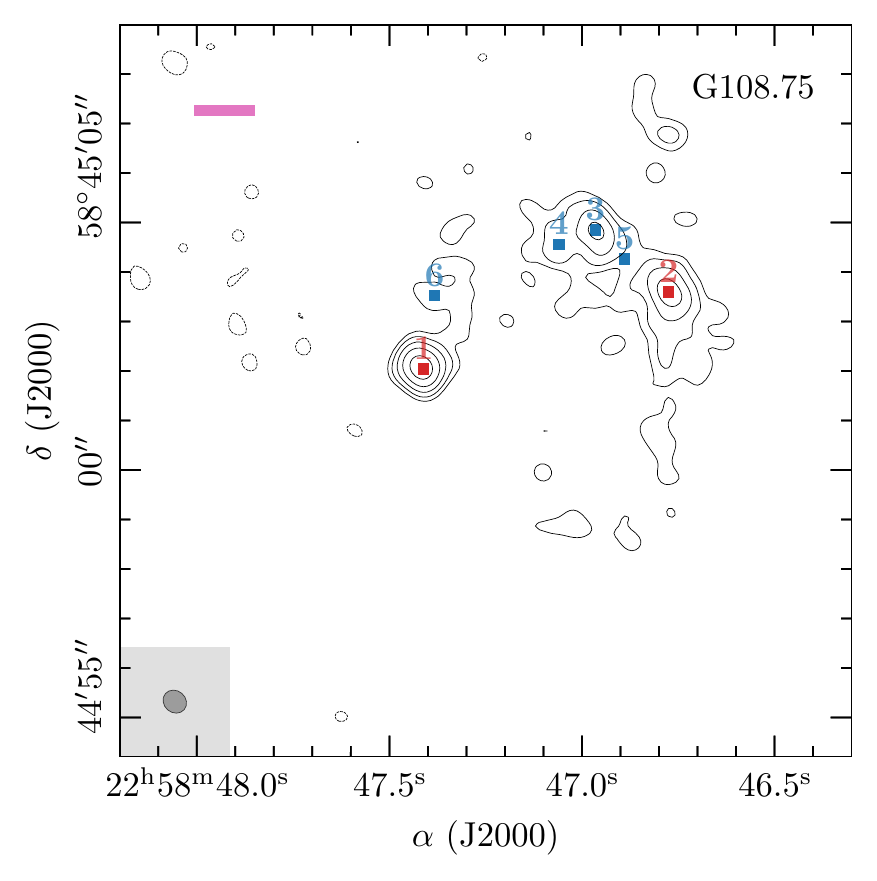}
\includegraphics[]{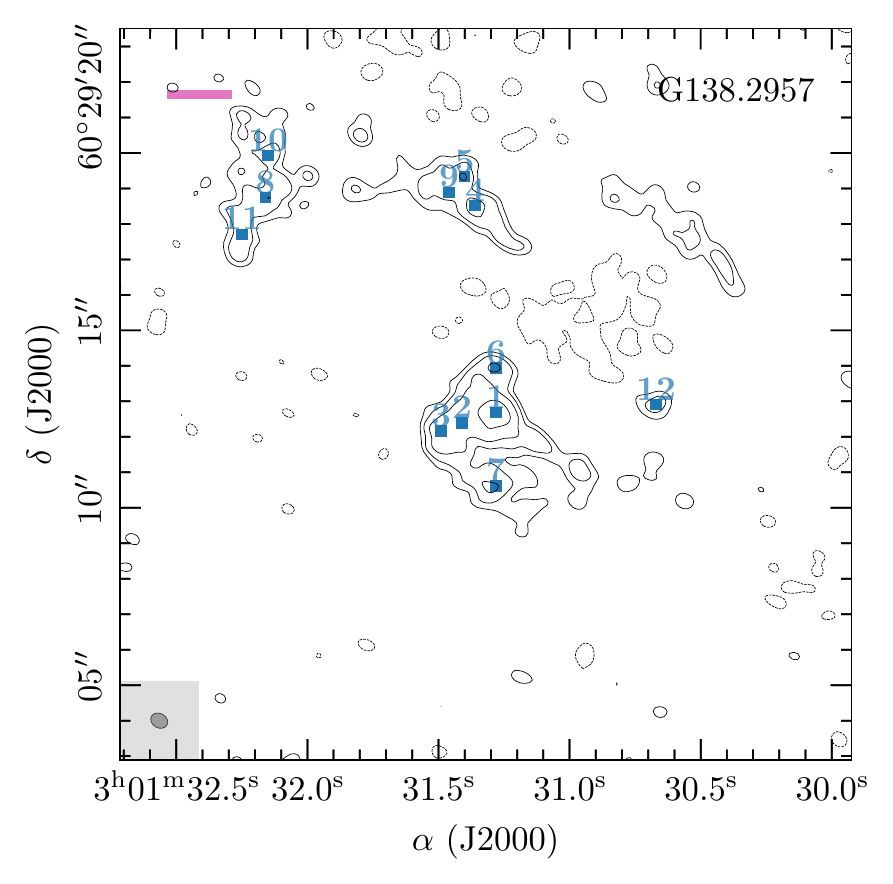}
\includegraphics[]{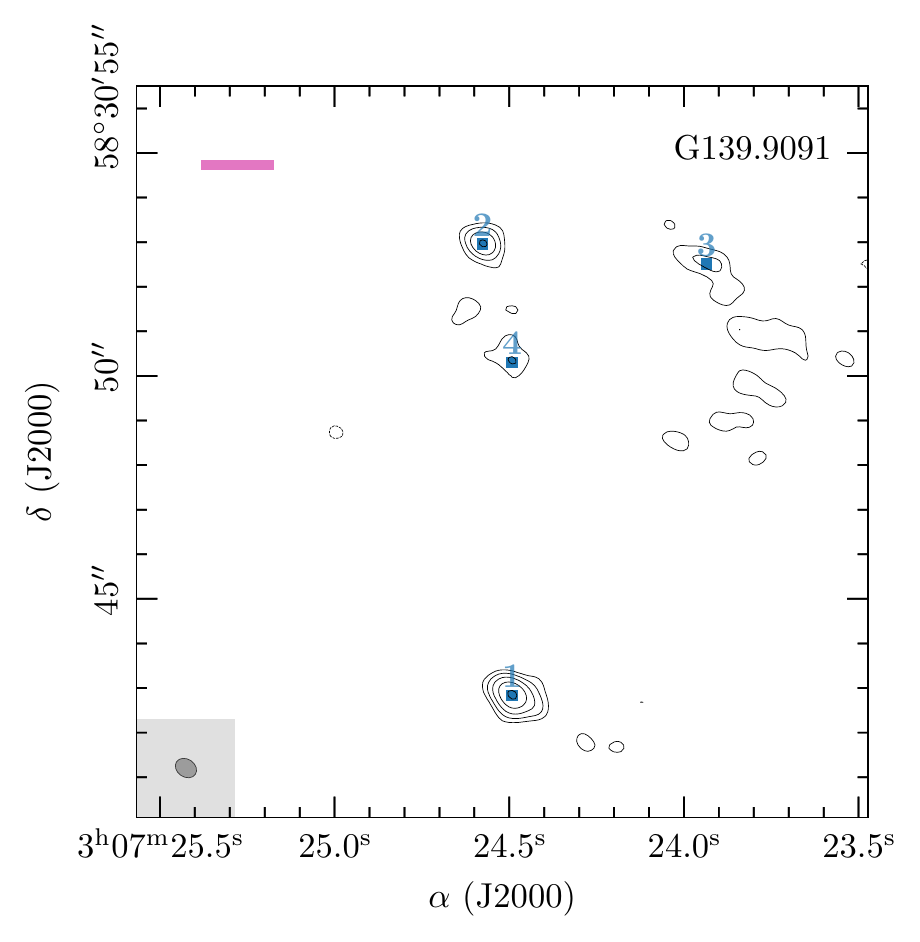}
\includegraphics[]{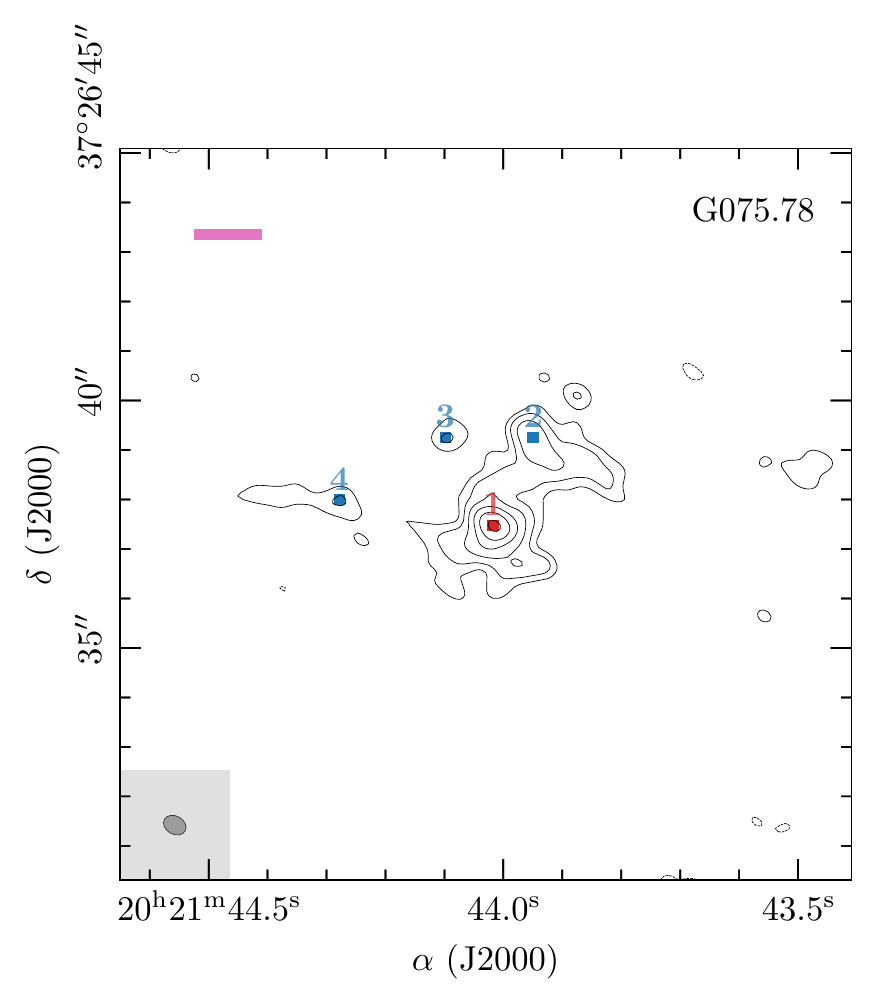}
\caption{1.37\,mm continuum emission. The selected positions analyzed in this paper (summarized in Table \ref{tab:positions}) are marked and labeled in red (core positions) and blue (all remaining positions). The dashed black contours show the $-5\sigma_\mathrm{cont}$ emission and the solid black contours start at 5$\sigma_\mathrm{cont}$ with contour steps increasing by a factor of 2 (e.g., $-$5, 5, 10, 20, 40, 80,...$\sigma_\mathrm{cont}$, see Table \ref{tab:dataproducts} for values of $\sigma_\mathrm{cont}$ for each region). In each panel, the beam size is shown in the bottom left corner and the pink bar in the top left corner indicates a linear spatial scale of 5\,000\,au. The field of view in each region is adjusted to only show the area with significant emission ($\geq 5\sigma_\mathrm{cont}$), and for regions with wide-spread emission, the primary beam is indicated by a black circle.}
\end{figure*}

\begin{figure*}
\ContinuedFloat
\captionsetup{list=off,format=cont}
\centering
\includegraphics[]{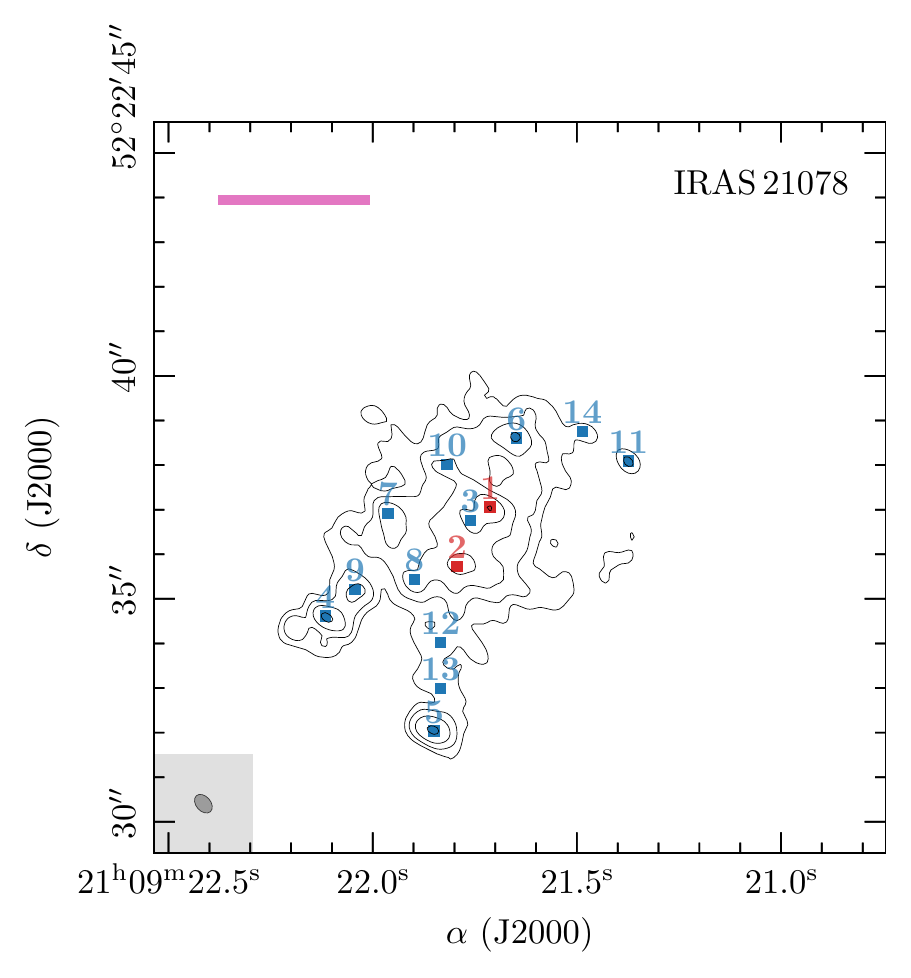}
\includegraphics[]{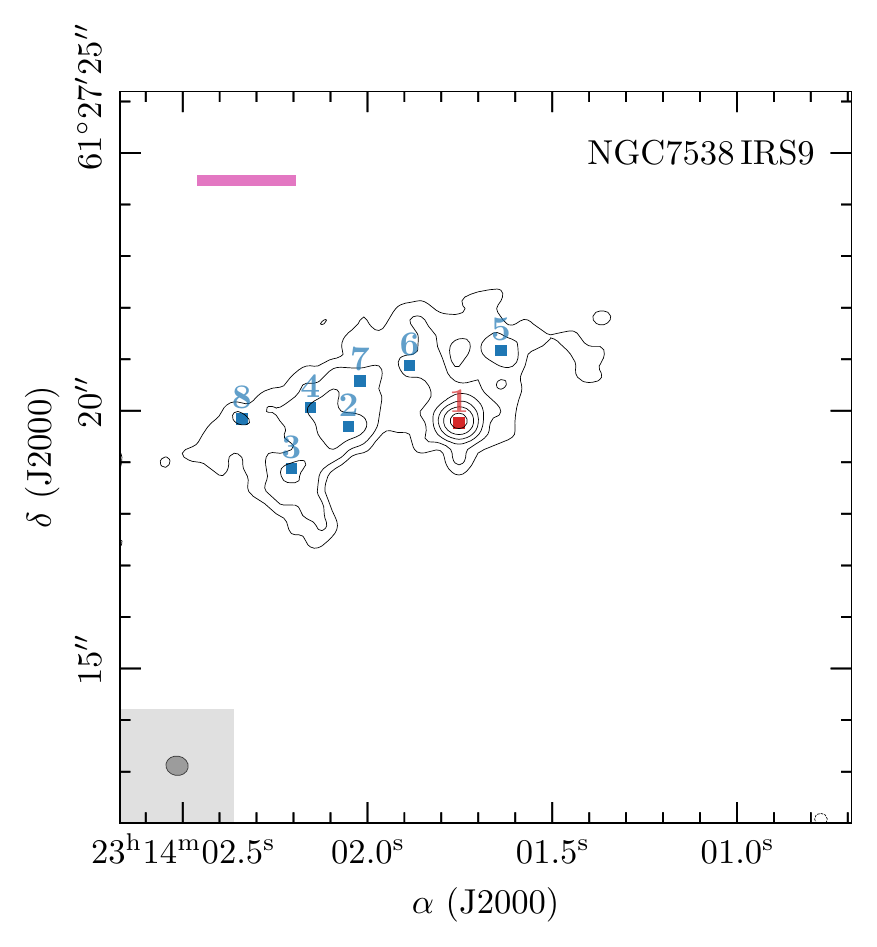}
\includegraphics[]{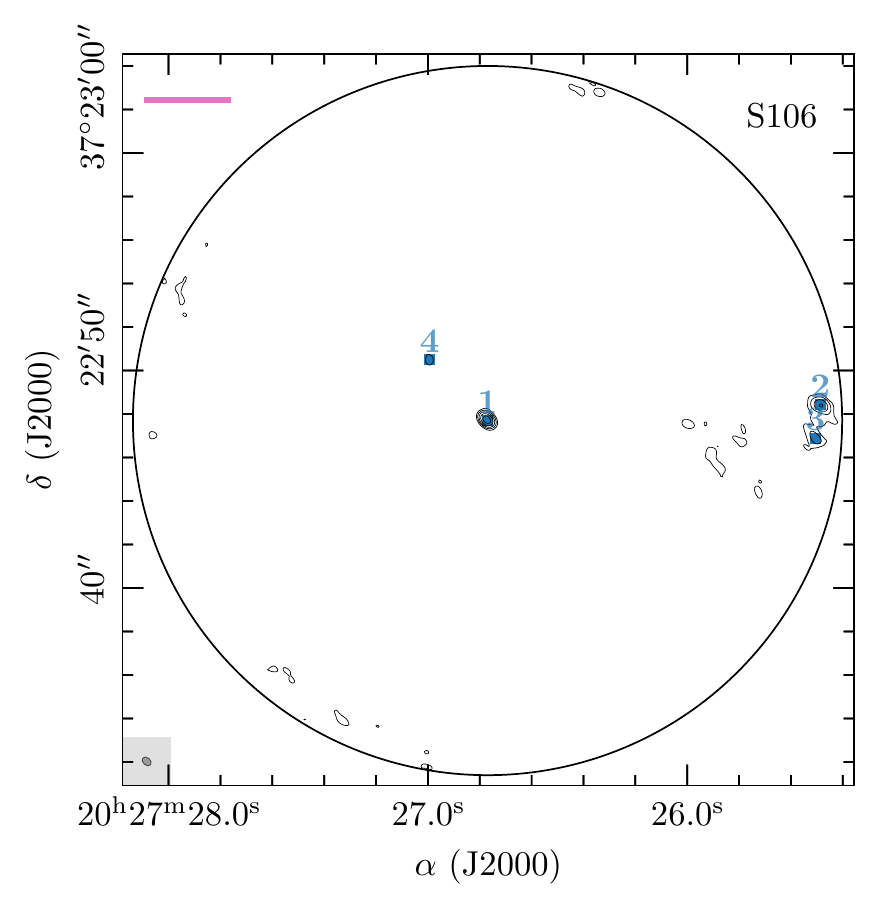}
\includegraphics[]{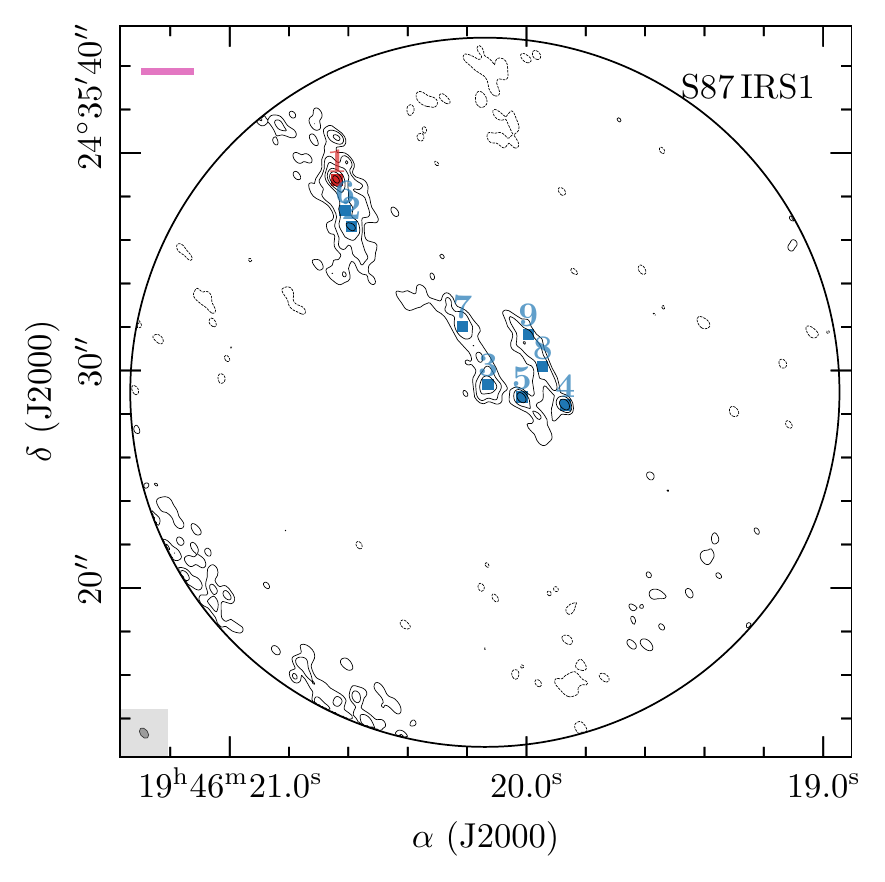}
\caption{1.37\,mm continuum emission. The selected positions analyzed in this paper (summarized in Table \ref{tab:positions}) are marked and labeled in red (core positions) and blue (all remaining positions). The dashed black contours show the $-5\sigma_\mathrm{cont}$ emission and the solid black contours start at 5$\sigma_\mathrm{cont}$ with contour steps increasing by a factor of 2 (e.g., $-$5, 5, 10, 20, 40, 80,...$\sigma_\mathrm{cont}$, see Table \ref{tab:dataproducts} for values of $\sigma_\mathrm{cont}$ for each region). In each panel, the beam size is shown in the bottom left corner and the pink bar in the top left corner indicates a linear spatial scale of 5\,000\,au. The field of view in each region is adjusted to only show the area with significant emission ($\geq 5\sigma_\mathrm{cont}$), and for regions with wide-spread emission, the primary beam is indicated by a black circle.}
\end{figure*}

\begin{figure*}
\ContinuedFloat
\captionsetup{list=off,format=cont}
\centering
\includegraphics[]{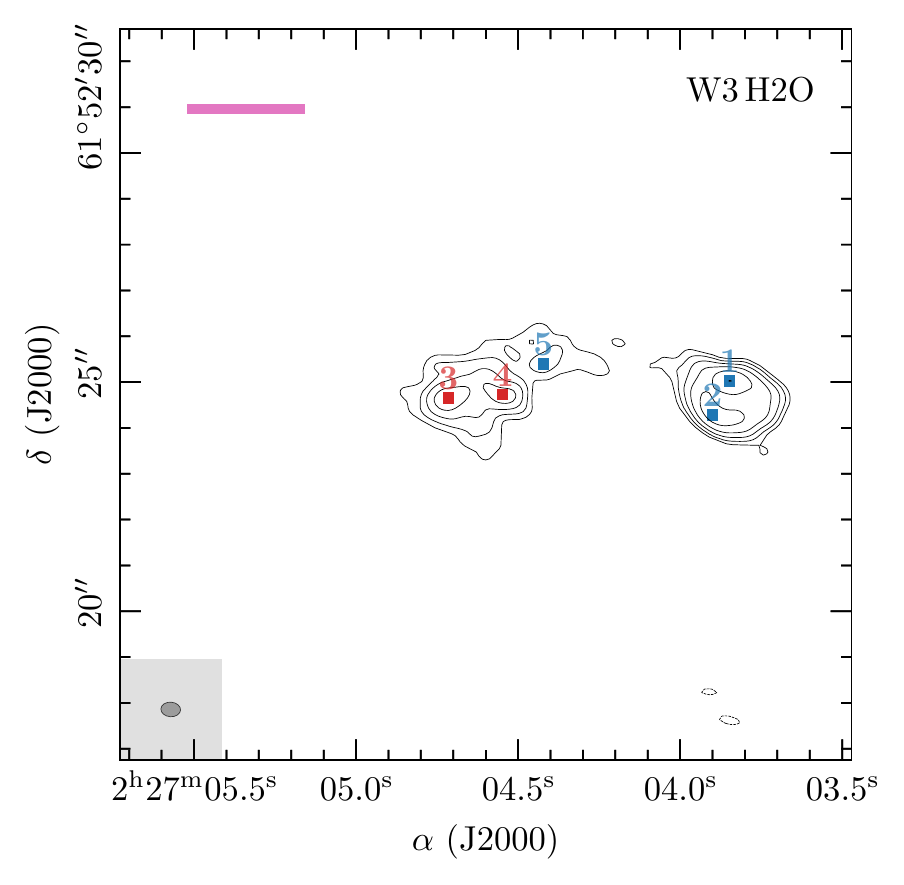}
\includegraphics[]{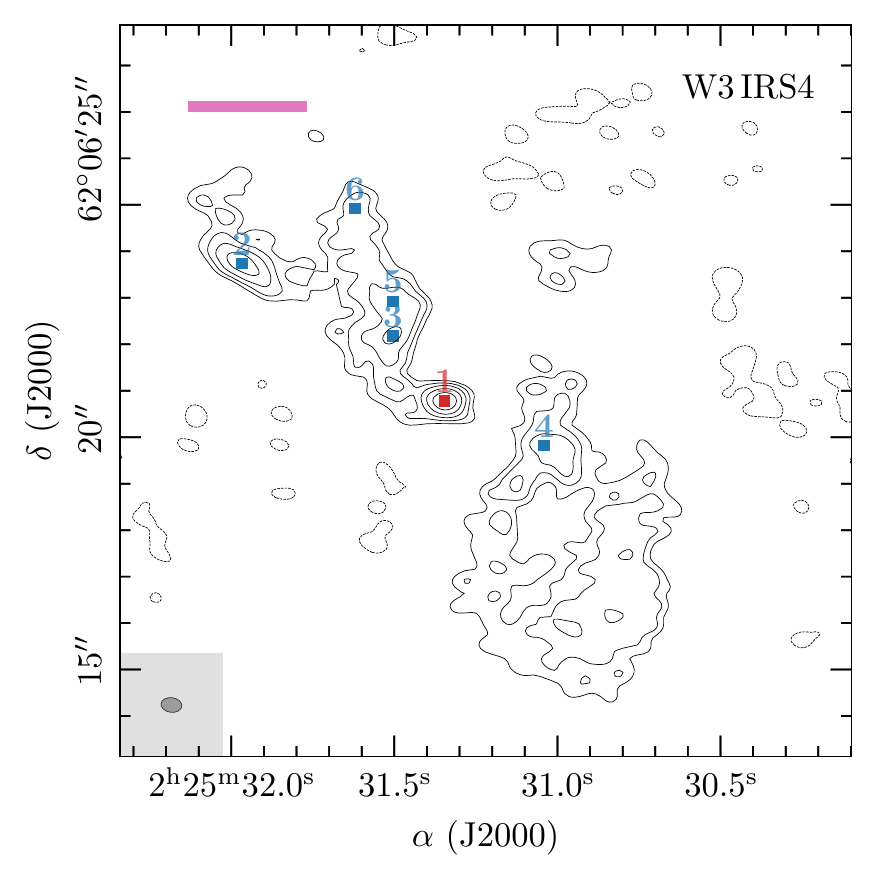}
\caption{1.37\,mm continuum emission. The selected positions analyzed in this paper (summarized in Table \ref{tab:positions}) are marked and labeled in red (core positions) and blue (all remaining positions). The dashed black contours show the $-5\sigma_\mathrm{cont}$ emission and the solid black contours start at 5$\sigma_\mathrm{cont}$ with contour steps increasing by a factor of 2 (e.g., $-$5, 5, 10, 20, 40, 80,...$\sigma_\mathrm{cont}$, see Table \ref{tab:dataproducts} for values of $\sigma_\mathrm{cont}$ for each region). In each panel, the beam size is shown in the bottom left corner and the pink bar in the top left corner indicates a linear spatial scale of 5\,000\,au. The field of view in each region is adjusted to only show the area with significant emission ($\geq 5\sigma_\mathrm{cont}$), and for regions with wide-spread emission, the primary beam is indicated by a black circle.}
\end{figure*}
	
	The NOrthern Extended Millimeter Array (NOEMA) large program ``Fragmentation and disk formation during high-mass star formation - CORE'' is a high angular resolution survey ($\sim$0\as4 at 1.37\,mm) designed to study the fragmentation, kinematic, and chemical properties of a homogeneous sample of 18 HMSFRs. The observations consist of spectral line and continuum data in Band 3 (1\,mm). An overview of the project and results of the analysis of the dust continuum and fragmentation properties are presented in \citet{Beuther2018}. 
	
	In combination with this paper, we release all data products of the CORE project with and without self-calibration conducted with \texttt{GILDAS}. The data can be found on the CORE collaboration website: \url{https://www.mpia.de/core}. The CLEANed data products as well as the calibrated $uv$-tables are available.
	
\subsection{Sample}\label{sec:sample}

	The sample of the CORE project consists of well-studied HMSFRs on the Northern hemisphere. They were selected based on their bolometric luminosity ($L > 10^4 L_{\odot}$) and distance ($d < 6$\,kpc) and allow us to study the evolution at early stages of HMSF in the HMPO/HMC stage. An overview of the properties of the regions is shown in Table \ref{tab:sample} in this paper and additional properties, such as the luminosity $L$ and mass $M$, are listed in Table 1 in \citet{Beuther2018}. With declinations higher than $+20^{\circ}$, most of the regions are difficult or impossible to observe with the Atacama Large Millimeter/submillimeter Array (ALMA), and therefore can only be studied with NOEMA at a high angular resolution ($< 1 ''$) at mm wavelengths. The angular resolution for all regions is homogeneous ($\sim$0\as4), but the resolved linear spatial scales vary from 300\,au to 2\,300\,au, as the regions are located at distances between 0.7\,kpc and 5.5\,kpc.

	The regions show a large diversity of fragmentation properties \citep{Beuther2018}: while some regions contain mainly a single isolated core (e.g., AFGL\,2591), other regions fragment into up to 20 cores (e.g., IRAS\,21078). Individual case-studies were carried out on some of the regions, e.g., the kinematic properties of W3\,H2O \citep{Ahmadi2018}, IRAS\,23385 \citep{Cesaroni2019}, IRAS\,23033 \citep{Bosco2019}, W3\,IRS4 \citep{Mottram2020}, and IRAS\,21078 \citep{Moscadelli2021}; the chemical composition of the pilot regions NGC7538\,S and NGC7538\,IRS1 \citep{Feng2016}, and AFGL\,2591 \citep{Gieser2019}. A multi-wavelength modeling study of AFGL\,2591 is presented in \citet{Olguin2020}. In Ahmadi et al. (in prep.) the kinematic analysis of the complete CORE sample will be covered, while in this study we focus on the physical structure (temperature and density) and chemical composition of the molecular gas. We do not include the pilot regions NGC7538\,S and NGC7538\,IRS1 presented in \citet{Beuther2018} in our analysis, as only for the remaining 18 regions we have a homogeneous multi-configuration NOEMA data set. The pilot studies have no corresponding D array observations, so less $uv$-coverage, and could hence not be accurately compared. However, a detailed individual chemical analysis of the pilot regions is already presented in \citet{Feng2016}.

\subsection{NOEMA Observations}\label{sec:NOEMAobs}

	The sample was observed from 2014 to 2017 in the A (extended), the old B or new C (intermediate), and D (compact) array configurations. Two regions each were observed in track-sharing pairs to reduce calibration time. Spectral line data were obtained with the broad-band WideX correlator at a rest-frequency range of $217.2-220.8$\,GHz with a spectral resolution of $\sim$2.7\,km s$^{-1}$. In addition, eight narrow-band units were placed within this frequency range in order to obtain high spectral resolution data ($\sim$0.4\,km s$^{-1}$) of kinematically interesting lines such as CH$_{3}$CN \citep[a summary of the high spectral resolution units is summarized in Table 2 and 3 in][]{Ahmadi2018}. The interferometric data were calibrated with the \texttt{CLIC} package in \texttt{GILDAS}\footnote{\url{https://www.iram.fr/IRAMFR/GILDAS/}}. The 1.37\,mm continuum data were extracted from the broad-band WideX data by carefully selecting line-free channels in each region. 

\subsection{IRAM 30m Observations}\label{sec:IRAMobs}

	Interferometers spatially filter the extended emission. The shortest baseline of the NOEMA array is $\sim$20\,m, thus spatial scales larger than 16$''$ are not recovered by the interferometric observations at 1.37\,mm. All CORE regions were therefore observed with the IRAM 30m telescope using the Eight MIxer Receiver \citep[EMIR,][]{EMIR} in order to recover the missing flux in the spectral line data due to spatial filtering.
	EMIR has four basebands and each have a width of $\sim$4\,GHz and a spectral resolution of 200\,kHz corresponding to $\sim$0.3\,km s$^{-1}$ at 1.37\,mm). The lower inner baseband (LI) of our EMIR spectral setup covers the same spectral range as the NOEMA observations with the broad-band WideX correlator. The half power beam width (HPBW) of the IRAM 30m telescope is 11\as8 at 1.37\,mm. For a detailed description of the IRAM 30m data calibration we refer to Appendix A in \citet{Mottram2020}. 
	The continuum data have no complementary single-dish observations, so here spatial filtering can remain an issue as discussed in \citet{Beuther2018}.

\subsection{Self-calibration}\label{sec:selfcal}

	The majority of the continuum data have a high signal-to-noise ($S$/$N$) ratio in the \texttt{GILDAS} standard calibrated data (see Table \ref{tab:dataproducts}). However, the phase noise can be high due to an unstable atmosphere during the observations causing the flux to be scattered around the source. This issue can be improved by applying phase self-calibration to the interferometric data \citep[e.g.,][]{Pearson1984, Radcliffe2016}.

	The CORE continuum data presented in \citet{Beuther2018} were successfully phase self-calibrated using the Common Astronomy Software Applications package (\texttt{CASA}) through an iterative masking of the source. However, at that time it was not possible to apply the self-calibration solution to the spectral line data. The \texttt{CASA} phase self-calibration of the continuum data was performed by applying an interactive mask in each self-calibration loop starting with the strongest structures first and proceeding with the weaker structures \citep{Beuther2018}. Depending on the $S$/$N$ ratio of the region, solution intervals of 220, 100, or 45\,s were used.
	
	We have now used the self-calibration tool in \texttt{GILDAS} to phase self-calibrate the continuum as well as the spectral line data of the CORE project to provide homogeneously calibrated data products of the full CORE observations. The crucial point of phase self-calibration is to start with a good enough spatial model of the source. The \texttt{selfcal} procedure of the \texttt{GILDAS/MAPPING} package uses as a source model the first CLEAN components $n_{\mathrm{CLEAN}}$ found during a previous step of deconvolution. The basic idea is that the first CLEAN components deliver a model of the source with a high $S$/$N$ ratio whose spatial structure is not much affected by flux scattering. The number of CLEAN components $n_{\mathrm{CLEAN}}$ must be large enough to get a fair representation of the source structure and small enough to avoid deconvolving scattered flux that would be confused with noise. The visibilities are usually averaged in time to increase their signal-to-noise ratio during the first self-calibration iteration and this averaging time is progressively lowered to the minimum possible integration time in the following iterations. In practice, we started to deconvolve the continuum source with an absolute flux stopping criterion set to 3 times the continuum noise. This gives a number of CLEAN iterations $n_{\mathrm{iter}}$ that we used to iterate the self-calibration three times increasing $n_{\mathrm{CLEAN}}$ from $n_{\mathrm{iter}}$/4, to $n_{\mathrm{iter}}$/2, and to $n_{\mathrm{iter}}$, and decreasing the averaging time from 200\,s, to 100\,s, and to 45\,s. Only visibilities with a $S$/$N$ ratio $> 6$ are self-calibrated, but the remaining visibilities are kept in the proceeding CLEANing of the data in order to not loose visibilities on the longest baselines which would decrease the angular resolution. With this method, simple structures, as well as regions with a complicated morphology can be successfully phase self-calibrated. As the continuum data has the highest sensitivity, we used the solution from the continuum self-calibration and apply the gain solution to the broad-band and narrow-band spectral line data using the \texttt{UV\_CAL} task.

	An example on how self-calibration is increasing the quality of the continuum data quality is shown in Fig. \ref{fig:calibration_comparison_continuum} for W3\,IRS4 which has a complex morphology in the continuum data. While \texttt{GILDAS} standard calibrated data already reveals the complex structure of the region, the noise is high throughout the primary beam with many negative features. Applying the \texttt{CASA} and \texttt{GILDAS} self-calibration significantly lowers the noise and increases the peak intensity. The continuum noise and peak intensity are summarized for all regions in Table \ref{tab:dataproducts}. The mean continuum noise $\sigma_{\mathrm{cont}}$ is 0.74\,mJy\,beam$^{-1}$ in the standard calibrated data, while for the \texttt{CASA} and \texttt{GILDAS} self-calibrated data, the mean noise $\sigma_{\mathrm{cont}}$ is lowered by $\sim$25\% to 0.53\,mJy\,beam$^{-1}$ and 0.56\,mJy\,beam$^{-1}$, respectively. In general, $\sigma_{\mathrm{cont}}$ is higher toward the strong continuum sources S106, CepA\,HW2, and W3\,H2O. The mean $S$/$N$ ratio of the \texttt{GILDAS} standard calibrated continuum data is 74, while for the \texttt{CASA} and \texttt{GILDAS} self-calibrated continuum data, the mean $S$/$N$ is improved by a factor of two to 130 and 132, respectively.
	 
	Even though both \texttt{CASA} and \texttt{GILDAS} self-calibration methods improve the data quality, due to the different techniques to define a source model (interactive masking and defining number of clean components, respectively), there are differences in the resulting images. In Fig. \ref{fig:calibration_comparison_continuum}, a faint structure toward the NW of the continuum peak is seen in the \texttt{GILDAS} standard calibrated and self-calibrated image, but it is not recovered in the \texttt{CASA} self-calibrated image. Comparing the self-calibration results (Table \ref{tab:dataproducts}), the peak intensities are higher in the \texttt{GILDAS} self-calibrated data, while a lower noise is achieved in the \texttt{CASA} self-calibrated data. For very faint structures and a complex source morphology, a careful interpretation of the self-calibrated data is recommended, but overall all the main features are recovered with both self-calibration methods.
	
	A comparison between the standard and self-calibrated broad-band spectral line data is shown in Fig. \ref{fig:calibration_comparison_line} for the CH$_{3}$CN $12_{3}-11_{3}$ transition around the location of the 1.37\,mm continuum peak in the W3\,IRS4 region. The emission is less fuzzy and more compact in the self-calibrated data product. On individual spectra, self-calibration provides a significant increase of the line intensity which has a big impact, e.g., when deriving column densities.

\subsection{Data products and public release}\label{sec:dataproducts}

	The data merging of the interferometric and single-dish data and imaging is done in the \texttt{GILDAS/MAPPING} package. The WideX and the corresponding EMIR spectral-line data are smoothed to a common spectral resolution of 3.0\,km\,s$^{-1}$. The narrow-band and the corresponding EMIR spectral line data are smoothed to a common spectral resolution of 0.5\,km\,s$^{-1}$. In order to merge the NOEMA and IRAM 30m data, the task \texttt{UVSHORT} converts the short spacings into a pseudo $uv$-table and combines them with the NOEMA data.

	The deconvolution of the NOEMA continuum, NOEMA spectral line and merged (combined NOEMA + IRAM 30m) spectral line data is performed in \texttt{GILDAS/MAPPING} using the Clark CLEAN algorithm \citep{Clark1980} adopting three different weightings: robust weighting with a robust parameter of 0.1 ($\theta \sim 0\as4$), a robust parameter of 1.0 ($\theta \sim 0\as6$), and natural weighting ($\theta \sim 1\as0$). The stopping criterion for the continuum and broad-band spectral line data is set to $f_{\mathrm{res}} = 0.01$ which corresponds to a minimum fraction of 1\% of the peak intensity in the dirty image. The stopping criterion for the narrow-band spectral line data is either $n_{\mathrm{iter}} = 5\,000$ (maximum number of iterations) or $a_{\mathrm{res}} = 0.01$, which corresponds to a maximum intensity of 10\,mJy\,beam$^{-1}$ in the residual image.

	Molecules such as CO isotopologues and SO may have extended emission in the merged data products \citep{Mottram2020}. As the Clark algorithm assumes emission from point sources, the CLEANed map may have point-like artifacts. The SDI algorithm \citep{Steer1984} may improve the CLEANed image in such cases. A comparison between the Clark and SDI algorithms for imaging of the W3\,IRS4 region is presented in \citet{Mottram2020}. Final data products of the merged broad-band spectral line data of molecular lines with potential large-scale emission ($^{13}$CO $2-1$, SO $6_{5}-5_{4}$, H$_{2}$CO $3_{0,3}-2_{0,2}$, H$_{2}$CO $3_{2,2}-2_{2,1}$, H$_{2}$CO $3_{2,1}-2_{2,0}$) CLEANed using the SDI algorithm with robust weighting (robust parameter of 3, $\theta \sim 0\as6$) with a stopping criterion of $n_{\mathrm{iter}} = 25\,000$ are provided as well. Primary beam correction is applied to the continuum and spectral line data products.

	In this study we use the primary beam corrected NOEMA-only continuum and merged (NOEMA + IRAM 30m) broad-band spectral line data. Both data products are imaged with the Clark algorithm and a robust parameter of 0.1 resulting in the highest angular resolution ($\theta \sim 0\as4$). Table \ref{tab:dataproducts} summarizes the properties of the standard and self-calibrated data products for all regions including the synthesized beam (major axis $\theta_\mathrm{maj}$, minor axis $\theta_\mathrm{min}$, and position angle PA), noise of the continuum data $\sigma_{\mathrm{cont}}$, continuum peak intensity $I_{\mathrm{peak}}$, and noise in the merged (NOEMA + IRAM 30m) spectral line data $\sigma_{\mathrm{line,map}}$. The mean map noise of the merged spectral line data is 0.44\,K.
	
\section{Physical structure}\label{sec:physicalstructure}

\begin{figure*}[!htb]
\centering
\includegraphics[]{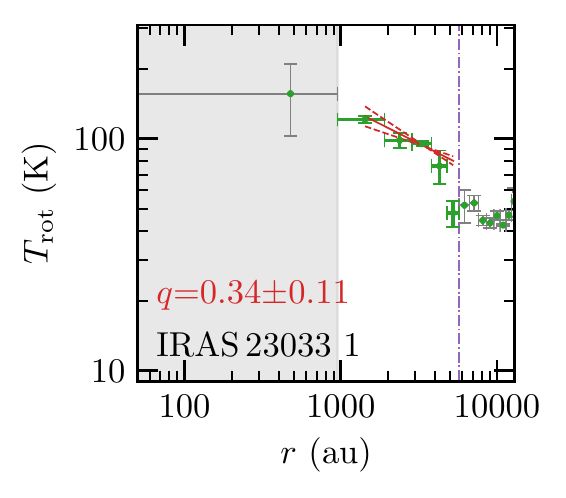}
\includegraphics[]{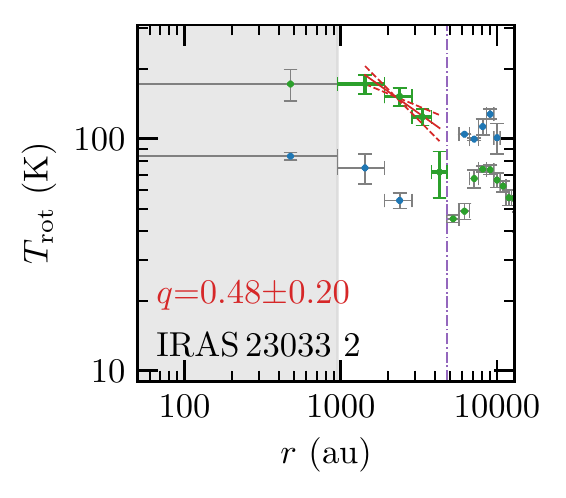}
\includegraphics[]{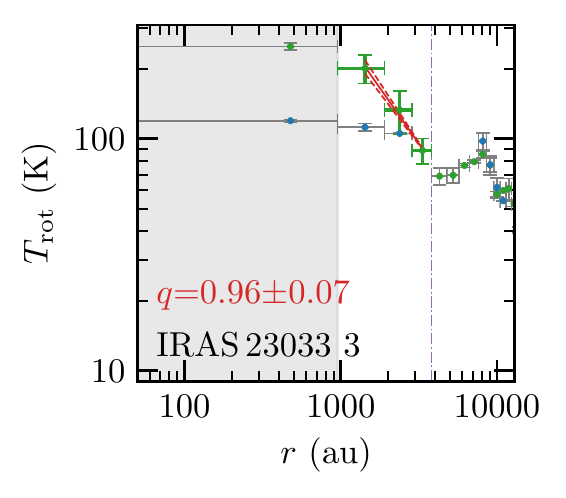}
\includegraphics[]{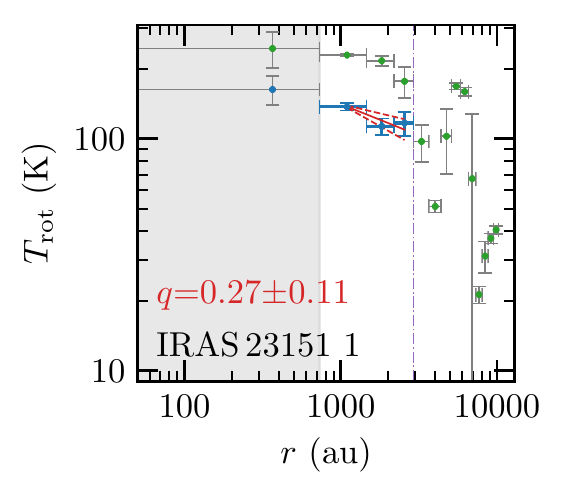}
\includegraphics[]{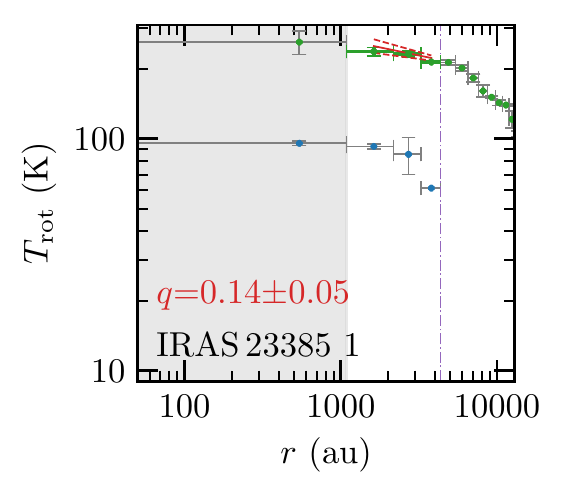}
\includegraphics[]{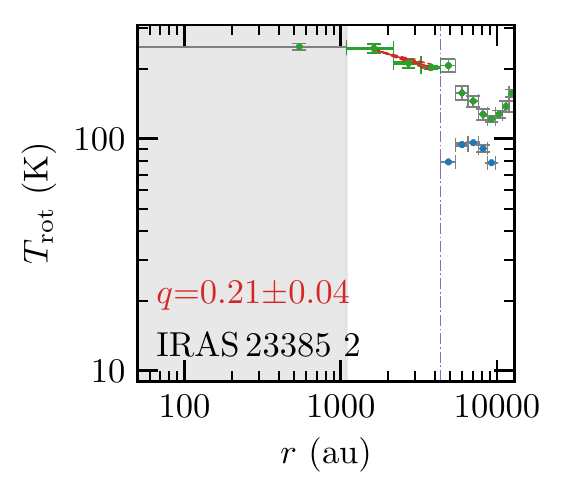}
\includegraphics[]{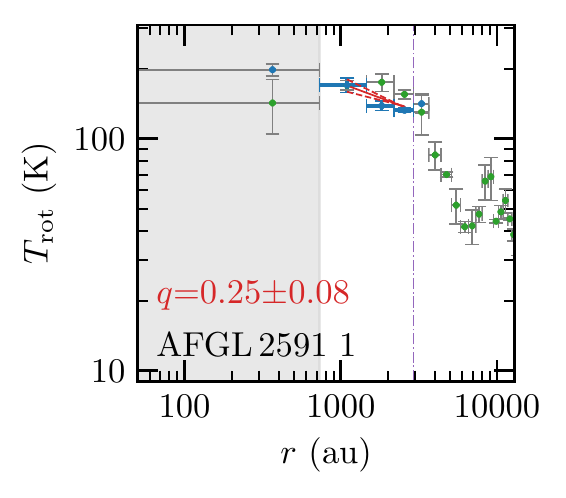}
\includegraphics[]{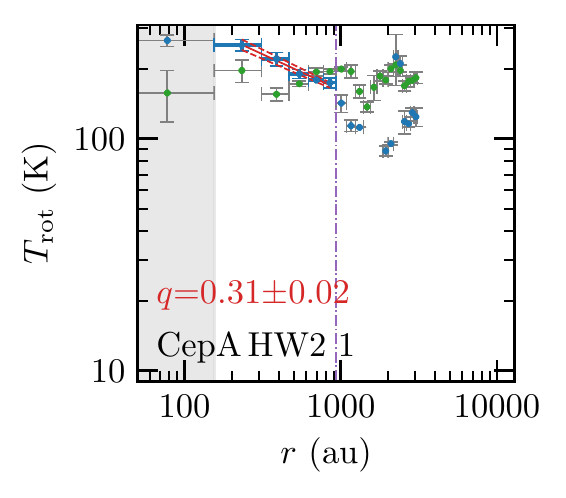}
\includegraphics[]{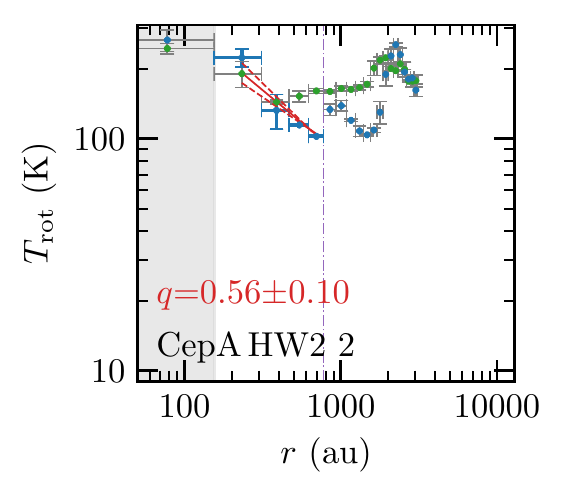}
\includegraphics[]{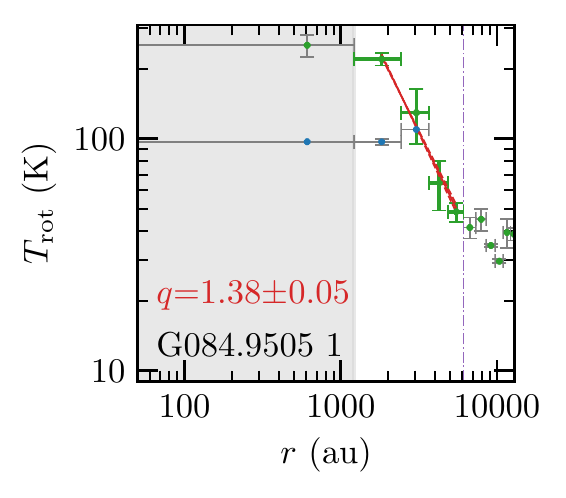}
\includegraphics[]{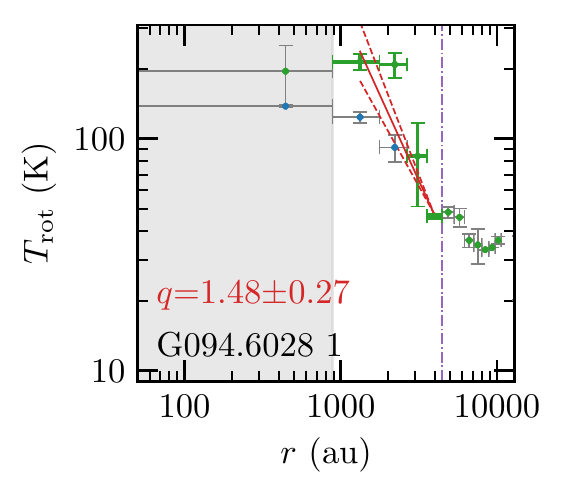}
\includegraphics[]{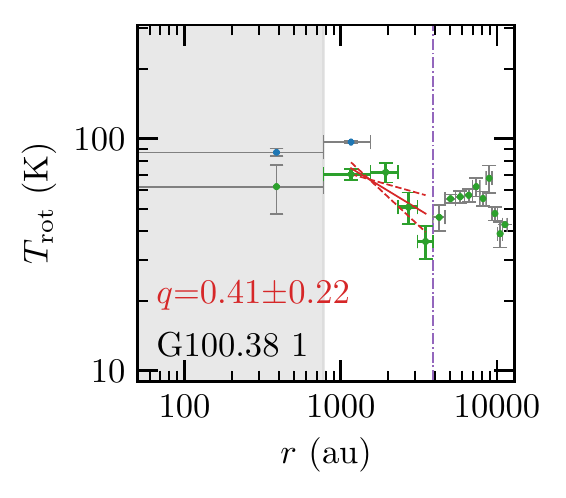}
\caption{Radial temperature profiles of the 22 cores. Each panel shows the binned radial temperature profile derived from the H$_2$CO (green) and CH$_3$CN (blue) temperature maps shown in Fig. \ref{fig:temperature_maps}. The data points used for the radial profile fit are shown by corresponding colored errorbars, the data points excluded from the fit are indicated by grey errorbars. The outer radius of the temperature fit is shown by the vertical purple dash-dotted line. The inner unresolved region is shown as a grey-shaded area. The linear fit and the $\pm1\sigma$ uncertainty are shown by the solid and dashed red lines, respectively.}
\label{fig:radialtemperatureprofile}
\end{figure*}

\begin{figure*}[!htb]
\ContinuedFloat
\captionsetup{list=off,format=cont}
\centering
\includegraphics[]{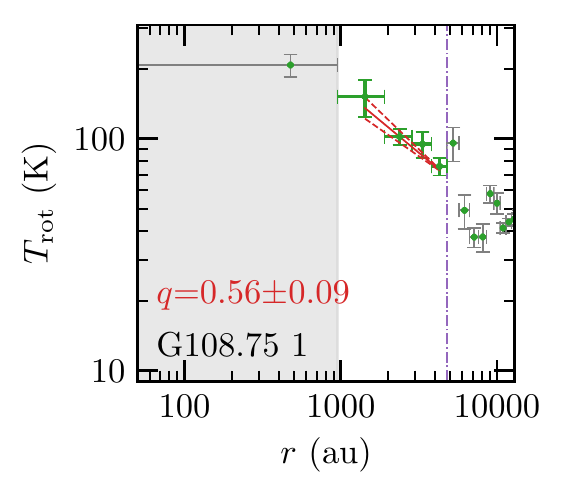}
\includegraphics[]{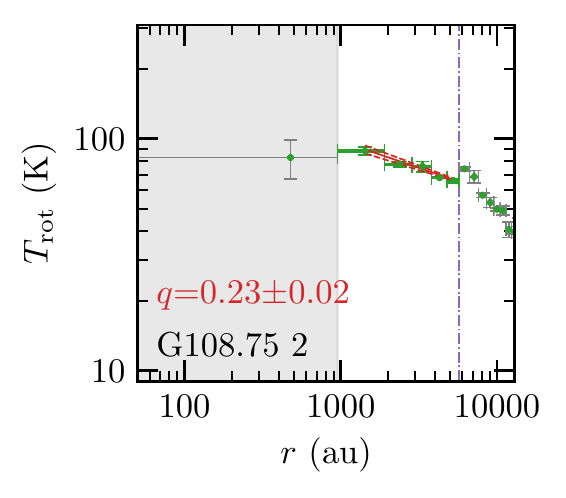}
\includegraphics[]{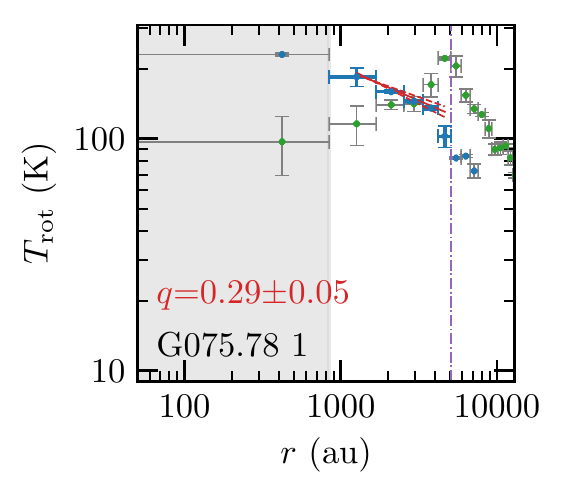}
\includegraphics[]{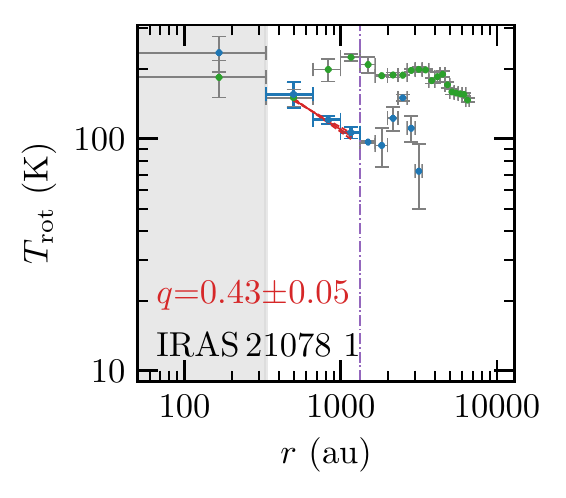}
\includegraphics[]{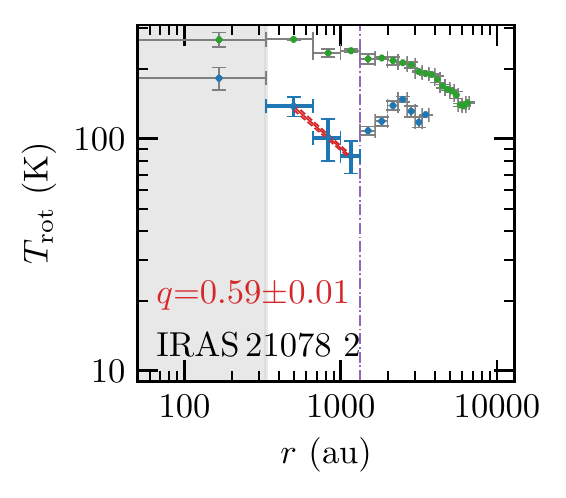}
\includegraphics[]{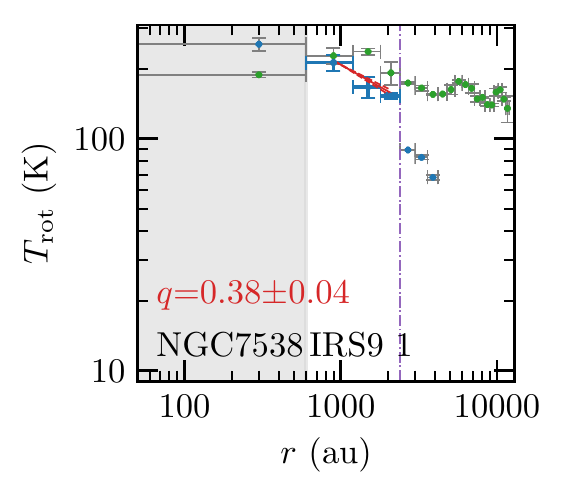}
\includegraphics[]{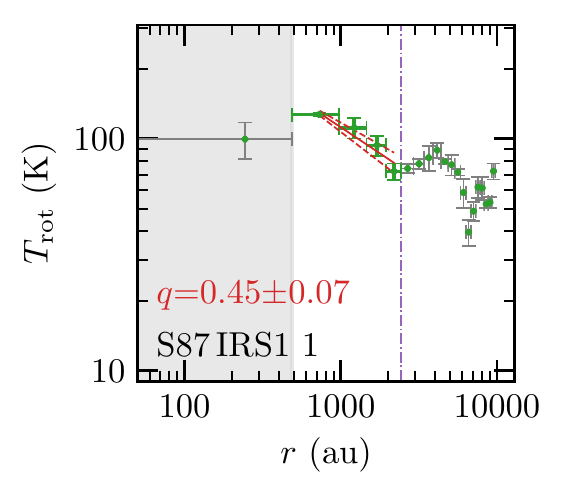}
\includegraphics[]{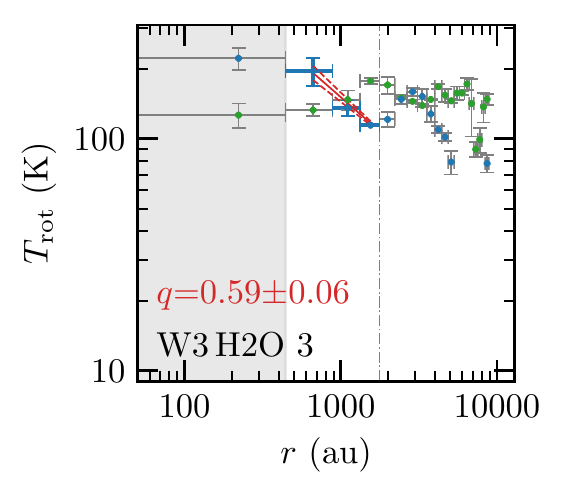}
\includegraphics[]{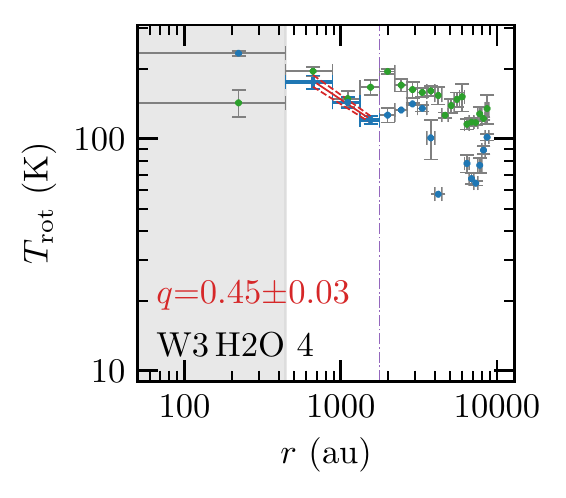}
\includegraphics[]{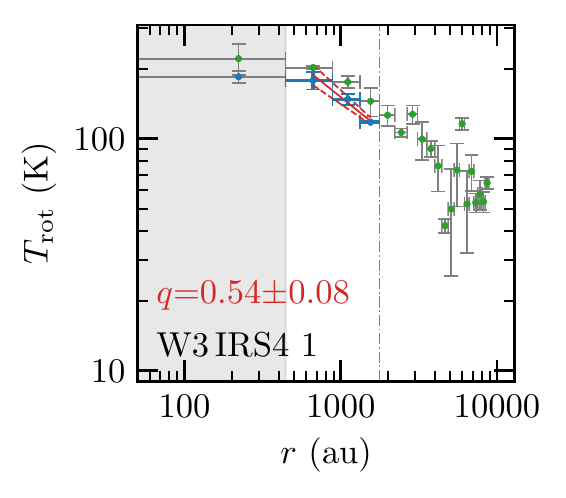}
\caption{Radial temperature profiles of the 22 cores. Each panel shows the binned radial temperature profile derived from the H$_2$CO (green) and CH$_3$CN (blue) temperature maps shown in Fig. \ref{fig:temperature_maps}. The data points used for the radial profile fit are shown by corresponding colored errorbars, the data points excluded from the fit are indicated by grey errorbars. The outer radius of the temperature fit is shown by the vertical purple dash-dotted line. The inner unresolved region is shown as a grey-shaded area. The linear fit and the $\pm1\sigma$ uncertainty are shown by the solid and dashed red lines, respectively.}
\end{figure*}

\begin{figure}[!htb]
\resizebox{\hsize}{!}{\includegraphics[]{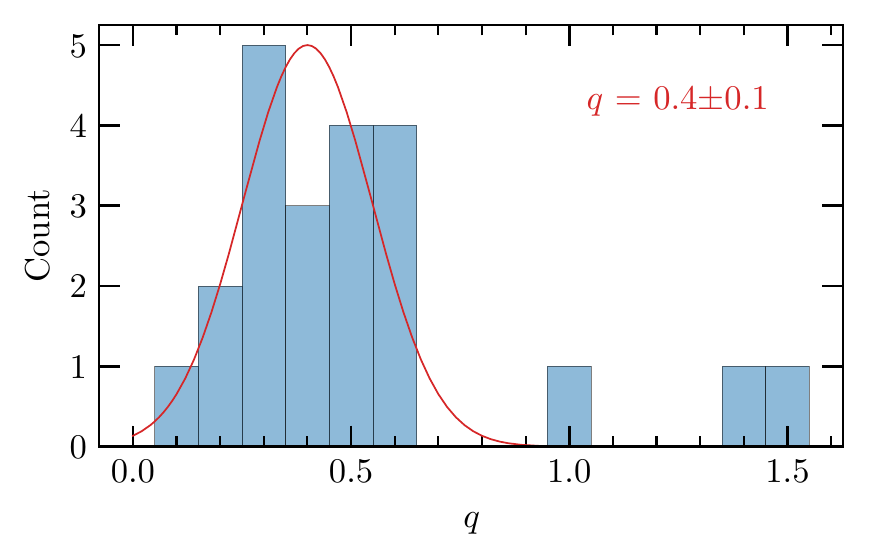}}
\caption{Histogram of the temperature power-law index $q$. The red line shows a Gaussian fit to the data points for $q = 0.1 - 0.6$.}
\label{fig:histogram_T}
\end{figure}

\begin{table*}
\caption{Results of the physical structure (Sect. \ref{sec:physicalstructure}) and estimated chemical ages (Sect. \ref{sec:chemicalmodeling}) of the cores. The density power-law index $p$ is calculated according to Eq. \eqref{eq:uvanalysis}. The core mass $M_\mathrm{core}$ is calculated according to Eq. \eqref{eq:Mcalc}. The estimate of the chemical timescale $\tau_\mathrm{chem}$ is explained in detail in Sect. \ref{sec:MUSCLEresults}.}
\label{tab:phys_struc}
\centering
\begin{tabular}{l l l l l l l l l}
\hline\hline
Region + & $r_\mathrm{in}$ & $r_\mathrm{out}$ & $T_{\mathrm{kin}}$($r_\mathrm{in}$) & $q$ & $\alpha$ & $p$ & $M_\mathrm{core}$ & $\tau_\mathrm{chem}$\\
Number & (au) & (au) & (K) & & & & ($M_{\odot}$) & (yrs) \\
\hline
IRAS\,23033 1 & 1\,837 & \,\,\,\,5\,720 & 114.9$\pm$\,\,\,8.2 & 0.34$\pm$0.11$^{\dagger}$ & $-0.44\pm$0.03 & 2.22$\pm$0.11 & \,\,\,6.06$\pm$1.29 & 3.4(4)$-$9.8(4)\\ 
IRAS\,23033 2 & 1\,837 & \,\,\,\,4\,767 & 167.2$\pm$\,\,\,6.6 & 0.48$\pm$0.20$^{\dagger}$ & $-0.73\pm$0.05 & 1.79$\pm$0.21 & \,\,\,7.81$\pm$1.59 & 1.8(4)\\ 
IRAS\,23033 3 & 1\,837 & \,\,\,\,3\,813 & 160.8$\pm$\,\,\,8.0 & 0.96$\pm$0.07$^{\dagger}$ & $-0.65\pm$0.05 & 1.39$\pm$0.09 & \,\,\,5.22$\pm$1.08 & 1.9(4)\\ 
IRAS\,23151 1 & 1\,392 & \,\,\,\,2\,926 & 129.2$\pm$\,\,\,4.7 & 0.27$\pm$0.11$^{\ast}$ & $-0.48\pm$0.01 & 2.25$\pm$0.11 & \,\,\,3.28$\pm$0.67 & 1.9(4)$-$8.4(4)\\ 
IRAS\,23385 1 & 2\,299 & \,\,\,\,4\,345 & 239.5$\pm$12.1 & 0.14$\pm$0.05$^{\dagger}$ & $-0.63\pm$0.05 & 2.23$\pm$0.07 & \,\,\,4.43$\pm$0.92 & 4.9(4)\\ 
IRAS\,23385 2 & 2\,299 & \,\,\,\,4\,345 & 226.0$\pm$\,\,\,2.3 & 0.21$\pm$0.04$^{\dagger}$ & $-0.39\pm$0.06 & 2.40$\pm$0.07 & \,\,\,2.30$\pm$0.46 & 9.5(4)\\ 
AFGL\,2591 1 & 1\,394 & \,\,\,\,2\,926 & 159.9$\pm$\,\,\,6.8 & 0.25$\pm$0.08$^{\ast}$ & $-0.63\pm$0.02 & 2.12$\pm$0.08 & \,\,\,7.43$\pm$1.52 & 8.4(4)\\ 
CepA\,HW2 1 & \,\,\,\,289 & \,\,\,\,\,\,\,931 & 238.2$\pm$11.4 & 0.31$\pm$0.02$^{\ast}$ & $-0.44\pm$0.02 & 2.25$\pm$0.03 & \,\,\,1.24$\pm$0.26 & 8.4(4)\\ 
CepA\,HW2 2 & \,\,\,\,289 & \,\,\,\,\,\,\,776 & 170.5$\pm$13.3 & 0.56$\pm$0.10$^{\ast}$ & $-0.45\pm$0.05 & 1.99$\pm$0.11 & \,\,\,0.28$\pm$0.06 & 1.9(4)$-$8.8(4)\\ 
G084.9505 1 & 2\,273 & \,\,\,\,6\,097 & 169.0$\pm$\,\,\,0.8 & 1.38$\pm$0.05$^{\dagger}$ & $-0.79\pm$0.03 & 0.83$\pm$0.06 & \,\,\,1.67$\pm$0.33 & 1.7(4)\\ 
G094.6028 1 & 1\,584 & \,\,\,\,4\,434 & 183.8$\pm$45.4 & 1.48$\pm$0.27$^{\dagger}$ & $-0.75\pm$0.04 & 0.77$\pm$0.27 & \,\,\,2.35$\pm$0.76 & 5.7(4)$-$8.4(4)\\ 
G100.38 1 & 1\,452 & \,\,\,\,3\,880 & \,\,\,68.2$\pm$\,\,\,0.7 & 0.41$\pm$0.22$^{\dagger}$ & $-0.76\pm$0.05 & 1.83$\pm$0.23 & \,\,\,1.85$\pm$0.37 & 8.4(4)\\ 
G108.75 1 & 2\,044 & \,\,\,\,4\,767 & 111.2$\pm$\,\,\,8.2 & 0.56$\pm$0.09$^{\dagger}$ & $-0.30\pm$0.03 & 2.14$\pm$0.09 & \,\,\,2.58$\pm$0.55 & 2.0(4)\\ 
G108.75 2 & 2\,044 & \,\,\,\,5\,720 & \,\,\,82.9$\pm$\,\,\,2.6 & 0.23$\pm$0.02$^{\dagger}$ & $-0.87\pm$0.04 & 1.90$\pm$0.04 & \,\,\,3.73$\pm$0.76 & 1.1(5)\\ 
G075.78 1 & 1\,642 & \,\,\,\,5\,055 & 176.8$\pm$\,\,\,0.3 & 0.29$\pm$0.05$^{\ast}$ & $-0.64\pm$0.04 & 2.07$\pm$0.06 & \,\,\,9.21$\pm$1.84 & 8.4(4)\\ 
IRAS\,21078 1 & \,\,\,\,612 & \,\,\,\,1\,330 & 135.5$\pm$\,\,\,0.2 & 0.43$\pm$0.05$^{\ast}$ & $-0.90\pm$0.05 & 1.67$\pm$0.07 & \,\,\,1.60$\pm$0.32 & 5.5(4)$-$9.2(4)\\ 
IRAS\,21078 2 & \,\,\,\,612 & \,\,\,\,1\,330 & 121.9$\pm$\,\,\,3.4 & 0.59$\pm$0.01$^{\ast}$ & $-1.10\pm$0.05 & 1.31$\pm$0.05 & \,\,\,1.70$\pm$0.34 & 4.9(4)\\ 
NGC7538\,IRS9 1 & 1\,114 & \,\,\,\,2\,394 & 200.5$\pm$\,\,\,0.9 & 0.38$\pm$0.04$^{\ast}$ & $-0.37\pm$0.04 & 2.25$\pm$0.06 & \,\,\,1.60$\pm$0.32 & 8.4(4)\\ 
S87\,IRS1 1 & 1\,005 & \,\,\,\,2\,439 & 112.0$\pm$\,\,\,5.2 & 0.45$\pm$0.07$^{\dagger}$ & $-0.52\pm$0.04 & 2.03$\pm$0.08 & \,\,\,2.15$\pm$0.44 & 5.9(4)\\ 
W3\,H2O 3 & \,\,\,\,770 & \,\,\,\,1\,774 & 176.6$\pm$10.9 & 0.59$\pm$0.06$^{\ast}$ & $-0.65\pm$0.09 & 1.76$\pm$0.11 & 11.61$\pm$2.44 & 8.4(4)\\ 
W3\,H2O 4 & \,\,\,\,770 & \,\,\,\,1\,774 & 166.4$\pm$\,\,\,9.0 & 0.45$\pm$0.03$^{\ast}$ & $-0.89\pm$0.09 & 1.66$\pm$0.09 & 11.22$\pm$2.33 & 8.6(4)\\ 
W3\,IRS4 1 & \,\,\,\,780 & \,\,\,\,1\,774 & 173.0$\pm$15.4 & 0.54$\pm$0.08$^{\ast}$ & $-0.52\pm$0.05 & 1.94$\pm$0.09 & \,\,\,1.14$\pm$0.25 & 2.0(4)$-$8.6(4)\\ 
\hline
\end{tabular}
\tablefoot{a(b) = a$\times$10$^{\mathrm{b}}$. Fit results for $r_\mathrm{in}$, $r_\mathrm{out}$, $T_{\mathrm{kin}}$($r_\mathrm{in}$), and $q$ derived from either the H$_{2}$CO or CH$_{3}$CN radial temperature profiles (Fig. \ref{fig:radialtemperatureprofile}) are indicated by either a $\dagger$ or $\ast$, respectively.}
\end{table*}

\begin{figure*}
\centering
\includegraphics[]{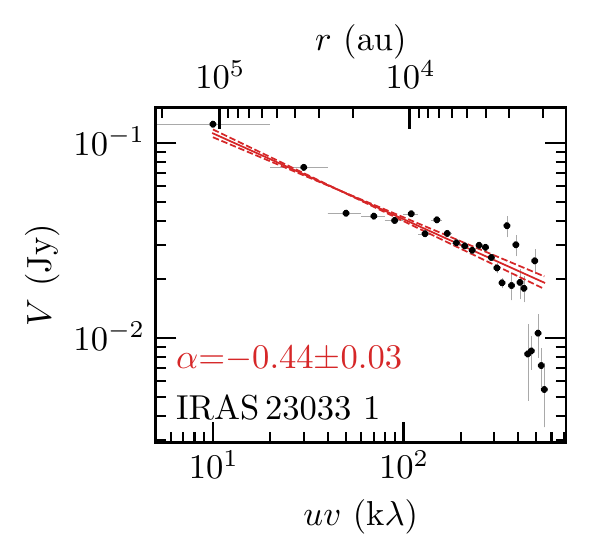}
\includegraphics[]{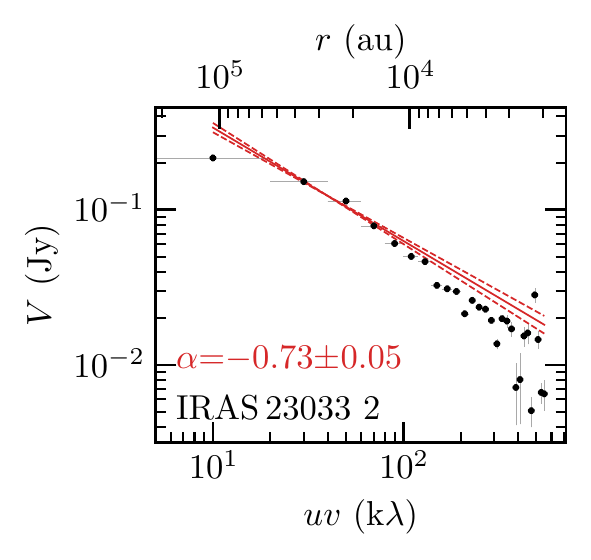}
\includegraphics[]{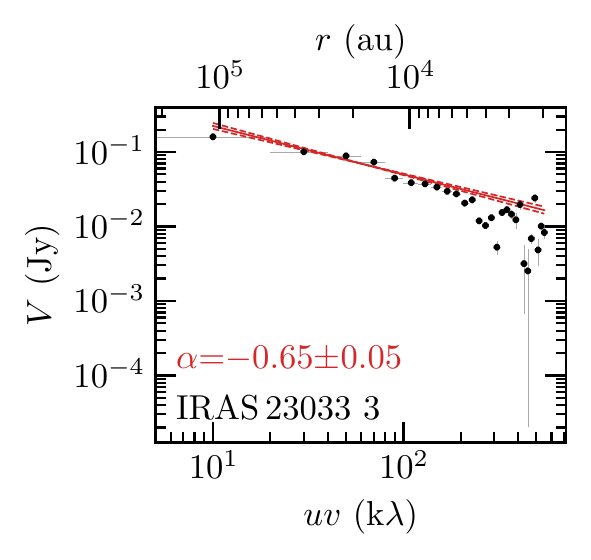}
\includegraphics[]{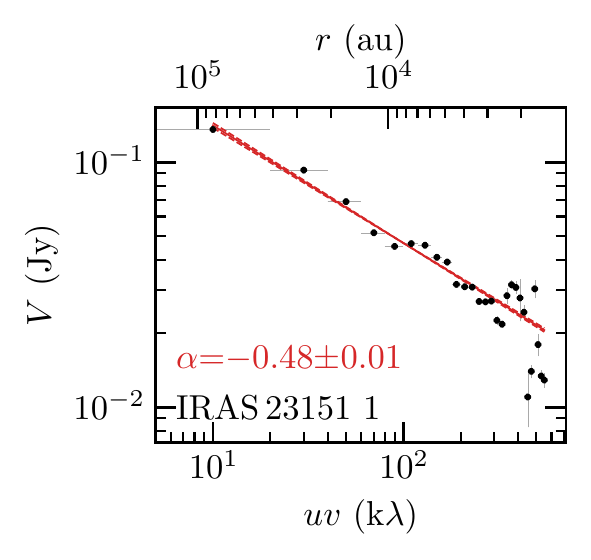}
\includegraphics[]{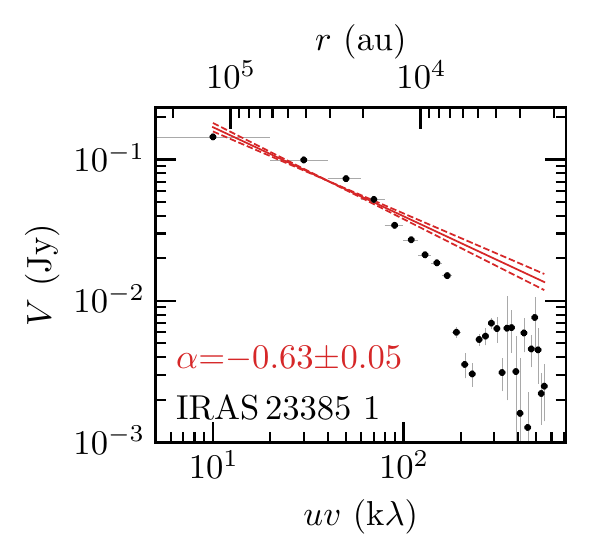}
\includegraphics[]{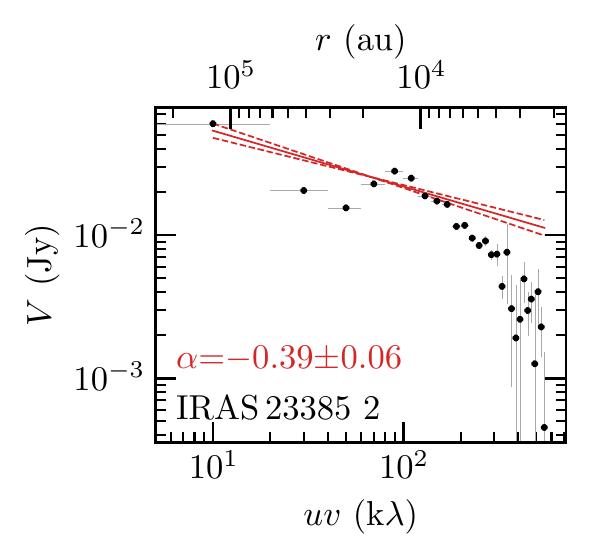}
\includegraphics[]{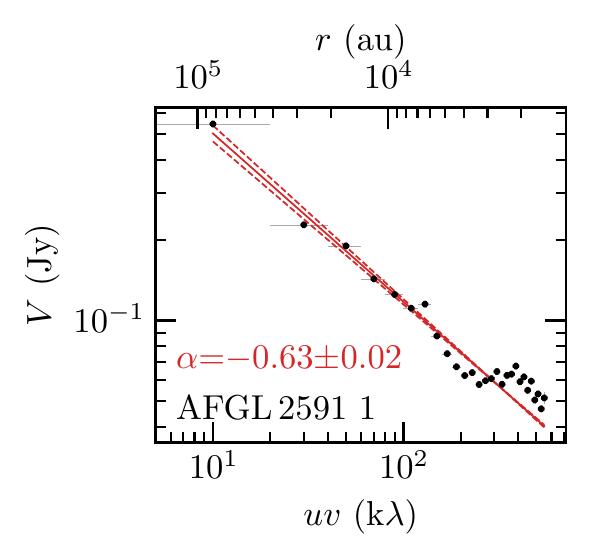}
\includegraphics[]{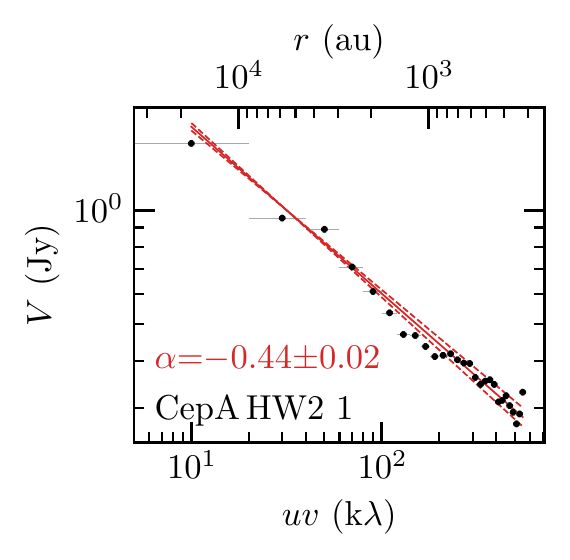}
\includegraphics[]{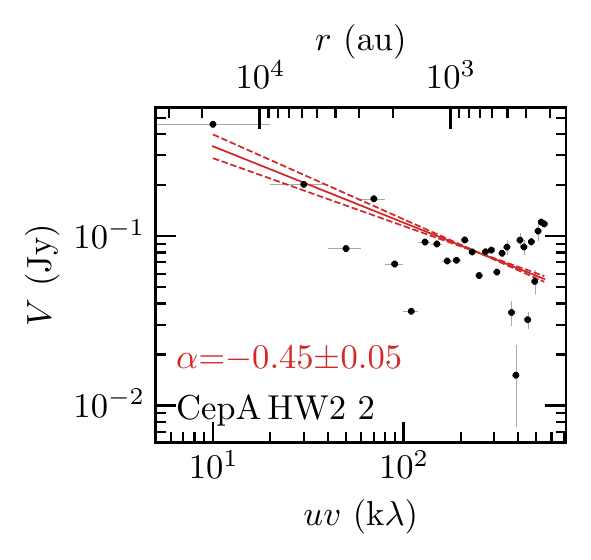}
\includegraphics[]{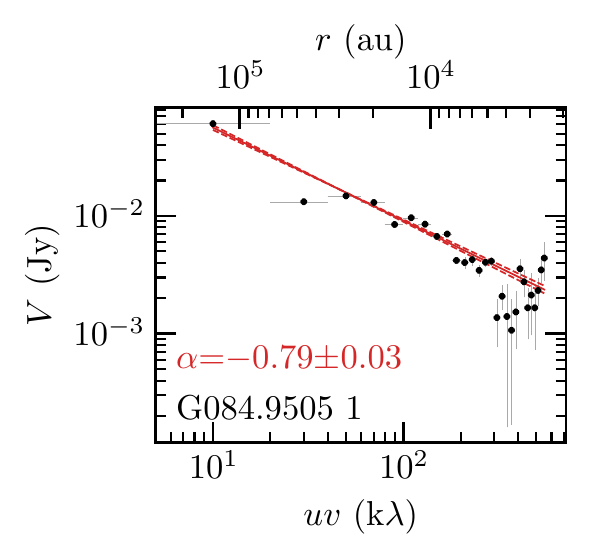}
\includegraphics[]{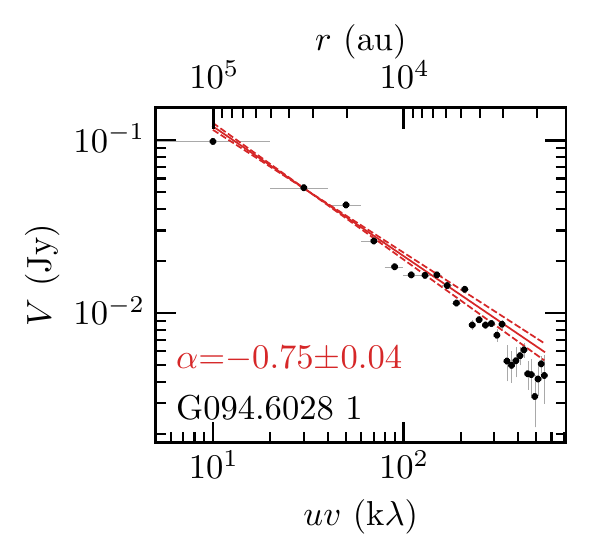}
\includegraphics[]{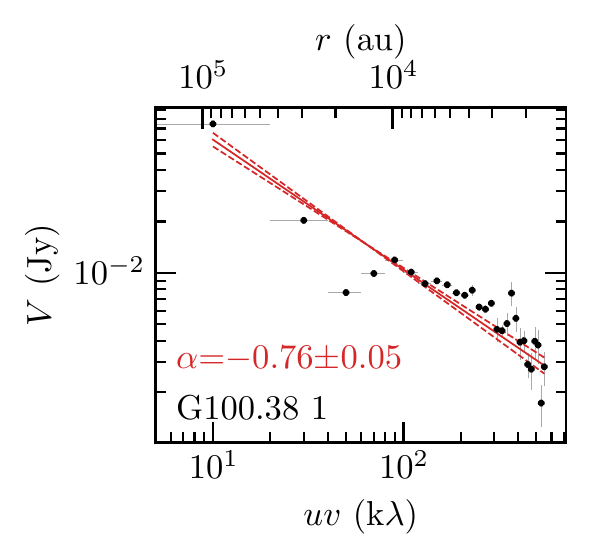}
\caption{Radial visibility profiles of the 1.37\,mm continuum emission of the 22 cores. The black data points show the radial profile of the averaged complex visibilities of the 1.37\,mm continuum as a function of $uv$ distance (bottom x-axis) and of the corresponding linear scale (top x-axis). The linear fit and the $\pm1\sigma$ uncertainties are indicated by the solid and dashed red line, respectively.}
\label{fig:visibilityanalysis}
\end{figure*}

\begin{figure*}
\ContinuedFloat
\captionsetup{list=off,format=cont}
\centering
\includegraphics[]{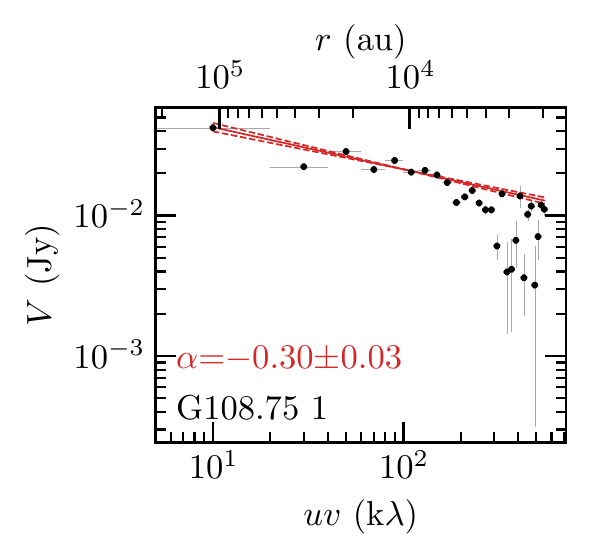}
\includegraphics[]{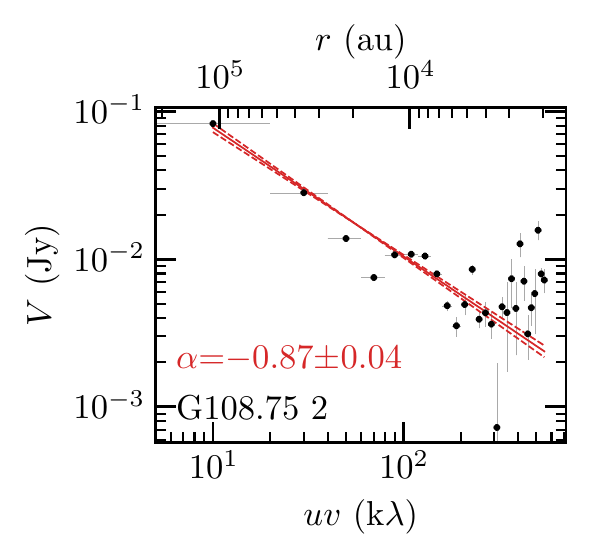}
\includegraphics[]{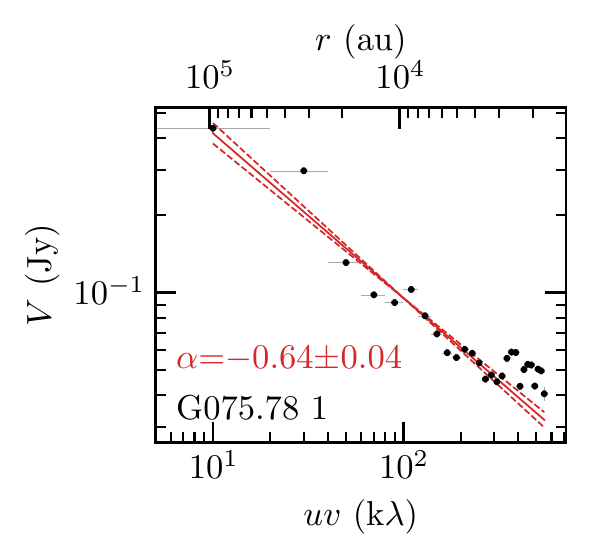}
\includegraphics[]{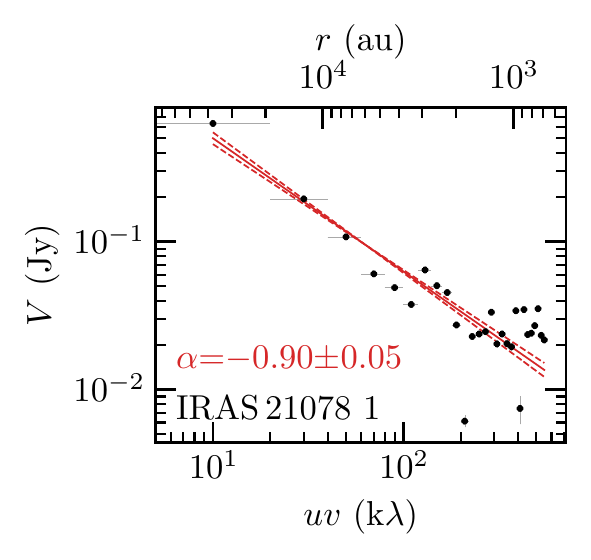}
\includegraphics[]{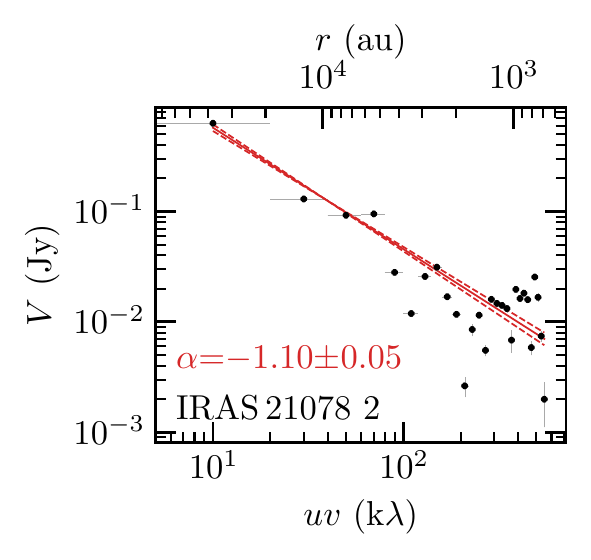}
\includegraphics[]{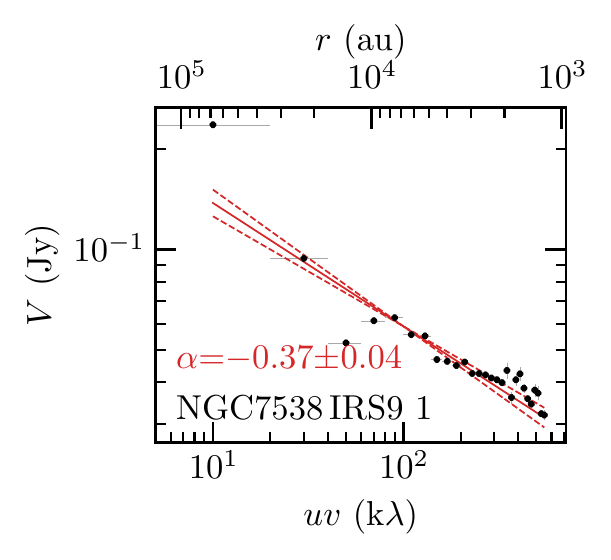}
\includegraphics[]{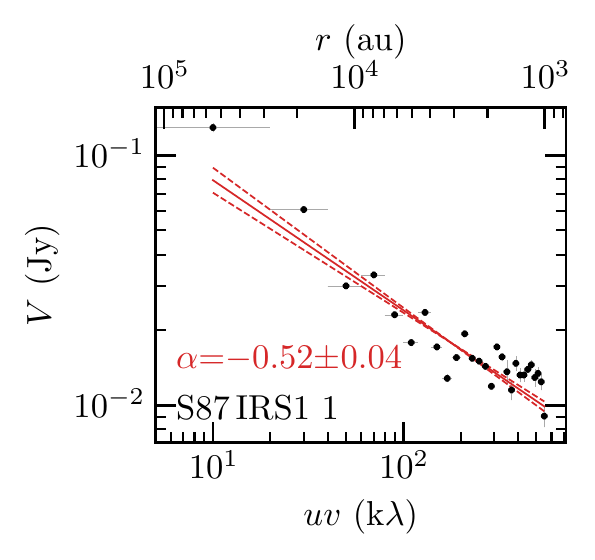}
\includegraphics[]{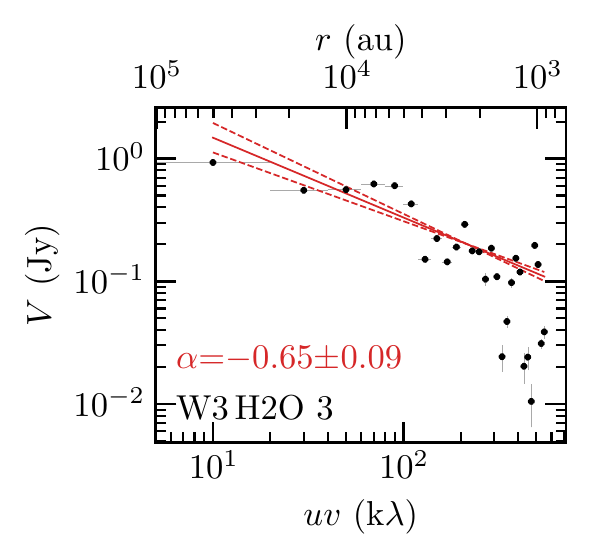}
\includegraphics[]{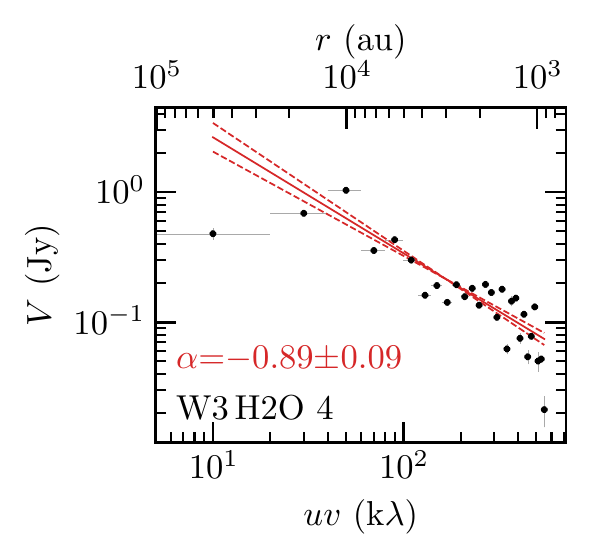}
\includegraphics[]{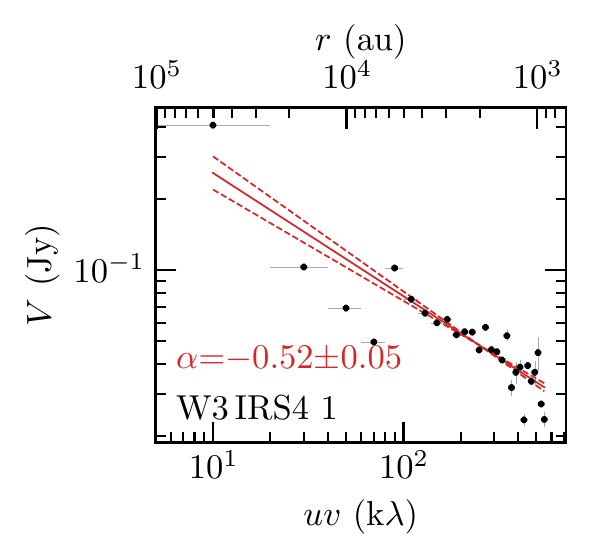}
\caption{Radial visibility profiles of the 1.37\,mm continuum emission of the 22 cores. The black data points show the radial profile of the averaged complex visibilities of the 1.37\,mm continuum as a function of $uv$ distance (bottom x-axis) and of the corresponding linear scale (top x-axis). The linear fit and the $\pm1\sigma$ uncertainties are indicated by the solid and dashed red line, respectively.}
\end{figure*}

\begin{figure}
\resizebox{\hsize}{!}{\includegraphics[]{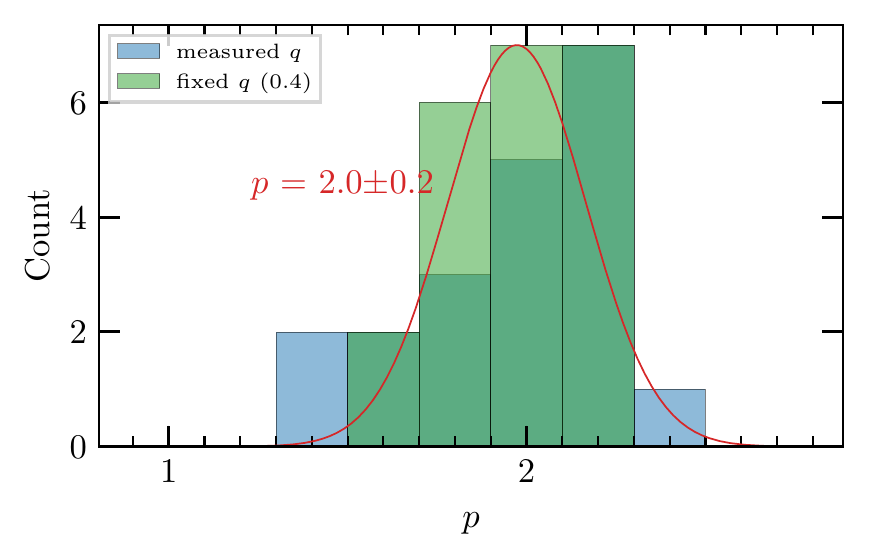}}
\caption{Histogram of the density power-law index $p$. The density power-law index derived with the measured values of $q$ for each core are shown in blue. In green, the results for $p$ calculated with a fixed temperature index ($q = 0.4$) for all cores are shown. The red line shows a Gaussian fit to the green histogram.}
\label{fig:histogram_p}
\end{figure}

	The linear spatial resolution of the CORE sample ranges from 300$-$2\,300\,au. At this spatial resolution it is not possible to resolve potential disks surrounding the protostars but these can be studied and characterized through the kinematic analysis of the line profiles \citep{Ahmadi2019}. However, we do resolve the gas and dust envelopes around individual cores which can be approximated as spherically symmetric objects for which the radial temperature profile $T(r)$ and radial density profile $n(r)$ of the envelope gas can be described by power-laws \citep[e.g.,][]{vanderTak2000,Beuther2002, Palau2014}:
\begin{equation}
\label{eq:temperatureprofile}
T(r) = T_{\mathrm{in}} \times \bigg(\frac{r}{r_{\mathrm{in}}}\bigg)^{-q}
\end{equation}
\begin{equation}
\label{eq:densityprofile}
n(r) = n_{\mathrm{in}} \times \bigg(\frac{r}{r_{\mathrm{in}}}\bigg)^{-p}
\end{equation}
	with $T_{\mathrm{in}} = T(r_{\mathrm{in}})$ and $n_{\mathrm{in}} = n(r_{\mathrm{in}})$ at an arbitrary radius $r_{\mathrm{in}}$. The temperature power-law index $q$ and density power-law index $p$ of the core envelopes are important properties of HMSFRs. By studying these quantities with observations, theoretical analytical models on how massive stars form can be constrained.

\subsection{Continuum emission}\label{sec:continuum}

	A detailed analysis of the continuum data is presented in \citet{Beuther2018}. The updated \texttt{GILDAS} self-calibrated continuum data are shown in contours in Fig. \ref{fig:continuum}. While the sample was selected to be largely homogeneous in luminosity, at an angular resolution of $\sim$0\as4 a variety of structures can be identified. While some regions appear as isolated single objects (IRAS\,23151, AFGL\,2591), others show spatially separated cores (IRAS\,23033, G108.75, G139.9091, S106, CepA\,HW2, W3\,H2O). There are regions in which fragmentation is observed within an embedded envelope (IRAS\,23385, IRAS\,21078). Many of the regions have a complex morphology such as filamentary structures and extended envelopes (G084.9505, G094.6028, G100.38, G138.2957, G75.78, NGC7538\,IRS9, S87\,IRS1, W3\,IRS4).

	In this study we aim to analyze the physical properties and chemical variation across the regions, and have therefore selected a number of positions throughout the different regions. Deriving the column density of all detected species in every single spectrum in each region is computationally expensive and we restrict this method to the analysis of the H$_{2}$CO and CH$_{3}$CN temperature maps (Sect. \ref{sec:temperaturestructure}). In order to study the cores and envelopes around forming protostars we select positions which have a clear spherically symmetric core-like morphology in the 1.37\,mm continuum data which is the case for most regions toward the continuum peak position, but also multiple spherically symmetric objects within a region can be identified (e.g., toward IRAS\,23033). In addition, we select positions in the extended envelopes to study potential differences in the chemical abundances. The in total 120 selected positions are summarized in Table \ref{tab:positions} and marked in Fig. \ref{fig:continuum}. The nomenclature of each position is denoted by the region name and increasing number with decreasing 1.37\,mm continuum intensity. For each region, position 1 is toward the 1.37\,mm continuum emission peak.
	
	In order to estimate the H$_{2}$ column density (Sect. \ref{sec:molecularhydrogen}), we require that all positions are detected in the 1.37\,mm continuum ($I_{\mathrm{1.37mm}} \geq 5\sigma_\mathrm{cont}$). The number of selected positions is higher toward regions with extended envelopes and complex morphologies (e.g., IRAS\,21078 and G138.2957) compared to compact regions (e.g., G139.9091 and S106). In this study, we define in Sect. \ref{sec:temperaturestructure} a ``core'' as an object that shows a radially decreasing temperature profile over at least the width of two beams. In Sect. \ref{sec:chemicalmodeling}, we will apply a 1D physical-chemical model to these defined cores using the observed temperature and density structure analyzed in Sect. \ref{sec:temperaturestructure} and Sect. \ref{sec:densitystructure}, respectively and observed molecular column densities (Sect. \ref{sec:molecularcontent}). These positions are labeled as ``C'' (core) in Table \ref{tab:positions}. The remaining positions are locations in the dust envelope and environment around the cores. There are a few positions with a clear spherically symmetric core morphology in the dust emission, but no temperature profile can be derived (e.g., position 1 in S106). Cores with unresolved radial temperature profiles are also not included in this approach. In total we select 120 positions including 22 core positions. The broad-band spectra of the 120 positions are shown in Fig. \ref{fig:spectrum}. Table \ref{tab:positions} lists the noise of the broad-band spectrum extracted from these positions and the systemic velocity $\varv_{\mathrm{LSR}}$ determined from the C$^{18}$O $2-1$ line, which may differ from the average region $\varv_{\mathrm{LSR}}$ listed in Table \ref{tab:sample} due to velocity gradients within the region (see Sect. \ref{sec:XCLASSfitting}). In contrast to this core definition, in the analysis by \citet{Beuther2018}, ``cores'' are defined as fragmented objects with emission $\geq 10\sigma_\mathrm{cont}$ using the \texttt{clumpfind} algorithm \citep{clumpfind} and thus more cores are found in their analysis.
	

\subsection{Temperature structure}\label{sec:temperaturestructure}

	
	To reliably determine the rotation temperature $T_{\mathrm{rot}}$, it is required to observe multiple transitions of a molecule. We use formaldehyde (H$_{2}$CO) and methyl cyanide (CH$_{3}$CN) as thermometers to determine the temperature structure of the regions, as both have multiple strong and optically thin lines in our spectral setup. The spectral line properties of these molecules are listed in Table \ref{tab:spectrallineproperties}.

	We model the spectral line emission of these molecules using the eXtended \texttt{CASA} Line Analysis Software Suite \citep[\texttt{XCLASS},][]{XCLASS}\footnote{\url{https://xclass.astro.uni-koeln.de/}}. With \texttt{XCLASS}, molecular lines can be modeled and fitted by solving the 1D radiative transfer equation. In the calculation of the Gaussian line profiles optical depth effects are included. For each molecule, the properties of all transitions are taken from an embedded SQlite3 database containing entries from the Cologne Database for Molecular Spectroscopy \citep[CDMS,][]{CDMS} and the Jet Propulsion Laboratory \citep[JPL,][]{JPL} using the Virtual Atomic and Molecular Data Centre \citep[VAMDC,][]{Endres2016}. 
	
	The broad-band spectral setup includes three strong formaldehyde lines, two of which have the same upper energy level, and one weak transition from the H$_{2}^{13}$CO isotopologue. H$_{2}$CO and H$_{2}^{13}$CO are fitted simultaneously in \texttt{XCLASS} using an isotopic ratio calculated from \citet{Wilson1994}: $^{12}$C/$^{13}$C$\approx 7.5 \times d_\mathrm{gal} + 7.6$. The $^{12}$C/$^{13}$C ratio is listed in Table \ref{tab:sample} for each region. While formaldehyde is a good low-temperature gas tracer at $T_{\mathrm{kin}} < 100$\,K \citep{Mangum1993}, at high densities and temperatures the $3_{0,3}-2_{0,2}$ transition is optically thick and a reliable temperature cannot be derived anymore from the line ratios with this method \citep{Rodon2012,Gieser2019}. To determine temperatures at higher densities, we use nine methyl cyanide lines ($J = 12-11$ and $K = 0 - 8$, $E_\mathrm{u}/k_\mathrm{B} = 69 - 526$\,K). In addition, 7 lines of the CH$_{3}^{13}$CN isotopologue are fitted simultaneously ($J = 12-11$ and $K = 0 - 6$, $E_\mathrm{u}/k_\mathrm{B} = 69 - 326$\,K). A detailed discussion of the line optical depth is given in Sect. \ref{sec:XCLASSfitting}.

	In \texttt{XCLASS}, each molecule can be described by a number of emission and absorption components. The fit parameter set of each component consists of the source size $\theta_\mathrm{source}$, the rotation temperature $T_\mathrm{rot}$, the column density $N$, the linewidth $\Delta v$, and the velocity offset from the systemic velocity $\varv_{\mathrm{off}}$. One can choose a variety of algorithms that can also be combined in an algorithm chain in order to find the best-fit parameters by minimizing the $\chi^{2}$ value. We adopt an algorithm chain with the Genetic and Levenberg-Marquardt (LM) methods optimizing toward global and local minima, respectively. 
	
	For each region, we use the \texttt{myXCLASSMapFit} function to fit the H$_{2}$CO and CH$_{3}$CN lines in each pixel within the primary beam. All lines with a peak intensity $> 10\sigma_\mathrm{line,map}$ are fitted with one emission component. We chose a high threshold of $10\sigma_\mathrm{line,map}$ so multiple transitions have a high signal-to-noise ratio to accurately determine the rotation temperature. Only a single value can be set as the threshold in the \texttt{myXCLASSMapFit} function, therefore toward the edges of the primary beam the temperature estimates are less reliable due to an increase of the noise.

	Under the assumption of local thermal equilibrium (LTE), the kinetic temperature of the gas can be estimated from the rotation temperature $T_\mathrm{kin} \approx T_\mathrm{rot}$. As HMSFRs have high densities $n > 10^{5}$\,cm$^{-3}$ toward the locations of the protostars, LTE can be assumed here \citep{Mangum2015}. The critical densities $n_{\mathrm{crit}} = \frac{A_{\mathrm{ul}}}{C_{\mathrm{ul}}}$ for the CH$_{3}$CN and H$_{2}$CO lines are $\sim$4$\times$10$^{6}$\,cm$^{-3}$ and $\sim$3$\times$10$^{6}$\,cm$^{-3}$, respectively ($A_{\mathrm{ul}}$ is the Einstein $A$ coefficient and $C_{\mathrm{ul}}$ is the collisional rate coefficient). Here, we use $C_{\mathrm{ul}}$ measured at 140\,K taken from the Leiden Atomic and Molecular Database \citep[LAMDA,][]{Schoier2005}. The H$_{2}$CO and CH$_{3}$CN temperature maps of each region are shown in Fig. \ref{fig:temperature_maps}. The H$_{2}$CO lines trace the extended low temperature gas at $10 - 100$\,K. The CH$_{3}$CN maps mostly show spatially compact emission tracing the higher temperature gas at $>$100\,K.

	For each selected position (Table \ref{tab:positions}), we extract the H$_{2}$CO and CH$_{3}$CN radial temperature profile which are binned in steps of half a beam ($\Delta r$). In each bin, the uncertainties are computed from the standard deviation of the mean. After a first visual inspection of all radial profiles of the 120 positions, we select all positions with a radially decreasing temperature profile along at least two beams in at least one temperature tracer. With this method, we are able to extract in total 22 positions, which we define as ``cores''. The radial temperature profiles of the cores are shown in Fig. \ref{fig:radialtemperatureprofile}. In cases where both temperature tracers are detected toward the central core (e.g., IRAS\,23033 2), the observed H$_{2}$CO temperature profile has significantly higher values compared to CH$_{3}$CN. As discussed previously, this is due to the fact that due to the high optical depth in these dense regions, the fit algorithm converges toward high temperatures. The line optical depth is discussed further in Sect. \ref{sec:XCLASSfitting}.
	
	We fit the profiles with a power-law according to Eq. \eqref{eq:temperatureprofile} using the minimum $\chi^{2}$ method to derive the temperature power-law index $q$. We use the CH$_{3}$CN temperature profile for the fit if it is detected along at least two beams and the H$_{2}$CO temperature profile otherwise. The inner radius is the temperature at a radius of half the beam size $r_{\mathrm{in}}$ = $\Delta r$. The outer radius $r_{\mathrm{out}}$ is chosen as a local minimum, when $T_{\mathrm{rot}}$($r_{\mathrm{out}}$) $<$ $T_{\mathrm{rot}}$($r_{\mathrm{out}} + \Delta r$) and $T_{\mathrm{rot}}$($r_{\mathrm{out}}$) $<$ $T_{\mathrm{rot}}$($r_{\mathrm{out}} + 2\Delta r$). In case the outer radii of nearby cores would overlap or the 2D temperature distribution becomes highly asymmetric, we reduce the outer radius. The outer radii are also marked in Fig. \ref{fig:radialtemperatureprofile}. The fit results for $q$, $r_{\mathrm{in}}$, $r_{\mathrm{out}}$, and $T_{\mathrm{kin}}$($r_{\mathrm{in}}$) are summarized in Table \ref{tab:phys_struc} for each core.
	
	A histogram of the temperature power-law index $q$ is shown in Fig. \ref{fig:histogram_T}. Most of the data points are distributed between 0.1 and 0.6, but three outliers are located at $q > 0.9$. A Gaussian fit to the data with $q = 0.1 - 0.6$ yields an average value of $q = 0.4 \pm 0.1$, which is in very good agreement with theoretical predictions \citep{Emerson1988, Osorio1999, vanderTak2000} and observations \citep[e.g.,][]{Palau2014, Palau2020}. The issue of optical depth of the H$_{2}$CO lines and CH$_{3}$CN being only detected toward the densest parts result in the high uncertainties and spread in $q$. We observe a broad range of $q$ with shallow profiles ($q = 0.1$) up to steep profiles ($q = 1.5$). In order to study if $q$ is constant for all high-mass cores or if the range of $q$ is real, observations of more cores are required. \citet{Osorio1999} and \citet{vanderTak2000} suggest steeper values for $q$ on scales $< 2\,000$\,au. In addition, the temperature maps and radial profiles are 2D projected, and the real 3D profile may be more complicated.
	
\subsection{Density structure}\label{sec:densitystructure}

	In contrast to the merged (NOEMA + IRAM 30m) spectral line data, the 1.37\,mm continuum NOEMA data suffer from missing flux due to the lack of short-spacing information. Therefore, it is difficult to reliably derive radial intensity profiles in the final images that are required to determine the density structure. This issue can be minimized by analyzing instead the complex visibilities of the 1.37\,mm continuum data in the $uv$-plane \citep{Adams1991, Looney2003, Beuther2007B, Zhang2009}. Assuming spherical symmetry, the power-law index of the complex visibility profile $\alpha$ is related to the density power-law index of the radial density distribution $p$ by \citep{Looney2003}:
\begin{equation}
\label{eq:uvanalysis}
\alpha = p + q - 3.
\end{equation}

	For each core position, the phase center is shifted to this location. The remaining cores within a region are fitted as a point source + circular Gaussian (to account both for the compact and the extended emission) and subtracted using the \texttt{UV\_FIT} task in \texttt{GILDAS}. With this approach blending of nearby cores in the visibility profiles within a single region can be minimized. The azimuthally averaged complex visibilities computed using the \texttt{UVAMP} task in \texttt{MIRIAD} \citep{miriad} are binned in steps of 20\,k$\lambda$. The radial visibility profiles are shown in Fig. \ref{fig:visibilityanalysis}. The $uv$ distances range from $\sim 20 - 800$\,m corresponding to spatial scales of $\sim 10 ^{3}-10^{5}$\,au at distances of a few kpc. We apply a power-law fit to the data in order to derive the power-law index $\alpha$. The density power-law index $p$ is calculated according to Eq. \eqref{eq:uvanalysis} taking into account the temperature power-law index $q$ derived in Sect. \ref{sec:temperaturestructure}. The results for $\alpha$ and $p$ for all cores are summarized in Table \ref{tab:phys_struc}.
	
	Most of the visibility profiles (Fig. \ref{fig:visibilityanalysis}) can be described by a single power-law. The higher scatter at large $uv$ distances is due to the fact that less data is available at long baselines. For IRAS\,23385, the visibility profiles of core 1 and 2 do not follow a simple power-law. Toward smaller spatial scales ($< 10^{4}$\,au), the profiles are steeper. This could be due to the fact, that even though the contribution of nearby cores is minimized by subtracting a point source + circular Gaussian, in the case of IRAS\,23385 it is not possible to clearly distinguish the contribution of both cores, which are embedded within a common dust envelope.

	A histogram of the density index $p$ is shown in Fig. \ref{fig:histogram_p}. Using the observationally derived values of $q$, the power-law index ranges from $1.4 - 2.4$. Fixing the temperature power-law index $q$ to the mean value of 0.4 and calculating $p$ with the results from the $uv$-analysis, the distribution of $p$ gets narrower, peaking at $p = 2.0 - 2.1$. A Gaussian fit to this distribution yields a mean of $p = 2.0 \pm 0.2$. The results of the derived physical structure of these 22 cores are discussed in detail in Sect. \ref{sec:discussphys}.

\section{Molecular gas content}\label{sec:molecularcontent}

\begin{figure*}[!htb]
\centering
\includegraphics[]{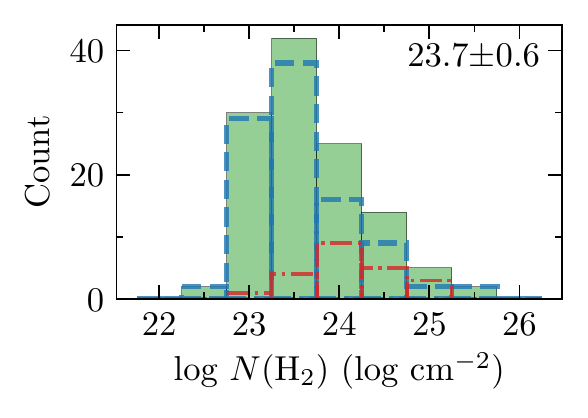}
\includegraphics[]{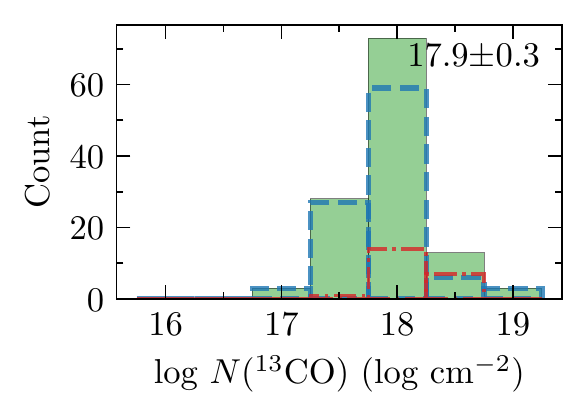}
\includegraphics[]{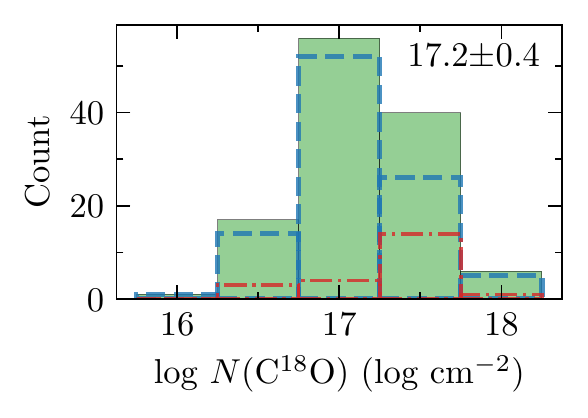}
\includegraphics[]{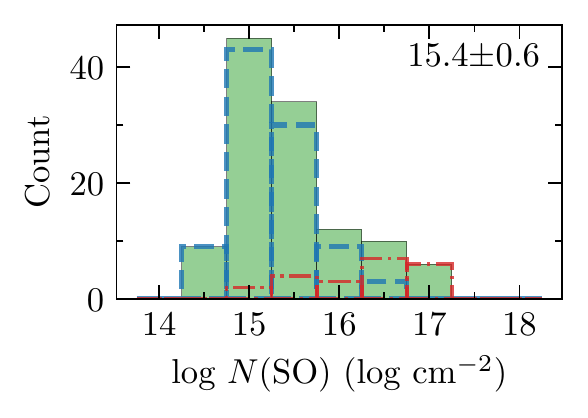}
\includegraphics[]{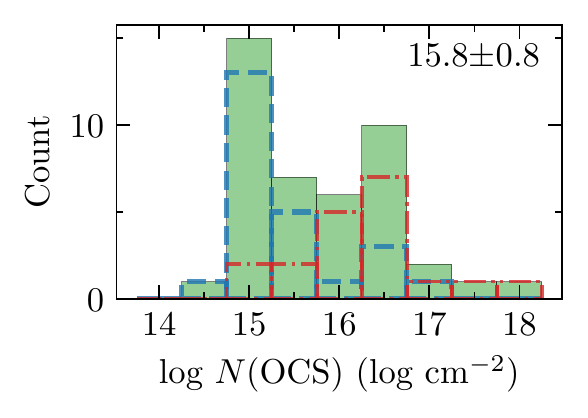}
\includegraphics[]{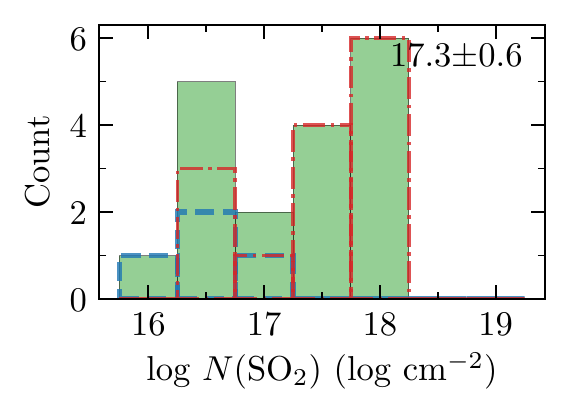}
\includegraphics[]{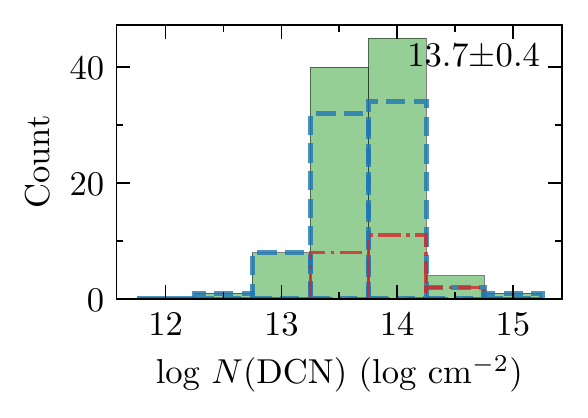}
\includegraphics[]{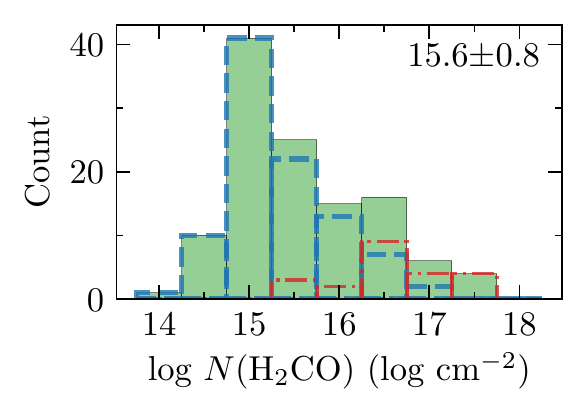}
\includegraphics[]{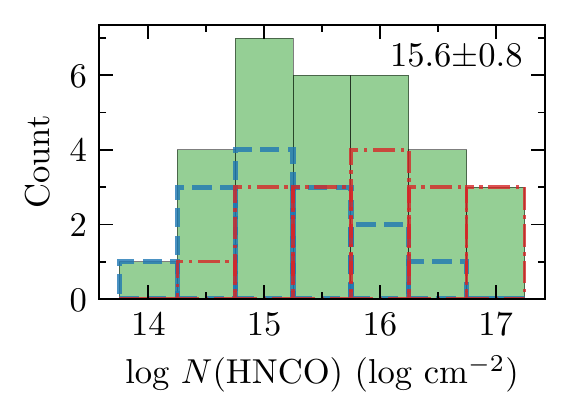}
\includegraphics[]{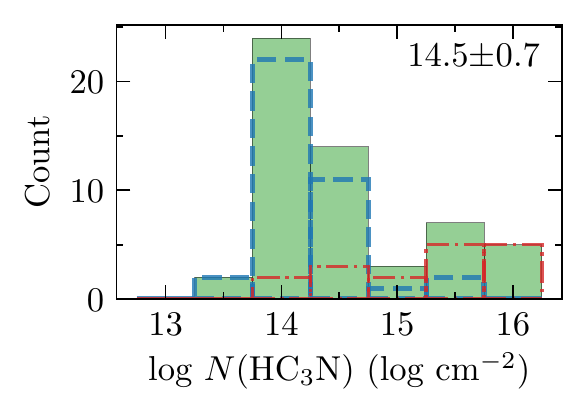}
\includegraphics[]{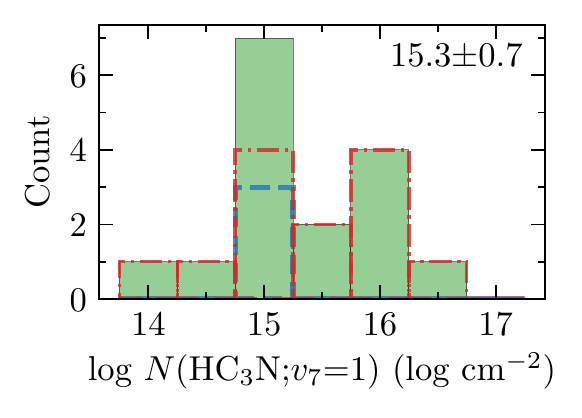}
\includegraphics[]{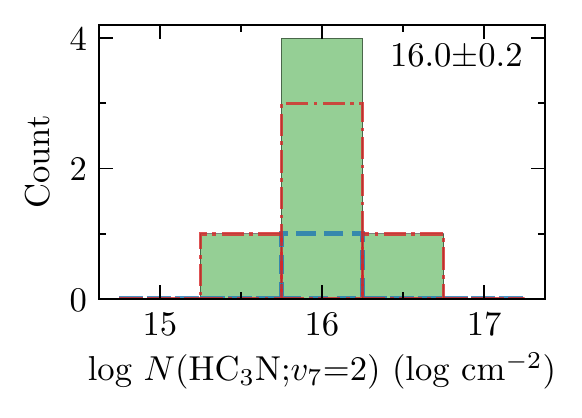}
\includegraphics[]{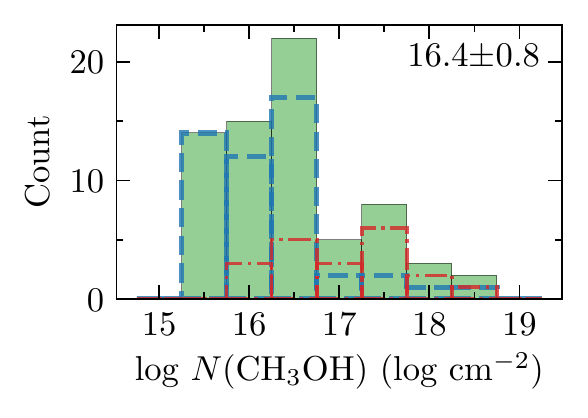}
\includegraphics[]{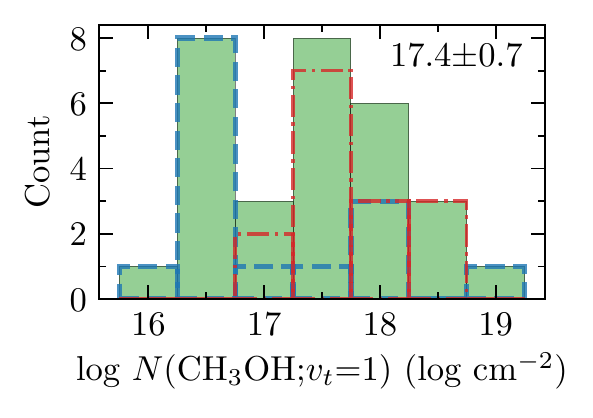}
\includegraphics[]{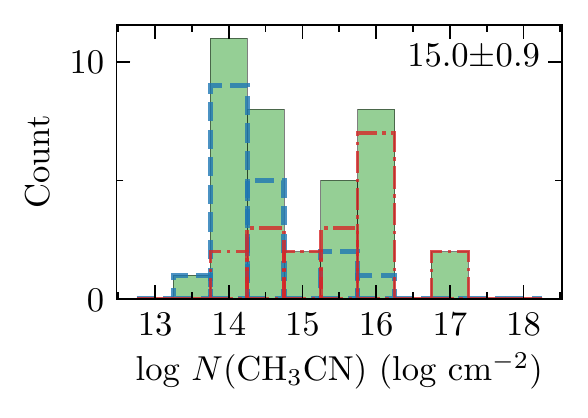}
\caption{Column density histograms. The H$_{2}$ column density is derived from the 1.37\,mm continuum emission and the remaining molecular column densities are derived with \texttt{XCLASS}. Column density histograms of all 120 positions are shown in green bars (upper limits are not included). Separate column density histograms of the core and the remaining positions are indicated by the dash-dotted red and dashed blue lines, respectively.}
\label{fig:Nhisto}
\end{figure*}

	In this section, we analyze the chemical content of the molecular gas toward the 18 CORE HMSFRs by studying the molecular column densities derived toward the 120 positions. The column density of molecular hydrogen $N$(H$_{2}$) is derived from the 1.37\,mm dust continuum emission (Sect. \ref{sec:molecularhydrogen}). In addition, molecular column densities are derived from the merged spectral line data with \texttt{XCLASS} by fitting the emission lines assuming LTE conditions (Sect \ref{sec:XCLASSfitting}). Spectra extracted toward the selected positions are shown in Fig. \ref{fig:spectrum}. In total, we consider 11 species among in total 16 isotopologues commonly detected within the 4\,GHz spectral bandwidth of the broad-band WideX correlator: $^{13}$CO, C$^{18}$O, SO, OCS, SO$_{2}$, DCN, H$_{2}$CO, HNCO, HC$_{3}$N, CH$_{3}$OH, CH$_{3}$CN. A spectral resolution of 3\,km\,s$^{-1}$ is not sufficient to study the line widths and kinematic properties in detail, but sufficient to derive molecular column densities $N$.
	
\subsection{Molecular hydrogen and core mass estimate}\label{sec:molecularhydrogen}

	The beam-convolved molecular hydrogen column density $N$(H$_{2}$) toward all 120 positions and mass calculation of the 22 cores $M_\mathrm{core}$ can be derived from the continuum intensity $I_{\nu}$ assuming that the emission is optically thin \citep{Hildebrand1983}:
	
\begin{equation}
\label{eq:H2calc}
N(\mathrm{H}_2) = \frac{I_{\nu} \eta }{\mu m_{\mathrm{H}} \Omega \kappa_{\nu} B_{\nu}(T)},
\end{equation}
\begin{equation}
\label{eq:Mcalc}
M_\mathrm{core} = \frac{F_{\nu} \eta d^2}{\kappa_{\nu} B_{\nu}(T)},
\end{equation}
	with a gas-to-dust mass ratio $\eta = 150$ \citep[$\eta = \frac{M_{\mathrm{gas}}}{M_{\mathrm{H}}}$/$\frac{M_{\mathrm{dust}}}{M_{\mathrm{H}}}$, with $\frac{M_{\mathrm{gas}}}{M_{\mathrm{H}}} = 1.4$ and $\frac{M_{\mathrm{dust}}}{M_{\mathrm{H}}} = 0.0091$, see Table 1.4 and 23.1 in][respectively]{Draine2011}, mean molecular weight $\mu = 2.8$ \citep{Kauffmann2008}, hydrogen mass $m_{\mathrm{H}}$, beam solid angle $\Omega$, dust opacity $\kappa_{\nu} = 0.9$\,cm$^{2}$\,g$^{-1}$ for dust grains with a thin icy mantle at a gas density of 10$^6$\,cm$^{-3}$ at 1.3\,mm \citep{Ossenkopf1994}, the Planck function $B_{\nu}(T)$ and distance $d$. $F_{\nu}$ is the integrated intensity of each core within the outer radius $r_\mathrm{out}$.

	We use $T_{\mathrm{kin}}$ for the temperature $T$ in the Planck function taken from the thermometers H$_{2}$CO and CH$_{3}$CN, assuming LTE conditions ($T_{\mathrm{kin}} \approx T_{\mathrm{rot}}$). If the spectrum is extracted from a core position, $T_{\mathrm{kin}}$ is taken from the radial temperature fit described in Sect. \ref{sec:temperaturestructure} with $T = T_{\mathrm{kin}}$($r_\mathrm{in}$) (see Table \ref{tab:phys_struc}). If the spectrum is extracted from a position not corresponding to a core, the kinetic temperature is computed from $T_{\mathrm{rot}}$(CH$_{3}$CN) if detected or from $T_{\mathrm{rot}}$(H$_{2}$CO) otherwise. If there is no temperature tracer detected toward the position, we adopt a lower limit of $T_{\mathrm{kin}} = 20\pm10$\,K, as the lowest derived rotation temperatures range between 10$-$30\,K. 
	
	In order to validate that the assumption of optically thin dust emission is valid, the continuum optical depth $\tau_{\nu}^{\mathrm{cont}}$ is computed for each position \citep[for a derivation of the equation, see Appendix A in][]{Frau2010}:
	\begin{equation}
	\label{eq:opticaldepth}
	\tau_{\nu}^{\mathrm{cont}} = -\mathrm{ln}\bigg( 1 - \frac{I_{\nu}}{\Omega B_{\nu}(T)} \bigg).
	\end{equation}
	
	The kinetic temperature $T_{\mathrm{kin}}$, molecular hydrogen column density $N$(H$_{2}$), and continuum optical depth $\tau_{\nu}^{\mathrm{cont}}$ are listed in Table \ref{tab:positions} for all 120 positions. The uncertainties of $N$(H$_{2}$) and $M_\mathrm{core}$ are calculated assuming Gaussian error propagation and include the uncertainty of the continuum intensity with an estimated 20\% flux calibration uncertainty \citep{Beuther2018} and uncertainty of the derived rotation temperature $\Delta T_{\mathrm{kin}}$ (listed in Table \ref{tab:positions}). The optical depth $\tau_{\nu}^{\mathrm{cont}}$ is $\ll 1$ toward most positions, so optically thin dust emission can be assumed here and the H$_{2}$ column density and core mass calculation provide reliable results. The only exceptions with $\tau_{\nu}^{\mathrm{cont}} > 1$ are the positions 1 and 2 of the W3\,H2O region which corresponds to the W3\,OH UCH{\sc ii} region \citep[see also][]{Ahmadi2018}. The molecular hydrogen column density $N$(H$_{2}$) has a mean value of $1.5\times10^{24}$\,cm$^{-2}$, ranging from $2.7\times10^{22}$\,cm$^{-2}$ up to $2.8\times10^{25}$\,cm$^{-2}$. The core masses $M_\mathrm{core}$ are listed in Table \ref{tab:phys_struc}. We find a mean core mass of 4.1\,$M_\odot$ in a range between 0.3\,$M_\odot$ and 11.6\,$M_\odot$. As discussed previously, due to missing short-spacing information, both $N$(H$_{2}$) and $M_\mathrm{core}$ should be considered as lower limits. However, the high core masses indicate that they are harboring protostars which will eventually end up as massive stars.

\subsection{Spectral line modeling with \texttt{XCLASS}}\label{sec:XCLASSfitting}
		
	We use the spectral line data of the CORE project to derive molecular column densities of 11 different species using the \texttt{XCLASS} software. A description of the \texttt{XCLASS} software is presented in Sect. \ref{sec:temperaturestructure}. Using the \texttt{myXCLASSFit} function, individual spectra are fitted species by species with one emission component to derive the molecular column density $N$.

	A spectrum is extracted from the merged spectral line data for each position listed in Table \ref{tab:positions}. The noise $\sigma_{\mathrm{line}}$ in each spectrum is computed in a line-free range from 219.00\,GHz to 219.13\,GHz. Compared to the average map noise $\sigma_{\mathrm{line,map}}$ listed in Table \ref{tab:sample}, the noise in each spectrum $\sigma_{\mathrm{line}}$ may have higher or lower noise values since the noise distribution is not uniform within the primary beam. The systemic velocity $\varv_{\mathrm{LSR}}$ is determined by fitting the C$^{18}$O $2-1$ transition and may differ from the global $\varv_{\mathrm{LSR}}$ listed in Table \ref{tab:sample} due to velocity gradients in the region, hence it allows us to employ a narrow parameter range for $\varv_{\mathrm{off}}$ so fitting strong nearby emission lines is avoided. The systemic velocity $\varv_{\mathrm{LSR}}$ and noise $\sigma_{\mathrm{line}}$ are listed in Table \ref{tab:positions} for each position. 

	All molecules and their corresponding transitions that are fitted with \texttt{XCLASS} are listed in Table \ref{tab:spectrallineproperties}. Lines which are blended with transitions of other detected species (at a resolution of 3\,km\,s$^{-1}$) are also listed, but excluded from the fit. For most molecules, only the rotational ground-state level $\varv = 0$ can be detected in our spectral setup. For SO$_{2}$ and HC$_{3}$N vibrationally excited levels ($\varv_{x}>0$) are present and for CH$_{3}$OH torsionally excited lines are detected ($\varv_{t} = 1$). The following species for which we observe multiple isotopologues are fitted simultaneously: OCS and O$^{13}$CS, SO$_{2}$ and $^{34}$SO$_{2}$, H$_{2}$CO and H$_{2}^{13}$CO, HC$_{3}$N and HCC$^{13}$CN, CH$_{3}$CN and CH$_{3}^{13}$CN. The isotopologue ratios are summarized in Table \ref{tab:sample} and are calculated either from \citet{Wilson1994} ($^{12}$C/$^{13}$C $\approx 7.5 \times d_\mathrm{gal} + 7.6$) or taken from references within ($^{32}$S/$^{34}$S $\approx$ 22). The uncertainties of these isotopic ratios are high due to a large scatter of the data points. However, we do not observe a sufficient number of strong isotopic lines to measure it more precisely. We do not fit $^{13}$CO and C$^{18}$O simultaneously, because the $^{13}$CO $2 - 1$ line has a high optical depth (see Table \ref{tab:spectrallineproperties}).

	We use an algorithm chain with the Genetic and Levenberg-Marquardt (LM) methods and to estimate the uncertainties of the fit parameters, we use the MCMC error estimation algorithm afterward. The following criteria are applied to the fitted model spectrum and column density $N^{+\Delta N_\mathrm{upp}}_{- \Delta N_\mathrm{low}}$ of each species in order to determine bad fits: 1) model spectra with a peak intensity $< 3\sigma_{\mathrm{line}}$; 2) the upper error of the column density $\Delta N_\mathrm{upp}$ is converging towards high values ($\Delta N_\mathrm{upp}> 10 \times N$); and 3) the lower error of the column density is not constrained $\Delta N_\mathrm{low} = 0$\,cm$^{-2}$. With these three criteria, weak and unresolved lines are automatically discarded and we use the best-fit value of $N$ only as an upper limit.

	The column densities $N$ for all species fitted with \texttt{XCLASS} and their uncertainties derived with the MCMC error estimation algorithm are summarized in Tables \ref{tab:XCLASSresults1} - \ref{tab:XCLASSresults3} for all positions. Histograms of the logarithmic column density distributions including $N$(H$_{2}$) are shown in Fig. \ref{fig:Nhisto} with the mean and standard variation of the column density noted in each panel (upper limits are not included). The logarithmic bin width is set to 0.5. Separate histograms of the core and non core populations are shown in the same panels. However, as discussed in Sect. \ref{sec:observations}, for the spectral line data we have short-spacing information, while we do not have this for the continuum data, hence we systematically underestimate the H$_{2}$ column density. 
	
	There are species with a distribution having a clear column density peak (e.g., H$_{2}$, $^{13}$CO, C$^{18}$O, SO, DCN, H$_{2}$CO, HNCO). Other species have a double-peaked distribution with a clear separation between the core and the remaining positions (e.g., OCS, HC$_{3}$N, CH$_{3}$OH, CH$_{3}$OH;$\varv_{t} = 1$, and CH$_{3}$CN). In these cases, the column density is enhanced by a factor of $\sim 10 - 100$ toward the core positions. There are not enough data points for the SO$_{2}$, HC$_{3}$N;$\varv_{7} = 1$, HC$_{3}$N;$\varv_{7} = 2$ to draw any conclusions about the distribution, however, they are detected mostly in the densest regions toward core positions. In general, high column densities are found toward core positions and low column densities are found toward the remaining positions.
	
	To account for the fact that toward the core positions the column density is expected to be higher in a higher density region, we show relative abundance $N$(X)/$N$(C$^{18}$O) histograms in Fig. \ref{fig:abundratiohisto} (upper limits are not included). Assuming that both species trace the same emission region, the relative abundances are independent of the absolute column density value, which differ from region to region and from core to core. Normalized to $N$(C$^{18}$O), most species have a single-peaked distribution. Exceptions are OCS, HNCO, HC$_{3}$N, CH$_{3}$OH, and CH$_{3}$CN with double-peaked distributions indicating that high temperature gas-phase chemistry has a big impact on these N-bearing molecules by increasing their abundances. However, in general there is still a clear difference between core positions (high abundance) and non core positions (low abundance). The fact that larger molecules have a clearer distinction between the core and non core positions, while for simpler species it is less obvious (e.g., $^{13}$CO and DCN), hints that the emission of COMs is associated with the cores while simple molecules are abundant in envelope as well. The difference of the core and the remaining positions indicates that the high densities and possible energetic processes around the protostars have a strong impact on the molecular abundances in the gas-phase (e.g. through protostellar outflows, shocks, disk accretion, and strong radiation from the protostars). Correlations of the derived column densities are discussed in Sect. \ref{sec:discusscorr}.

	Observed and \texttt{XCLASS} modeled spectra are shown in Fig. \ref{fig:spectrum} for all positions. The computed optical depth for all fitted lines as a function of rest-frequency is shown as well. Even though the sample was selected to be at a similar evolutionary stage (HMPO/HMC), the number of emission lines in the observed spectra vary from region to region, but also within a region. Typical hot core spectra are observed for the positions AFGL\,2591 1, CepA\,HW2 1, G75.78 1, W3\,H2O 1, and W3\,H2O 2. Many weak emission features are detected in the spectra at a $\sim 2 - $3$ \sigma_{\mathrm{line}}$ level originating from COMs. These COM emission features are difficult to fit as the transitions are weak, have similar upper energy levels $E_\mathrm{u}/k_\mathrm{B}$, and are blended at a spectral resolution of 3\,km\,s$^{-1}$, so they were excluded from this analysis. A detailed study of the integrated line emission (including line stacking of weak COM emission lines) will be presented in a future study (Gieser et al. in prep.). Fewer species are detected in spectra toward the non core positions. In contrast to the line-rich sources, there are several sources that show only a handful of emission lines mainly from CO-isotopologues, SO, and H$_{2}$CO even toward the continuum peak positions. Line-poor regions are G139.9091, G138.2957, S87\,IRS1, and S106. In the case of IRAS\,23033, core 1 has a line-poor spectrum, while core 2 and 3, which are embedded in a common envelope, significantly more emission lines are detected. These cores either could be at different evolutionary stages or are embedded in an inhomogeneous local radiation field. The former can be investigated by applying a chemical model to the observed column densities and by estimating the chemical ages of the regions, which we investigate in the next section.
	
	The mean and maximum line optical depth, $\tau_\mathrm{mean}^\mathrm{line}$ and $\tau_\mathrm{max}^\mathrm{line}$, computed for each fitted transition with \texttt{XCLASS} are summarized in Table \ref{tab:spectrallineproperties}. The $^{13}$CO $2-1$ transition has the highest mean optical depth with $\tau_\mathrm{mean}^\mathrm{line} = 1.5$, but for most other species and transitions, the mean line optical depth is $< 1$, so the column density and temperature determination should be reliable, especially when many transitions of a molecule are fitted simultaneously. The mean optical depth of the H$_{2}$CO $3_{0,3}-2_{0,2}$ transition is a factor of 4 higher than the remaining two transitions. When the $3_{0,3}-2_{0,2}$ transition is optically thick, temperature estimates depending on the line ratios are difficult to determine with the remaining two optically thin lines as they have similar upper energy levels (68\,K). In 23 out of the 118 spectra where H$_{2}$CO was detected and fitted with \texttt{XCLASS}, the calculated optical depth of the $3_{0,3}-2_{0,2}$ transition is $\tau^\mathrm{line} > 1$.

\section{Physical-chemical modeling of the cores}\label{sec:chemicalmodeling}

\begin{table}
\caption{\texttt{MUSCLE} input parameters.}
\label{tab:modelinput}
\centering
\begin{tabular}{ll}
\hline \hline
Parameter & Value \\
\hline
\textbf{Radiation field:}&\\
CR ionization rate $\zeta_{\mathrm{CR}}$ & 1.8($-16$)\,s$^{-1}$\, \tablefootmark{a} \\
Extinction $A_\mathrm{v}$ at $r_{\mathrm{out}}$ & 10$^{\mathrm{mag}}$\\
UV photodesorption yield &1($-$05)\,\tablefootmark{b,c}\\
\hline
\textbf{Grain properties:} &\\
Grain radius $r_{\mathrm{g}}$ &0.1\,$\mu$m\,\tablefootmark{d}\\
Dust density $\rho_{\mathrm{d}}$ & 3\,g\,cm$^{-3}$\,\tablefootmark{d}\\
Gas-to-dust mass ratio $\eta$ & 150\,\tablefootmark{e}\\
Surface diffusivity $E_{\mathrm{Diff}}$/$E_{\mathrm{Bind}}$ & 0.4\,\tablefootmark{f}\\
Mantle composition & olivine\,\tablefootmark{d}\\
\hline
\textbf{Initial Chemical Abundances} \\
\textbf{HMPO} model & best-fit IRDC stage\,\tablefootmark{g}\\
Inner radius $r_{\mathrm{in}}$ &$12\,700$\,au\\
Outer radius $r_{\mathrm{out}}$ &$0.5$\,pc\\
Temperature $T_{\mathrm{in}}$ at $r_{\mathrm{in}}$ &$11.3$\,K\\
Temperature power-law index $q$ & 0.0 (isothermal)\\
Density $n_{\mathrm{in}}$ at $r_{\mathrm{in}}$ & 1.4(5)\,cm$^{-3}$\\
Density power-law index $p$ & 1.5\\
Stage lifetime $\tau_\mathrm{IRDC}$ & $16\,500$\,years\\
\textbf{HMC} model & best-fit HMPO stage\,\tablefootmark{g}\\
Inner radius $r_{\mathrm{in}}$ &$103$\,au\\
Outer radius $r_{\mathrm{out}}$ &$0.5$\,pc\\
Temperature $T_{\mathrm{in}}$ at $r_{\mathrm{in}}$ &$75.8$\,K\\
Temperature power-law index $q$ & 0.4\\
Density $n_{\mathrm{in}}$ at $r_{\mathrm{in}}$ & 1.5(9)\,cm$^{-3}$\\
Density power-law index $p$ & 1.8\\
Stage lifetime $\tau_\mathrm{HMPO}$ & $32\,000$\,years\\
\textbf{UCH{\sc ii}} model & best-fit HMC stage\,\tablefootmark{g}\\
Inner radius $r_{\mathrm{in}}$ &$1\,140$\,au\\
Outer radius $r_{\mathrm{out}}$ &$0.5$\,pc\\
Temperature $T_{\mathrm{in}}$ at $r_{\mathrm{in}}$ &$162.9$\,K\\
Temperature power-law index $q$ & 0.4\\
Density $n_{\mathrm{in}}$ at $r_{\mathrm{in}}$ & 1.3(8)\,cm$^{-3}$\\
Density power-law index $p$ & 2.0\\
Stage lifetime $\tau_\mathrm{HMC}$ & $35\,000$\,years\\
\hline
\end{tabular}
\tablefoot{a(b) = a$\times$10$^{\mathrm{b}}$.\\
\tablefoottext{a}{\citet{Indriolo2015};}
\tablefoottext{b}{\citet{CruzDiaz2016};}
\tablefoottext{c}{\citet{Bertin2016};}
\tablefoottext{d}{\citet{Gerin2013};}
\tablefoottext{e}{\citet{Draine2011};}
\tablefoottext{f}{\citet{Cuppen2017};}
\tablefoottext{g}{Table A.4, A.5, and A.6 in \citet{Gerner2015}.}
}
\end{table}

\begin{figure*}[!htb]
\centering
\includegraphics[]{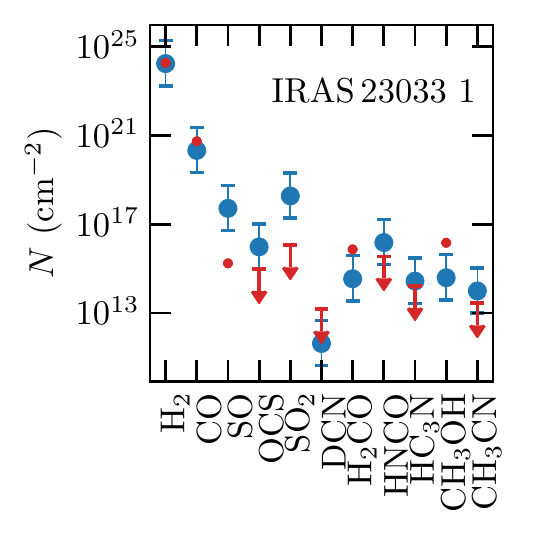}
\includegraphics[]{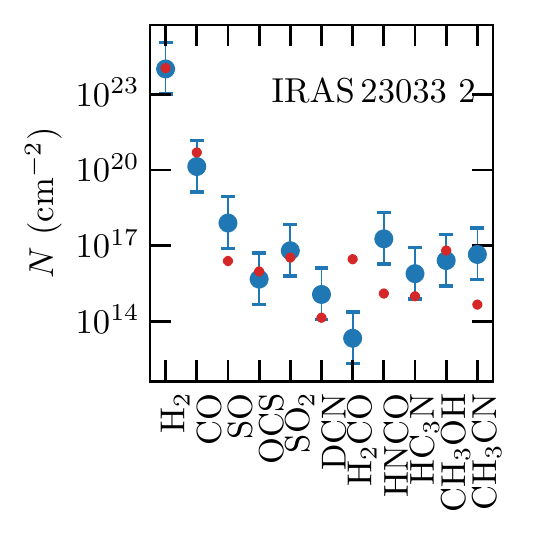}
\includegraphics[]{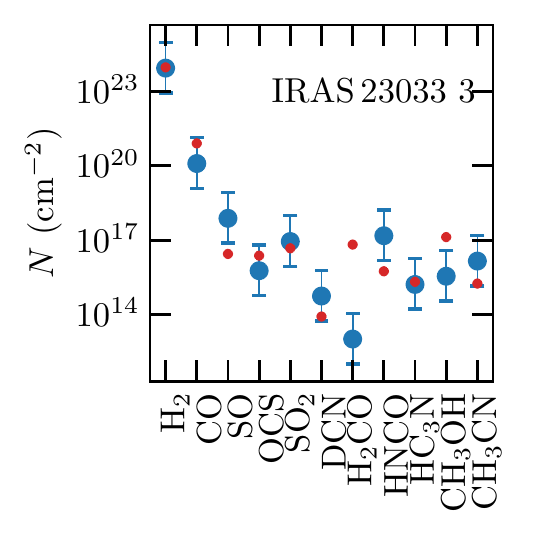}
\includegraphics[]{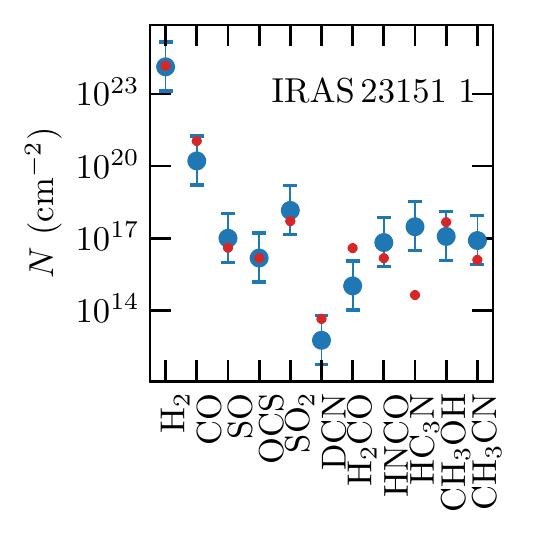}
\includegraphics[]{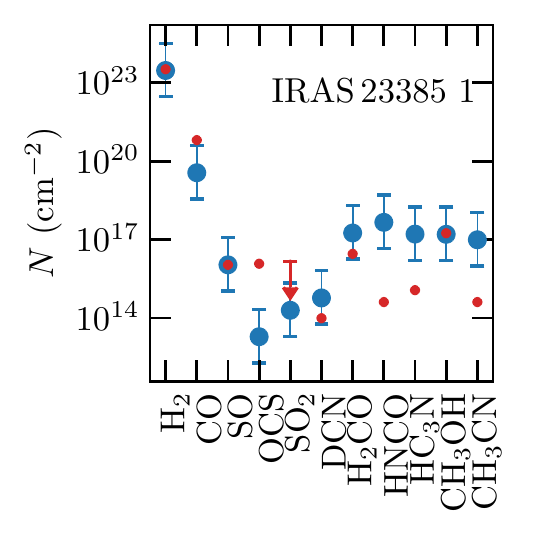}
\includegraphics[]{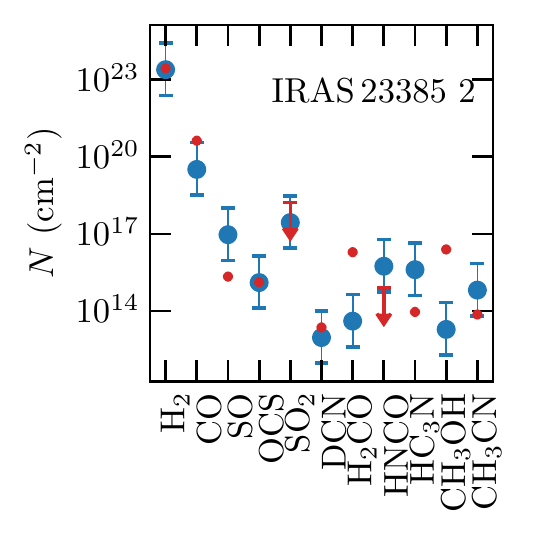}
\includegraphics[]{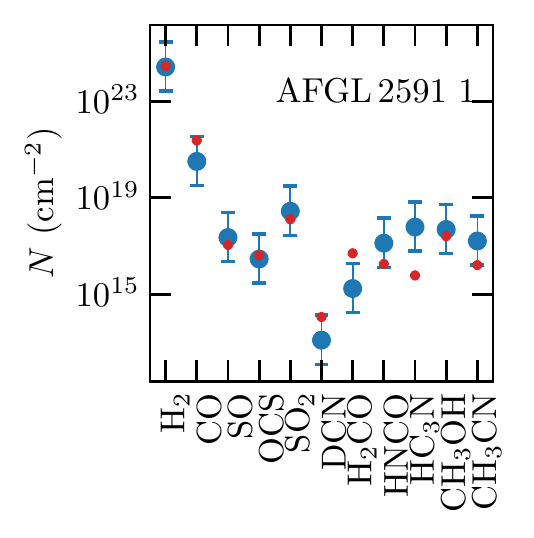}
\includegraphics[]{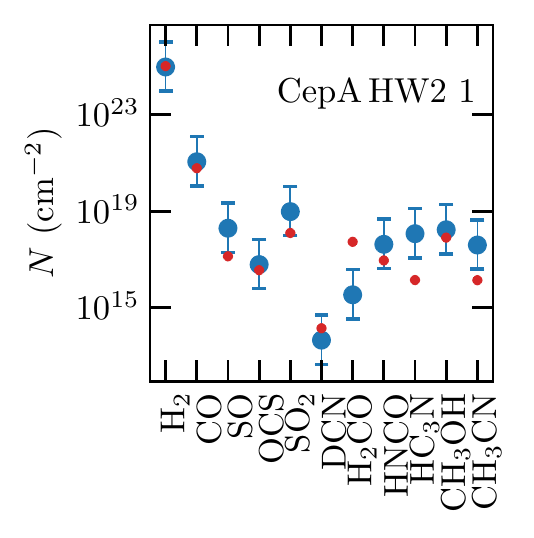}
\includegraphics[]{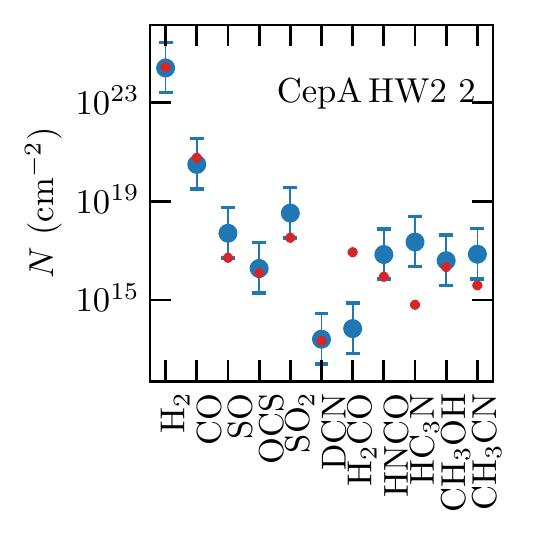}
\includegraphics[]{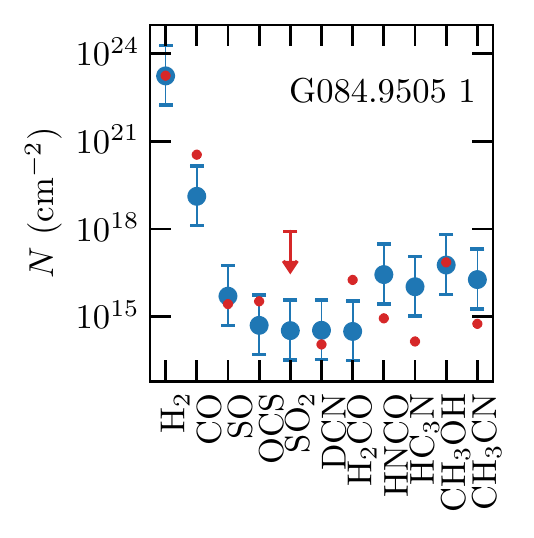}
\includegraphics[]{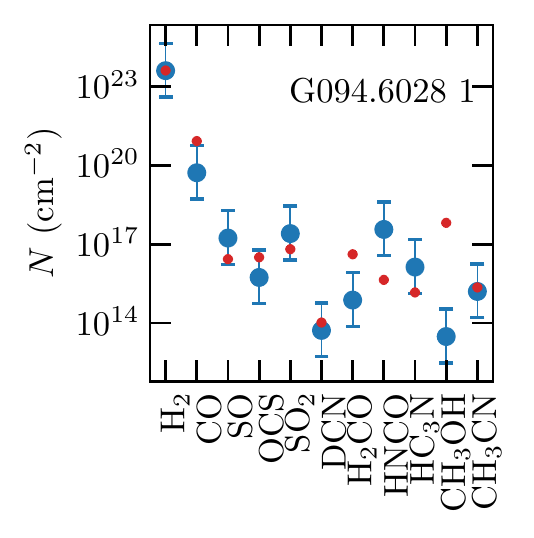}
\includegraphics[]{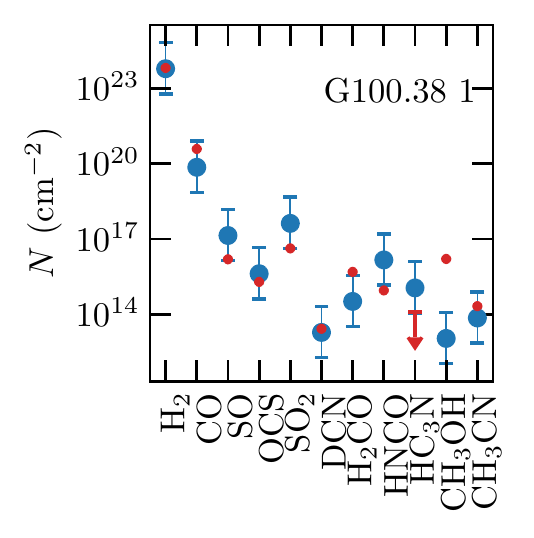}
\caption{Comparison of the observed and modeled column densities shown in red and blue, respectively. Upper limits are indicated by an arrow.}
\label{fig:model}
\end{figure*}

\begin{figure*}[!htb]
\ContinuedFloat
\captionsetup{list=off,format=cont}
\centering
\includegraphics[]{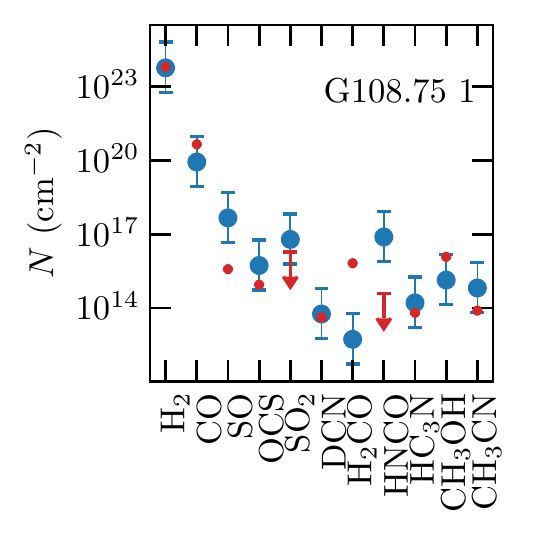}
\includegraphics[]{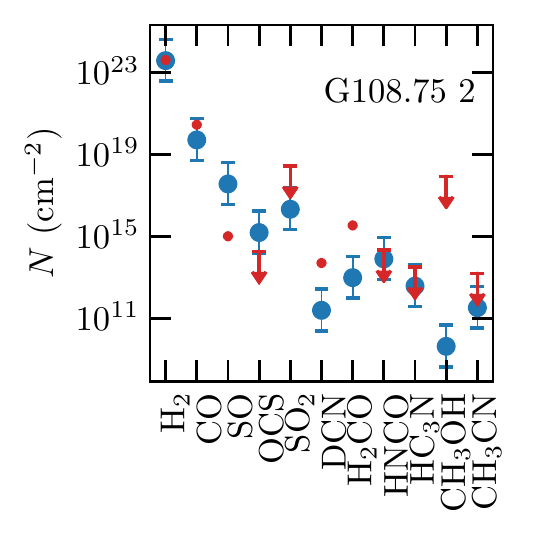}
\includegraphics[]{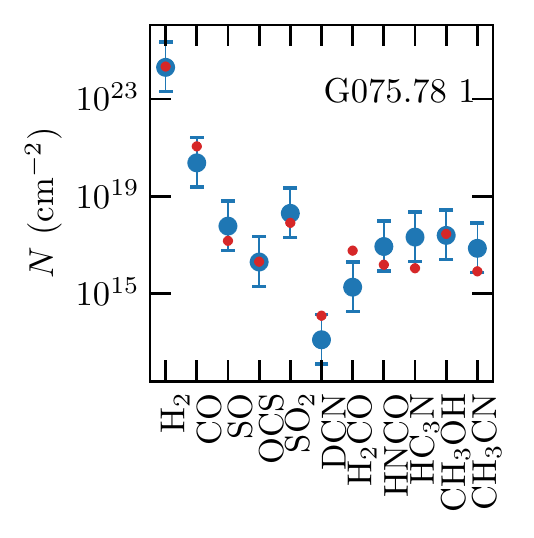}
\includegraphics[]{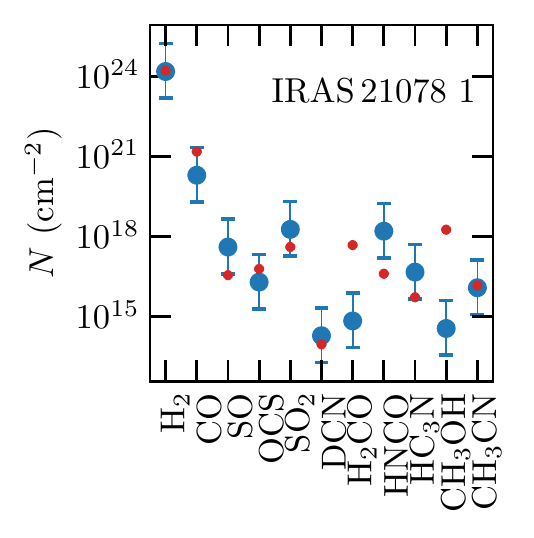}
\includegraphics[]{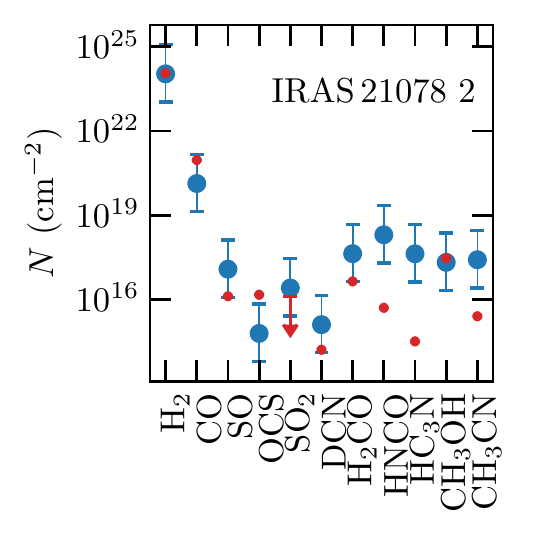}
\includegraphics[]{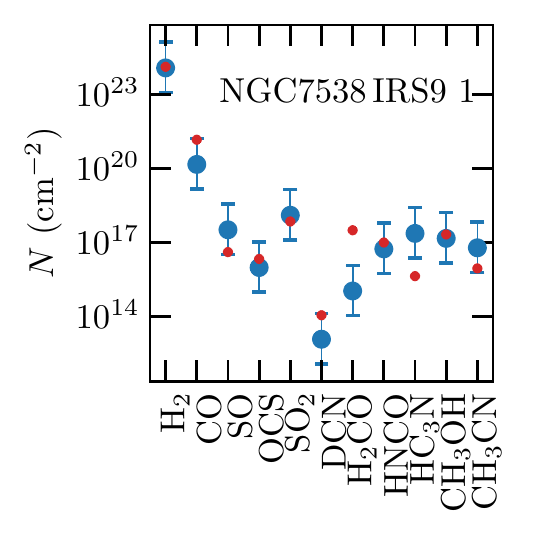}
\includegraphics[]{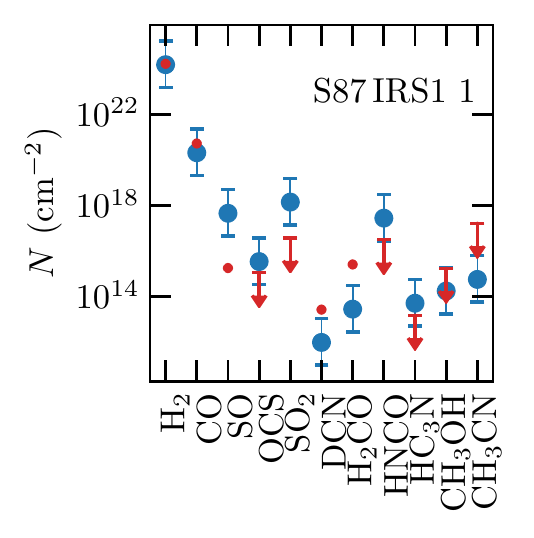}
\includegraphics[]{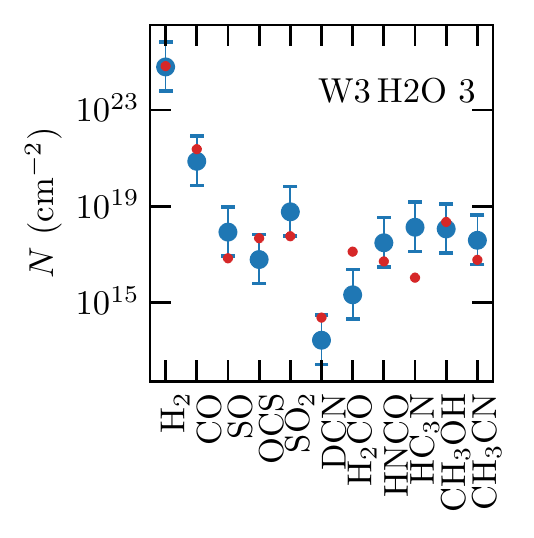}
\includegraphics[]{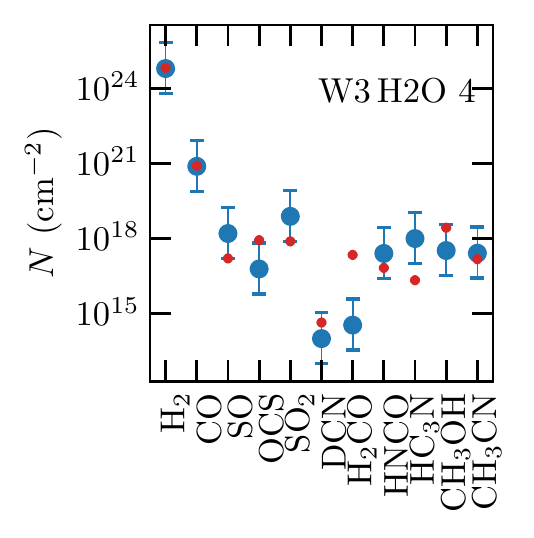}
\includegraphics[]{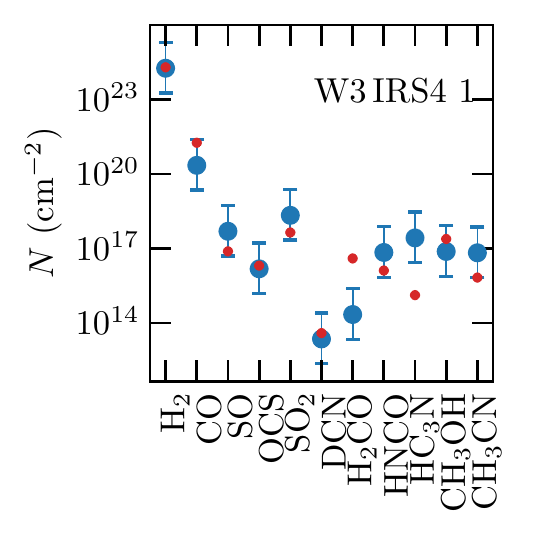}
\caption{Comparison of the observed and modeled column densities shown in red and blue, respectively. Upper limits are indicated by an arrow.}
\end{figure*}

	The continuum data of the CORE sample show a large diversity in fragmentation properties \citep{Beuther2018} and our analysis in Sect. \ref{sec:molecularcontent} showed that the composition of the molecular gas varies within as well as between the regions: some have a rich plethora of molecular lines, while others have line-poor spectra. This diversity of physical and chemical properties could be explained by a number of reasons. Magnetic fields and/or different initial density structures could explain the variety in fragmentation properties \citep{Beuther2018}. Different initial conditions in, e.g., the large-scale kinematics and mass distribution might also have an effect on the molecular abundances. To investigate if the observed variation of the physical and chemical properties of the cores may be due to the fact that the CORE regions have different ages, we model the chemical evolution of the 22 cores in the following.

\subsection{\texttt{MUSCLE} setup}\label{sec:MUSCLEsetup}

	A physical-chemical model is applied to the physical properties and molecular column densities of each core determined from the CORE 1\,mm observations in order to estimate the chemical ages. \texttt{MUSCLE} (MUlti Stage ChemicaL codE) was already successfully applied to the CORE pilot regions NGC7538\,S and NGC7538\,IRS1 \citep{Feng2016} and the CORE region AFGL\,2591 \citep{Gieser2019}. The model comprises spherically symmetric physical structures. The temperature and density profiles of the cores are described by power-laws up to the outer radius $r_{\mathrm{out}}$ with index $q$ and $p$, respectively, see Eq. \eqref{eq:temperatureprofile} and \eqref{eq:densityprofile}. At an inner radius $r_{\mathrm{in}}$ and further in, the density and temperature reach a constant value. We adopt 40 logarithmic grid points for the radial profiles.
	
	On top of this static physical structure, the time-dependent gas-grain chemical network ALCHEMIC \citep{Semenov2010} computes the abundances of hundreds of atomic and molecular species using thousands of reactions. A detailed description of \texttt{MUSCLE} can be found in \citet{Gerner2014,Gerner2015}. We adopt most of the model parameters from the AFGL\,2591 case-study described in \citet{Gieser2019} which yielded a good estimate of the chemical age of this hot core compared to literature estimates. A summary of the input parameters is listed in Table \ref{tab:modelinput}. In contrast to the AFGL\,2591 case-study, we use a higher value for the cosmic ionization rate $\zeta_{\mathrm{CR}}$ based on a study of multiple HMSFRs by \citet{Indriolo2015}. These authors find that $\zeta_{\mathrm{CR}}$ is constant at a Galactic radius $> 5$\,kpc and with all the CORE regions at Galactic distances $> 7$\,kpc (Table \ref{tab:sample}), we use a constant value of 1.8$\times$10$^{-16}$\,s$^{-1}$. By setting the extinction at $r_{\mathrm{out}}$ to $A_\mathrm{v} = 10^{\mathrm{mag}}$, the core is shielded from the interstellar ultraviolet radiation field.
	
	For each of the 22 cores, we run a physical-chemical model with \texttt{MUSCLE}. The input are the H$_{2}$ column density (Sect. \ref{sec:molecularhydrogen}) and all molecular column densities derived with \texttt{XCLASS} (Sect. \ref{sec:XCLASSfitting}). The CO column density $N$(CO) is calculated from $N$(C$^{18}$O) as C$^{18}$O is less optically thick than $^{13}$CO (Table \ref{tab:spectrallineproperties}) and hence more reliably fitted in \texttt{XCLASS}. For each region, we calculate the $^{16}$O/$^{18}$O isotopic ratio according to \citet{Wilson1994}: $^{16}$O/$^{18}$O $\approx 58.8 d_\mathrm{gal} + 37.1$. The $^{16}$O/$^{18}$O ratio is listed in Table \ref{tab:sample} for each region. For HC$_{3}$N and CH$_{3}$OH we compute the mean column density of the rotational ground state and vibrationally/torsionally excited states for the \texttt{MUSCLE} input. Molecular column densities, for which only upper limits could be determined, are also set as upper limits in \texttt{MUSCLE}. We set the temperature structure of the model core to the observed temperature profile (Sect. \ref{sec:temperaturestructure}) and use the density power-law index derived from the continuum visibility analysis (Sect. \ref{sec:densitystructure}).
	
	Two undetermined model parameters remain. First, we do not know how evolved the cores are, described by the chemical age $\tau_{\mathrm{chem}}$, and second, what the initial chemical composition of the parental molecular cloud/clump was. Due to the fact that the CORE regions are far more evolved than typical cold IRDCs and the physical structure of each model stage is static, one has to define a sensible initial condition for the chemical composition. The initial conditions we apply are based on a study of 69 HMSFRs using single-dish observations \citep{Gerner2014,Gerner2015}. These HMSFRs were classified according to their evolutionary stage (IRDCs, HMPOs, HMCs, and UCH{\sc ii} regions) and a template was created from the average column densities for each evolutionary stage. The four template stages were modeled using \texttt{MUSCLE} to create average abundances for all molecular and atomic species and to estimate a mean chemical age of each evolutionary stage. The properties of their template IRDC, HMPO, and HMC model are summarized in Table \ref{tab:modelinput}. Following the convention by \citet{Gerner2014,Gerner2015}, the chemical age $\tau_{\mathrm{chem}}$ is 0\,yrs when the gas density reaches 10$^{4}$\,cm$^{-3}$.
	
	Based on the temperature profiles of the 22 cores, we can assume that they lie somewhere between the HMPO and early UCH{\sc ii} stage, since the average temperatures around cores are too high to be classified as IRDCs ($T \sim 20$\,K). There are a few known UCH{\sc ii} regions with strong free-free emission at cm wavelengths resolved in the CORE data. In W3\,H2O, the Western part (around position 1 and 2 in Fig. \ref{fig:continuum}) is the UCH{\sc ii} region W3\,OH. In W3\,IRS4, the Southern ring-like structure is an UCH{\sc ii} region as well \citep{Mottram2020}. However, for the W3\,OH UCH{\sc ii} region we do not find a clear radial decreasing temperature profile and toward the W3\,IRS4 UCH{\sc ii} region no H$_{2}$CO or CH$_{3}$CN line emission is detected at a 10$\sigma_{\mathrm{line,map}}$ level to estimate the kinetic temperature. The S106 region is a UCH{\sc ii} as well for which we do not detect neither H$_{2}$CO nor CH$_{3}$CN emission around the compact core. This already suggests that toward this later stage the molecular richness in these regions is decreased.
	
	To test which initial conditions (see Table \ref{tab:modelinput}) fit best to the observed molecular column densities, we model each of the 22 cores with initial abundances after an initial IRDC phase (referred to as the HMPO model), after an initial HMPO phase (referred to as the HMC model), and after an initial HMC phase (referred to as the UCH{\sc ii} model). While most cores are unlikely to have formed a strong UCH{\sc ii} region yet, we include the UCH{\sc ii} model, as the observations in \citet{Gerner2014,Gerner2015} have large beam sizes (11$''$ and 29$''$), and UCH{\sc ii} regions may have contamination from less evolved line-rich objects, which are blended into the single-pointing spectra. It is not our aim to classify the cores into these evolutionary stages, but find sensible initial chemical conditions as an input for our physical-chemical modeling. For example, an evolved HMC, that is more evolved than the template HMC from \citet{Gerner2015}, will be described best by the UCH{\sc ii} model in our nomenclature.
	
	For each model, the chemical evolution $\tau_\mathrm{model}$ runs up to 100\,000\,years in 100 logarithmic time steps. In each time step, the computed radial abundance profiles are converted into beam-convolved column densities with the beam size fixed to the mean synthesized beam of the observations. The best-fit model is determined by a minimum $\chi^2$ analysis by comparing the modeled and observed column densities in each time step and for all three adopted initial condition models. Applying this physical-chemical model allows us to estimate the chemical age.
	
\subsection{Chemical ages}\label{sec:MUSCLEresults}

	The best-fit chemical age $\tau_\mathrm{chem}$, $\chi^{2}$, and percentage of well modeled molecules are shown in Table \ref{tab:MUSCLE_results} for each initial abundance model and core. \citet{Gerner2014} estimate that chemical ages are uncertain by a factor of $2-3$ and that the modeled column densities are uncertain by a factor of $10$. But this depends on the number of modeled molecules, but also cores embedded in complex dynamic environments are harder to fit with our model. The chemical age $\tau_\mathrm{chem}$ is the sum of the time of the initial abundance model and $\tau_\mathrm{model}$, so for the HMPO model: $\tau_\mathrm{chem} = \tau_\mathrm{IRDC} + \tau_\mathrm{model}$, for the HMC model: $\tau_\mathrm{chem} = \tau_\mathrm{IRDC} + \tau_\mathrm{HMPO} + \tau_\mathrm{model}$, and for the UCH{\sc ii} model: $\tau_\mathrm{chem} = \tau_\mathrm{IRDC} + \tau_\mathrm{HMPO}+ \tau_\mathrm{HMC} + \tau_\mathrm{model}$. 
	
	For some cores, multiple initial condition models have a similarly low $\chi^{2}$ (e.g., the HMPO and UCH{\sc ii} model for core 1 in IRAS\,23033). In these cases, we cannot constrain the chemical age well. Comparing the lowest $\chi^{2}$ model with the remaining initial condition models, if the $\chi^{2}$ difference is less than 5\%, only chemical age ranges spanning over these models are further considered. Table \ref{tab:phys_struc} shows the chemical age $\tau_\mathrm{chem}$ for models with a clear lowest $\chi^{2}$ initial condition model or a time range in chemical age for cores with multiple best-fit initial condition models. The estimated chemical timescales $\tau_\mathrm{chem}$ of the cores vary between $\sim$20\,000$-$100\,000\,yrs within the regions of the CORE sample with a mean of $\sim$60\,000\,yrs. The youngest core being G084.9505 1 and the oldest core being G108.75 2. 
	
	A comparison of the best-fit modeled and observed column densities for all cores is shown in Fig. \ref{fig:model}. For most cores, the model underestimates the H$_{2}$CO and CH$_{3}$OH column densities compared to the observed values. This can be explained by the fact that the quasi-static model does not sufficiently take into account the warm-up stage from $30-80$\,K where surface chemistry on the dust grains plays an important role and where these two molecules are formed by subsequent hydrogenation of CO. These discrepancies between modeled and observed H$_{2}$CO and CH$_{3}$OH column densities have already been noticed by \citet{Gerner2014} in their template HMPO stage modeled with \texttt{MUSCLE}. They explain that this is due to the fact that the formation route of H$_{2}$CO consists of grain-surface as well as gas-phase chemistry which are both time-dependent and not correctly implemented in the chemical models. This results in the over and underproduction of these species, which is also the case in our modeling results shown in Fig. \ref{fig:model}.
	
	 Large discrepancies between the modeled and observed column density exist also for the SO molecule for which the model overproduces the SO column density by a factor $>10$ for many cores (e.g., IRAS\,23033 core 1, 2, and 3). This overproduction in SO is seen in all three initial condition models, but in most cases, other modeled S-bearing species (OCS and SO$_{2}$) are modeled well. The applied initial chemical conditions based on the \citet{Gerner2015} models also included S-bearing species (SO, CS, and OCS). Their initial IRDC stage model started with elemental abundances and only H$_{2}$ in molecular form taken from the low metals set of \citet{Lee1998}. But in order to fit the IRDC phase accurately, \citet{Gerner2015} had to increase the initial elemental S abundance from $8 \times 10^{-8}$ to $8 \times 10^{-7}$ (w.r.t. H). However, an overproduction of SO is also seen in their best-fit HMPO, HMC and UCH{\sc ii} models. This might be connected to a poorly understood chemistry of the reactive SO molecule, as also in their models the remaining S-bearing species can be reproduced properly. In addition, as only one SO transition is covered in our spectral setup, which can be typically optically thick (Table \ref{tab:spectrallineproperties} and Fig. \ref{fig:spectrum}), we may underestimate the observed SO column density. This might partially explain the differences between the modeled and observed SO column density.
	
	With multiple cores resolved within a region, it is possible to study how the chemical timescale $\tau_\mathrm{chem}$ varies across small spatial scales. In the IRAS\,23033 region, core 1 seems to be more evolved ($\sim 30\,000-100\,000$\,yrs), even though the spectrum is line-poor (see Fig. \ref{fig:spectrum}) compared to the spectra of core 2 and 3 which are embedded in a common envelope (see Fig. \ref{fig:continuum}) and for which we estimate similar chemical timescales of $\sim$20\,000\,yrs. Core 1 and 2 in CepA\,HW2 have a chemical age of $\sim$80\,000\,yrs and $\sim 20\,000 - 90\,000$\,yrs, respectively. The cores are very close ($\sim$2\,300\,au), but within our sensitivity limit, these cores are not embedded in a common envelope, but have very steep density profiles ($p \gtrsim 2$). In IRAS\,21078, core 1 and 2 have a chemical age of $\sim 60\,000 - 90\,000$\,yrs and $\sim$50\,000\,yrs, respectively, suggesting a small age gradient. The cores are embedded within a common envelope and have small projected separations. Core 3 and 4 in W3\,H2O have a chemical age of $\sim$90\,000\,yrs and $\sim 20\,000 - 90\,000$\,yrs, respectively. In the IRAS\,23385 region, core 1 is younger ($\sim$50\,000\,yrs), while core 2 is estimated to be much older ($\sim$100\,000\,yrs). In G108.75 a large difference between the chemical ages of core 1 ($\sim$20\,000\,yrs) and core 2 ($\sim$110\,000\,yrs) is estimated. The cores have a separation of $\sim$20\,000\,au, but have the same systemic velocity (Table \ref{tab:positions}). A strong external radiation field or complex dynamics could be the reason for this large chemical age difference.
	
	One of the limitations of \texttt{MUSCLE} is that the physical structure (radial temperature and density profiles) is static within each evolutionary stage (IRDC, HMPO, HMC and UCH{\sc ii}). In reality these properties do change on timescales smaller than the chemical timescales derived here, and also the dynamics (e.g., gas inflow) are important factors to consider. Currently, 3D time-dependent physical models in combination with a full chemical network are computationally expensive. Therefore, we use the approach of our quasi-static physical model by considering the four different evolutionary stages. More sophisticated physical-chemical models in the future are required to include 3D gas dynamics and the evolution of the density and temperature structure. In addition, a larger number of molecular column densities would better constrain the model parameter space.

\section{Discussion}\label{sec:discussion}

\subsection{Physical structure of high-mass star-forming cores}\label{sec:discussphys}
	
\begin{figure}
\centering
\includegraphics[]{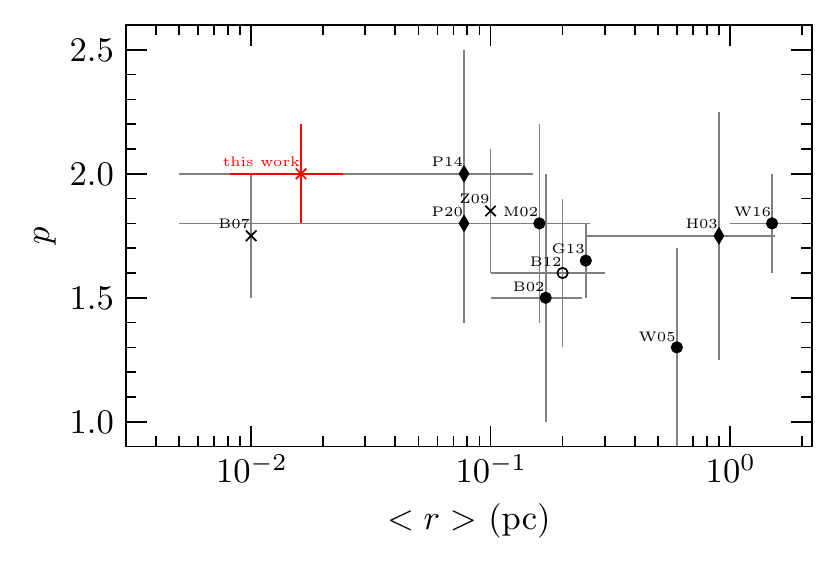}
\caption{Literature comparison of the density index $p$ at different core/clump sizes $<r>$. Studies based on interferometric observations are marked by a ``$\times$'', (sub)mm single-dish observations by a ``$\bullet$'', multi-wavelength observations by a ``$\blacklozenge$'', and mid-infrared observations by a ``$\circ$''. Notes. M02: \citet{Mueller2002}, B02: \citet{Beuther2002}, H03: \citet{Hatchell2003}, W05: \citet{Williams2005}, B07: \citet{Beuther2007B}, Z09: \citet{Zhang2009}, B12: \citet{Butler2012}, G13: \citet{Giannetti2013}, P14: \citet{Palau2014}, W16: \citet{Wyrowski2016}, P20: \citet{Palau2020}.}
\label{fig:density_lit}
\end{figure}

\begin{figure}
\centering
\includegraphics[]{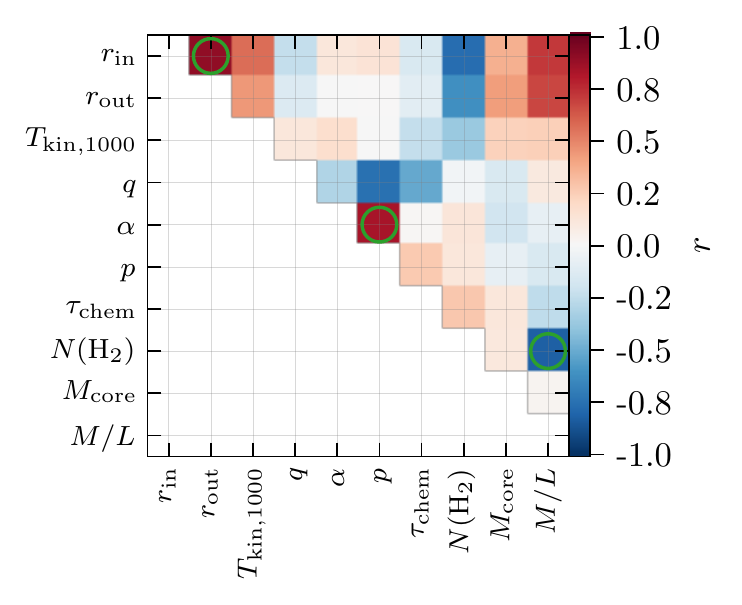}
\caption{Spearman correlation coefficient $r$ for the derived physical and chemical core parameters listed in Table \ref{tab:phys_struc}. Values higher than 0.8 are marked by a green circle.}
\label{fig:correlationcoefficient_core}
\end{figure}

\begin{figure}
\centering
\includegraphics[]{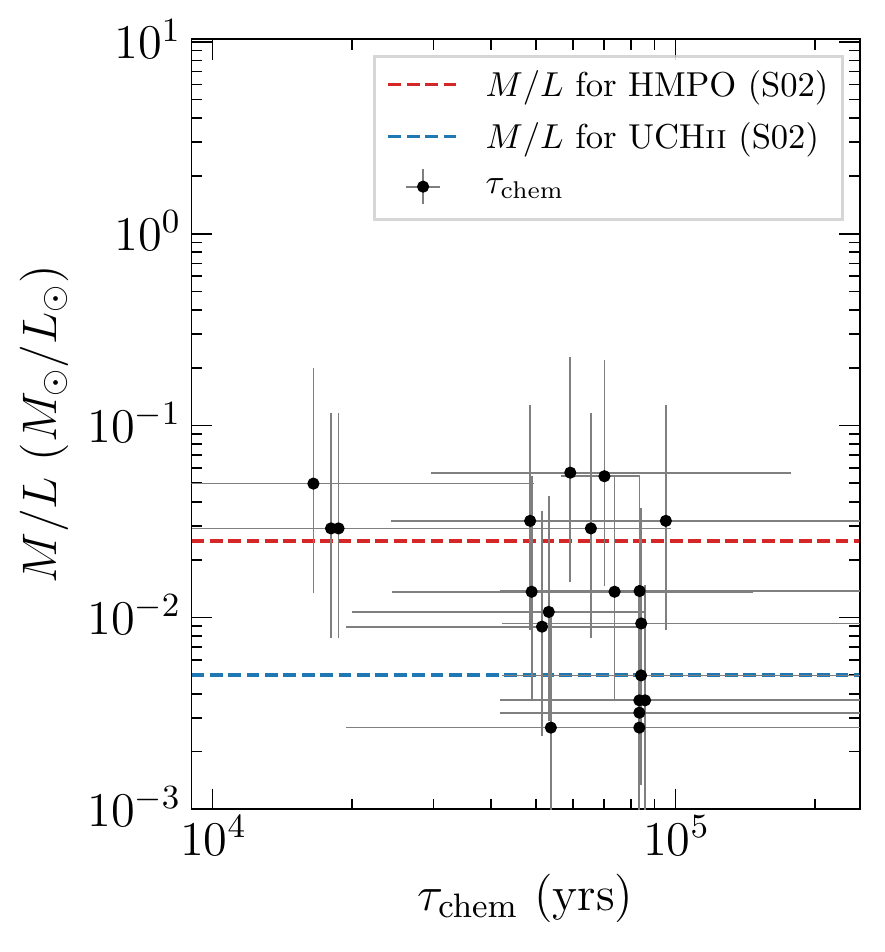}
\caption{The $M$/$L$ ratio and chemical ages $\tau_\mathrm{chem}$. The mass $M$ and luminosity $L$ for each region are taken from \citet{Beuther2018}. The horizontal dashed lines correspond to average $M$/$L$ ratios for HMPOs (red) and UCH{\sc ii} regions (blue) taken from \citet{Sridharan2002}.}
\label{fig:agevsluminositymass}
\end{figure}
	
	Various methods were applied in the literature to observationally derive the density profiles of envelopes in HMSFRs \citep[e.g., summarized in Table 6 in][]{Gieser2019}. Some of these studies are based on observations with single-dish telescopes with beam sizes $> 10''$ tracing the clump-scale envelope, while interferometric observations trace the core-scale envelope. We select studies in the literature for which the density structure was determined for a sample of cores or clumps and also extract the typical sizes $<r>$ from their studies. The results in comparison with our study are shown in Fig. \ref{fig:density_lit}. At scales of 1\,pc down to 0.01\,pc, it seems that the density index $p$ lies between 1.7 and 2.0, which is close to the values inferred in low-mass star-forming regions \citep{Motte2001}. To investigate this further, we have observed the CORE regions with the NIKA2 instrument at the IRAM 30m telescope and an analysis of the density structure at clump-scales will follow (Beuther et al. in prep.).

	The observationally-derived density and temperature profiles ($q = 0.4 \pm 0.1$ and $p = 2.0 \pm 0.2$) are in agreement with theoretical studies of HMSF, but the physics of how massive stars form is not fully understood yet. Currently theoretical models propose the formation of high-mass stars through: a monolithic collapse of turbulent cores \citep{McKee2002,McKee2003}; protostellar collisions and coalescence in dense clusters \citep{Bonnell1998,Bonnell2002}; or competitive accretion in clusters \citep{Bonnell2001,Smith2009,Hartmann2012,Murray2012}. The density and temperature structure are important parameters of the initial cloud and proceeding clump and core scales. For example, early star formation models by \citet{Shu1977} and \citet{Shu1987} that model the gravitational collapse of an isothermal sphere find that $p = 2$ in the outer envelope and $p = 1.5$ in the inner region where the gas is free-falling onto the central region. \citet{McLaughlin1996,McLaughlin1997} used a logatropic equation of state and a non-isothermal sphere and find that at an initial density profile of $p = 1$, the profile steepens to $p = 1.5$ after the collapse. In the turbulent core model by \citet{McKee2002,McKee2003} the authors assume $p = 1.5$ based on observational constraints. In \citet{Bonnell1998} the density profile in the outer region has the form $p = 2$ and a shallower, near-uniform profile in the central region. \citet{Murray2012} explore their models by varying $p$ from 0 (uniform), 1, and 2 (isothermal). 
	
	The density structure is important for the physical and chemical evolution of HMSFRs. It is therefore important to quantify the initial density profile on cloud to clump and cores scales and how it changes with time. While theoretical models usually do not predict, but rather assume a given density profile, observations of HMSFRs on different scales can help to narrow down the parameter space (see Fig. \ref{fig:density_lit}). Hydrodynamic simulations reported by \citet{Chen2020} investigate how changes of $p$ in giant molecular clouds with an initial radius of 20\,pc affect massive star cluster formation. They find that for steep density profiles, $p = 2$, there is a centrally-concentrated cluster, while for shallower profiles hierarchical fragmentation occurs. Hydrodynamic simulations by \citet{Girichidis2011} show that massive protostars form only in clouds with a density index of $p = 1.5$ or $p = 2.0$, while for uniform or Bonnor-Ebert-like profiles a large fraction of low-mass stars form. They found that turbulence and the initial density profile are important aspects for the evolution of the cloud and the formation of clusters.
	
	We study correlations of all core properties shown in Table \ref{tab:phys_struc} using the Spearman correlation coefficient $r$. This statistical tool can be used to check if two data sets have a positive correlation ($r = 1$), negative correlation ($r = -1$), or no correlation ($r = 0$). We define that a high correlation exists if $r > 0.8$. For example, \citet{Feng2020} finds a negative correlation of the H$_{2}$ column density and dust temperature for cold high-mass clumps using the Spearman correlation coefficient $r$. A big advantage compared to the Pearson correlation coefficient is that linear, as well as nonlinear correlations are considered in the calculation of $r$. In addition, we add the $M$/$L$ ratio of the region listed in Table 1 in \citet{Beuther2018} as a parameter for each core. However, the interpretation is difficult as multiple cores within a region have the same $M$/$L$ ratio. A mean chemical age is used in the computation of $r$ for cores for which only a time range can be estimated (Table \ref{tab:phys_struc}).
	
	The results for the correlation coefficient $r$ are shown in Fig. \ref{fig:correlationcoefficient_core}, where all pairs with a correlation $> 0.8$ are highlighted. Unfortunately, a small sample of only 22 cores does not allow us to find many strong correlations. Observations of many HMSFRs at core-scales are required to study these relations in a better statistical way, which will be possible, e.g., with the ALMAGAL survey, an ongoing ALMA large program observing more than 1\,000 HMSFRs. A high correlation is found between the inner and outer radius, which is due to the fact that we are resolution limited and the regions are located at different distances. The correlation between $p$ and $\alpha$ is due to Eq. \eqref{eq:uvanalysis}. However, we find a strong negative correlation between the $M$/$L$ ratio and the H$_{2}$ column density of the cores, so more evolved cores have a higher beam-convolved H$_{2}$ column density. The $M$/$L$ ratio, proposed to be a good tracer of the evolutionary stage, will be investigated in the following section.
	
\subsection{M/L ratio as a tracer of evolutionary trend}\label{sec:discussML}

	\citet{Sridharan2002} found that the distance-independent ratio $M$/$L$ of UCH{\sc ii} regions is lower ($\sim$0.005) than the ratio of HMPOs ($\sim$0.025), as UCH{\sc ii} regions are more evolved and thus more luminous. This has also been confirmed by observations of large samples of HMSFRs, e.g., by \citet{Molinari2008, Molinari2010, Maud2015, Molinari2016, Urquhart2018, Molinari2019}. In Fig. \ref{fig:agevsluminositymass} we plot the region-average $M$/$L$ ratio taken from Table 1 in \citet{Beuther2018} against the estimated chemical ages $\tau_\mathrm{chem}$. The chemical timescale and a factor 2 uncertainty is shown for cores with a clear best-fit model. For cores with estimated chemical age ranges, the mean chemical age are shown with error bars spanning over the time range. The luminosities $L$ have uncertainties on the order of $\sim$30\% \citep{Mottram2011b} and the masses $M$ calculated in \citet{Beuther2018} are expected to be uncertain within a factor of 3. We also show the average $M$/$L$ ratio for HMPOs and UCH{\sc ii} regions derived by \citet{Sridharan2002}. We corrected the $M$/$L$ ratios of \citet{Sridharan2002} by a factor of 0.5, as $M$ was taken from \citet{Beuther2002} where the reported values for $M$ were lower by a factor of 2 \citep{Beuther2005}. The cores G108.75 1 and 2 are excluded here, because no consistent continuum data is available to reliably derive the mass \citep{Beuther2018}. There is a tendency such that the older cores have a lower $M$/$L$ ratio. However, as the $M$/$L$ ratios are determined on much larger scales and we resolve multiple cores within each region, it is difficult to compare these properties.
	
	Estimating the chemical ages with \texttt{MUSCLE} shows that a line-rich spectrum does not have to imply that a core is more evolved compared to a core with a line-poor spectrum. For example, the line-poor core 1 in IRAS\,23033 is estimated to be older than the line-richer cores 2 and 3. When a hot core evolves to become an UCH{\sc ii} region, the strong protostellar radiation destroys molecules. Hence line-poor spectra can be found at early and at late evolutionary stages. Observations of more species would be helpful to better constrain the chemical model.
	
	We are not able to derive radial temperature profiles from the temperature maps of G138.2957, G139.9091, and S106 (Fig. \ref{fig:temperature_maps}). In G139.9091 and S106 there is no emission at the adopted $10\sigma_{\mathrm{line,map}}$ level. There is diffuse emission of H$_{2}$CO in G138.2957, but it is too diffuse to derive a radial temperature profile. The spectra of these regions are line-poor and only simple species are detected ($^{13}$CO, C$^{18}$O, SO, H$_{2}$CO, see Fig. \ref{fig:spectrum}). Based on the shape of the 1.37\,mm continuum emission, G139.9091 and S106 have isolated cores with no significant envelope emission, while G138.2957 has diffuse dust emission (Fig. \ref{fig:continuum}). Therefore one may conclude that G139.9091 and S106 are already more evolved and probably have $\tau_\mathrm{chem} > 100\,000$\,yrs. S106 is a well studied bipolar H{\sc ii} region \citep[e.g.,][]{Roberts1995, Schneider2007}. G139.9091 is associated with a H{\sc ii} region as well \citep[e.g.,][]{Kurtz1994,Purser2017, Obonyo2019}. G138.2957 could be in a very young strongly depleted phase with $\tau_\mathrm{chem} < 20\,000$\,yrs or could be an evolved region and the diffuse emission is due to the disruption by the protostellar radiation. With observations of G138.2957 at 5.8\,cm and 20\,cm, a core component with an associated synchrotron jet is seen toward the location of the 1.37\,mm continuum peak and the position has an associated infrared source \citep{Obonyo2019}. This suggests that G138.2957 is an evolved embedded cluster and not a young region. Deep observations at radio wavelengths ($\sim$5$-$50\,GHz) would also provide information on the presence of UCH{\sc ii} regions within the CORE sample. We find that for the known H{\sc ii} regions, we only detect simple species such as CO isotopologues, H$_{2}$CO suggesting that a large fraction of the larger molecules are already destroyed in this stage.

\subsection{Correlations between chemical species}\label{sec:discusscorr}

\begin{figure}
\centering
\includegraphics[]{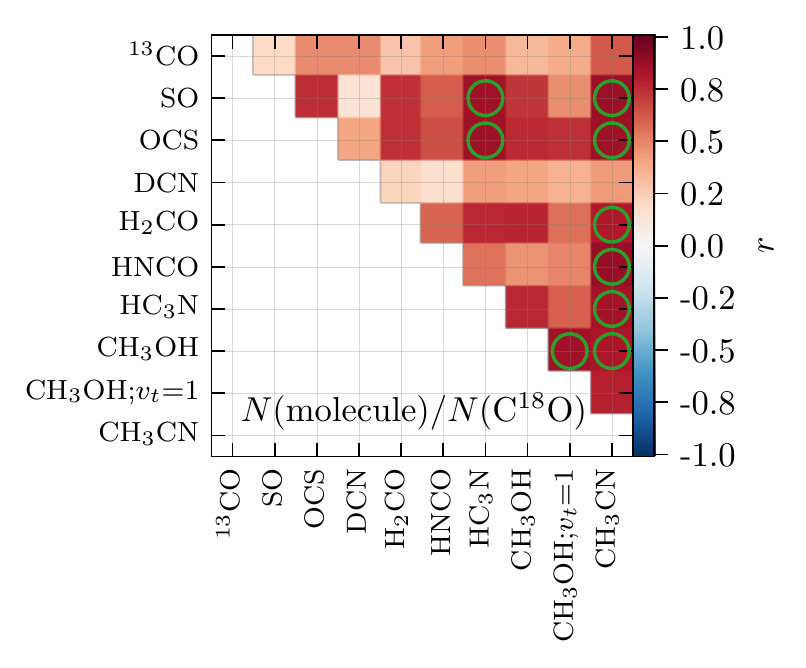}
\caption{Spearman correlation coefficient $r$ for pairs of molecular column densities relative to N(C$^{18}$O). Values higher than 0.8 are marked by a green circle.}
\label{fig:correlationcoefficient_N}
\end{figure}

	In this section, we aim to analyze which molecules show correlations by chemical links or temperature effects. We compute the Spearman correlation coefficient $r$ for the molecular column density pair combinations relative to N(C$^{18}$O). SO$_{2}$, HC$_{3}$N;$\varv_{7} = 1$, and HC$_{3}$N;$\varv_{7} = 2$ are excluded from this analysis, as there are less than 20 column density data points. Ideally, we would compare the correlations of relative abundance pairs relative to N(H$_{2}$), but due to the missing flux issue, we use N(C$^{18}$O) which is also detected toward all positions and is optically thin (Table \ref{tab:spectrallineproperties}). Comparing the correlation between column density pairs, one faces the issue that toward higher densities the column density is also higher as discussed in Sect. \ref{sec:XCLASSfitting}. The core and non core positions were equally considered in the calculation of $r$ and the results for all column density pairs are shown in Fig. \ref{fig:correlationcoefficient_N}.
	
	In general, all pairs show a positive correlation, $r > 0$. High correlation coefficients ($r > 0.8$) are found between pairs of the following molecules: HC$_{3}$N--SO, HC$_{3}$N--OCS, CH$_{3}$CN--SO, CH$_{3}$CN--OCS, CH$_{3}$CN--H$_{2}$CO, CH$_{3}$CN--HNCO, CH$_{3}$CN--HC$_{3}$N, CH$_{3}$OH--CH$_{3}$OH;$\varv_{t}=1$, CH$_{3}$CN--CH$_{3}$OH. The lowest correlations are found for $^{13}$CO and DCN where they show no correlation with any species. In the case of $^{13}$CO this is most likely due to a high optical depth (Table \ref{tab:spectrallineproperties}), so the column density cannot be reliably derived. The case of DCN is more puzzling, but a more detailed study of the deuteration would be required. Unfortunately, the CORE spectral setup covers only the DCN 3$-$2 line, so follow-up observations of deuterated species are necessary to study how deuterium chemistry behaves on such scales.
	
	Methyl cyanide (CH$_{3}$CN) shows a strong positive correlation with most other species, including S-bearing and N-bearing species. A correlation of HC$_{3}$N-CH$_{3}$CN can be explained by gas-phase N-chemistry in the envelope gas \citep{Bergner2017}. We find that OCS is correlated with other dense gas tracers (HC$_{3}$N and CH$_{3}$CN). A correlation exists between CH$_{3}$OH and CH$_{3}$CN even though, there is no chemical link between methanol and methyl cyanide. Such a correlation has also been found toward low-mass star-forming regions \citep{Bergner2017,Belloche2020} and toward the massive star-forming region G10.6$-$0.4 \citep{Law2021}. \citet{Belloche2020} argued that this is a temperature effect caused by chemically unrelated species being evaporated from icy grain mantles by energetic processes. Additional high angular resolution observations of SiO would be helpful to study the impact of shocks in more detail. In the 1D physical-chemical modeling of HMSFRs with \texttt{MUSCLE} by \citet{Gerner2014}, CH$_{3}$CN and CH$_{3}$OH are co-spatial in radial abundance profiles toward the inner hot core region. \citet{Urquhart2019} find that the line integrated ratios of S- and N-bearing species are positively correlated with the dust temperature in a sample of high-mass star-forming clumps. \citet{Gratier2013} found that CH$_{3}$CN is much more abundant in the photo-dissociation region (PDR) of the Horsehead nebula than the associated cold and dense core. This is consistent with the fact that \citet{Purcell2006} detect 3\,mm lines of CH$_{3}$CN in 58 candidate hot molecular cores on a sample of 83 methanol maser-selected star-forming regions. They detect CH$_{3}$CN in isolated methanol maser sites and find that CH$_{3}$CN is more prevalent and brighter when an UCH{\sc ii} region is present, independent of the distance to the source. \citet{Guzman2014} propose that correlated abundances of CH$_{3}$OH and CH$_{3}$CN could be related with photochemistry, e.g., by photodesorption.
	
	The column densities relative to N(C$^{18}$O) show in general positive correlations. N-bearing and S-bearing species seem to be chemically related by high temperature gas-phase chemistry. The strong non-correlation of DCN with any observed species needs high angular resolution follow-up observations of deuterated molecules toward these HMSFRs (e.g., N2D$^{+}$, DCO$^{+}$). The correlation between CH$_{3}$CN and CH$_{3}$OH can be due to a mutual evaporation temperature and/or due to photo-desorption.
	
	Computing the Spearman correlation coefficient $r$ of the chemical timescales $\tau_\mathrm{chem}$ (Table \ref{tab:phys_struc}) with the observed column densities $N$ (Fig. \ref{fig:Nhisto}) and with the relative abundances $N$/$N$(C$^{18}$O) (Fig. \ref{fig:abundratiohisto}) for all 22 cores, we do not find strong correlations for most of the molecular column densities and relative abundances with the chemical age. We use the mean chemical age for cores for which only a time range can be estimated. Positive correlations of the column density and relative abundance with the chemical timescale ($r = 0.5 - 0.9$) are found for SO$_{2}$, HNCO, HC$_{3}$N;$\varv_{7}=1$, and CH$_{3}$OH;$\varv_{t}=1$. We only detect HC$_{3}$N;$\varv_{7}=2$ in cores which have a chemical age estimated to be $> 80\,000$\,yrs. This suggests that vibrationally and torsionally excited states of molecules are good indicators for a more evolved region, as these transitions have higher upper energy levels (Table \ref{tab:spectrallineproperties}) and are only excited at a high kinetic temperature. We propose that SO$_{2}$ and HNCO are also good tracers of the evolutionary stage, however, this has to be investigated in a larger statistical sample.
	
	\subsection{Comparison of physical and chemical timescales}
	
\begin{figure}
\centering
\includegraphics[]{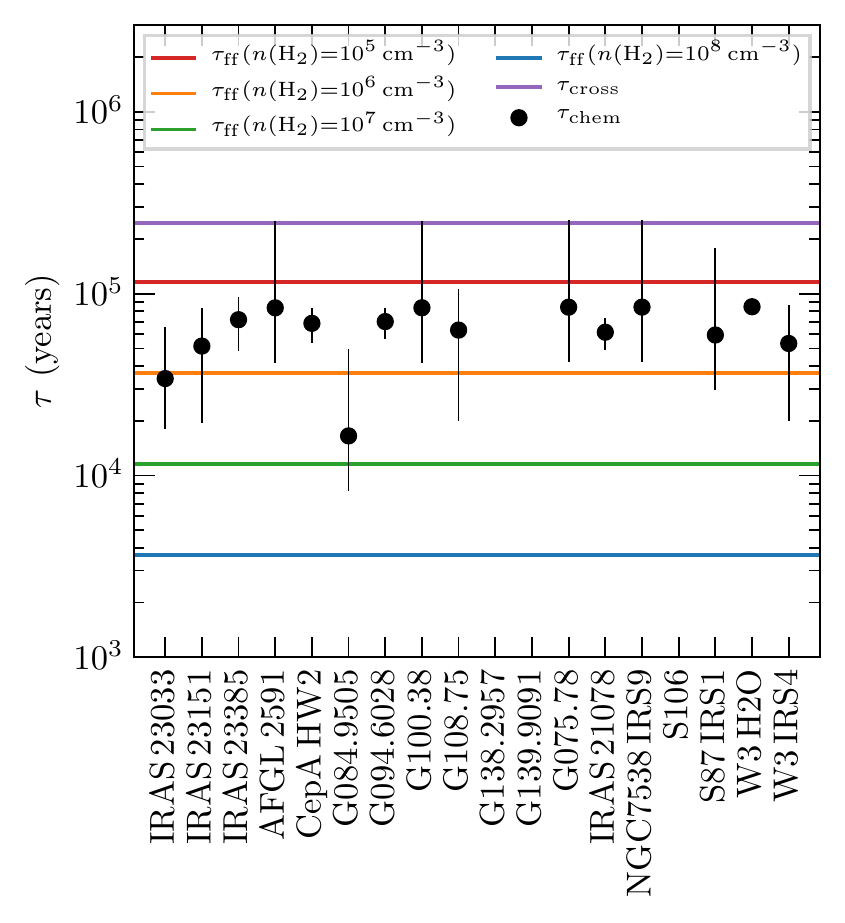}
\caption{Comparison of the free-fall timescale $\tau_{\mathrm{ff}}$ and crossing timescale $\tau_{\mathrm{cross}}$ (colored lines), and chemical timescales $\tau_{\mathrm{chem}}$ (black data points).}
\label{fig:timescales}
\end{figure}

	In the following, we compare the estimated chemical timescales with commonly applied physical timescales. The free-fall timescale $\tau_{\mathrm{ff}}$ is the time it takes for a spherical object to fully collapse under the influence of gravity with no additional forces. It only depends on the initial density $\rho$:
\begin{equation}
\tau_{\mathrm{ff}} = \sqrt{\frac{3\pi}{32G\rho\mathrm{(H}_{2}\mathrm{)}}},
\end{equation}
	where $G$ is the gravitational constant and $\rho$(H$_{2}$) the initial H$_{2}$ mass density. 
	
	The crossing timescale is the time it takes to cross the system once:
\begin{equation}
\tau_{\mathrm{cross}} = \frac{R}{\varv},
\end{equation}
	with clump radius $R$ and velocity dispersion $\varv$.

	A comparison between the free-fall timescales for corresponding H$_{2}$ number densities of $10^{5}$\,cm$^{-3}$, $10^{6}$\,cm$^{-3}$, $10^{7}$\,cm$^{-3}$, and $10^{8}$\,cm$^{-3}$, crossing timescale \citep[assuming $R = 1$\,pc and $\varv = 4$\,km\,s$^{-1}$,][]{Elmegreen2000} and the chemical ages of the regions estimated with \texttt{MUSCLE} is shown in Fig. \ref{fig:timescales}. A mean chemical age is used for cores for which only a time range can be determined. We then calculate the mean chemical age for regions associated with multiple cores. In these cases, the errorbars indicate the range of chemical ages. For regions associated with only one core, we assume a factor of 2 uncertainty for cores with a clear best-fit chemical age or the errorbars indicate the estimated time range for cores for which we can only determine a time range. The derived chemical ages are in agreement with an initial clump density of $10^{5}-10^{6}$\,cm$^{-3}$. Indeed, observations suggest that IRDCs have densities $n$(H$_{2}$)$ > 10^{5}$\,cm$^{-3}$ \citep[e.g.,][]{Carey1998}. The estimated crossing time is an upper limit, in agreement with the scenario that suggests that star formation occurs within a few crossing times \citep{Elmegreen2000}. The agreement of the physical timescales with the chemical ages shows that even though the physical-chemical model is pseudo time-dependent, the estimates are feasible, especially when a homogeneous data set and analysis is used. 
	
	Other forces such as magnetic fields and turbulence can slow down or prevent collapse and are important in the inital diffuse ISM \citep{BallesterosParedes2007,Burge2016}. However, in the diffuse ISM, molecules, with the exception of H$_{2}$, have not formed yet. Our zero point for the chemical timescales is when the density reaches $10^{4}$\,cm$^{-3}$ and for the free-fall timescale we also assume densities $\geqslant 10^{4}$\,cm$^{-3}$. In \citet{Beuther2018}, we showed that the fragmentation scales we observe is in agreement with thermal Jeans fragmentation and turbulent Jeans fragmentation is not important. This is fully consistent with the results found in a different sample by \citet{Palau2015}. The role of magnetic fields in the CORE regions are being investigated by polarization observations with the Submillimeter Array (SMA).

\section{Conclusions}\label{sec:conclusions}

	We study the physical and chemical structure of 18 high-mass star-forming regions using the CORE 1\,mm continuum and spectral line data. We quantify the chemical content for in total 120 positions including in total 22 cores. Combining the CORE observations with the physical-chemical model \texttt{MUSCLE}, we estimate the chemical timescales of the cores. Our main conclusions are summarized as follows:

\begin{enumerate}
\item Using H$_{2}$CO and CH$_{3}$CN line emission, we derive temperature maps of the regions. We identify ``cores'' as objects having a clear radially decreasing temperature profile. We fit these radial temperature profiles and obtain an average power-law index $q = 0.4 \pm 0.1$ excluding three outliers with $q \geq 1$, which is in agreement with theoretical predictions and calculations.
\item The 1.37\,mm continuum visibility profiles of the cores reveal a mean density power-law index of $p = 2.0 \pm 0.2$. Comparing these high-resolution density profiles to previous single-dish and interferometric studies, the density profiles appear to stay roughly constant between $p = 1.7 - 2.0$ from scales of 1\,pc down to 1\,000\,au. The molecular hydrogen column density $N$(H$_{2}$) has a mean of $1.5\times10^{24}$\,cm$^{-2}$ toward the 120 positions and the core mass $M_\mathrm{core}$ of the 22 cores has a mean of 4.1\,$M_{\odot}$.
\item We derive molecular column densities of 11 species by fitting their spectral lines with \texttt{XCLASS}. Spearman correlation coefficients are evaluated for all molecule pairs. We find high correlations between N-bearing and S-bearing species that are chemically related through high temperature gas-phase chemistry. High correlations are also found for molecules that are not chemically related (CH$_{3}$CN and CH$_{3}$OH), but for which the correlation results from a common evaporation temperature and/or photo-desorption.
\item We apply the physical-chemical model \texttt{MUSCLE} to the observed column densities of each core to estimate the chemical age $\tau_\mathrm{chem}$. We find a spread in age from $\tau_\mathrm{chem} \sim 20\,000 - 100\,000$\,yrs and a mean chemical age of $\sim$60\,000\,yrs. Multiple cores within a region show that there can be an age gradient for largely separated cores (e.g. toward IRAS\,23033 and G108.75), while close cores have a similar chemical age (e.g. toward IRAS\,21078 and W3\,H2O) suggesting a sequential star formation.
\item A strong correlation is found between the peak column density $N$(H$_{2}$) of the 22 cores and the $M$/$L$ ratio of the region. In addition, we find a trend that older cores have a lower $M$/$L$ ratio. However, a larger sample is required to study the physical and chemical properties on core-scales in more detail.
\item We compare the chemical age with physical timescales. We find that the chemical age is consistent with a free-fall timescale of the initial clump at a density of $10^{5}$\,cm$^{-3}$ and $10^{6}$\,cm$^{-3}$ consistent with density values typically found toward IRDCs. The chemical ages are smaller than a crossing time of the parental clump.
\item We improve the quality (both intensity and noise) of the NOEMA continuum and spectral line data by applying self-calibration. All of the standard and self-calibrated CORE data are publically available at \url{https://www.mpia.de/core}.
\end{enumerate}
 
	The CORE data reveal a large physical and chemical diversity on scales down to $\sim$300\,au. Here, we use the molecular column densities of 120 positions toward the 18 high-mass star-forming regions to quantify the chemical content, with an emphasis on 22 cores. The case-study of AFGL\,2591 already revealed complex structures around a single hot core \citep{Gieser2019} using the high resolution interferometric observations. In a next step, we will study the spatial morphology of the molecular emission in order to study spatial correlations of the species. 

\begin{acknowledgements}
	The authors would like to thank the anonymous referee whose comments helped improve the clarity of this paper. This work is based on observations carried out under project number L14AB with the IRAM NOEMA Interferometer and the IRAM 30m telescope. IRAM is supported by INSU/CNRS (France), MPG (Germany), and IGN (Spain). We thank the IRAM staff, especially R. Neri and S. Bardeau, for their help in obtaining, calibrating, and imaging the NOEMA and IRAM 30m data. C. G., H. B., and S. S. acknowledge support from the European Research Council under the Horizon 2020 Framework Programme via the ERC Consolidator Grant CSF-648505. D. S. acknowledges support by the Deutsche Forschungsgemeinschaft through SPP 1833: ``Building a Habitable Earth (SE 1962/6-1)''. R. G.-M. acknowledges support from UNAM-PAPIIT project IN104319. R. K. acknowledges financial support via the Emmy Noether Research Group on Accretion Flows and Feedback in Realistic Models of Massive Star Formation funded by the German Research Foundation (DFG) under grant no. KU 2849/3-1 and KU 2849/3-2. A. S.-M. research is partially carried out within the Collaborative Research Centre 956 (subproject A6), funded by the Deutsche Forschungsgemeinschaft (DFG) - project ID 184018867. A. P. acknowledges financial support from CONACyT and UNAM-PAPIIT IN113119 grant, M{\'e}xico. J. P. acknowledges support by the Programme National “Physique et Chimie du Milieu Interstellaire” (PCMI) of CNRS/INSU with INC/INP co-funded by CEA and CNES. This research made use of Astropy\footnote{\url{http://www.astropy.org}}, a community-developed core Python package for Astronomy \citep{Astropy2013, Astropy2018}.
\end{acknowledgements}

\bibliographystyle{aa} 
\bibliography{paper_CORE_chemistry_arxiv} 

\begin{appendix}
\FloatBarrier
\onecolumn
\setlength{\tabcolsep}{5pt}
\section{Position properties}\label{app:continuum}

	Table \ref{tab:positions} summarizes the properties of all 120 positions extracted within the 18 CORE regions which were analyzed in this study. The positions are also marked in Fig. \ref{fig:continuum} showing the 1.37\,mm continuum emission. Positions which show a clear temperature profile are labeled as ``C'' (further explained in Sect. \ref{sec:temperaturestructure}).

\begin{longtable}{lcccccccc}
\caption{Overview of selected positions in each region. The systemic velocity is derived from the C$^{18}$O $2-1$ line (Sect. \ref{sec:XCLASSfitting}). The noise of the spectrum $\sigma_{\mathrm{line}}$ is computed in a line-free range from a rest-frequency of 219.00\,GHz to 219.13\,GHz. The continuum optical depth $\tau_{\nu}^{\mathrm{cont}}$ is calculated according to Eq. \eqref{eq:opticaldepth} and the H$_{2}$ column density $N$(H$_{2}$) is calculated according to Eq. \eqref{eq:H2calc} using the continuum intensity $I_{1.37\mathrm{mm}}$ and $T_\mathrm{kin}$ (Sect. \ref{sec:molecularhydrogen}). As the H$_{2}$ column density is derived from the 1.37\,mm dust continuum emission and we do not have short-spacing information this should be considered as an lower limit due to potential missing flux.}
\label{tab:positions}
\\
\hline\hline
Position & $\alpha$ & $\delta$ & $\varv_{\mathrm{LSR}}$ & $\sigma_{\mathrm{line}}$ & $T_\mathrm{kin}$ & $I_{1.37\mathrm{mm}}$ & $\tau_{\nu}^{\mathrm{cont}}$ & $N$(H$_{2}$)\\
 & J(2000) & J(2000) & (km\,s$^{-1}$) & (K) & (K) & (mJy\,beam$^{-1}$) & & (cm$^{-2}$)\\
\hline
\endfirsthead
\caption{continued.}\\
\hline\hline
Position & $\alpha$ & $\delta$ & $\varv_{\mathrm{LSR}}$ & $\sigma_{\mathrm{line}}$ & $T_\mathrm{kin}$ & $I_{1.37\mathrm{mm}}$ & $\tau_{\nu}^{\mathrm{cont}}$ & $N$(H$_{2}$)\\
 & J(2000) & J(2000) & (km\,s$^{-1}$) & (K) & (K) & (mJy\,beam$^{-1}$) & & (cm$^{-2}$)\\
\hline
\endhead
\hline
Notes. a(b) = a$\times$10$^{\mathrm{b}}$.
\endfoot
IRAS\,23033 1 C & 23:05:24.63 & +60:08:09.2 & $-53.6$ & $0.30$ & 114.9$\pm$\,\,\,8.2 & \,\,\,38.13 & 5.5($-2$) & 1.9(24)$\pm$4.0(23)\\ 
IRAS\,23033 2 C & 23:05:25.04 & +60:08:15.8 & $-53.4$ & $0.20$ & 167.2$\pm$\,\,\,6.6 & \,\,\,32.55 & 3.1($-2$) & 1.1(24)$\pm$2.2(23)\\ 
IRAS\,23033 3 C & 23:05:24.92 & +60:08:13.9 & $-54.6$ & $0.20$ & 160.8$\pm$\,\,\,8.0 & \,\,\,27.04 & 2.7($-2$) & 9.5(23)$\pm$2.0(23)\\ 
IRAS\,23033 4 & 23:05:24.97 & +60:08:15.0 & $-53.7$ & $0.19$ & \,\,\, 84.9$\pm$\,\,\,6.7 & \,\,\,\,\,\,7.14 & 1.4($-2$) & 4.9(23)$\pm$1.1(23)\\ 
IRAS\,23033 5 & 23:05:25.03 & +60:08:17.3 & $-52.9$ & $0.21$ & \,\,\, 15.3$\pm$\,\,\,2.5 & \,\,\,\,\,\,6.30 & 9.5($-2$) & 3.2(24)$\pm$9.7(23)\\ 
\hline 
IRAS\,23151 1 C & 23:17:20.89 & +59:28:47.6 & $-54.5$ & $0.22$ & 129.2$\pm$\,\,\,4.7 & \,\,\,33.78 & 4.3($-2$) & 1.5(24)$\pm$3.1(23)\\ 
IRAS\,23151 2 & 23:17:20.84 & +59:28:47.0 & $-54.2$ & $0.22$ & 168.8$\pm$78.9 & \,\,\,\,\,\,3.06 & 2.9($-3$) & 1.0(23)$\pm$5.4(22)\\ 
IRAS\,23151 3 & 23:17:20.82 & +59:28:48.4 & $-56.2$ & $0.23$ & 171.7$\pm$42.2 & \,\,\,\,\,\,2.96 & 2.8($-3$) & 9.8(22)$\pm$3.2(22)\\ 
IRAS\,23151 4 & 23:17:20.88 & +59:28:49.0 & $-56.4$ & $0.33$ & 204.1$\pm$35.5 & \,\,\,\,\,\,2.35 & 1.8($-3$) & 6.5(22)$\pm$1.7(22)\\ 
IRAS\,23151 5 & 23:17:20.96 & +59:28:46.2 & $-53.7$ & $0.29$ & \,\,\, 69.0$\pm$37.7 & \,\,\,\,\,\,1.87 & 4.5($-3$) & 1.6(23)$\pm$1.0(23)\\ 
\hline 
IRAS\,23385 1 C & 23:40:54.51 & +61:10:28.0 & $-50.9$ & $0.20$ & 239.5$\pm$12.1 & \,\,\,17.84 & 9.3($-3$) & 3.3(23)$\pm$6.8(22)\\ 
IRAS\,23385 2 C & 23:40:54.72 & +61:10:28.5 & $-51.3$ & $0.25$ & 226.0$\pm$\,\,\,2.3 & \,\,\,14.36 & 7.9($-3$) & 2.8(23)$\pm$5.6(22)\\ 
IRAS\,23385 3 & 23:40:54.44 & +61:10:27.7 & $-50.8$ & $0.17$ & \,\,\, 72.4$\pm$37.5 & \,\,\,13.79 & 2.5($-2$) & 8.9(23)$\pm$5.2(23)\\ 
IRAS\,23385 4 & 23:40:54.42 & +61:10:27.0 & $-49.9$ & $0.21$ & \,\,\, 21.5$\pm$\,\,\,4.7 & \,\,\,\,\,\,4.12 & 3.0($-2$) & 1.1(24)$\pm$3.7(23)\\ 
IRAS\,23385 5 & 23:40:54.32 & +61:10:28.2 & $-50.1$ & $0.22$ & 262.2$\pm$11.6 & \,\,\,\,\,\,3.99 & 1.9($-3$) & 6.7(22)$\pm$1.4(22)\\ 
IRAS\,23385 6 & 23:40:54.59 & +61:10:27.3 & $-51.0$ & $0.18$ & \,\,\, 51.1$\pm$32.1 & \,\,\,\,\,\,3.28 & 8.7($-3$) & 3.1(23)$\pm$2.2(23)\\ 
IRAS\,23385 7 & 23:40:54.49 & +61:10:29.4 & $-50.7$ & $0.20$ & 164.2$\pm$42.2 & \,\,\,\,\,\,2.49 & 1.9($-3$) & 6.8(22)$\pm$2.2(22)\\ 
IRAS\,23385 8 & 23:40:54.34 & +61:10:27.2 & $-50.3$ & $0.21$ & 116.4$\pm$36.6 & \,\,\,\,\,\,0.92 & 1.0($-3$) & 3.6(22)$\pm$1.4(22)\\ 
\hline 
AFGL\,2591 1 C & 20:29:24.88 & +40:11:19.4 & $-\,\,\, 5.0$ & $0.47$ & 159.9$\pm$\,\,\,6.8 & \,\,\,85.10 & 8.6($-2$) & 2.9(24)$\pm$6.0(23)\\ 
AFGL\,2591 2 & 20:29:24.47 & +40:11:24.0 & $-\,\,\, 7.2$ & $0.36$ & \,\,\, 39.3$\pm$\,\,\,6.4 & \,\,\,\,\,\,5.76 & 2.5($-2$) & 9.0(23)$\pm$2.4(23)\\ 
AFGL\,2591 3 & 20:29:25.03 & +40:11:21.3 & $-\,\,\, 5.7$ & $0.40$ & \,\,\, 61.6$\pm$14.8 & \,\,\,\,\,\,5.68 & 1.5($-2$) & 5.4(23)$\pm$1.8(23)\\ 
AFGL\,2591 4 & 20:29:24.80 & +40:11:20.5 & $-\,\,\, 5.8$ & $0.35$ & \,\,\, 94.3$\pm$20.4 & \,\,\,\,\,\,5.01 & 8.4($-3$) & 3.0(23)$\pm$9.1(22)\\ 
\hline 
CepA\,HW2 1 C & 22:56:17.98 & +62:01:49.6 & $-\,\,\, 7.7$ & $0.97$ & 238.2$\pm$11.4 & 448.50 & 3.4($-1$) & 1.0(25)$\pm$2.1(24)\\ 
CepA\,HW2 2 C & 22:56:17.96 & +62:01:46.2 & $-\,\,\, 9.8$ & $0.39$ & 170.5$\pm$13.3 & \,\,\,84.21 & 8.0($-2$) & 2.7(24)$\pm$5.9(23)\\ 
CepA\,HW2 3 & 22:56:17.87 & +62:01:50.3 & $-\,\,\, 7.4$ & $0.48$ & 124.9$\pm$\,\,\,8.7 & \,\,\,28.63 & 3.7($-2$) & 1.3(24)$\pm$2.7(23)\\ 
CepA\,HW2 4 & 22:56:17.35 & +62:01:54.8 & $-11.1$ & $0.44$ & \,\,\, 20.0$\pm$10.0 & \,\,\,20.45 & 2.3($-1$) & 7.2(24)$\pm$4.9(24)\\ 
CepA\,HW2 5 & 22:56:17.89 & +62:01:48.1 & $-\,\,\, 9.4$ & $0.42$ & 273.8$\pm$\,\,\,8.5 & \,\,\,14.19 & 8.0($-3$) & 2.8(23)$\pm$5.7(22)\\ 
\hline 
G084.9505 1 C & 20:55:32.50 & +44:06:10.2 & $-34.3$ & $0.15$ & 169.0$\pm$\,\,\,0.8 & \,\,\,\,\,\,5.19 & 5.0($-3$) & 1.8(23)$\pm$3.5(22)\\ 
G084.9505 2 & 20:55:32.22 & +44:06:07.9 & $-34.5$ & $0.17$ & \,\,\, 81.8$\pm$\,\,\,3.6 & \,\,\,\,\,\,3.38 & 6.9($-3$) & 2.4(23)$\pm$5.0(22)\\ 
G084.9505 3 & 20:55:32.13 & +44:06:08.8 & $-34.2$ & $0.14$ & \,\,\, 77.0$\pm$22.5 & \,\,\,\,\,\,2.18 & 4.7($-3$) & 1.7(23)$\pm$6.2(22)\\ 
G084.9505 4 & 20:55:32.30 & +44:06:08.0 & $-34.0$ & $0.15$ & 111.2$\pm$36.4 & \,\,\,\,\,\,2.05 & 3.0($-3$) & 1.1(23)$\pm$4.3(22)\\ 
G084.9505 5 & 20:55:32.41 & +44:06:08.5 & $-34.2$ & $0.14$ & \,\,\, 53.7$\pm$42.5 & \,\,\,\,\,\,1.68 & 5.4($-3$) & 1.9(23)$\pm$1.7(23)\\ 
G084.9505 6 & 20:55:32.16 & +44:06:09.2 & $-34.2$ & $0.17$ & 111.2$\pm$37.9 & \,\,\,\,\,\,1.27 & 1.9($-3$) & 6.7(22)$\pm$2.7(22)\\ 
G084.9505 7 & 20:55:32.31 & +44:06:09.0 & $-34.7$ & $0.17$ & \,\,\, 33.3$\pm$\,\,\,4.2 & \,\,\,\,\,\,1.15 & 6.4($-3$) & 2.3(23)$\pm$5.6(22)\\ 
G084.9505 8 & 20:55:32.38 & +44:06:06.6 & $-33.9$ & $0.17$ & \,\,\, 57.2$\pm$27.2 & \,\,\,\,\,\,1.04 & 3.1($-3$) & 1.1(23)$\pm$6.2(22)\\ 
\hline 
G094.6028 1 C & 21:39:58.27 & +50:14:21.0 & $+29.0$ & $0.26$ & 183.8$\pm$45.4 & \,\,\,12.84 & 1.2($-2$) & 4.2(23)$\pm$1.3(23)\\ 
G094.6028 2 & 21:39:58.16 & +50:14:20.5 & $+30.0$ & $0.18$ & \,\,\, 39.1$\pm$17.7 & \,\,\,\,\,\,2.18 & 1.0($-2$) & 3.7(23)$\pm$2.1(23)\\ 
G094.6028 3 & 21:39:58.11 & +50:14:26.1 & $+28.4$ & $0.25$ & \,\,\, 50.9$\pm$\,\,\,0.0 & \,\,\,\,\,\,2.11 & 7.5($-3$) & 2.7(23)$\pm$5.3(22)\\ 
G094.6028 4 & 21:39:58.31 & +50:14:19.6 & $+28.3$ & $0.19$ & \,\,\, 37.8$\pm$\,\,\,8.5 & \,\,\,\,\,\,2.05 & 1.0($-2$) & 3.6(23)$\pm$1.2(23)\\ 
G094.6028 5 & 21:39:58.13 & +50:14:21.6 & $+29.0$ & $0.17$ & \,\,\, 21.7$\pm$\,\,\,5.0 & \,\,\,\,\,\,2.00 & 1.9($-2$) & 6.9(23)$\pm$2.4(23)\\ 
G094.6028 6 & 21:39:58.05 & +50:14:26.0 & $+28.2$ & $0.29$ & \,\,\, 65.8$\pm$38.7 & \,\,\,\,\,\,1.86 & 5.0($-3$) & 1.8(23)$\pm$1.2(23)\\ 
G094.6028 7 & 21:39:58.22 & +50:14:21.7 & $+29.0$ & $0.20$ & \,\,\, 26.7$\pm$11.7 & \,\,\,\,\,\,1.70 & 1.3($-2$) & 4.5(23)$\pm$2.6(23)\\ 
G094.6028 8 & 21:39:58.59 & +50:14:20.5 & $+28.9$ & $0.30$ & \,\,\, 16.7$\pm$\,\,\,4.4 & \,\,\,\,\,\,1.32 & 1.8($-2$) & 6.4(23)$\pm$2.6(23)\\ 
\hline 
G100.38 1 C & 22:16:10.37 & +52:21:34.1 & $-37.4$ & $0.13$ & \,\,\, 68.2$\pm$\,\,\,0.7 & \,\,\,\,\,\,7.28 & 1.8($-2$) & 6.5(23)$\pm$1.3(23)\\ 
G100.38 2 & 22:16:10.48 & +52:21:36.7 & $-37.3$ & $0.16$ & \,\,\, 37.4$\pm$\,\,\,0.0 & \,\,\,\,\,\,1.21 & 5.9($-3$) & 2.1(23)$\pm$4.2(22)\\ 
G100.38 3 & 22:16:10.42 & +52:21:37.1 & $-37.4$ & $0.18$ & \,\,\, 42.3$\pm$37.6 & \,\,\,\,\,\,1.20 & 5.1($-3$) & 1.8(23)$\pm$1.9(23)\\ 
G100.38 4 & 22:16:10.58 & +52:21:33.3 & $-37.6$ & $0.18$ & \,\,\, 14.7$\pm$\,\,\,6.1 & \,\,\,\,\,\,0.99 & 1.6($-2$) & 5.5(23)$\pm$3.4(23)\\ 
G100.38 5 & 22:16:10.62 & +52:21:36.1 & $-37.5$ & $0.20$ & \,\,\, 20.0$\pm$10.0 & \,\,\,\,\,\,0.67 & 7.0($-3$) & 2.5(23)$\pm$1.7(23)\\ 
\hline 
G108.75 1 C & 22:58:47.41 & +58:45:02.0 & $-51.0$ & $0.15$ & 111.2$\pm$\,\,\,8.2 & \,\,\,16.01 & 1.8($-2$) & 6.2(23)$\pm$1.3(23)\\ 
G108.75 2 C & 22:58:46.78 & +58:45:03.6 & $-50.7$ & $0.14$ & \,\,\, 82.9$\pm$\,\,\,2.6 & \,\,\,\,\,\,7.81 & 1.2($-2$) & 4.2(23)$\pm$8.4(22)\\ 
G108.75 3 & 22:58:46.97 & +58:45:04.9 & $-50.9$ & $0.16$ & \,\,\, 39.1$\pm$17.5 & \,\,\,\,\,\,6.24 & 2.1($-2$) & 7.6(23)$\pm$4.2(23)\\ 
G108.75 4 & 22:58:47.06 & +58:45:04.6 & $-50.8$ & $0.16$ & \,\,\, 22.9$\pm$\,\,\,5.9 & \,\,\,\,\,\,2.14 & 1.4($-2$) & 4.9(23)$\pm$1.8(23)\\ 
G108.75 5 & 22:58:46.89 & +58:45:04.3 & $-50.8$ & $0.13$ & 155.7$\pm$75.4 & \,\,\,\,\,\,0.97 & 7.5($-4$) & 2.7(22)$\pm$1.4(22)\\ 
G108.75 6 & 22:58:47.38 & +58:45:03.5 & $-50.7$ & $0.14$ & \,\,\, 28.8$\pm$17.2 & \,\,\,\,\,\,0.97 & 4.7($-3$) & 1.7(23)$\pm$1.2(23)\\ 
\hline 
G138.2957 1 & 03:01:31.28 & +60:29:12.7 & $-38.9$ & $0.16$ & \,\,\, 31.3$\pm$\,\,\,6.7 & \,\,\,\,\,\,6.28 & 3.0($-2$) & 1.1(24)$\pm$3.4(23)\\ 
G138.2957 2 & 03:01:31.41 & +60:29:12.4 & $-39.2$ & $0.19$ & \,\,\, 34.8$\pm$16.4 & \,\,\,\,\,\,3.78 & 1.6($-2$) & 5.6(23)$\pm$3.3(23)\\ 
G138.2957 3 & 03:01:31.49 & +60:29:12.2 & $-39.3$ & $0.19$ & 149.3$\pm$28.4 & \,\,\,\,\,\,3.13 & 2.7($-3$) & 9.6(22)$\pm$2.7(22)\\ 
G138.2957 4 & 03:01:31.36 & +60:29:18.5 & $-38.0$ & $0.20$ & \,\,\, 19.3$\pm$\,\,\,5.7 & \,\,\,\,\,\,2.64 & 2.3($-2$) & 8.1(23)$\pm$3.5(23)\\ 
G138.2957 5 & 03:01:31.40 & +60:29:19.3 & $-38.3$ & $0.26$ & \,\,\, 35.1$\pm$12.8 & \,\,\,\,\,\,2.28 & 9.4($-3$) & 3.3(23)$\pm$1.6(23)\\ 
G138.2957 6 & 03:01:31.28 & +60:29:13.9 & $-38.6$ & $0.16$ & \,\,\, 74.9$\pm$14.1 & \,\,\,\,\,\,2.27 & 4.1($-3$) & 1.4(23)$\pm$4.1(22)\\ 
G138.2957 7 & 03:01:31.28 & +60:29:10.6 & $-39.4$ & $0.16$ & 117.1$\pm$41.7 & \,\,\,\,\,\,2.19 & 2.4($-3$) & 8.7(22)$\pm$3.7(22)\\ 
G138.2957 8 & 03:01:32.16 & +60:29:18.7 & $-38.6$ & $0.29$ & \,\,\, 41.1$\pm$11.1 & \,\,\,\,\,\,2.02 & 7.0($-3$) & 2.5(23)$\pm$9.1(22)\\ 
G138.2957 9 & 03:01:31.46 & +60:29:18.9 & $-37.3$ & $0.21$ & \,\,\, 63.6$\pm$14.3 & \,\,\,\,\,\,1.87 & 4.0($-3$) & 1.4(23)$\pm$4.5(22)\\ 
G138.2957 10 & 03:01:32.15 & +60:29:19.9 & $-37.2$ & $0.26$ & \,\,\, 20.8$\pm$\,\,\,4.9 & \,\,\,\,\,\,1.85 & 1.4($-2$) & 5.1(23)$\pm$1.8(23)\\ 
G138.2957 11 & 03:01:32.25 & +60:29:17.7 & $-38.5$ & $0.35$ & \,\,\, 20.0$\pm$\,\,\,2.6 & \,\,\,\,\,\,1.80 & 1.5($-2$) & 5.3(23)$\pm$1.4(23)\\ 
G138.2957 12 & 03:01:30.67 & +60:29:12.9 & $-36.9$ & $0.22$ & \,\,\, 27.4$\pm$14.6 & \,\,\,\,\,\,1.44 & 8.0($-3$) & 2.8(23)$\pm$1.9(23)\\ 
\hline 
G139.9091 1 & 03:07:24.49 & +58:30:42.8 & $-39.8$ & $0.27$ & 257.5$\pm$59.1 & \,\,\,13.49 & 6.7($-3$) & 2.4(23)$\pm$7.4(22)\\ 
G139.9091 2 & 03:07:24.58 & +58:30:53.0 & $-39.7$ & $0.21$ & 298.3$\pm$\,\,\,0.0 & \,\,\,\,\,\,6.63 & 2.8($-3$) & 1.0(23)$\pm$2.0(22)\\ 
G139.9091 3 & 03:07:23.94 & +58:30:52.5 & $-39.8$ & $0.24$ & \,\,\, 57.2$\pm$61.6 & \,\,\,\,\,\,1.79 & 4.3($-3$) & 1.5(23)$\pm$1.8(23)\\ 
G139.9091 4 & 03:07:24.49 & +58:30:50.3 & $-40.8$ & $0.30$ & \,\,\, 55.3$\pm$43.1 & \,\,\,\,\,\,1.59 & 4.0($-3$) & 1.4(23)$\pm$1.2(23)\\ 
\hline 
G075.78 1 C & 20:21:44.02 & +37:26:37.5 & $-\,\,\, 6.9$ & $0.38$ & 176.8$\pm$\,\,\,0.3 & \,\,\,71.67 & 6.2($-2$) & 2.1(24)$\pm$4.3(23)\\ 
G075.78 2 & 20:21:43.95 & +37:26:39.3 & $-\,\,\, 6.0$ & $0.30$ & \,\,\, 73.1$\pm$28.6 & \,\,\,13.88 & 3.0($-2$) & 1.0(24)$\pm$4.9(23)\\ 
G075.78 3 & 20:21:44.10 & +37:26:39.3 & $-\,\,\, 6.7$ & $0.31$ & \,\,\, 64.7$\pm$23.2 & \,\,\,\,\,\,4.20 & 1.0($-2$) & 3.6(23)$\pm$1.6(23)\\ 
G075.78 4 & 20:21:44.28 & +37:26:38.0 & $-\,\,\, 7.1$ & $0.30$ & 276.8$\pm$\,\,\,3.2 & \,\,\,\,\,\,4.16 & 2.2($-3$) & 7.8(22)$\pm$1.6(22)\\ 
\hline 
IRAS\,21078 1 C & 21:09:21.71 & +52:22:37.1 & $-\,\,\, 6.5$ & $0.40$ & 135.5$\pm$\,\,\,0.2 & \,\,\,36.69 & 4.7($-2$) & 1.6(24)$\pm$3.3(23)\\ 
IRAS\,21078 2 C & 21:09:21.79 & +52:22:35.7 & $-\,\,\, 6.8$ & $0.25$ & 121.9$\pm$\,\,\,3.4 & \,\,\,22.95 & 3.3($-2$) & 1.1(24)$\pm$2.3(23)\\ 
IRAS\,21078 3 & 21:09:21.76 & +52:22:36.8 & $-\,\,\, 5.4$ & $0.29$ & 120.5$\pm$15.1 & \,\,\,21.21 & 3.0($-2$) & 1.1(24)$\pm$2.6(23)\\ 
IRAS\,21078 4 & 21:09:22.12 & +52:22:34.6 & $-\,\,\, 3.9$ & $0.21$ & \,\,\, 46.3$\pm$14.3 & \,\,\,20.09 & 8.3($-2$) & 2.8(24)$\pm$1.1(24)\\ 
IRAS\,21078 5 & 21:09:21.85 & +52:22:32.0 & $-\,\,\, 5.4$ & $0.27$ & 249.0$\pm$\,\,\,9.1 & \,\,\,19.24 & 1.3($-2$) & 4.6(23)$\pm$9.3(22)\\ 
IRAS\,21078 6 & 21:09:21.65 & +52:22:38.6 & $-\,\,\, 6.3$ & $0.25$ & \,\,\, 61.8$\pm$\,\,\,3.8 & \,\,\,19.06 & 5.6($-2$) & 2.0(24)$\pm$4.1(23)\\ 
IRAS\,21078 7 & 21:09:21.96 & +52:22:36.9 & $-\,\,\, 4.6$ & $0.25$ & 206.9$\pm$23.3 & \,\,\,16.49 & 1.3($-2$) & 4.8(23)$\pm$1.1(23)\\ 
IRAS\,21078 8 & 21:09:21.90 & +52:22:35.4 & $-\,\,\, 5.2$ & $0.26$ & \,\,\, 77.7$\pm$33.6 & \,\,\,13.98 & 3.2($-2$) & 1.1(24)$\pm$5.6(23)\\ 
IRAS\,21078 9 & 21:09:22.04 & +52:22:35.2 & $-\,\,\, 5.5$ & $0.26$ & 229.5$\pm$18.5 & \,\,\,10.08 & 7.4($-3$) & 2.6(23)$\pm$5.7(22)\\ 
IRAS\,21078 10 & 21:09:21.82 & +52:22:38.0 & $-\,\,\, 6.0$ & $0.25$ & 115.7$\pm$36.4 & \,\,\,\,\,\,9.99 & 1.5($-2$) & 5.3(23)$\pm$2.0(23)\\ 
IRAS\,21078 11 & 21:09:21.37 & +52:22:38.1 & $-\,\,\, 6.0$ & $0.32$ & 229.1$\pm$59.2 & \,\,\,\,\,\,5.16 & 3.8($-3$) & 1.3(23)$\pm$4.4(22)\\ 
IRAS\,21078 12 & 21:09:21.83 & +52:22:34.0 & $-\,\,\, 5.3$ & $0.28$ & \,\,\, 41.4$\pm$\,\,\,8.6 & \,\,\,\,\,\,4.03 & 1.8($-2$) & 6.4(23)$\pm$2.0(23)\\ 
IRAS\,21078 13 & 21:09:21.83 & +52:22:33.0 & $-\,\,\, 5.3$ & $0.33$ & 298.1$\pm$\,\,\,0.9 & \,\,\,\,\,\,3.96 & 2.2($-3$) & 7.9(22)$\pm$1.6(22)\\ 
IRAS\,21078 14 & 21:09:21.49 & +52:22:38.8 & $-\,\,\, 5.8$ & $0.28$ & 286.2$\pm$\,\,\,2.2 & \,\,\,\,\,\,3.29 & 1.9($-3$) & 6.8(22)$\pm$1.4(22)\\ 
\hline 
NGC7538\,IRS9 1 C & 23:14:01.75 & +61:27:19.8 & $-56.5$ & $0.64$ & 200.5$\pm$\,\,\,0.9 & \,\,\,46.09 & 3.7($-2$) & 1.3(24)$\pm$2.6(23)\\ 
NGC7538\,IRS9 2 & 23:14:02.05 & +61:27:19.7 & $-57.7$ & $0.26$ & 189.4$\pm$18.4 & \,\,\,\,\,\,6.12 & 5.2($-3$) & 1.8(23)$\pm$4.1(22)\\ 
NGC7538\,IRS9 3 & 23:14:02.21 & +61:27:18.9 & $-58.4$ & $0.27$ & 194.6$\pm$21.8 & \,\,\,\,\,\,4.65 & 3.8($-3$) & 1.4(23)$\pm$3.1(22)\\ 
NGC7538\,IRS9 4 & 23:14:02.15 & +61:27:20.1 & $-57.9$ & $0.27$ & 104.2$\pm$52.4 & \,\,\,\,\,\,4.44 & 7.0($-3$) & 2.5(23)$\pm$1.4(23)\\ 
NGC7538\,IRS9 5 & 23:14:01.64 & +61:27:21.2 & $-58.5$ & $0.22$ & 263.4$\pm$\,\,\,7.7 & \,\,\,\,\,\,3.27 & 2.0($-3$) & 7.0(22)$\pm$1.4(22)\\ 
NGC7538\,IRS9 6 & 23:14:01.89 & +61:27:20.9 & $-58.8$ & $0.32$ & \,\,\, 84.5$\pm$22.5 & \,\,\,\,\,\,2.81 & 5.5($-3$) & 2.0(23)$\pm$6.8(22)\\ 
NGC7538\,IRS9 7 & 23:14:02.02 & +61:27:20.6 & $-58.5$ & $0.27$ & 203.8$\pm$29.4 & \,\,\,\,\,\,2.71 & 2.1($-3$) & 7.6(22)$\pm$1.9(22)\\ 
NGC7538\,IRS9 8 & 23:14:02.34 & +61:27:19.8 & $-58.0$ & $0.19$ & 134.8$\pm$70.6 & \,\,\,\,\,\,2.33 & 2.8($-3$) & 1.0(23)$\pm$5.8(22)\\ 
\hline 
S87\,IRS1 1 C & 19:46:20.64 & +24:35:38.8 & $+22.9$ & $0.61$ & 112.0$\pm$\,\,\,5.2 & \,\,\,37.31 & 4.8($-2$) & 1.7(24)$\pm$3.4(23)\\ 
S87\,IRS1 2 & 19:46:20.59 & +24:35:36.6 & $+22.5$ & $0.38$ & \,\,\, 84.4$\pm$27.0 & \,\,\,\,\,\,8.24 & 1.4($-2$) & 5.0(23)$\pm$2.0(23)\\ 
S87\,IRS1 3 & 19:46:20.13 & +24:35:29.4 & $+22.0$ & $0.32$ & \,\,\, 63.5$\pm$40.1 & \,\,\,\,\,\,6.77 & 1.6($-2$) & 5.5(23)$\pm$4.0(23)\\ 
S87\,IRS1 4 & 19:46:19.87 & +24:35:28.4 & $+22.2$ & $0.28$ & \,\,\, 28.4$\pm$\,\,\,7.4 & \,\,\,\,\,\,6.60 & 3.9($-2$) & 1.3(24)$\pm$5.0(23)\\ 
S87\,IRS1 5 & 19:46:20.02 & +24:35:28.8 & $+21.2$ & $0.25$ & \,\,\, 53.2$\pm$\,\,\,9.9 & \,\,\,\,\,\,5.92 & 1.7($-2$) & 5.9(23)$\pm$1.7(23)\\ 
S87\,IRS1 6 & 19:46:20.61 & +24:35:37.4 & $+22.9$ & $0.48$ & 100.3$\pm$52.1 & \,\,\,\,\,\,5.42 & 7.7($-3$) & 2.7(23)$\pm$1.6(23)\\ 
S87\,IRS1 7 & 19:46:20.22 & +24:35:32.0 & $+22.7$ & $0.24$ & \,\,\, 69.9$\pm$16.4 & \,\,\,\,\,\,3.31 & 6.9($-3$) & 2.4(23)$\pm$7.9(22)\\ 
S87\,IRS1 8 & 19:46:19.94 & +24:35:30.2 & $+21.5$ & $0.24$ & 107.7$\pm$60.5 & \,\,\,\,\,\,3.22 & 4.2($-3$) & 1.5(23)$\pm$9.3(22)\\ 
S87\,IRS1 9 & 19:46:19.99 & +24:35:31.7 & $+21.8$ & $0.35$ & 186.6$\pm$43.0 & \,\,\,\,\,\,2.58 & 1.9($-3$) & 6.8(22)$\pm$2.1(22)\\ 
\hline 
S106 1 & 20:27:26.77 & +37:22:47.7 & $+\,\,\, 0.5$ & $0.24$ & \,\,\, 64.3$\pm$63.4 & 150.47 & 5.2($-1$) & 1.4(25)$\pm$1.6(25)\\ 
S106 2 & 20:27:25.49 & +37:22:48.4 & $-\,\,\, 0.5$ & $0.57$ & 124.8$\pm$30.8 & \,\,\,50.13 & 6.9($-2$) & 2.4(24)$\pm$7.7(23)\\ 
S106 3 & 20:27:25.51 & +37:22:46.9 & $-\,\,\, 0.8$ & $0.75$ & \,\,\, 64.2$\pm$12.5 & \,\,\,19.13 & 5.3($-2$) & 1.8(24)$\pm$5.3(23)\\ 
S106 4 & 20:27:26.99 & +37:22:50.5 & $-\,\,\, 2.4$ & $0.25$ & \,\,\, 96.7$\pm$38.8 & \,\,\,\,\,\,8.05 & 1.4($-2$) & 5.0(23)$\pm$2.3(23)\\ 
\hline 
W3\,H2O 1 & 02:27:03.85 & +61:52:25.0 & $-47.1$ & $0.54$ & 124.2$\pm$28.3 & 456.29 & 1.3($0$) & 2.5(25)$\pm$7.9(24)\\ 
W3\,H2O 2 & 02:27:03.90 & +61:52:24.3 & $-46.8$ & $0.41$ & \,\,\, 81.0$\pm$\,\,\,9.1 & 317.19 & 1.5($0$) & 2.8(25)$\pm$6.5(24)\\ 
W3\,H2O 3 C & 02:27:04.71 & +61:52:24.7 & $-51.7$ & $0.97$ & 176.6$\pm$10.9 & 170.41 & 2.0($-1$) & 6.6(24)$\pm$1.4(24)\\ 
W3\,H2O 4 C & 02:27:04.55 & +61:52:24.7 & $-49.7$ & $1.27$ & 166.4$\pm$\,\,\,9.0 & 161.32 & 2.1($-1$) & 6.6(24)$\pm$1.4(24)\\ 
W3\,H2O 5 & 02:27:04.42 & +61:52:25.4 & $-48.1$ & $0.42$ & 109.0$\pm$21.4 & \,\,\,42.75 & 8.0($-2$) & 2.7(24)$\pm$7.8(23)\\ 
\hline 
W3\,IRS4 1 C & 02:25:31.35 & +62:06:20.8 & $-44.7$ & $0.44$ & 173.0$\pm$15.4 & \,\,\,51.50 & 5.6($-2$) & 1.9(24)$\pm$4.3(23)\\ 
W3\,IRS4 2 & 02:25:31.97 & +62:06:23.7 & $-46.6$ & $0.63$ & \,\,\, 48.2$\pm$10.6 & \,\,\,20.76 & 9.0($-2$) & 3.1(24)$\pm$9.6(23)\\ 
W3\,IRS4 3 & 02:25:31.50 & +62:06:22.2 & $-43.9$ & $0.46$ & \,\,\, 61.3$\pm$19.4 & \,\,\,16.36 & 5.3($-2$) & 1.9(24)$\pm$7.4(23)\\ 
W3\,IRS4 4 & 02:25:31.04 & +62:06:19.8 & $-44.6$ & $0.45$ & 251.5$\pm$\,\,\,9.5 & \,\,\,10.46 & 7.6($-3$) & 2.7(23)$\pm$5.5(22)\\ 
W3\,IRS4 5 & 02:25:31.50 & +62:06:22.9 & $-44.5$ & $0.43$ & \,\,\, 27.9$\pm$\,\,\,3.1 & \,\,\,10.02 & 8.1($-2$) & 2.8(24)$\pm$6.7(23)\\ 
W3\,IRS4 6 & 02:25:31.62 & +62:06:24.9 & $-44.6$ & $0.39$ & \,\,\, 84.8$\pm$29.0 & \,\,\,\,\,\,5.91 & 1.3($-2$) & 4.7(23)$\pm$2.0(23)\\ 
\end{longtable}

\section{Spectral line properties}\label{app:lineproperties}

	 The line properties of the detected spectral lines analyzed in this study are summarized in Table \ref{tab:spectrallineproperties}.

\begin{table*}
\caption{Properties of the analyzed spectral lines taken from Splatalogue \citep{Splatalogue}. The entries are taken from the CDMS and JPL catalogs \citep{CDMS,JPL}. Blended lines are indicated by a $\ddagger$. The mean and maximum line optical depth for each transition fitted with \texttt{XCLASS} are computed considering all 120 positions (see Sect. \ref{sec:XCLASSfitting}).}
\label{tab:spectrallineproperties}
\centering
\begin{tabular}{l l l l l l l l}
\hline\hline
 & & & Einstein & Upper & & \multicolumn{2}{c}{Line} \\
 & & Frequency & coefficient & Energy Level & & \multicolumn{2}{c}{Optical Depth}\\
Molecule & Quantum Numbers & $\nu$ & log($A_{\mathrm{ul}}$) & $E_\mathrm{u}/k_\mathrm{B}$ & Catalog & $\tau^{\mathrm{line}}_{\mathrm{mean}}$ & $\tau^{\mathrm{line}}_{\mathrm{max}}$\\
 & & (GHz) & (log s$^{-1}$) & (K) & & \\
\hline
$^{13}$CO & $2-1$ & 220.399 & $-6.22$ & \,\,\,\,\,\,15.9 & JPL & $1.5(0)$ & $2.1(2)$\\ 
\hline 
C$^{18}$O & $2-1$ & 219.560 & $-6.22$ & \,\,\,\,\,\,15.8 & JPL & $1.1(-1)$ & $5.4(0)$\\ 
\hline 
SO & $6_{5}-5_{4}$ & 219.949 & $-3.87$ & \,\,\,\,\,\,35.0 & JPL & $2.7(-1)$ & $8.3(0)$\\ 
\hline 
OCS & $18-17$ & 218.903 & $-4.52$ & \,\,\,\,\,\,99.8 & JPL & $4.3(-1)$ & $2.4(1)$\\ 
\hline 
O$^{13}$CS & $18-17$ & 218.199 & $-4.52$ & \,\,\,\,\,\,99.5 & JPL & $2.0(-3)$ & $3.4(-1)$\\ 
\hline 
SO$_{2}$ & $22_{7,15}-23_{6,18}$ & 219.276 & $-4.67$ & \,\,\,352.8 & JPL & $2.8(-2)$ & $5.6(-1)$\\ 
\hline 
$^{34}$SO$_{2}$ & $11_{1,11}-10_{0,10}$ & 219.355 & $-3.96$ & \,\,\,\,\,\,60.2 & JPL & $1.1(-2)$ & $1.1(-1)$\\ 
\hline 
DCN & $3-2$ & 217.239 & $-3.34$ & \,\,\,\,\,\,20.9 & CDMS & $9.7(-2)$ & $2.2(1)$\\ 
\hline 
H$_{2}$CO & $3_{0,3}-2_{0,2}$ & 218.222 & $-3.55$ & \,\,\,\,\,\,21.0 & JPL & $6.2(-1)$ & $2.5(1)$\\ 
H$_{2}$CO & $3_{2,2}-2_{2,1}$ & 218.476 & $-3.80$ & \,\,\,\,\,\,68.1 & JPL & $1.5(-1)$ & $5.8(0)$\\ 
H$_{2}$CO & $3_{2,1}-2_{2,0}$ & 218.760 & $-3.80$ & \,\,\,\,\,\,68.1 & JPL & $1.3(-1)$ & $8.5(0)$\\ 
\hline 
H$_{2}^{13}$CO & $3_{1,2}-2_{1,1}$ & 219.909 & $-3.59$ & \,\,\,\,\,\,32.9 & JPL & $1.5(-2)$ & $7.9(-1)$\\ 
\hline 
HNCO & $10_{1,10}-9_{1,9}$ & 218.981 & $-3.85$ & \,\,\,101.1 & CDMS & $6.5(-2)$ & $2.3(0)$\\ 
HNCO & $10_{3,8}-9_{3,7}$ & 219.657 & $-3.92$ & \,\,\,433.0 & CDMS & $1.0(-2)$ & $1.7(-1)$\\ 
HNCO & $10_{3,7}-9_{3,6}$ & 219.657 & $-3.92$ & \,\,\,433.0 & CDMS & $1.0(-2)$ & $1.7(-1)$\\ 
HNCO & $10_{2,9}-9_{2,8}$ & 219.734 & $-3.87$ & \,\,\,228.3 & CDMS & $5.8(-2)$ & $9.9(-1)$\\ 
HNCO & $10_{2,8}-9_{2,7}$ & 219.737 & $-3.87$ & \,\,\,228.3 & CDMS & $6.5(-2)$ & $9.9(-1)$\\ 
HNCO & $10_{0,10}-9_{0,9}$ & 219.798 & $-3.83$ & \,\,\,\,\,\,58.0 & CDMS & $1.3(-1)$ & $3.9(0)$\\ 
HNCO & $10_{1,9}-9_{1,8}$ & 220.585 & $-3.84$ & \,\,\,101.5 & CDMS & $9.1(-2)$ & $2.3(0)$\\ 
\hline 
HC$_{3}$N & $24-23$ & 218.325 & $-3.08$ & \,\,\,131.0 & JPL & $8.3(-2)$ & $1.3(0)$\\ 
\hline 
HC$_{3}$N;~$\varv_{7}=1$$^{\ddagger}$ & $24-23$,~$l=1e$ & 218.861 & $-3.08$ & \,\,\,452.1 & CDMS & - & -\\ 
HC$_{3}$N;~$\varv_{7}=1$ & $24-23$,~$l=1f$ & 219.174 & $-3.08$ & \,\,\,452.3 & CDMS & $2.5(-2)$ & $5.8(-1)$\\ 
\hline 
HC$_{3}$N;~$\varv_{7}=2$ & $24-23$,~$l=0$ & 219.675 & $-3.08$ & \,\,\,773.5 & CDMS & $3.4(-3)$ & $6.8(-2)$\\ 
HC$_{3}$N;~$\varv_{7}=2$ & $24-23$,~$l=2e$ & 219.707 & $-3.08$ & \,\,\,776.8 & CDMS & $4.2(-3)$ & $6.6(-2)$\\ 
HC$_{3}$N;~$\varv_{7}=2$$^{\ddagger}$ & $24-23$,~$l=2f$ & 219.742 & $-3.08$ & \,\,\,776.8 & CDMS & - & -\\ 
\hline 
HCC$^{13}$CN & $24-23$ & 217.420 & $-3.09$ & \,\,\,130.4 & CDMS & $1.1(-3)$ & $1.8(-2)$\\ 
\hline 
CH$_{3}$OH & $20_{1,19}-20_{0,20}E$ & 217.887 & $-4.47$ & \,\,\,508.4 & CDMS & $1.4(-2)$ & $5.3(-1)$\\ 
CH$_{3}$OH & $4_{2,3}-3_{1,2}E$ & 218.440 & $-4.33$ & \,\,\,\,\,\,45.5 & CDMS & $5.7(-1)$ & $4.0(1)$\\ 
CH$_{3}$OH & $25_{3,23}-24_{4,20}E$ & 219.984 & $-4.69$ & \,\,\,802.2 & CDMS & $1.5(-3)$ & $9.6(-2)$\\ 
CH$_{3}$OH & $23_{5,18}-22_{6,17}E$ & 219.994 & $-4.76$ & \,\,\,775.9 & CDMS & $1.5(-3)$ & $9.4(-2)$\\ 
CH$_{3}$OH & $8_{0,8}-7_{1,6}E$ & 220.079 & $-4.60$ & \,\,\,\,\,\,96.6 & CDMS & $9.5(-2)$ & $5.0(0)$\\ 
CH$_{3}$OH$^{\ddagger}$ & $10_{5,6}-11_{4,8}E$ & 220.401 & $-4.95$ & \,\,\,251.6 & CDMS & - & -\\ 
\hline 
CH$_{3}$OH;~$\varv_{t}=1$ & $6_{1,5}-7_{2,5}A$ & 217.299 & $-4.37$ & \,\,\,373.9 & CDMS & $4.1(-2)$ & $1.6(0)$\\ 
CH$_{3}$OH;~$\varv_{t}=1$ & $15_{6,9}-16_{5,11}A$ & 217.643 & $-4.72$ & \,\,\,745.6 & CDMS & $1.0(-2)$ & $1.5(-1)$\\ 
CH$_{3}$OH;~$\varv_{t}=1$ & $15_{6,10}-16_{5,12}A$ & 217.643 & $-4.72$ & \,\,\,745.6 & CDMS & $1.0(-2)$ & $1.5(-1)$\\ 
\hline 
CH$_{3}$CN & $12_{8}-11_{8}$ & 220.476 & $-3.45$ & \,\,\,525.6 & JPL & $7.4(-3)$ & $4.0(-1)$\\ 
CH$_{3}$CN & $12_{7}-11_{7}$ & 220.539 & $-3.38$ & \,\,\,418.6 & JPL & $2.2(-2)$ & $7.7(-1)$\\ 
CH$_{3}$CN & $12_{6}-11_{6}$ & 220.594 & $-3.32$ & \,\,\,325.9 & JPL & $9.2(-2)$ & $2.7(0)$\\ 
CH$_{3}$CN & $12_{5}-11_{5}$ & 220.641 & $-3.28$ & \,\,\,247.4 & JPL & $5.9(-2)$ & $2.1(0)$\\ 
CH$_{3}$CN & $12_{4}-11_{4}$ & 220.679 & $-3.25$ & \,\,\,183.1 & JPL & $1.1(-1)$ & $2.7(0)$\\ 
CH$_{3}$CN & $12_{3}-11_{3}$ & 220.709 & $-3.22$ & \,\,\,133.2 & JPL & $2.7(-1)$ & $8.2(0)$\\ 
CH$_{3}$CN & $12_{2}-11_{2}$ & 220.730 & $-3.21$ & \,\,\,\,\,\,97.4 & JPL & $2.1(-1)$ & $4.7(0)$\\ 
CH$_{3}$CN & $12_{1}-11_{1}$ & 220.743 & $-3.20$ & \,\,\,\,\,\,76.0 & JPL & $4.5(-1)$ & $6.7(0)$\\ 
CH$_{3}$CN & $12_{0}-11_{0}$ & 220.747 & $-3.20$ & \,\,\,\,\,\,68.9 & JPL & $3.3(-1)$ & $6.7(0)$\\ 
\hline 
CH$_{3}^{13}$CN & $12_{6}-11_{6}$ & 220.486 & $-3.16$ & \,\,\,325.9 & JPL & $9.8(-4)$ & $3.1(-2)$\\ 
CH$_{3}^{13}$CN & $12_{5}-11_{5}$ & 220.532 & $-3.12$ & \,\,\,247.4 & JPL & $5.7(-3)$ & $4.8(-1)$\\ 
CH$_{3}^{13}$CN & $12_{4}-11_{4}$ & 220.570 & $-3.09$ & \,\,\,183.1 & JPL & $1.3(-3)$ & $3.9(-2)$\\ 
CH$_{3}^{13}$CN & $12_{3}-11_{3}$ & 220.600 & $-3.06$ & \,\,\,133.1 & JPL & $1.5(-2)$ & $6.4(-1)$\\ 
CH$_{3}^{13}$CN & $12_{2}-11_{2}$ & 220.621 & $-3.05$ & \,\,\,\,\,\,97.4 & JPL & $2.0(-3)$ & $6.3(-2)$\\ 
CH$_{3}^{13}$CN & $12_{1}-11_{1}$ & 220.634 & $-3.04$ & \,\,\,\,\,\,76.0 & JPL & $1.4(-2)$ & $1.0(0)$\\ 
CH$_{3}^{13}$CN & $12_{0}-11_{0}$ & 220.638 & $-3.03$ & \,\,\,\,\,\,68.8 & JPL & $6.4(-2)$ & $2.1(0)$\\ 
\hline
\end{tabular}
\tablefoot{a(b) = a$\times$10$^{\mathrm{b}}$.}
\end{table*}

\section{Temperature maps}\label{app:temperaturemaps}

	Figure \ref{fig:temperature_maps} shows the H$_{2}$CO and CH$_{3}$CN temperature maps derived with \texttt{XCLASS} for each region (Sect. \ref{sec:temperaturestructure}). Each core position is marked in red and the derived outer radii from the radial temperature profiles are indicated by dashed red circles.
	
\begin{figure*}[!htb]
\centering
\includegraphics[]{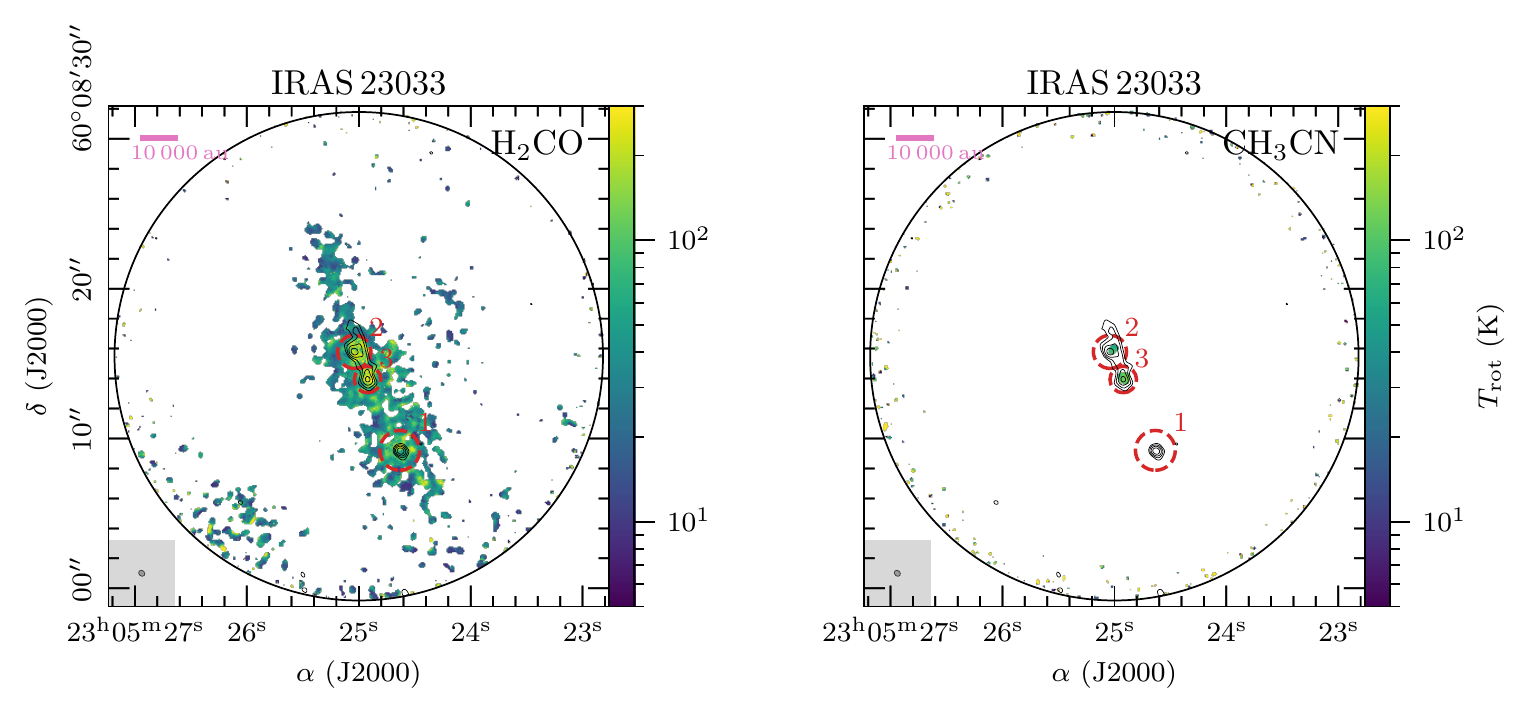}
\includegraphics[]{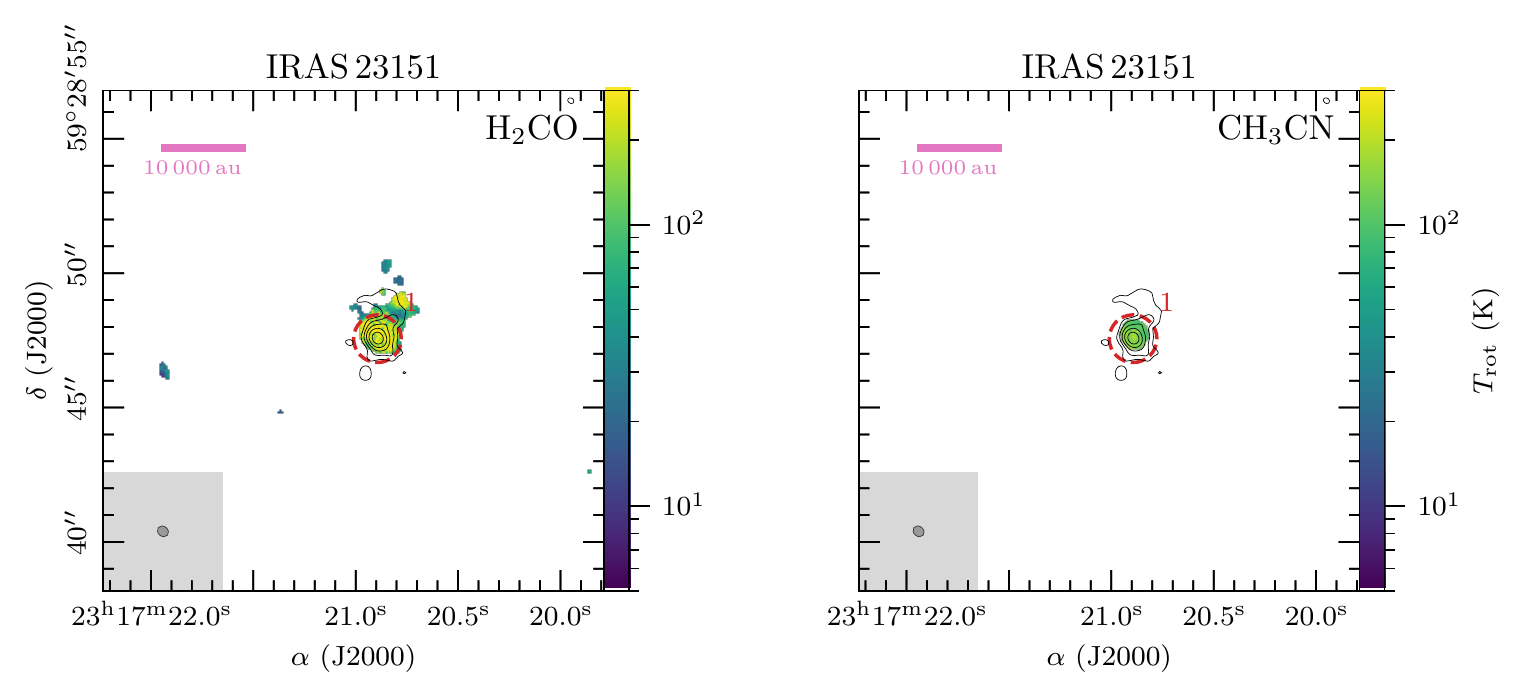}
\includegraphics[]{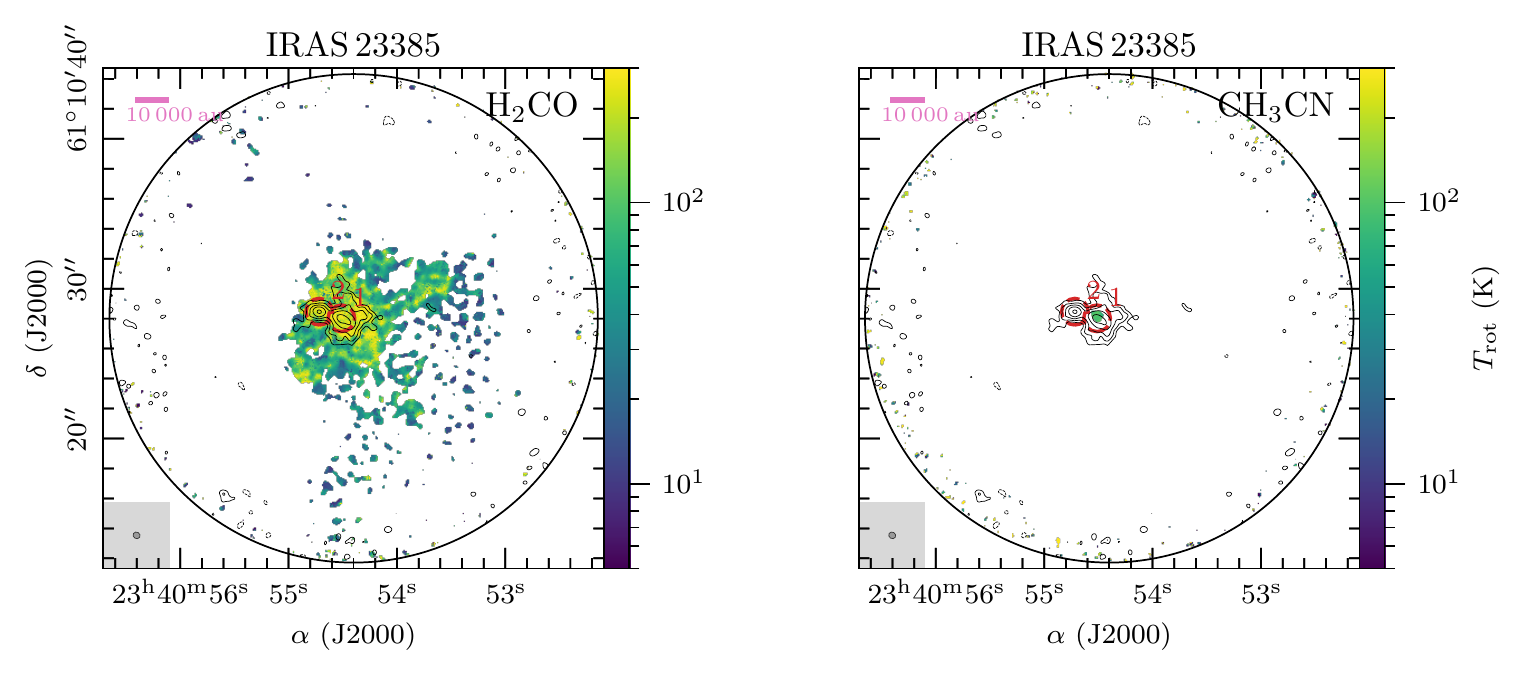}
\caption{Temperature maps derived with \texttt{XCLASS}. Each panel shows in color the temperature map (left: H$_2$CO, right: CH$_3$CN) and in black contours the 1.37\,mm continuum emission. The dashed black contours show the $-5\sigma_\mathrm{cont}$ emission and the solid black contours start at 5$\sigma_\mathrm{cont}$ with steps increasing by a factor of 2 (see Table \ref{tab:dataproducts} for values of $\sigma_\mathrm{cont}$ for each region). Each core is marked in red and the dashed red circle indicates the outer radius of the radial temperature fit (Sect. \ref{sec:temperaturestructure}). The beam size is shown in the bottom left corner in each panel. The pink bar in the top left corner indicates a linear spatial scale of 10\,000\,au. The primary beam size is indicated by a black circle and for regions with no extended H$_{2}$CO temperature map a smaller field of view is shown.}
\label{fig:temperature_maps}
\end{figure*}

\begin{figure*}[!htb]
\ContinuedFloat
\captionsetup{list=off,format=cont}
\centering
\includegraphics[]{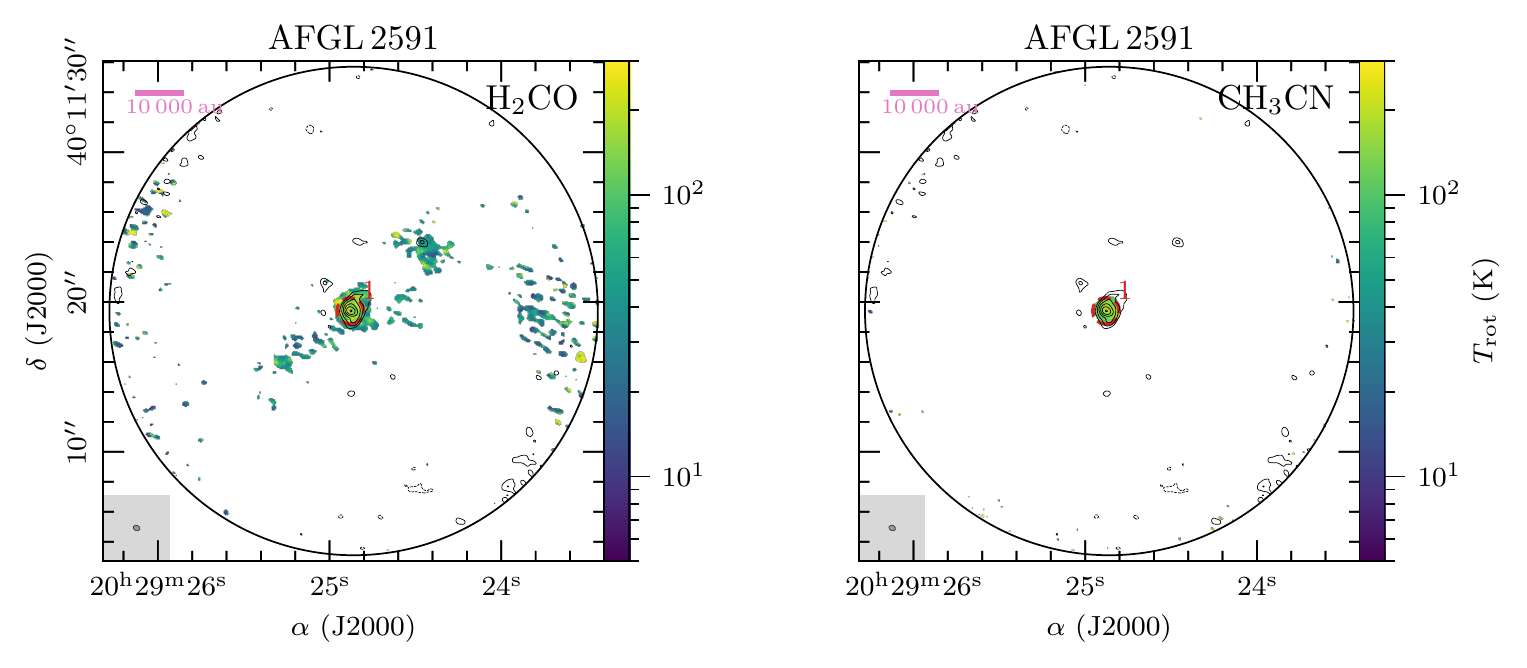}
\includegraphics[]{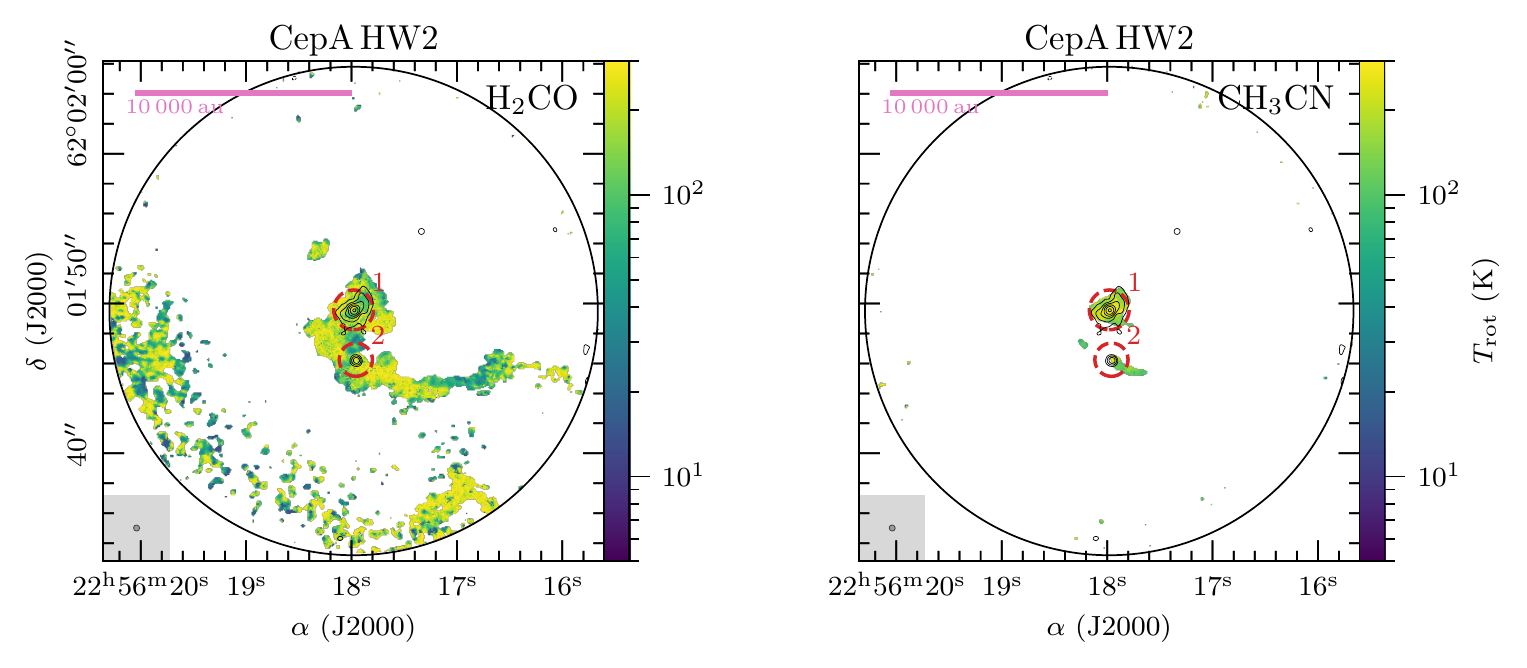}
\includegraphics[]{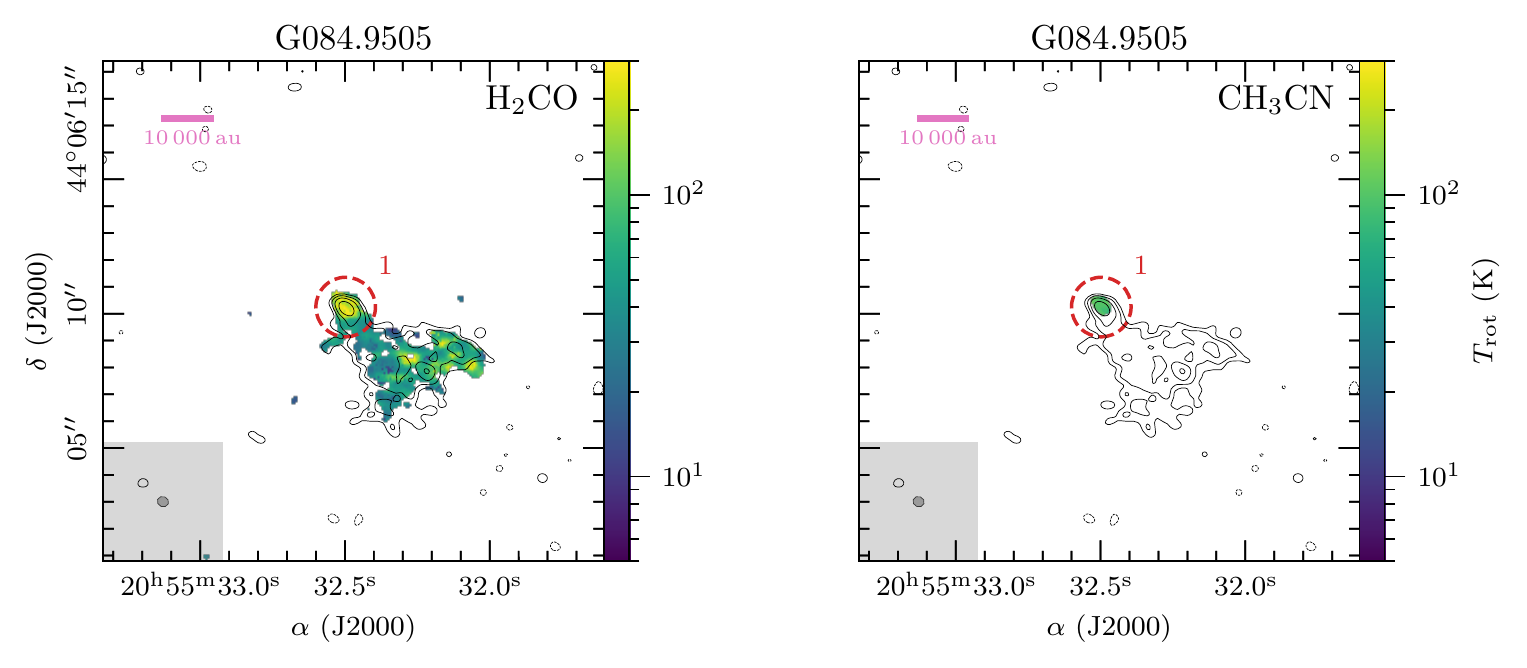}
\caption{Temperature maps derived with \texttt{XCLASS}. Each panel shows in color the temperature map (left: H$_2$CO, right: CH$_3$CN) and in black contours the 1.37\,mm continuum emission. The dashed black contours show the $-5\sigma_\mathrm{cont}$ emission and the solid black contours start at 5$\sigma_\mathrm{cont}$ with steps increasing by a factor of 2 (see Table \ref{tab:dataproducts} for values of $\sigma_\mathrm{cont}$ for each region). Each core is marked in red and the dashed red circle indicates the outer radius of the radial temperature fit (Sect. \ref{sec:temperaturestructure}). The beam size is shown in the bottom left corner in each panel. The pink bar in the top left corner indicates a linear spatial scale of 10\,000\,au. The primary beam size is indicated by a black circle and for regions with no extended H$_{2}$CO temperature map a smaller field of view is shown.}
\end{figure*}

\begin{figure*}[!htb]
\ContinuedFloat
\captionsetup{list=off,format=cont}
\centering
\includegraphics[]{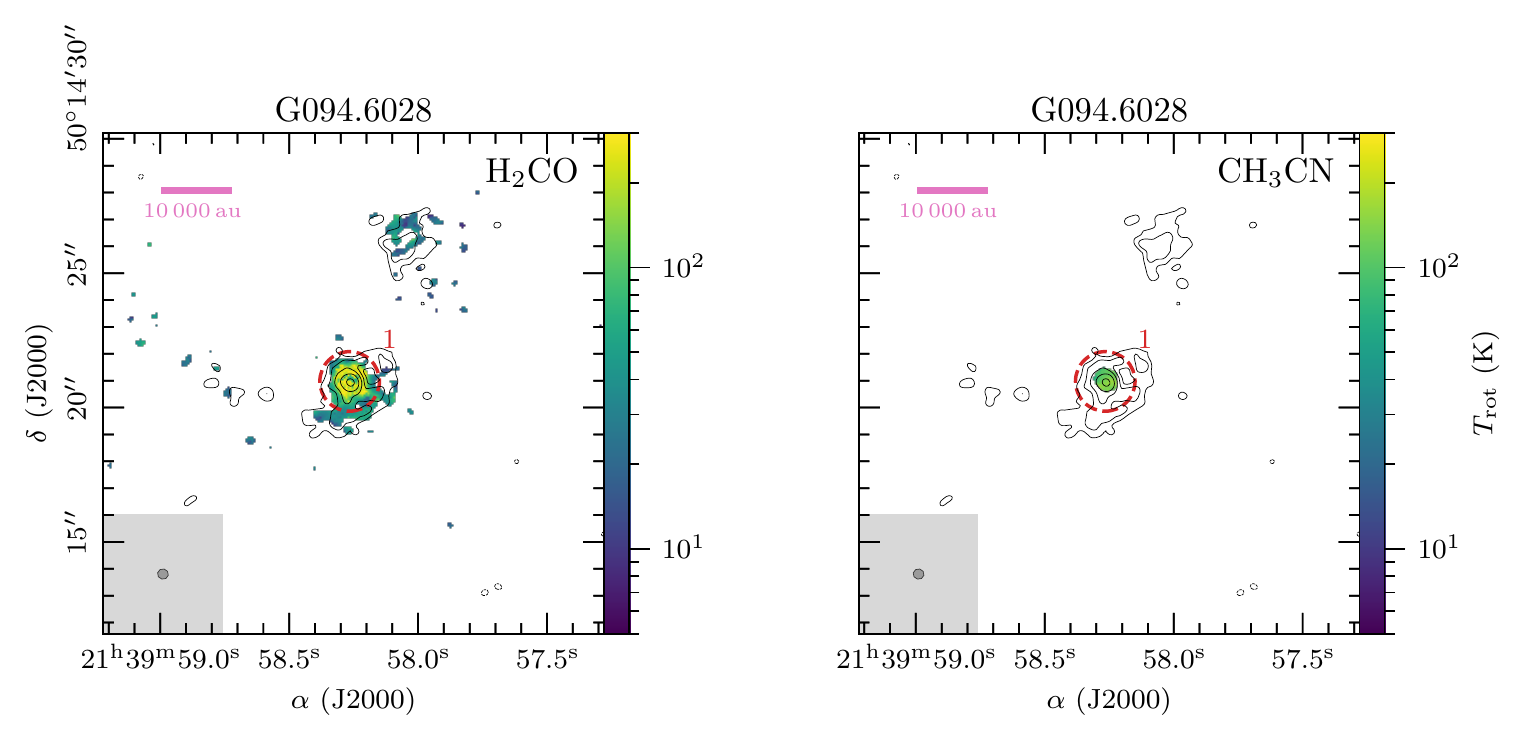}
\includegraphics[]{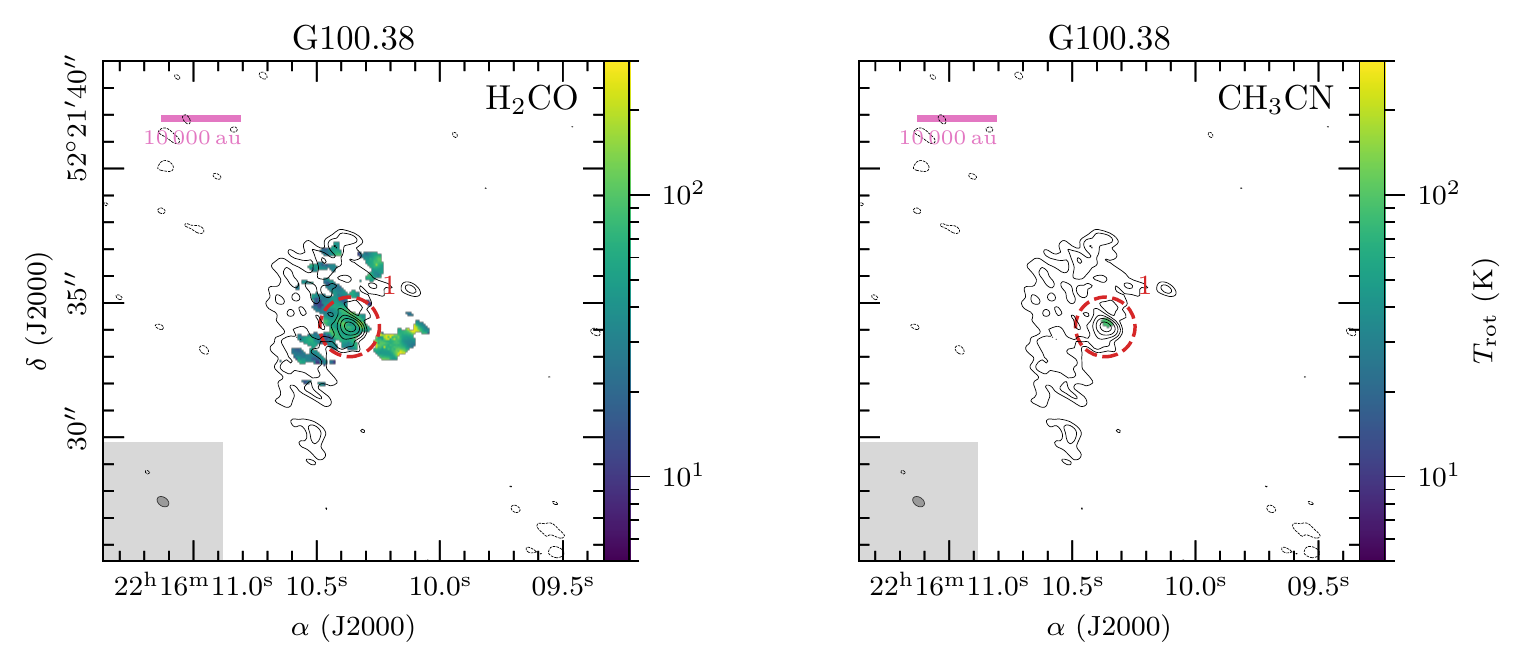}
\includegraphics[]{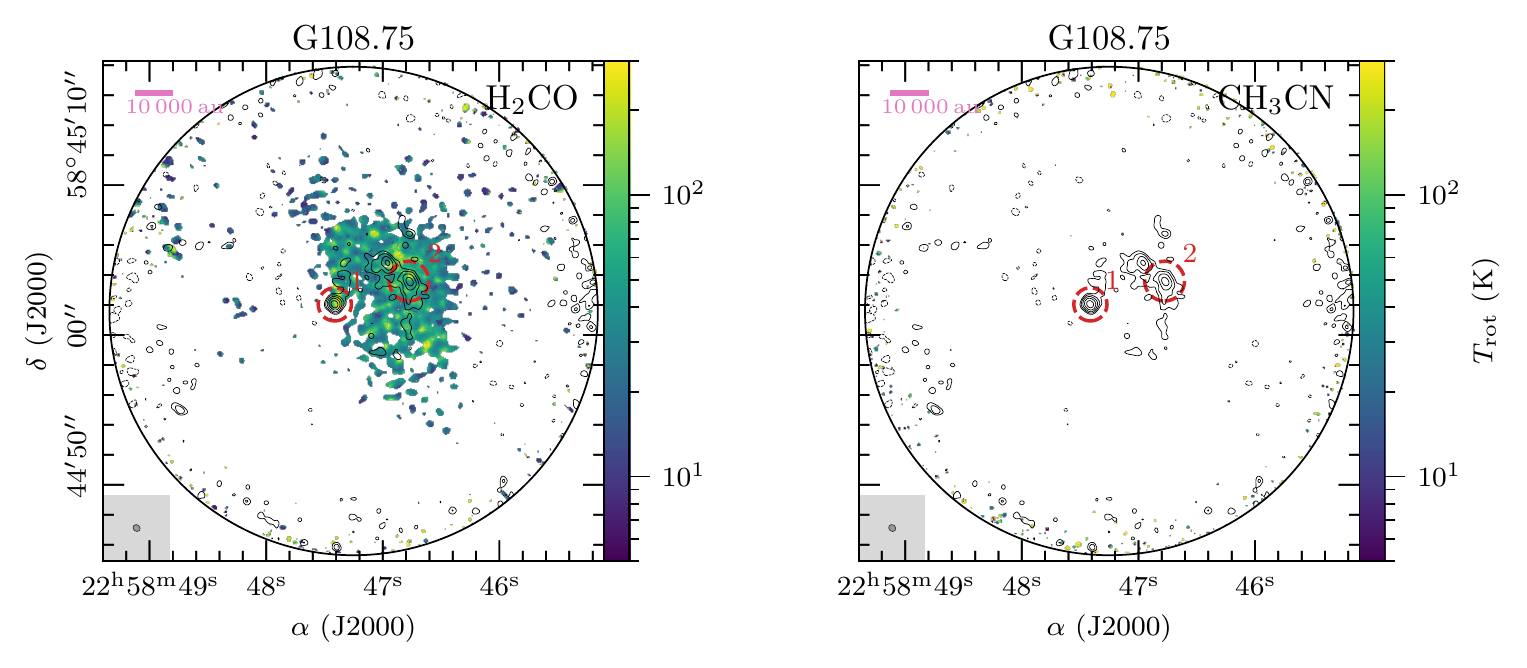}
\caption{Temperature maps derived with \texttt{XCLASS}. Each panel shows in color the temperature map (left: H$_2$CO, right: CH$_3$CN) and in black contours the 1.37\,mm continuum emission. The dashed black contours show the $-5\sigma_\mathrm{cont}$ emission and the solid black contours start at 5$\sigma_\mathrm{cont}$ with steps increasing by a factor of 2 (see Table \ref{tab:dataproducts} for values of $\sigma_\mathrm{cont}$ for each region). Each core is marked in red and the dashed red circle indicates the outer radius of the radial temperature fit (Sect. \ref{sec:temperaturestructure}). The beam size is shown in the bottom left corner in each panel. The pink bar in the top left corner indicates a linear spatial scale of 10\,000\,au. The primary beam size is indicated by a black circle and for regions with no extended H$_{2}$CO temperature map a smaller field of view is shown.}
\end{figure*}

\begin{figure*}[!htb]
\ContinuedFloat
\captionsetup{list=off,format=cont}
\centering
\includegraphics[]{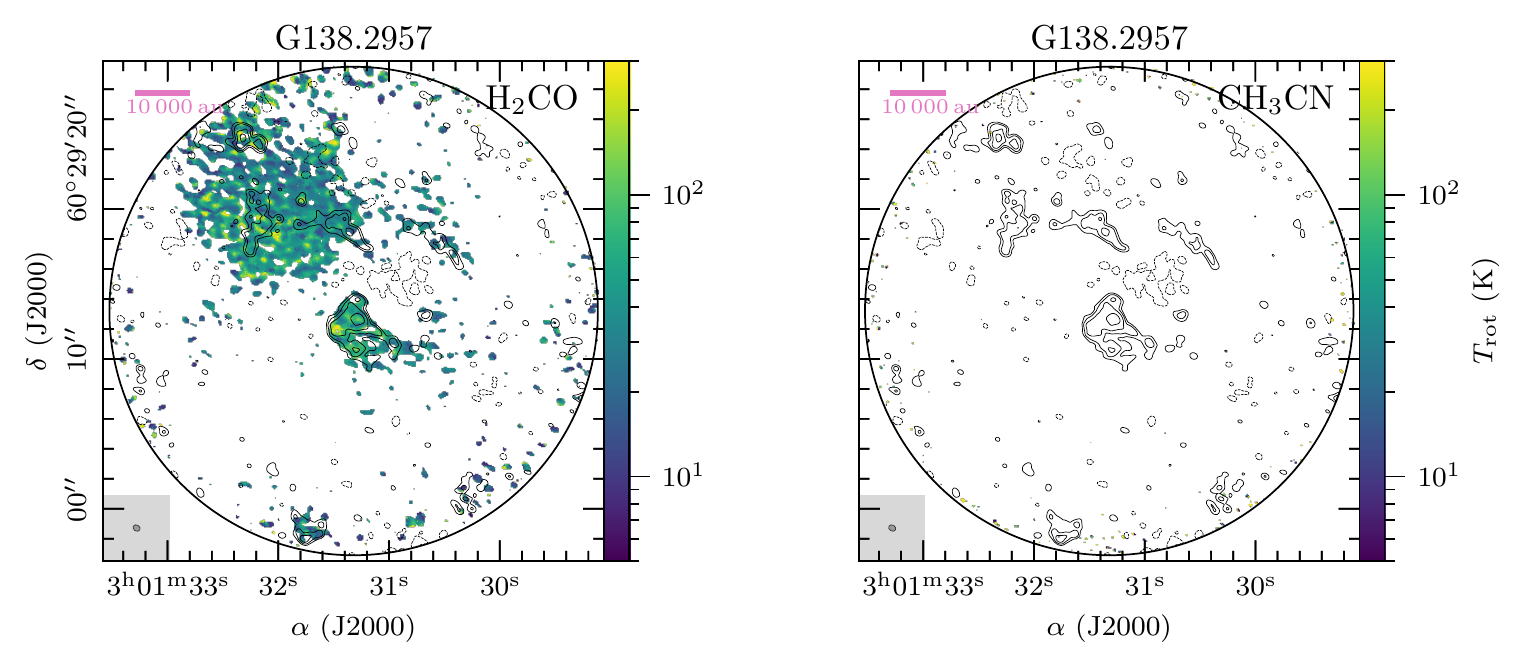}
\includegraphics[]{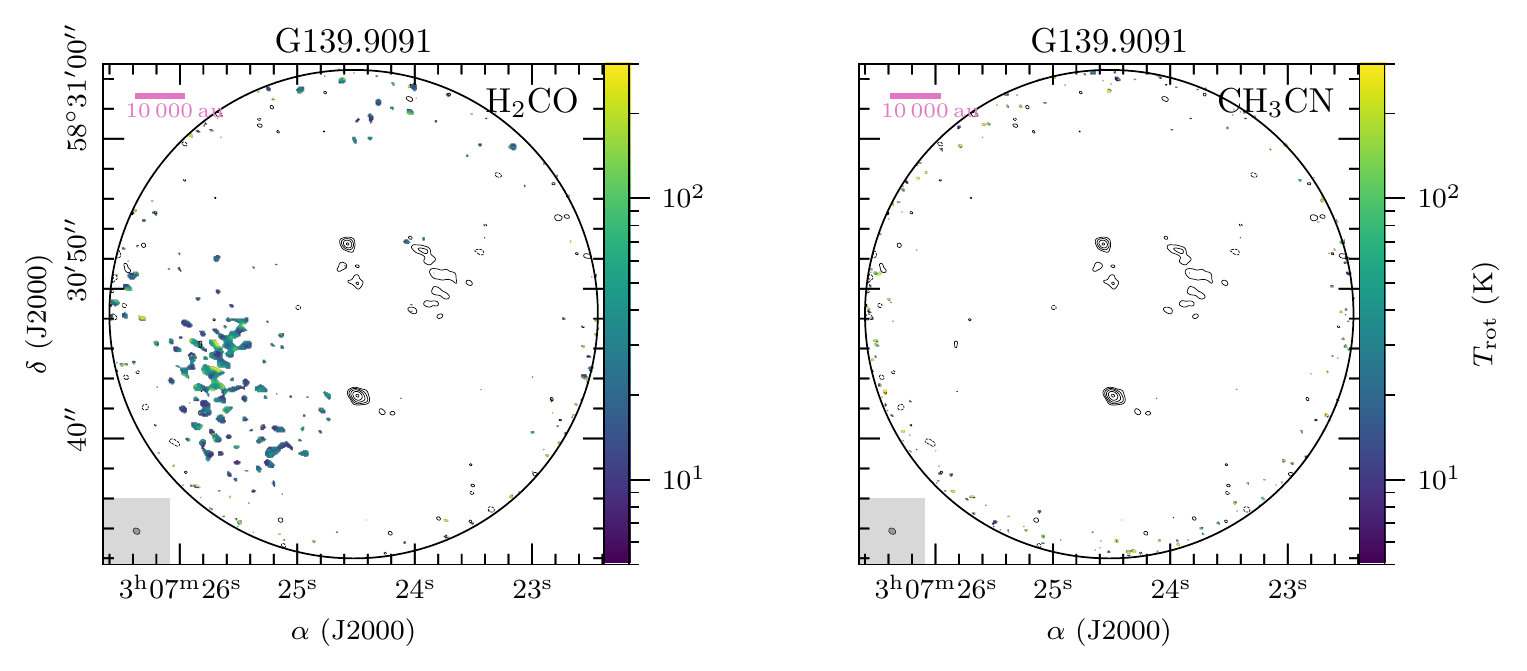}
\includegraphics[]{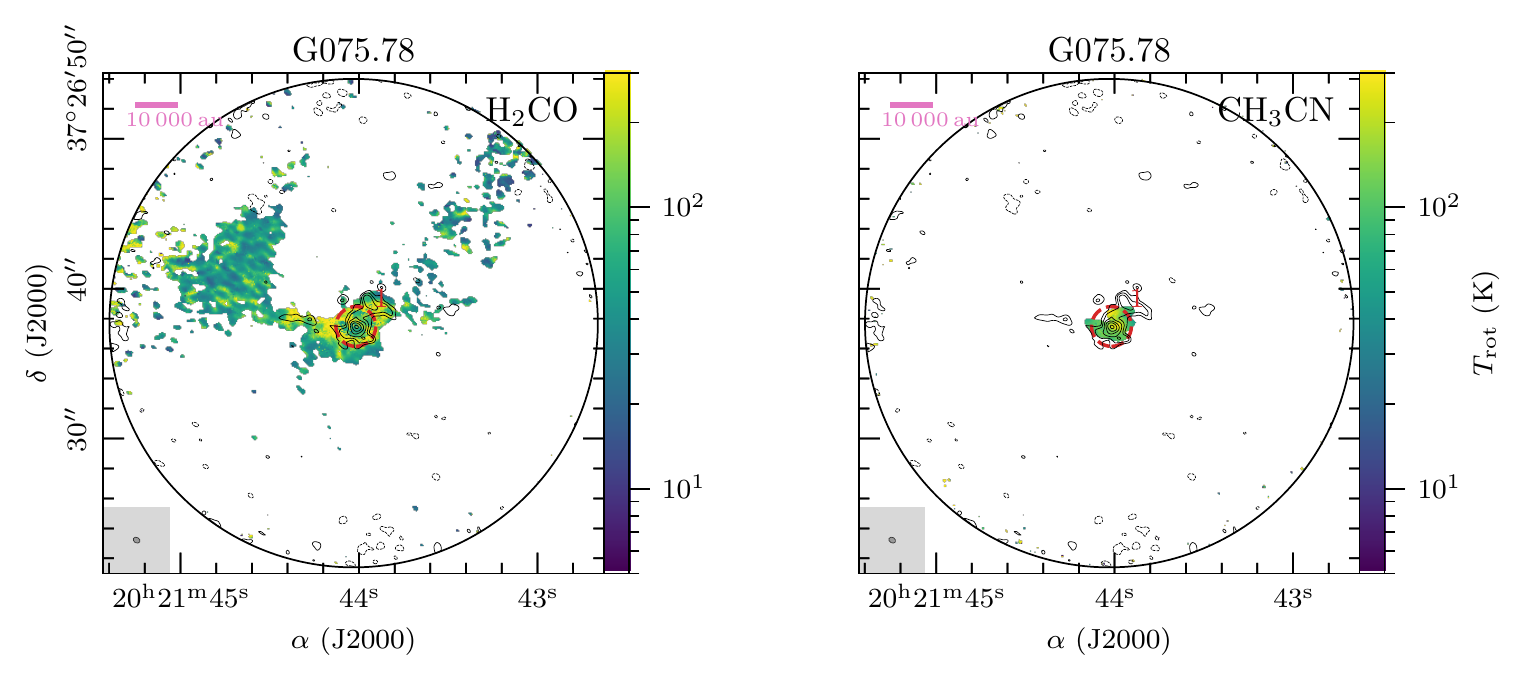}
\caption{Temperature maps derived with \texttt{XCLASS}. Each panel shows in color the temperature map (left: H$_2$CO, right: CH$_3$CN) and in black contours the 1.37\,mm continuum emission. The dashed black contours show the $-5\sigma_\mathrm{cont}$ emission and the solid black contours start at 5$\sigma_\mathrm{cont}$ with steps increasing by a factor of 2 (see Table \ref{tab:dataproducts} for values of $\sigma_\mathrm{cont}$ for each region). Each core is marked in red and the dashed red circle indicates the outer radius of the radial temperature fit (Sect. \ref{sec:temperaturestructure}). The beam size is shown in the bottom left corner in each panel. The pink bar in the top left corner indicates a linear spatial scale of 10\,000\,au. The primary beam size is indicated by a black circle and for regions with no extended H$_{2}$CO temperature map a smaller field of view is shown.}
\end{figure*}

\begin{figure*}[!htb]
\ContinuedFloat
\captionsetup{list=off,format=cont}
\centering
\includegraphics[]{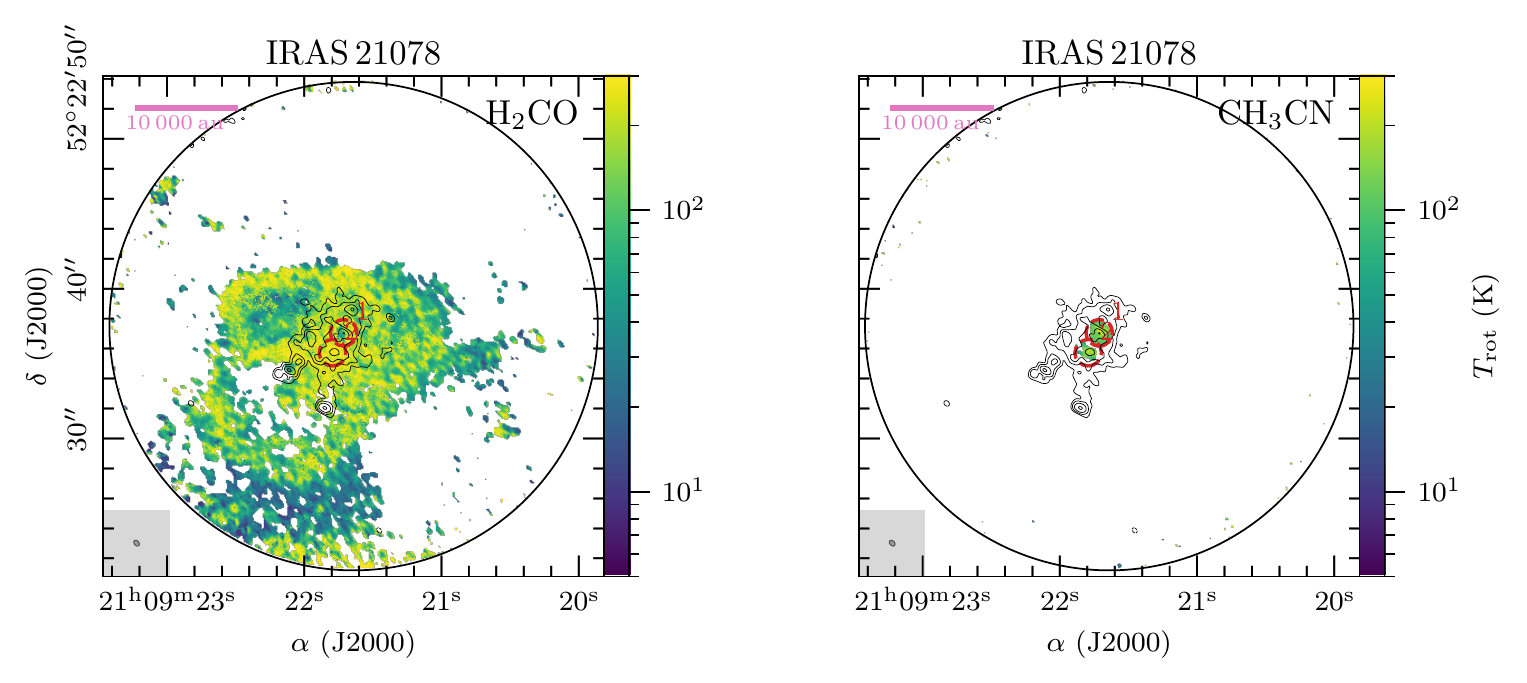}
\includegraphics[]{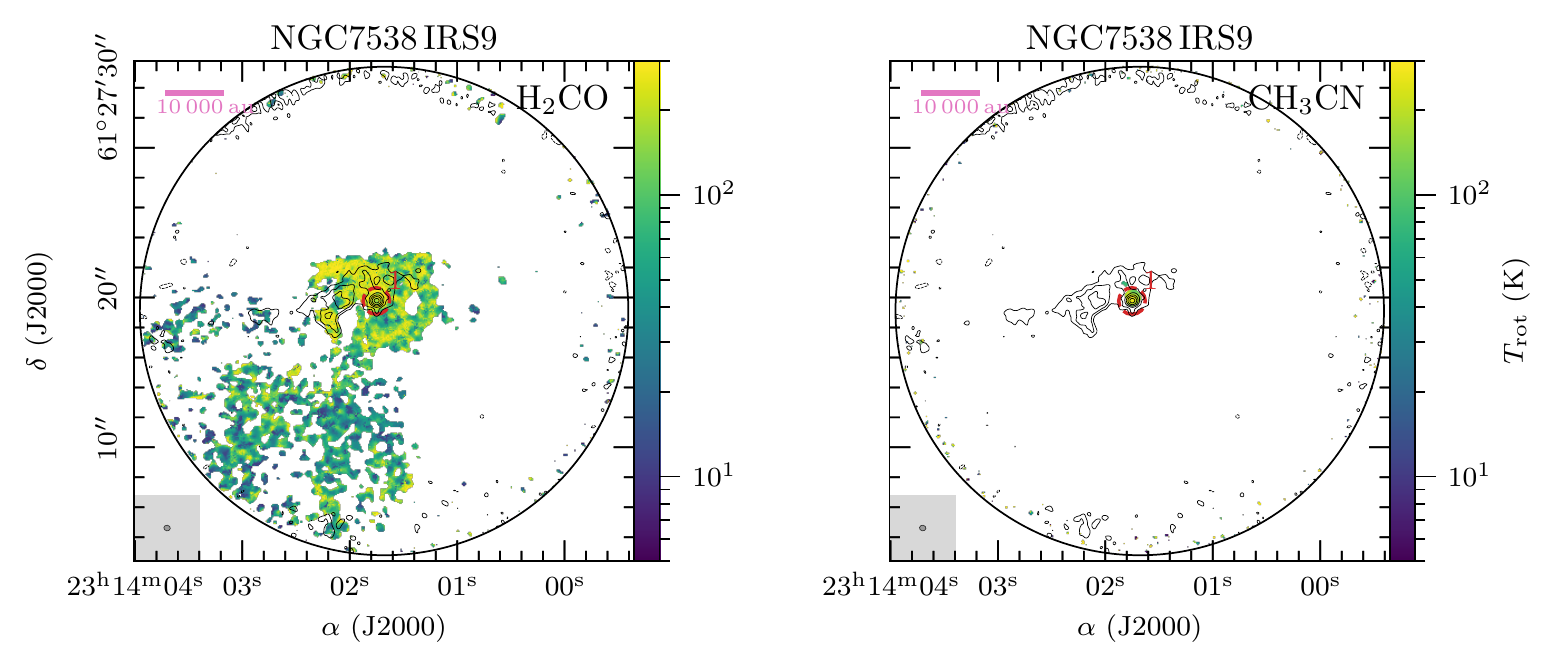}
\includegraphics[]{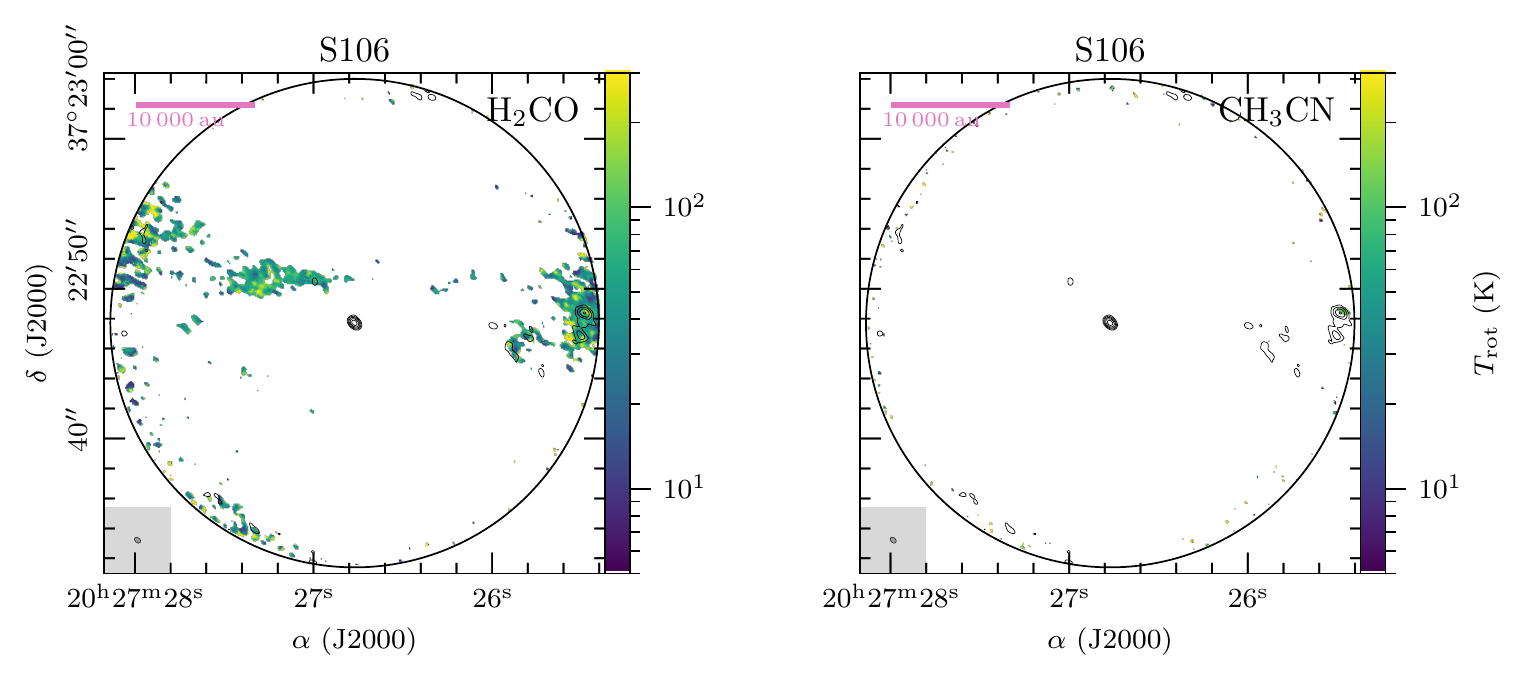}
\caption{Temperature maps derived with \texttt{XCLASS}. Each panel shows in color the temperature map (left: H$_2$CO, right: CH$_3$CN) and in black contours the 1.37\,mm continuum emission. The dashed black contours show the $-5\sigma_\mathrm{cont}$ emission and the solid black contours start at 5$\sigma_\mathrm{cont}$ with steps increasing by a factor of 2 (see Table \ref{tab:dataproducts} for values of $\sigma_\mathrm{cont}$ for each region). Each core is marked in red and the dashed red circle indicates the outer radius of the radial temperature fit (Sect. \ref{sec:temperaturestructure}). The beam size is shown in the bottom left corner in each panel. The pink bar in the top left corner indicates a linear spatial scale of 10\,000\,au. The primary beam size is indicated by a black circle and for regions with no extended H$_{2}$CO temperature map a smaller field of view is shown.}
\end{figure*}

\begin{figure*}[!htb]
\ContinuedFloat
\captionsetup{list=off,format=cont}
\centering
\includegraphics[]{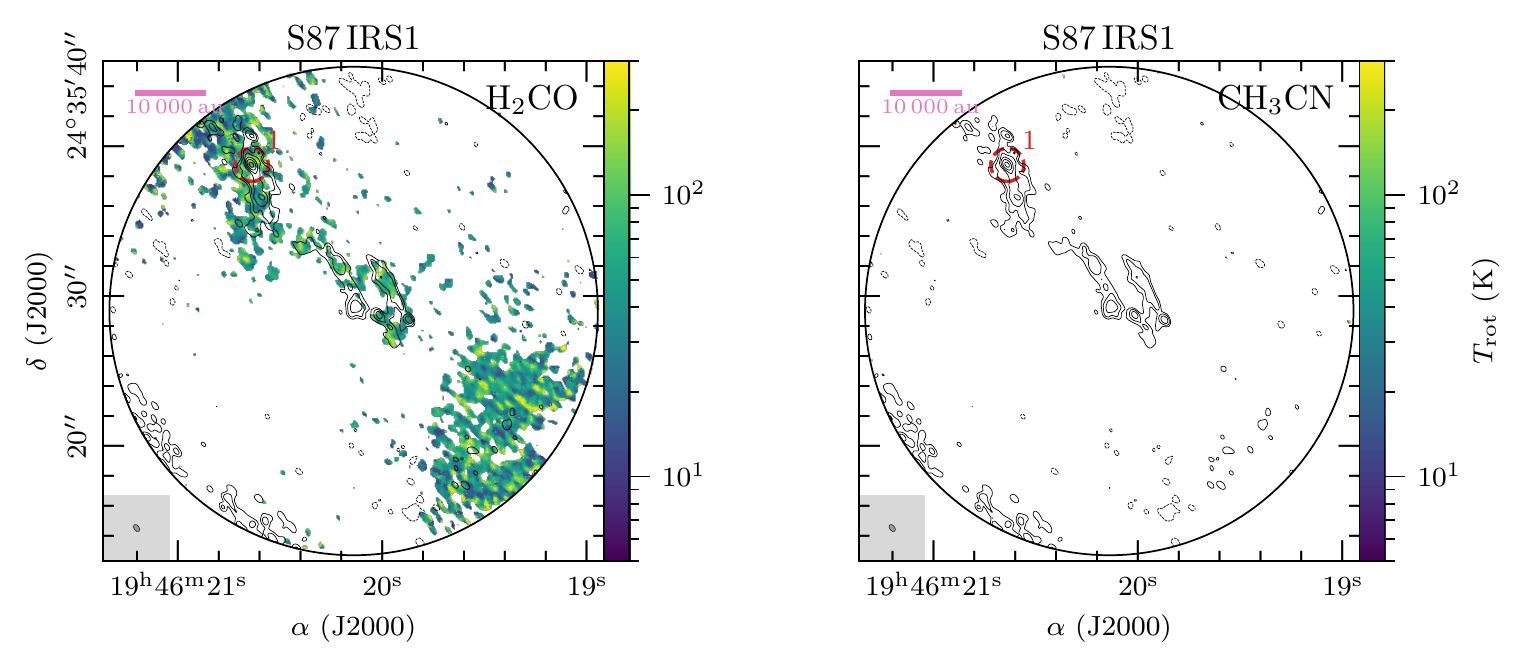}
\includegraphics[]{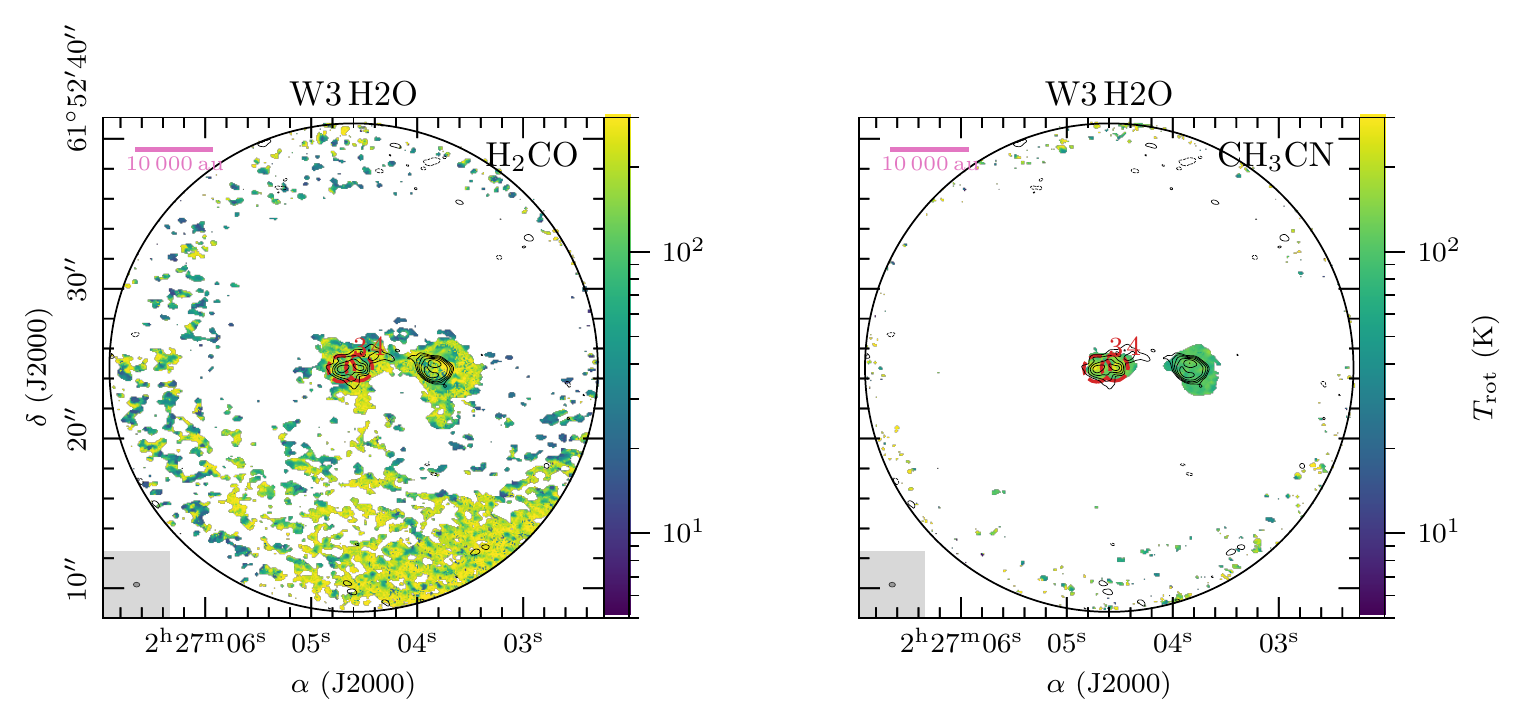}
\includegraphics[]{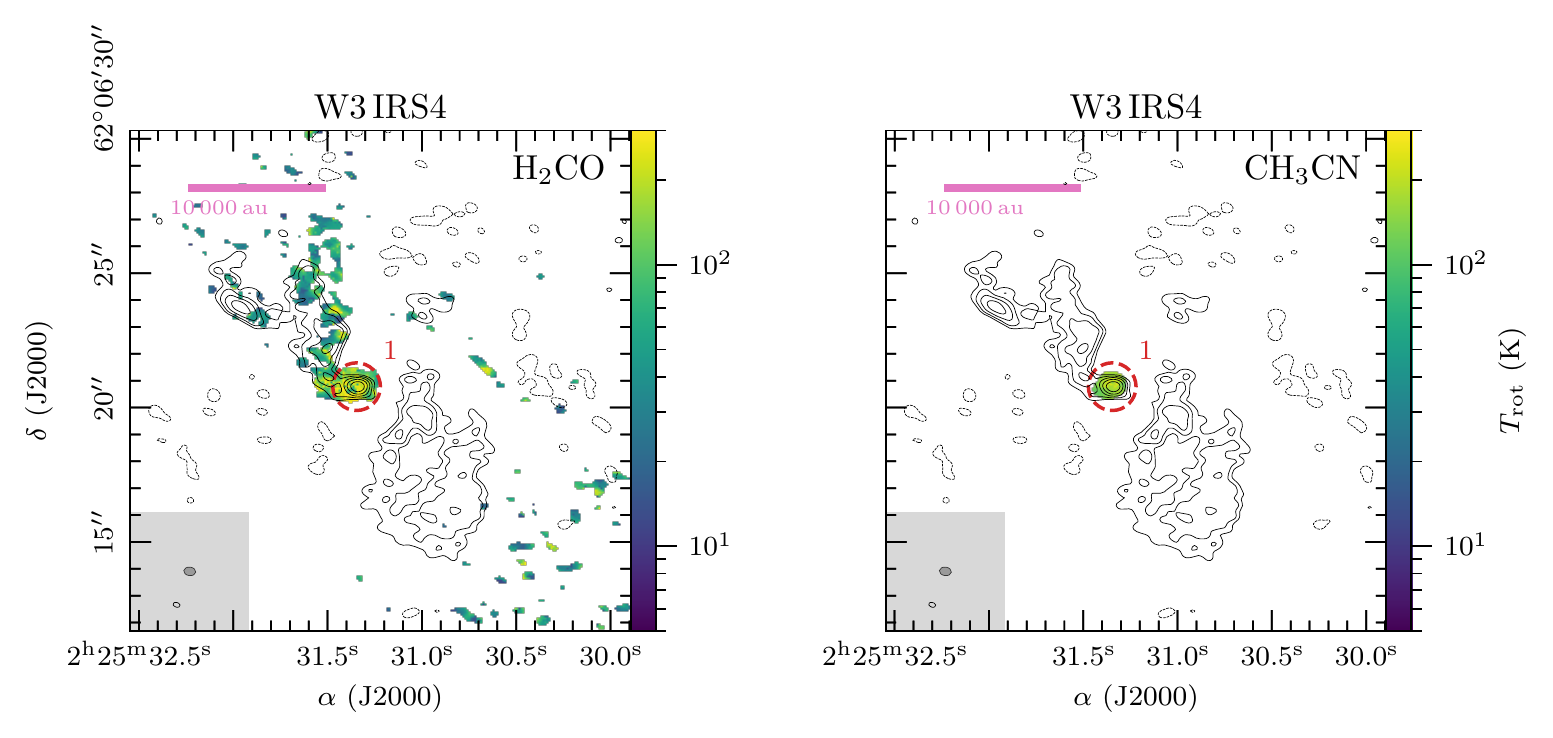}
\caption{Temperature maps derived with \texttt{XCLASS}. Each panel shows in color the temperature map (left: H$_2$CO, right: CH$_3$CN) and in black contours the 1.37\,mm continuum emission. The dashed black contours show the $-5\sigma_\mathrm{cont}$ emission and the solid black contours start at 5$\sigma_\mathrm{cont}$ with steps increasing by a factor of 2 (see Table \ref{tab:dataproducts} for values of $\sigma_\mathrm{cont}$ for each region). Each core is marked in red and the dashed red circle indicates the outer radius of the radial temperature fit (Sect. \ref{sec:temperaturestructure}). The beam size is shown in the bottom left corner in each panel. The pink bar in the top left corner indicates a linear spatial scale of 10\,000\,au. The primary beam size is indicated by a black circle and for regions with no extended H$_{2}$CO temperature map a smaller field of view is shown.}
\end{figure*}

\section{Abundance histograms}

\begin{figure*}
\centering
\includegraphics[]{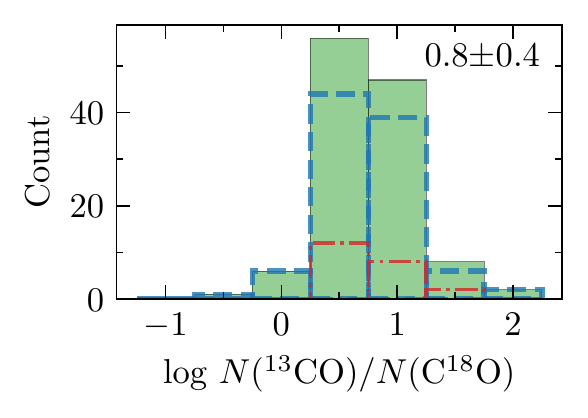}
\includegraphics[]{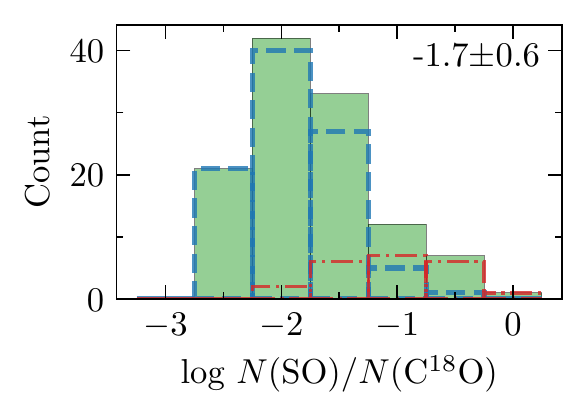}
\includegraphics[]{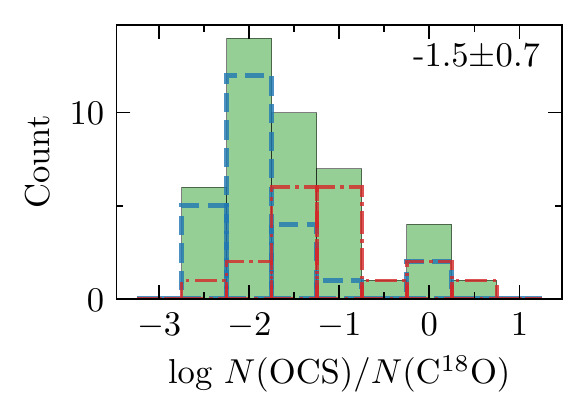}
\includegraphics[]{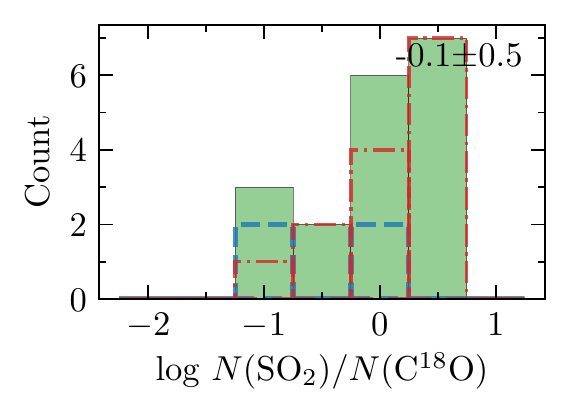}
\includegraphics[]{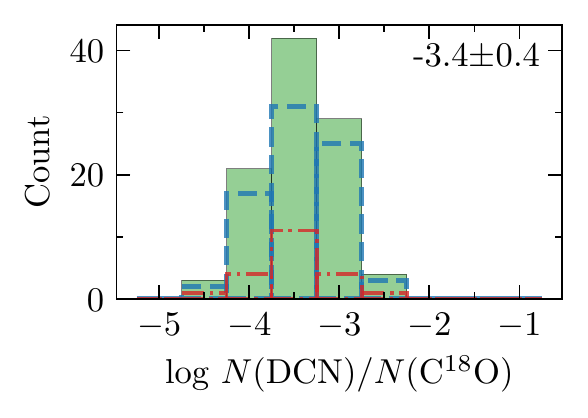}
\includegraphics[]{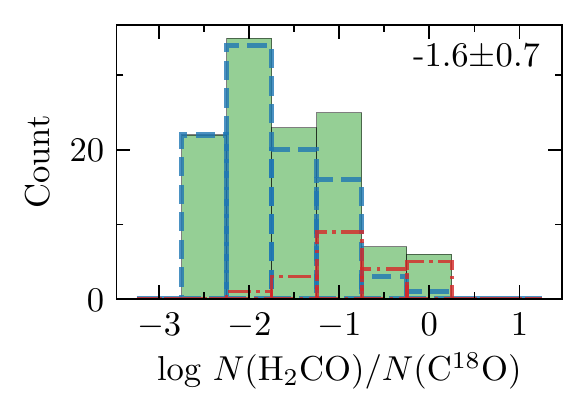}
\includegraphics[]{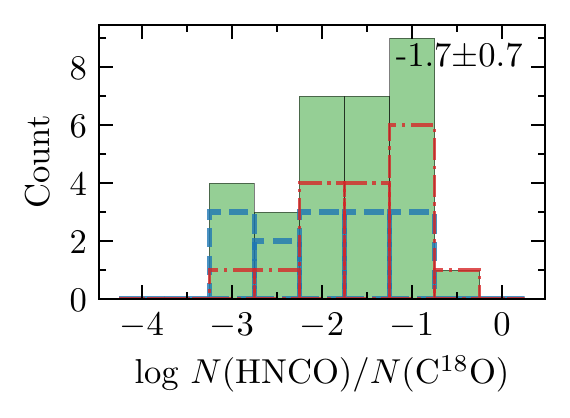}
\includegraphics[]{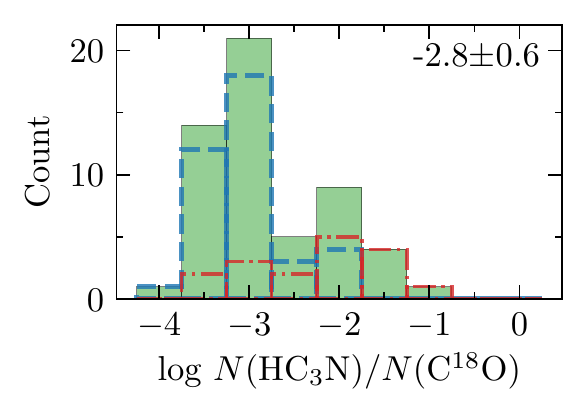}
\includegraphics[]{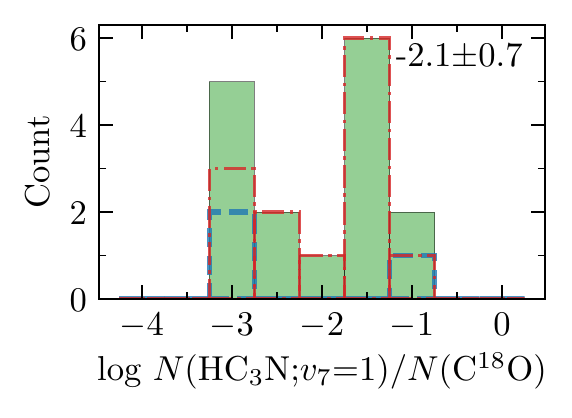}
\includegraphics[]{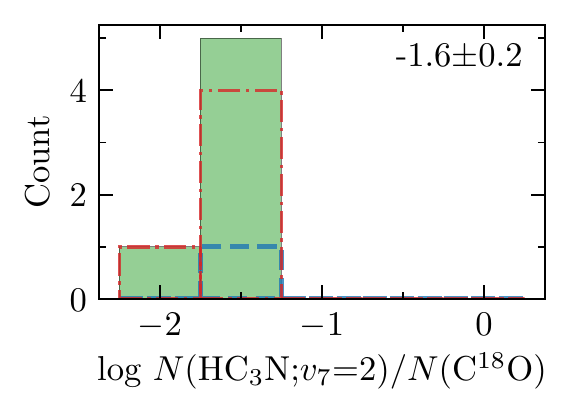}
\includegraphics[]{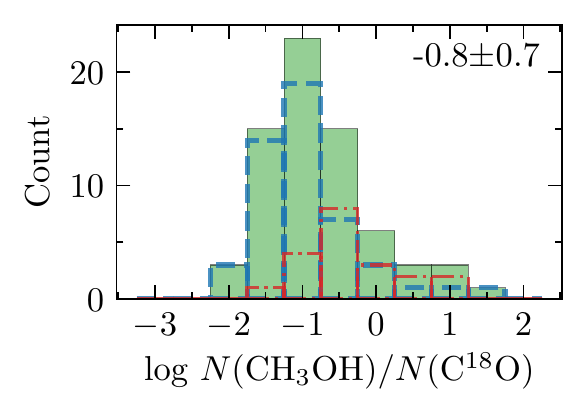}
\includegraphics[]{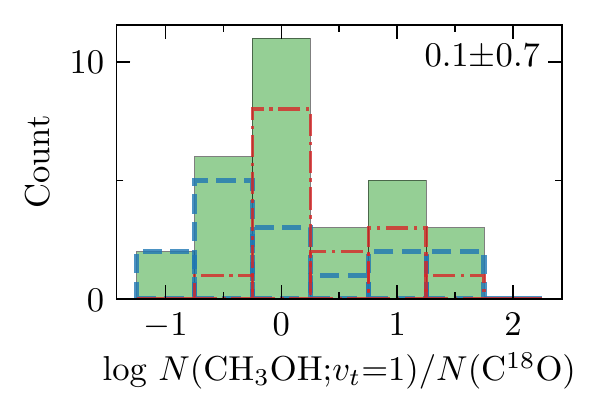}
\includegraphics[]{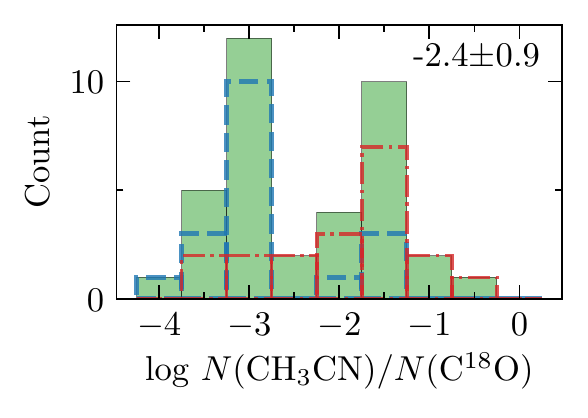}
\caption{Abundance ratio histograms. Abundance ratio histograms of all 120 positions are shown in green (upper limits are not included). Separate abundance ratio histograms of the core and non core positions are indicated by the dash-dotted red and dashed blue lines, respectively.}
\label{fig:abundratiohisto}
\end{figure*}

	Figure \ref{fig:abundratiohisto} shows a histogram of the abundance ratio relative to $N$(C$^{18}$O) for each molecule complementary to the column density histograms discussed in Sect. \ref{sec:XCLASSfitting}.
	
\section{Column density results}

	Table \ref{tab:XCLASSresults1}, \ref{tab:XCLASSresults2}, and \ref{tab:XCLASSresults3} show the derived column densities for all molecules fitted with \texttt{XCLASS} (Sect. \ref{sec:XCLASSfitting}). The upper and lower errors are estimated using the MCMC error estimation algorithm in \texttt{XCLASS}. If the fitting results do not match our defined criteria for a good fit (further explained in Sect. \ref{sec:XCLASSfitting}), the results are considered as upper limits.
	
\begin{longtable}{llllll}
\caption{Molecular column densities derived with \texttt{XCLASS}.}
\label{tab:XCLASSresults1}
\\
\hline\hline
Position & $^{13}$CO & C$^{18}$O & SO & OCS & SO$_{2}$ \\
\hline
\endfirsthead
\caption{continued.}\\
\hline\hline
Position & $^{13}$CO & C$^{18}$O & SO & OCS & SO$_{2}$ \\
\hline
\endhead
\hline
Notes. a(b) = a$\times$10$^{\mathrm{b}}$.
\endfoot
\hline
IRAS\,23033 1 & 8.4(17)$^{+9.1(16)}_{-1.2(17)}$ & 5.3(16)$^{+5.8(16)}_{-1.7(16)}$ & 1.7(15)$^{+1.7(15)}_{-3.5(14)}$ & $<$9.5(14) & $<$1.1(16)\\ 
IRAS\,23033 2 & 7.6(17)$^{+1.8(17)}_{-8.4(16)}$ & 1.9(17)$^{+1.0(17)}_{-3.9(16)}$ & 2.5(16)$^{+1.1(16)}_{-4.6(15)}$ & 9.5(15)$^{+5.0(15)}_{-1.3(15)}$ & 3.4(16)$^{+3.1(16)}_{-1.4(16)}$\\ 
IRAS\,23033 3 & 1.2(18)$^{+1.7(17)}_{-1.0(17)}$ & 2.6(17)$^{+5.4(16)}_{-4.5(16)}$ & 2.8(16)$^{+1.6(15)}_{-1.5(15)}$ & 2.4(16)$^{+4.3(15)}_{-3.7(15)}$ & 4.7(16)$^{+3.2(16)}_{-1.4(16)}$\\ 
IRAS\,23033 4 & 8.2(17)$^{+2.8(17)}_{-9.6(16)}$ & 2.4(17)$^{+6.4(16)}_{-6.0(16)}$ & 2.3(15)$^{+2.1(15)}_{-8.0(14)}$ & $<$1.8(14) & $<$1.4(16)\\ 
IRAS\,23033 5 & 8.0(17)$^{+2.0(17)}_{-7.4(16)}$ & 1.3(17)$^{+1.8(17)}_{-3.2(16)}$ & 6.3(14)$^{+1.2(14)}_{-5.0(14)}$ & $<$3.9(14) & $<$1.3(16)\\ 
\hline 
IRAS\,23151 1 & 1.8(18)$^{+2.5(17)}_{-1.5(17)}$ & 6.9(17)$^{+1.0(17)}_{-7.4(16)}$ & 4.0(16)$^{+1.5(15)}_{-5.5(15)}$ & 1.5(16)$^{+3.3(15)}_{-1.5(15)}$ & 5.1(17)$^{+3.8(16)}_{-1.9(16)}$\\ 
IRAS\,23151 2 & 6.4(17)$^{+6.9(16)}_{-1.1(17)}$ & 2.1(17)$^{+7.5(16)}_{-3.2(16)}$ & 7.9(15)$^{+1.8(15)}_{-1.2(15)}$ & 7.6(14)$^{+8.2(13)}_{-6.5(14)}$ & $<$1.8(16)\\ 
IRAS\,23151 3 & 6.7(17)$^{+1.2(17)}_{-1.0(17)}$ & 4.0(17)$^{+5.9(16)}_{-6.2(16)}$ & 8.3(15)$^{+5.2(15)}_{-1.9(15)}$ & 2.0(15)$^{+1.1(16)}_{-7.5(14)}$ & $<$1.5(16)\\ 
IRAS\,23151 4 & 2.8(17)$^{+1.1(17)}_{-5.6(16)}$ & 2.3(17)$^{+9.5(16)}_{-3.2(16)}$ & 3.3(15)$^{+1.1(15)}_{-3.8(14)}$ & 7.1(14)$^{+1.5(14)}_{-5.8(14)}$ & $<$2.7(16)\\ 
IRAS\,23151 5 & 9.5(18)$^{+4.8(16)}_{-8.9(16)}$ & 1.3(17)$^{+5.1(16)}_{-4.4(16)}$ & 4.3(15)$^{+1.2(15)}_{-9.4(14)}$ & $<$3.7(14) & $<$1.5(16)\\ 
\hline 
IRAS\,23385 1 & 9.3(17)$^{+2.3(17)}_{-1.5(17)}$ & 2.6(17)$^{+1.4(17)}_{-3.5(16)}$ & 1.1(16)$^{+3.8(15)}_{-1.6(15)}$ & 1.2(16)$^{+1.7(15)}_{-2.1(15)}$ & $<$1.5(16)\\ 
IRAS\,23385 2 & 6.2(17)$^{+2.2(17)}_{-1.1(17)}$ & 2.1(17)$^{+1.6(17)}_{-2.8(16)}$ & 2.2(15)$^{+1.9(15)}_{-7.5(14)}$ & 1.3(15)$^{+6.2(15)}_{-6.9(14)}$ & $<$1.7(18)\\ 
IRAS\,23385 3 & 4.3(17)$^{+5.9(16)}_{-1.1(17)}$ & 2.3(17)$^{+1.5(17)}_{-3.5(16)}$ & 5.6(15)$^{+2.1(15)}_{-2.4(15)}$ & 9.6(14)$^{+1.9(15)}_{-4.2(14)}$ & $<$2.0(16)\\ 
IRAS\,23385 4 & 1.0(18)$^{+1.9(17)}_{-1.9(17)}$ & 9.5(16)$^{+1.2(17)}_{-2.8(16)}$ & 4.6(15)$^{+1.8(15)}_{-1.3(15)}$ & 8.9(14)$^{+1.8(14)}_{-7.2(14)}$ & $<$1.6(16)\\ 
IRAS\,23385 5 & 5.8(17)$^{+2.1(17)}_{-1.6(17)}$ & 7.1(16)$^{+8.2(16)}_{-3.3(16)}$ & 4.3(15)$^{+1.0(15)}_{-1.4(15)}$ & $<$6.1(14) & $<$1.2(16)\\ 
IRAS\,23385 6 & 6.5(17)$^{+1.5(17)}_{-1.7(17)}$ & 1.2(17)$^{+8.6(16)}_{-4.6(16)}$ & 3.5(15)$^{+1.7(15)}_{-8.0(14)}$ & $<$2.6(14) & $<$1.1(16)\\ 
IRAS\,23385 7 & 1.3(18)$^{+3.3(17)}_{-3.7(17)}$ & 1.2(17)$^{+7.4(16)}_{-5.3(16)}$ & 1.4(16)$^{+6.9(15)}_{-7.4(15)}$ & $<$8.8(14) & $<$1.0(16)\\ 
IRAS\,23385 8 & 3.8(18)$^{+5.4(17)}_{-6.8(17)}$ & 8.4(16)$^{+5.0(16)}_{-4.8(16)}$ & 1.3(15)$^{+4.9(14)}_{-3.6(14)}$ & 1.6(15)$^{+1.2(16)}_{-6.7(14)}$ & $<$3.8(16)\\ 
\hline 
AFGL\,2591 1 & 4.6(18)$^{+3.1(17)}_{-2.4(17)}$ & 4.1(17)$^{+6.0(16)}_{-3.6(16)}$ & 1.1(17)$^{+3.8(15)}_{-6.8(15)}$ & 4.1(16)$^{+5.7(15)}_{-4.2(15)}$ & 1.3(18)$^{+2.2(17)}_{-1.3(17)}$\\ 
AFGL\,2591 2 & 6.6(17)$^{+1.5(17)}_{-6.8(16)}$ & 6.0(17)$^{+9.7(16)}_{-5.9(16)}$ & 1.3(15)$^{+7.1(14)}_{-6.1(14)}$ & $<$1.6(15) & $<$1.3(16)\\ 
AFGL\,2591 3 & 6.3(17)$^{+2.0(17)}_{-1.0(17)}$ & 2.7(17)$^{+3.1(16)}_{-4.9(16)}$ & 4.8(15)$^{+2.0(16)}_{-2.8(15)}$ & $<$1.7(17) & $<$2.4(16)\\ 
AFGL\,2591 4 & 9.8(17)$^{+1.7(17)}_{-9.9(16)}$ & 1.9(17)$^{+8.3(16)}_{-2.2(16)}$ & 6.3(15)$^{+1.6(15)}_{-1.2(15)}$ & $<$1.5(15) & $<$2.1(16)\\ 
\hline 
CepA\,HW2 1 & 1.1(18)$^{+4.8(17)}_{-2.4(17)}$ & 5.6(17)$^{+1.4(17)}_{-6.2(16)}$ & 1.4(17)$^{+2.3(16)}_{-1.9(16)}$ & 3.6(16)$^{+9.3(15)}_{-2.0(15)}$ & 1.3(18)$^{+1.2(17)}_{-1.1(17)}$\\ 
CepA\,HW2 2 & 1.1(18)$^{+1.4(17)}_{-1.5(17)}$ & 5.6(17)$^{+4.8(16)}_{-9.9(16)}$ & 5.1(16)$^{+2.1(16)}_{-5.5(15)}$ & 1.2(16)$^{+2.3(16)}_{-2.1(15)}$ & 3.3(17)$^{+1.2(17)}_{-5.2(16)}$\\ 
CepA\,HW2 3 & 2.2(18)$^{+2.6(17)}_{-4.1(17)}$ & 7.5(17)$^{+5.5(16)}_{-5.0(16)}$ & 3.0(16)$^{+6.8(15)}_{-4.6(15)}$ & 2.0(16)$^{+5.5(15)}_{-2.5(15)}$ & 9.1(16)$^{+1.5(16)}_{-1.3(16)}$\\ 
CepA\,HW2 4 & 9.9(18)$^{+2.3(16)}_{-2.3(16)}$ & 1.2(17)$^{+5.3(16)}_{-5.9(16)}$ & $<$1.1(14) & $<$1.0(14) & $<$1.7(16)\\ 
CepA\,HW2 5 & 9.9(17)$^{+1.3(17)}_{-1.1(17)}$ & 2.9(17)$^{+1.1(17)}_{-4.6(16)}$ & 1.2(16)$^{+1.4(15)}_{-2.7(15)}$ & $<$7.5(14) & $<$1.0(16)\\ 
\hline 
G084.9505 1 & 5.9(17)$^{+2.3(17)}_{-1.2(17)}$ & 1.1(17)$^{+9.3(16)}_{-3.1(16)}$ & 2.7(15)$^{+2.9(14)}_{-7.2(14)}$ & 3.4(15)$^{+2.0(15)}_{-7.4(14)}$ & $<$8.2(17)\\ 
G084.9505 2 & 2.5(17)$^{+1.3(17)}_{-1.1(16)}$ & 4.1(16)$^{+1.4(16)}_{-2.9(16)}$ & 1.0(15)$^{+5.9(14)}_{-3.2(14)}$ & $<$4.7(14) & $<$1.2(16)\\ 
G084.9505 3 & 1.1(17)$^{+1.2(17)}_{-1.4(16)}$ & 5.1(16)$^{+4.7(16)}_{-2.5(16)}$ & 1.4(15)$^{+1.2(15)}_{-4.5(14)}$ & $<$1.5(14) & $<$1.0(16)\\ 
G084.9505 4 & 1.5(17)$^{+4.3(16)}_{-4.7(16)}$ & 2.6(16)$^{+2.5(16)}_{-4.1(15)}$ & 1.6(15)$^{+6.7(14)}_{-7.6(14)}$ & $<$1.9(14) & $<$3.5(18)\\ 
G084.9505 5 & 6.8(17)$^{+4.1(17)}_{-1.5(17)}$ & 9.3(16)$^{+1.6(17)}_{-2.7(16)}$ & 1.8(15)$^{+6.2(14)}_{-1.2(15)}$ & $<$2.3(14) & $<$1.2(16)\\ 
G084.9505 6 & 2.2(17)$^{+1.7(17)}_{-2.1(16)}$ & 5.3(16)$^{+2.4(16)}_{-2.7(16)}$ & 2.0(15)$^{+1.5(15)}_{-1.2(15)}$ & 8.9(14)$^{+3.0(15)}_{-4.2(14)}$ & $<$4.0(16)\\ 
G084.9505 7 & 6.7(17)$^{+4.0(17)}_{-1.8(17)}$ & 1.1(17)$^{+7.4(16)}_{-5.2(16)}$ & 2.0(15)$^{+2.5(15)}_{-9.6(14)}$ & 8.3(14)$^{+1.7(15)}_{-4.9(14)}$ & $<$1.0(16)\\ 
G084.9505 8 & 4.1(17)$^{+1.4(17)}_{-1.2(17)}$ & 4.9(16)$^{+4.3(16)}_{-2.3(16)}$ & 4.2(14)$^{+1.7(14)}_{-3.1(14)}$ & $<$5.5(14) & $<$3.0(16)\\ 
\hline 
G094.6028 1 & 1.5(18)$^{+3.5(17)}_{-1.0(17)}$ & 3.1(16)$^{+1.8(16)}_{-9.5(15)}$ & 2.7(16)$^{+4.0(15)}_{-3.5(15)}$ & 3.2(16)$^{+8.4(15)}_{-6.6(15)}$ & 6.5(16)$^{+2.9(16)}_{-2.8(16)}$\\ 
G094.6028 2 & 3.4(17)$^{+1.5(17)}_{-6.0(16)}$ & 6.4(16)$^{+1.9(16)}_{-4.1(16)}$ & 1.0(15)$^{+1.0(14)}_{-8.4(14)}$ & $<$2.9(14) & $<$1.3(16)\\ 
G094.6028 3 & 1.1(18)$^{+2.3(17)}_{-4.9(17)}$ & 1.2(17)$^{+8.0(16)}_{-4.1(16)}$ & 4.8(14)$^{+1.2(15)}_{-1.6(14)}$ & $<$1.3(14) & $<$1.3(16)\\ 
G094.6028 4 & 4.0(17)$^{+1.5(17)}_{-6.2(16)}$ & 6.2(16)$^{+4.0(16)}_{-2.9(16)}$ & 1.3(15)$^{+4.6(14)}_{-6.7(14)}$ & $<$1.3(15) & $<$1.2(16)\\ 
G094.6028 5 & 4.7(17)$^{+3.6(16)}_{-1.2(17)}$ & 5.3(16)$^{+6.4(16)}_{-2.5(16)}$ & 2.8(14)$^{+4.1(14)}_{-4.1(13)}$ & $<$4.0(14) & $<$1.0(16)\\ 
G094.6028 6 & 1.3(18)$^{+1.4(17)}_{-2.8(17)}$ & 8.2(16)$^{+3.8(16)}_{-2.7(16)}$ & 1.0(15)$^{+5.9(14)}_{-3.8(14)}$ & $<$1.1(15) & $<$1.3(16)\\ 
G094.6028 7 & 5.3(17)$^{+1.1(17)}_{-9.4(16)}$ & 5.7(16)$^{+6.7(15)}_{-4.0(16)}$ & 5.3(14)$^{+4.2(13)}_{-4.1(14)}$ & 8.8(14)$^{+2.4(15)}_{-4.5(14)}$ & $<$1.2(16)\\ 
G094.6028 8 & 4.8(17)$^{+3.3(17)}_{-7.7(16)}$ & 4.8(16)$^{+4.4(16)}_{-2.0(16)}$ & $<$2.5(14) & $<$1.9(14) & $<$1.8(16)\\ 
\hline 
G100.38 1 & 6.6(17)$^{+1.5(17)}_{-1.2(17)}$ & 1.7(17)$^{+1.6(17)}_{-5.0(16)}$ & 1.6(16)$^{+2.7(15)}_{-2.4(15)}$ & 2.0(15)$^{+4.0(15)}_{-1.5(15)}$ & 4.3(16)$^{+1.9(16)}_{-1.1(16)}$\\ 
G100.38 2 & 5.5(17)$^{+1.9(17)}_{-1.0(17)}$ & 2.2(17)$^{+1.7(17)}_{-6.9(16)}$ & 1.8(15)$^{+2.5(15)}_{-6.7(14)}$ & $<$1.0(18) & $<$1.3(16)\\ 
G100.38 3 & 3.6(17)$^{+5.6(16)}_{-3.3(16)}$ & 1.4(17)$^{+7.8(16)}_{-3.5(16)}$ & 8.4(14)$^{+5.3(14)}_{-6.4(14)}$ & $<$6.7(16) & $<$3.0(16)\\ 
G100.38 4 & 7.1(16)$^{+4.5(16)}_{-3.7(16)}$ & 1.7(17)$^{+6.8(16)}_{-6.7(16)}$ & 2.6(15)$^{+4.4(15)}_{-1.8(15)}$ & $<$1.4(14) & $<$8.9(17)\\ 
G100.38 5 & 2.7(17)$^{+3.0(17)}_{-1.6(17)}$ & 9.0(16)$^{+7.7(16)}_{-4.8(16)}$ & 1.1(15)$^{+1.1(14)}_{-9.2(14)}$ & $<$4.4(14) & $<$1.2(16)\\ 
\hline 
G108.75 1 & 6.9(17)$^{+3.1(16)}_{-8.5(16)}$ & 2.1(17)$^{+1.8(17)}_{-2.9(16)}$ & 3.8(15)$^{+2.5(14)}_{-6.7(14)}$ & 8.9(14)$^{+4.6(14)}_{-7.2(14)}$ & $<$1.9(16)\\ 
G108.75 2 & 4.4(17)$^{+1.6(17)}_{-2.3(17)}$ & 5.2(16)$^{+1.4(16)}_{-3.2(16)}$ & 1.0(15)$^{+6.5(14)}_{-4.6(14)}$ & $<$1.9(14) & $<$2.7(18)\\ 
G108.75 3 & 4.3(17)$^{+3.8(16)}_{-6.3(16)}$ & 5.8(16)$^{+4.6(16)}_{-3.0(16)}$ & 6.0(14)$^{+6.6(14)}_{-3.1(14)}$ & $<$1.4(14) & $<$1.1(16)\\ 
G108.75 4 & 4.5(17)$^{+7.8(16)}_{-4.5(16)}$ & 1.3(17)$^{+1.5(17)}_{-6.1(16)}$ & 5.2(14)$^{+7.3(14)}_{-2.3(14)}$ & $<$5.4(15) & $<$1.0(16)\\ 
G108.75 5 & 4.8(17)$^{+5.6(17)}_{-1.4(17)}$ & 5.0(16)$^{+4.9(16)}_{-2.5(16)}$ & 1.2(15)$^{+3.1(15)}_{-6.9(14)}$ & 3.5(14)$^{+9.8(14)}_{-7.3(13)}$ & $<$1.0(16)\\ 
G108.75 6 & 8.8(17)$^{+2.7(17)}_{-1.2(17)}$ & 1.0(17)$^{+1.1(17)}_{-3.9(16)}$ & 1.2(15)$^{+2.8(15)}_{-6.7(14)}$ & $<$2.6(14) & $<$1.1(16)\\ 
\hline 
G138.2957 1 & 6.3(17)$^{+1.9(17)}_{-6.7(16)}$ & 6.1(16)$^{+4.1(16)}_{-2.6(16)}$ & 1.1(15)$^{+1.4(14)}_{-9.8(14)}$ & $<$2.1(14) & $<$1.2(16)\\ 
G138.2957 2 & 4.0(17)$^{+5.9(16)}_{-2.9(16)}$ & 1.5(17)$^{+7.0(16)}_{-4.0(16)}$ & 1.0(15)$^{+1.2(15)}_{-8.1(14)}$ & $<$9.9(17) & $<$1.0(16)\\ 
G138.2957 3 & 7.4(17)$^{+1.8(17)}_{-1.7(17)}$ & 1.7(17)$^{+1.1(17)}_{-5.7(16)}$ & 1.4(15)$^{+1.6(15)}_{-6.9(14)}$ & $<$5.5(14) & $<$2.4(18)\\ 
G138.2957 4 & 1.3(18)$^{+7.0(17)}_{-4.5(17)}$ & 1.9(17)$^{+1.0(17)}_{-8.0(16)}$ & 1.8(15)$^{+2.6(15)}_{-6.8(14)}$ & $<$5.0(14) & $<$1.5(16)\\ 
G138.2957 5 & 6.0(17)$^{+1.9(17)}_{-8.2(16)}$ & 1.7(17)$^{+9.6(16)}_{-6.1(16)}$ & 9.9(14)$^{+8.5(14)}_{-6.0(14)}$ & $<$1.0(14) & $<$2.4(16)\\ 
G138.2957 6 & 6.4(17)$^{+3.2(17)}_{-1.2(17)}$ & 4.2(16)$^{+5.9(16)}_{-1.3(16)}$ & 8.4(14)$^{+2.8(15)}_{-3.9(14)}$ & $<$4.1(14) & $<$6.3(18)\\ 
G138.2957 7 & 6.2(17)$^{+1.7(17)}_{-1.6(17)}$ & 6.1(16)$^{+3.9(16)}_{-1.2(16)}$ & 4.8(14)$^{+5.6(14)}_{-2.9(14)}$ & $<$1.7(14) & $<$1.1(16)\\ 
G138.2957 8 & 7.2(17)$^{+1.2(17)}_{-7.5(16)}$ & 1.0(17)$^{+8.8(16)}_{-4.7(16)}$ & 1.2(15)$^{+9.7(14)}_{-4.5(14)}$ & $<$6.1(14) & $<$2.0(16)\\ 
G138.2957 9 & 7.9(17)$^{+1.2(17)}_{-1.8(17)}$ & 1.9(17)$^{+1.2(17)}_{-6.3(16)}$ & 7.6(14)$^{+8.4(14)}_{-1.7(14)}$ & $<$8.7(14) & 1.5(16)$^{+9.8(15)}_{-1.3(15)}$\\ 
G138.2957 10 & 5.5(17)$^{+6.4(17)}_{-4.6(16)}$ & 7.1(16)$^{+4.8(16)}_{-3.8(16)}$ & 1.1(15)$^{+8.3(14)}_{-6.2(14)}$ & $<$7.2(14) & $<$1.7(16)\\ 
G138.2957 11 & 1.1(18)$^{+2.6(17)}_{-9.6(16)}$ & 1.1(17)$^{+2.0(17)}_{-3.9(16)}$ & 9.2(14)$^{+9.7(14)}_{-4.7(14)}$ & $<$9.2(15) & $<$8.9(18)\\ 
G138.2957 12 & 3.9(17)$^{+2.7(17)}_{-1.2(17)}$ & 1.1(17)$^{+7.9(16)}_{-4.0(16)}$ & 1.3(15)$^{+6.2(14)}_{-1.0(15)}$ & $<$3.2(14) & $<$1.7(18)\\ 
\hline 
G139.9091 1 & 3.4(17)$^{+1.4(17)}_{-5.6(16)}$ & 2.1(16)$^{+1.4(16)}_{-3.4(15)}$ & $<$3.8(14) & $<$1.1(14) & $<$8.3(18)\\ 
G139.9091 2 & 1.4(18)$^{+4.0(17)}_{-2.6(17)}$ & 5.3(16)$^{+4.6(16)}_{-2.1(16)}$ & 1.9(15)$^{+1.2(15)}_{-5.2(14)}$ & $<$3.0(14) & $<$1.6(16)\\ 
G139.9091 3 & 6.6(17)$^{+9.3(16)}_{-7.0(16)}$ & 5.6(16)$^{+3.1(16)}_{-2.7(16)}$ & 5.7(14)$^{+7.6(14)}_{-2.9(14)}$ & $<$1.0(18) & $<$1.0(16)\\ 
G139.9091 4 & 1.2(18)$^{+1.6(17)}_{-4.7(17)}$ & 1.0(17)$^{+2.6(16)}_{-4.3(16)}$ & 5.6(14)$^{+1.9(14)}_{-3.9(14)}$ & $<$1.9(14) & $<$1.0(16)\\ 
\hline 
G075.78 1 & 2.2(18)$^{+5.2(17)}_{-2.9(17)}$ & 3.1(17)$^{+6.9(16)}_{-3.9(16)}$ & 1.5(17)$^{+2.4(16)}_{-1.7(16)}$ & 2.0(16)$^{+3.8(15)}_{-8.6(14)}$ & 8.0(17)$^{+5.4(16)}_{-6.2(16)}$\\ 
G075.78 2 & 9.0(17)$^{+2.6(17)}_{-7.3(16)}$ & 5.6(17)$^{+8.3(16)}_{-6.0(16)}$ & 1.8(16)$^{+4.8(15)}_{-3.1(15)}$ & $<$1.4(14) & $<$1.1(16)\\ 
G075.78 3 & 6.4(17)$^{+2.9(17)}_{-7.4(16)}$ & 2.4(17)$^{+7.7(16)}_{-8.2(16)}$ & 2.1(15)$^{+8.8(14)}_{-6.6(14)}$ & $<$3.2(14) & $<$1.1(17)\\ 
G075.78 4 & 2.0(18)$^{+3.1(17)}_{-5.7(17)}$ & 1.6(17)$^{+5.7(16)}_{-3.5(16)}$ & 2.2(15)$^{+9.4(14)}_{-5.6(14)}$ & $<$3.5(14) & $<$2.7(16)\\ 
\hline 
IRAS\,21078 1 & 2.9(18)$^{+4.8(17)}_{-1.8(17)}$ & 3.1(17)$^{+9.1(16)}_{-4.0(16)}$ & 3.7(16)$^{+3.9(15)}_{-2.0(15)}$ & 6.2(16)$^{+7.9(15)}_{-3.9(15)}$ & 4.1(17)$^{+1.2(17)}_{-1.3(17)}$\\ 
IRAS\,21078 2 & 1.7(18)$^{+2.1(17)}_{-1.2(17)}$ & 3.6(17)$^{+1.2(17)}_{-6.1(16)}$ & 1.3(16)$^{+2.2(15)}_{-1.9(15)}$ & 1.5(16)$^{+7.9(15)}_{-3.7(15)}$ & $<$1.3(16)\\ 
IRAS\,21078 3 & 1.8(18)$^{+1.2(17)}_{-5.0(16)}$ & 4.8(17)$^{+7.4(16)}_{-5.3(16)}$ & 9.5(15)$^{+2.2(15)}_{-1.8(15)}$ & 2.1(16)$^{+1.8(15)}_{-3.0(15)}$ & $<$1.5(16)\\ 
IRAS\,21078 4 & 9.6(17)$^{+4.8(17)}_{-1.9(17)}$ & 5.7(16)$^{+1.9(16)}_{-3.5(16)}$ & 6.7(14)$^{+5.8(14)}_{-4.1(14)}$ & $<$3.4(14) & $<$1.9(16)\\ 
IRAS\,21078 5 & 5.3(17)$^{+2.5(17)}_{-1.4(17)}$ & 6.5(16)$^{+7.6(16)}_{-3.1(16)}$ & 1.4(15)$^{+1.0(15)}_{-3.6(14)}$ & 1.7(15)$^{+6.2(15)}_{-1.2(15)}$ & $<$1.2(16)\\ 
IRAS\,21078 6 & 9.2(17)$^{+8.7(16)}_{-1.2(17)}$ & 4.6(17)$^{+9.4(16)}_{-6.5(16)}$ & 2.8(15)$^{+1.3(15)}_{-1.1(14)}$ & 1.9(15)$^{+3.0(15)}_{-1.4(15)}$ & $<$1.1(16)\\ 
IRAS\,21078 7 & 8.7(17)$^{+5.9(17)}_{-1.1(17)}$ & 1.7(17)$^{+1.4(17)}_{-6.4(16)}$ & 1.9(15)$^{+3.3(14)}_{-4.8(14)}$ & $<$1.0(14) & $<$2.4(18)\\ 
IRAS\,21078 8 & 5.8(17)$^{+1.3(17)}_{-1.5(17)}$ & 1.7(17)$^{+1.1(17)}_{-6.7(16)}$ & 3.7(15)$^{+1.2(15)}_{-9.0(14)}$ & 1.0(15)$^{+1.2(14)}_{-8.8(14)}$ & $<$9.8(17)\\ 
IRAS\,21078 9 & 3.2(17)$^{+1.3(17)}_{-5.3(16)}$ & 8.5(16)$^{+1.4(16)}_{-4.9(16)}$ & 1.7(15)$^{+8.6(14)}_{-4.8(14)}$ & $<$1.9(14) & $<$6.5(16)\\ 
IRAS\,21078 10 & 7.1(17)$^{+1.5(17)}_{-1.2(17)}$ & 1.3(17)$^{+8.1(16)}_{-3.6(16)}$ & 7.6(15)$^{+4.8(15)}_{-1.9(15)}$ & 7.9(14)$^{+1.1(15)}_{-5.8(14)}$ & $<$1.2(16)\\ 
IRAS\,21078 11 & 7.1(17)$^{+2.7(17)}_{-1.1(17)}$ & 1.6(17)$^{+2.2(16)}_{-5.2(16)}$ & 4.0(15)$^{+2.1(15)}_{-8.1(14)}$ & $<$1.1(15) & $<$1.4(16)\\ 
IRAS\,21078 12 & 1.0(18)$^{+5.5(17)}_{-4.8(17)}$ & 1.5(17)$^{+1.7(17)}_{-7.9(16)}$ & 1.8(15)$^{+8.2(14)}_{-6.7(14)}$ & 2.3(15)$^{+5.6(15)}_{-1.8(15)}$ & $<$1.4(16)\\ 
IRAS\,21078 13 & 4.0(17)$^{+3.1(17)}_{-7.0(16)}$ & 3.0(16)$^{+1.8(16)}_{-2.0(16)}$ & 1.1(15)$^{+8.3(14)}_{-2.3(14)}$ & $<$2.2(14) & $<$1.0(19)\\ 
IRAS\,21078 14 & 1.7(18)$^{+5.7(17)}_{-2.4(17)}$ & 1.3(17)$^{+7.1(16)}_{-5.6(16)}$ & 1.2(15)$^{+1.5(14)}_{-1.6(14)}$ & $<$4.3(14) & $<$1.6(16)\\ 
\hline 
NGC7538\,IRS9 1 & 2.4(18)$^{+6.3(17)}_{-3.3(17)}$ & 2.2(17)$^{+1.1(17)}_{-5.3(16)}$ & 4.0(16)$^{+4.3(15)}_{-4.7(15)}$ & 2.1(16)$^{+4.7(15)}_{-1.7(15)}$ & 7.0(17)$^{+1.7(17)}_{-2.0(17)}$\\ 
NGC7538\,IRS9 2 & 9.1(17)$^{+2.2(17)}_{-2.1(17)}$ & 1.2(17)$^{+5.9(16)}_{-4.7(16)}$ & 2.1(15)$^{+6.8(14)}_{-5.2(14)}$ & 1.9(15)$^{+4.9(15)}_{-1.2(15)}$ & $<$1.5(16)\\ 
NGC7538\,IRS9 3 & 6.6(17)$^{+9.9(16)}_{-5.8(16)}$ & 1.3(17)$^{+9.8(16)}_{-4.0(16)}$ & 1.8(15)$^{+2.4(14)}_{-5.6(14)}$ & $<$3.0(14) & $<$1.4(18)\\ 
NGC7538\,IRS9 4 & 1.6(18)$^{+4.3(17)}_{-2.8(17)}$ & 7.4(16)$^{+4.1(16)}_{-1.6(16)}$ & 2.7(14)$^{+4.2(14)}_{-4.1(13)}$ & $<$3.5(14) & $<$1.0(16)\\ 
NGC7538\,IRS9 5 & 6.8(18)$^{+1.5(18)}_{-1.0(18)}$ & 2.6(17)$^{+6.0(16)}_{-6.7(16)}$ & 1.1(15)$^{+2.7(14)}_{-3.7(14)}$ & $<$2.8(14) & $<$6.3(18)\\ 
NGC7538\,IRS9 6 & 8.0(17)$^{+1.2(17)}_{-1.3(17)}$ & 2.6(17)$^{+1.8(17)}_{-6.3(16)}$ & 2.1(15)$^{+1.2(15)}_{-6.7(14)}$ & 1.6(15)$^{+3.4(15)}_{-9.7(14)}$ & $<$2.2(16)\\ 
NGC7538\,IRS9 7 & 6.9(17)$^{+3.3(17)}_{-1.3(17)}$ & 2.2(17)$^{+8.6(16)}_{-5.2(16)}$ & 2.0(15)$^{+5.1(14)}_{-4.1(14)}$ & 1.6(15)$^{+2.5(14)}_{-1.5(15)}$ & $<$1.7(16)\\ 
NGC7538\,IRS9 8 & 4.6(17)$^{+5.8(16)}_{-7.5(16)}$ & 1.6(17)$^{+2.5(16)}_{-5.6(16)}$ & 8.7(14)$^{+1.1(15)}_{-4.8(14)}$ & $<$1.0(14) & $<$1.4(16)\\ 
\hline 
S87\,IRS1 1 & 1.1(18)$^{+2.2(17)}_{-1.0(17)}$ & 1.7(17)$^{+1.4(17)}_{-5.5(16)}$ & 1.8(15)$^{+1.7(15)}_{-8.7(14)}$ & $<$1.1(15) & $<$3.8(16)\\ 
S87\,IRS1 2 & 5.4(17)$^{+5.8(16)}_{-7.1(16)}$ & 2.2(17)$^{+1.3(17)}_{-1.1(16)}$ & 5.8(14)$^{+2.0(14)}_{-2.8(14)}$ & $<$6.4(15) & $<$1.1(16)\\ 
S87\,IRS1 3 & 5.0(17)$^{+9.3(16)}_{-4.4(16)}$ & 1.7(17)$^{+3.7(16)}_{-6.3(16)}$ & 7.1(14)$^{+6.3(14)}_{-2.4(14)}$ & $<$2.5(14) & $<$1.3(16)\\ 
S87\,IRS1 4 & 6.8(17)$^{+3.9(17)}_{-3.3(16)}$ & 1.4(17)$^{+9.1(16)}_{-3.9(16)}$ & 8.4(14)$^{+1.2(15)}_{-4.9(14)}$ & $<$5.1(14) & $<$1.5(16)\\ 
S87\,IRS1 5 & 1.0(18)$^{+5.3(16)}_{-1.1(17)}$ & 1.1(17)$^{+5.5(16)}_{-5.1(16)}$ & 3.7(15)$^{+1.1(15)}_{-1.6(14)}$ & $<$3.8(14) & $<$2.7(16)\\ 
S87\,IRS1 6 & 1.9(18)$^{+6.3(17)}_{-7.9(17)}$ & 1.8(17)$^{+3.8(16)}_{-5.7(16)}$ & 6.6(14)$^{+7.3(14)}_{-2.6(14)}$ & $<$5.6(14) & $<$1.2(18)\\ 
S87\,IRS1 7 & 5.6(17)$^{+9.9(16)}_{-5.7(16)}$ & 1.4(17)$^{+5.5(16)}_{-7.9(16)}$ & 4.2(14)$^{+3.1(14)}_{-8.9(13)}$ & $<$1.1(14) & $<$2.4(16)\\ 
S87\,IRS1 8 & 3.9(18)$^{+7.6(17)}_{-7.6(17)}$ & 1.2(17)$^{+1.8(17)}_{-5.6(16)}$ & 6.3(14)$^{+2.4(14)}_{-3.7(14)}$ & $<$1.4(14) & $<$1.0(16)\\ 
S87\,IRS1 9 & 7.4(17)$^{+9.6(16)}_{-1.2(17)}$ & 1.6(17)$^{+1.3(17)}_{-6.2(16)}$ & 2.0(15)$^{+9.0(14)}_{-8.1(13)}$ & $<$1.3(14) & $<$1.6(18)\\ 
\hline 
S106 1 & 5.7(17)$^{+1.5(17)}_{-1.2(17)}$ & 3.8(16)$^{+1.8(16)}_{-2.3(16)}$ & $<$2.3(14) & $<$4.0(14) & $<$1.3(16)\\ 
S106 2 & 1.3(18)$^{+3.6(17)}_{-1.0(17)}$ & 3.3(17)$^{+1.6(17)}_{-5.1(16)}$ & 1.6(16)$^{+5.2(15)}_{-2.0(15)}$ & 5.2(15)$^{+2.4(15)}_{-1.6(15)}$ & $<$2.7(16)\\ 
S106 3 & 1.4(18)$^{+2.2(17)}_{-1.3(17)}$ & 7.6(17)$^{+4.3(16)}_{-4.0(16)}$ & 2.0(15)$^{+1.4(15)}_{-6.9(14)}$ & $<$1.6(15) & $<$5.7(16)\\ 
S106 4 & 6.8(17)$^{+1.3(17)}_{-1.3(17)}$ & 2.5(17)$^{+4.6(16)}_{-3.2(16)}$ & 9.0(14)$^{+4.8(14)}_{-7.4(14)}$ & $<$3.8(14) & $<$2.5(16)\\ 
\hline 
W3\,H2O 1 & 3.8(17)$^{+1.3(17)}_{-8.1(16)}$ & 5.3(16)$^{+4.0(16)}_{-2.3(16)}$ & 8.6(15)$^{+6.0(15)}_{-2.6(15)}$ & 3.2(16)$^{+4.8(15)}_{-3.4(15)}$ & 5.0(16)$^{+1.7(16)}_{-1.5(16)}$\\ 
W3\,H2O 2 & 5.7(17)$^{+5.3(16)}_{-4.9(16)}$ & 1.7(16)$^{+2.2(15)}_{-6.6(15)}$ & 4.0(15)$^{+9.8(14)}_{-9.4(14)}$ & 1.3(16)$^{+2.8(15)}_{-1.3(15)}$ & 2.9(16)$^{+2.1(16)}_{-1.2(16)}$\\ 
W3\,H2O 3 & 3.9(18)$^{+1.1(17)}_{-2.5(17)}$ & 4.4(17)$^{+3.6(16)}_{-4.9(16)}$ & 6.9(16)$^{+1.1(16)}_{-5.4(15)}$ & 4.7(17)$^{+5.5(16)}_{-4.9(16)}$ & 5.7(17)$^{+1.6(17)}_{-7.1(16)}$\\ 
W3\,H2O 4 & 1.3(18)$^{+1.3(17)}_{-1.2(17)}$ & 3.9(17)$^{+6.7(16)}_{-3.9(16)}$ & 1.6(17)$^{+1.6(16)}_{-2.1(16)}$ & 8.5(17)$^{+1.5(16)}_{-1.5(16)}$ & 7.7(17)$^{+6.8(16)}_{-7.6(16)}$\\ 
W3\,H2O 5 & 1.1(18)$^{+3.0(17)}_{-2.4(17)}$ & 8.1(17)$^{+3.6(16)}_{-5.1(16)}$ & 2.5(16)$^{+3.3(15)}_{-2.9(15)}$ & 1.4(17)$^{+5.5(16)}_{-2.6(16)}$ & $<$1.4(16)\\ 
\hline 
W3\,IRS4 1 & 3.0(18)$^{+1.6(17)}_{-2.8(17)}$ & 1.4(17)$^{+2.5(16)}_{-3.7(16)}$ & 7.6(16)$^{+5.5(15)}_{-2.2(15)}$ & 2.0(16)$^{+7.4(15)}_{-5.4(15)}$ & 4.4(17)$^{+2.8(16)}_{-1.8(16)}$\\ 
W3\,IRS4 2 & 7.5(17)$^{+1.6(17)}_{-5.8(16)}$ & 3.4(17)$^{+3.4(16)}_{-5.5(16)}$ & 1.7(15)$^{+8.6(14)}_{-6.8(14)}$ & $<$1.0(14) & $<$1.7(18)\\ 
W3\,IRS4 3 & 1.1(18)$^{+2.1(17)}_{-7.9(16)}$ & 5.0(17)$^{+7.7(16)}_{-6.7(16)}$ & 3.5(15)$^{+2.6(15)}_{-3.0(14)}$ & $<$9.0(14) & $<$1.5(16)\\ 
W3\,IRS4 4 & 1.5(18)$^{+2.8(17)}_{-1.9(17)}$ & 5.9(17)$^{+4.0(16)}_{-1.1(17)}$ & 1.3(15)$^{+8.1(14)}_{-4.3(14)}$ & $<$1.0(15) & $<$1.7(16)\\ 
W3\,IRS4 5 & 7.3(17)$^{+3.2(17)}_{-2.9(16)}$ & 5.5(17)$^{+5.8(16)}_{-6.3(16)}$ & 1.6(15)$^{+8.8(14)}_{-2.7(14)}$ & $<$1.6(15) & $<$2.1(16)\\ 
W3\,IRS4 6 & 1.2(18)$^{+7.0(16)}_{-1.2(17)}$ & 5.4(17)$^{+3.8(16)}_{-8.3(16)}$ & 1.2(15)$^{+6.8(14)}_{-2.1(14)}$ & $<$9.0(14) & $<$1.1(16)\\ 
\end{longtable}

\begin{longtable}{llllll}
\caption{Molecular column densities derived with \texttt{XCLASS}.}
\label{tab:XCLASSresults2}
\\
\hline\hline
Position & DCN & H$_{2}$CO & HNCO & HC$_{3}$N & HC$_{3}$N;$\varv_{7}$=1 \\
\hline
\endfirsthead
\caption{continued.}\\
\hline\hline
Position & DCN & H$_{2}$CO & HNCO & HC$_{3}$N & HC$_{3}$N;$\varv_{7}$=1 \\
\hline
\endhead
\hline
Notes. a(b) = a$\times$10$^{\mathrm{b}}$.
\endfoot
\hline
IRAS\,23033 1 & $<$1.5(13) & 7.3(15)$^{+2.4(15)}_{-1.5(15)}$ & $<$3.6(15) & $<$1.7(14) & $<$2.3(16)\\ 
IRAS\,23033 2 & 1.4(14)$^{+7.7(13)}_{-2.9(13)}$ & 2.9(16)$^{+2.3(15)}_{-4.7(15)}$ & 1.3(15)$^{+7.3(14)}_{-6.3(14)}$ & 9.9(14)$^{+6.2(14)}_{-1.9(14)}$ & $<$1.1(14)\\ 
IRAS\,23033 3 & 8.3(13)$^{+1.4(13)}_{-8.0(12)}$ & 6.6(16)$^{+1.4(16)}_{-5.1(15)}$ & 5.5(15)$^{+2.7(15)}_{-1.5(15)}$ & 3.3(15)$^{+5.8(14)}_{-5.9(14)}$ & 8.5(14)$^{+9.5(14)}_{-4.5(14)}$\\ 
IRAS\,23033 4 & 7.2(13)$^{+7.7(13)}_{-3.2(13)}$ & 3.4(15)$^{+1.2(15)}_{-9.2(14)}$ & 3.6(16)$^{+3.7(15)}_{-3.1(16)}$ & 1.1(14)$^{+2.3(14)}_{-6.4(13)}$ & $<$8.4(14)\\ 
IRAS\,23033 5 & 1.3(14)$^{+2.1(13)}_{-4.8(13)}$ & 2.2(15)$^{+1.1(15)}_{-1.3(15)}$ & $<$8.3(14) & 5.4(13)$^{+3.4(13)}_{-3.7(13)}$ & $<$1.1(14)\\ 
\hline 
IRAS\,23151 1 & 4.4(13)$^{+9.9(13)}_{-3.4(13)}$ & 3.8(16)$^{+5.4(15)}_{-6.8(15)}$ & 1.5(16)$^{+2.7(15)}_{-1.9(15)}$ & 4.3(14)$^{+1.2(15)}_{-2.2(14)}$ & $<$1.6(14)\\ 
IRAS\,23151 2 & 6.0(13)$^{+9.2(13)}_{-4.1(13)}$ & 9.1(15)$^{+1.6(15)}_{-3.5(15)}$ & 4.9(14)$^{+7.5(14)}_{-3.0(14)}$ & $<$1.0(13) & $<$3.9(15)\\ 
IRAS\,23151 3 & 2.1(13)$^{+1.2(14)}_{-1.2(13)}$ & 3.1(15)$^{+9.7(14)}_{-6.2(14)}$ & $<$2.2(14) & $<$3.0(13) & $<$1.1(15)\\ 
IRAS\,23151 4 & 2.8(13)$^{+1.6(13)}_{-1.6(13)}$ & 2.5(15)$^{+5.9(14)}_{-9.8(14)}$ & $<$1.7(14) & $<$3.1(13) & $<$6.4(15)\\ 
IRAS\,23151 5 & 7.1(13)$^{+2.2(14)}_{-3.8(13)}$ & 7.7(14)$^{+1.4(15)}_{-2.7(14)}$ & $<$1.1(14) & $<$3.2(13) & $<$3.3(15)\\ 
\hline 
IRAS\,23385 1 & 9.8(13)$^{+6.3(13)}_{-2.8(13)}$ & 2.8(16)$^{+4.7(15)}_{-2.8(15)}$ & 4.1(14)$^{+2.8(14)}_{-1.2(14)}$ & 1.2(15)$^{+2.7(14)}_{-3.1(14)}$ & $<$1.6(14)\\ 
IRAS\,23385 2 & 2.3(13)$^{+1.1(14)}_{-1.0(13)}$ & 1.9(16)$^{+6.0(15)}_{-4.6(15)}$ & $<$7.8(14) & 9.1(13)$^{+1.6(14)}_{-6.1(13)}$ & $<$1.0(14)\\ 
IRAS\,23385 3 & 1.3(14)$^{+8.6(13)}_{-3.7(13)}$ & 2.1(16)$^{+5.7(15)}_{-4.4(15)}$ & $<$3.1(14) & 2.3(14)$^{+1.5(14)}_{-8.5(13)}$ & $<$9.7(15)\\ 
IRAS\,23385 4 & 4.3(13)$^{+3.1(13)}_{-2.4(13)}$ & 2.4(16)$^{+7.2(15)}_{-4.7(15)}$ & $<$7.9(14) & 9.0(13)$^{+1.3(14)}_{-6.4(13)}$ & $<$3.6(14)\\ 
IRAS\,23385 5 & 4.4(13)$^{+1.0(14)}_{-3.1(13)}$ & 1.2(16)$^{+4.5(15)}_{-2.7(15)}$ & $<$3.5(14) & 1.4(14)$^{+9.6(14)}_{-5.6(13)}$ & $<$2.5(16)\\ 
IRAS\,23385 6 & 7.6(13)$^{+2.1(14)}_{-6.5(13)}$ & 1.3(16)$^{+5.3(15)}_{-1.4(15)}$ & $<$3.9(15) & 1.1(14)$^{+2.5(14)}_{-8.6(13)}$ & $<$9.7(15)\\ 
IRAS\,23385 7 & 3.1(13)$^{+3.8(13)}_{-1.5(13)}$ & 1.6(16)$^{+1.1(16)}_{-5.7(15)}$ & 7.4(15)$^{+1.9(16)}_{-7.0(15)}$ & 1.0(14)$^{+1.7(14)}_{-8.0(13)}$ & $<$1.0(14)\\ 
IRAS\,23385 8 & 5.7(13)$^{+3.5(14)}_{-1.7(13)}$ & 6.1(15)$^{+2.1(15)}_{-1.9(15)}$ & $<$6.9(14) & $<$4.1(13) & $<$1.5(16)\\ 
\hline 
AFGL\,2591 1 & 1.1(14)$^{+4.6(13)}_{-3.3(13)}$ & 5.0(16)$^{+7.9(15)}_{-3.8(15)}$ & 1.8(16)$^{+2.2(15)}_{-8.5(14)}$ & 3.9(15)$^{+4.2(14)}_{-3.8(14)}$ & 9.9(15)$^{+1.7(15)}_{-8.3(14)}$\\ 
AFGL\,2591 2 & $<$1.0(13) & 1.4(15)$^{+3.9(14)}_{-4.3(14)}$ & $<$1.8(14) & $<$1.2(14) & 7.0(14)$^{+8.1(14)}_{-5.3(14)}$\\ 
AFGL\,2591 3 & 4.7(13)$^{+4.4(13)}_{-1.2(13)}$ & 1.0(15)$^{+1.0(15)}_{-2.1(14)}$ & $<$1.0(14) & $<$1.5(13) & $<$9.7(15)\\ 
AFGL\,2591 4 & 2.8(13)$^{+1.1(14)}_{-1.8(13)}$ & 2.4(15)$^{+1.2(15)}_{-3.9(14)}$ & $<$2.5(14) & 1.7(14)$^{+6.1(14)}_{-8.3(13)}$ & $<$1.0(14)\\ 
\hline 
CepA\,HW2 1 & 1.4(14)$^{+4.1(13)}_{-6.4(13)}$ & 5.4(17)$^{+4.8(16)}_{-3.0(16)}$ & 9.1(16)$^{+1.9(15)}_{-1.8(15)}$ & 7.6(15)$^{+8.1(14)}_{-6.1(14)}$ & 1.3(16)$^{+6.4(14)}_{-6.1(14)}$\\ 
CepA\,HW2 2 & 2.2(13)$^{+1.1(13)}_{-1.9(13)}$ & 8.6(16)$^{+2.9(16)}_{-1.0(16)}$ & 8.7(15)$^{+9.9(14)}_{-2.4(15)}$ & $<$2.8(14) & 6.4(14)$^{+1.4(15)}_{-2.0(14)}$\\ 
CepA\,HW2 3 & 9.2(14)$^{+1.6(13)}_{-2.9(13)}$ & 5.4(16)$^{+1.2(16)}_{-6.8(15)}$ & 5.5(15)$^{+9.4(14)}_{-4.5(14)}$ & 9.3(14)$^{+4.4(14)}_{-2.1(14)}$ & 9.9(14)$^{+1.0(15)}_{-4.9(14)}$\\ 
CepA\,HW2 4 & $<$9.7(12) & $<$1.8(14) & $<$1.0(14) & $<$3.7(14) & $<$8.9(14)\\ 
CepA\,HW2 5 & 3.7(13)$^{+8.7(13)}_{-2.3(13)}$ & 8.8(15)$^{+2.8(15)}_{-7.8(14)}$ & $<$1.0(14) & $<$3.4(13) & $<$9.7(15)\\ 
\hline 
G084.9505 1 & 1.1(14)$^{+3.6(13)}_{-2.7(13)}$ & 1.8(16)$^{+6.5(15)}_{-1.3(15)}$ & 8.9(14)$^{+2.7(14)}_{-2.0(14)}$ & 1.9(14)$^{+7.9(13)}_{-1.5(14)}$ & 1.0(14)$^{+8.8(11)}_{-9.2(10)}$\\ 
G084.9505 2 & 6.2(12)$^{+3.2(12)}_{-5.1(12)}$ & 7.6(14)$^{+9.1(14)}_{-1.8(14)}$ & $<$1.2(14) & $<$5.2(13) & $<$1.0(14)\\ 
G084.9505 3 & 1.8(13)$^{+4.6(13)}_{-1.4(13)}$ & 3.4(15)$^{+4.4(15)}_{-8.2(14)}$ & $<$1.6(14) & $<$4.7(13) & $<$1.5(14)\\ 
G084.9505 4 & 1.4(13)$^{+6.9(13)}_{-5.0(12)}$ & 1.3(15)$^{+1.0(15)}_{-6.7(14)}$ & 6.9(14)$^{+7.1(14)}_{-5.1(14)}$ & 4.1(14)$^{+1.0(15)}_{-3.7(14)}$ & $<$1.3(16)\\ 
G084.9505 5 & 4.9(13)$^{+6.6(13)}_{-3.5(13)}$ & 6.7(14)$^{+5.8(14)}_{-3.6(14)}$ & $<$7.9(14) & $<$1.4(13) & $<$6.1(16)\\ 
G084.9505 6 & 1.1(13)$^{+1.3(12)}_{-9.6(12)}$ & 2.3(15)$^{+3.1(15)}_{-1.5(15)}$ & $<$1.0(14) & $<$1.9(13) & $<$6.7(14)\\ 
G084.9505 7 & 1.5(13)$^{+6.9(12)}_{-1.3(13)}$ & 6.0(14)$^{+5.5(14)}_{-3.7(14)}$ & $<$1.1(16) & $<$2.0(13) & $<$1.3(14)\\ 
G084.9505 8 & 2.6(12)$^{+3.8(12)}_{-1.2(12)}$ & 7.7(14)$^{+3.8(14)}_{-5.4(14)}$ & $<$3.6(14) & $<$1.0(17) & $<$1.1(14)\\ 
\hline 
G094.6028 1 & 1.0(14)$^{+6.0(13)}_{-1.5(13)}$ & 4.1(16)$^{+8.0(15)}_{-9.7(15)}$ & 4.4(15)$^{+3.2(14)}_{-6.7(14)}$ & 2.3(15)$^{+1.1(15)}_{-4.4(14)}$ & 6.5(14)$^{+8.0(14)}_{-5.0(14)}$\\ 
G094.6028 2 & 6.9(13)$^{+2.0(13)}_{-3.1(13)}$ & 5.9(14)$^{+5.4(14)}_{-8.9(13)}$ & $<$1.6(16) & $<$2.5(13) & $<$1.0(14)\\ 
G094.6028 3 & 3.6(13)$^{+8.9(12)}_{-3.2(13)}$ & 5.5(14)$^{+9.8(14)}_{-1.6(14)}$ & $<$2.8(14) & $<$1.0(13) & $<$5.7(14)\\ 
G094.6028 4 & 5.2(13)$^{+5.5(13)}_{-2.0(13)}$ & 7.1(14)$^{+6.1(14)}_{-2.8(14)}$ & $<$1.5(16) & 7.6(13)$^{+7.5(13)}_{-5.5(13)}$ & $<$3.3(16)\\ 
G094.6028 5 & 5.8(13)$^{+1.8(14)}_{-2.6(13)}$ & 5.5(14)$^{+9.8(13)}_{-4.2(14)}$ & 6.3(15)$^{+5.8(16)}_{-5.4(15)}$ & $<$2.2(13) & $<$1.2(15)\\ 
G094.6028 6 & 2.8(13)$^{+3.0(13)}_{-2.0(13)}$ & 7.1(14)$^{+6.5(14)}_{-2.9(14)}$ & $<$1.0(14) & $<$1.0(13) & $<$1.2(14)\\ 
G094.6028 7 & 1.5(13)$^{+9.4(13)}_{-4.4(12)}$ & 3.7(14)$^{+2.0(14)}_{-2.4(14)}$ & $<$3.6(14) & $<$2.6(13) & $<$2.4(14)\\ 
G094.6028 8 & $<$1.9(12) & 1.2(14)$^{+1.6(13)}_{-2.5(13)}$ & $<$4.6(14) & $<$1.6(13) & $<$1.4(14)\\ 
\hline 
G100.38 1 & 2.7(13)$^{+3.3(12)}_{-2.4(13)}$ & 4.9(15)$^{+1.8(15)}_{-1.3(15)}$ & 9.1(14)$^{+3.3(14)}_{-2.1(14)}$ & $<$1.2(14) & $<$2.2(15)\\ 
G100.38 2 & $<$6.5(12) & 4.3(14)$^{+4.0(14)}_{-2.1(14)}$ & $<$1.7(14) & $<$2.4(13) & $<$5.2(14)\\ 
G100.38 3 & $<$1.0(13) & 3.1(14)$^{+2.0(14)}_{-1.6(14)}$ & $<$2.8(14) & $<$1.9(13) & $<$1.7(15)\\ 
G100.38 4 & $<$9.0(12) & 1.2(15)$^{+3.5(14)}_{-1.1(15)}$ & $<$1.1(14) & $<$8.3(13) & $<$2.0(15)\\ 
G100.38 5 & $<$7.1(12) & $<$1.1(14) & $<$2.0(14) & $<$4.9(13) & $<$2.9(14)\\ 
\hline 
G108.75 1 & 4.1(13)$^{+3.7(14)}_{-2.2(13)}$ & 6.6(15)$^{+2.9(15)}_{-1.7(15)}$ & $<$3.8(14) & 6.3(13)$^{+1.3(14)}_{-1.5(13)}$ & $<$3.5(14)\\ 
G108.75 2 & 5.0(13)$^{+2.3(13)}_{-4.4(13)}$ & 3.5(15)$^{+3.4(15)}_{-1.1(15)}$ & $<$2.1(14) & $<$3.1(13) & $<$6.7(14)\\ 
G108.75 3 & 2.0(13)$^{+2.3(13)}_{-1.5(13)}$ & 8.4(14)$^{+4.3(14)}_{-4.5(14)}$ & $<$1.0(17) & $<$1.3(13) & $<$1.6(14)\\ 
G108.75 4 & 1.4(13)$^{+3.7(13)}_{-9.0(12)}$ & 1.7(15)$^{+3.1(15)}_{-3.9(14)}$ & $<$1.9(14) & $<$2.8(13) & $<$4.0(14)\\ 
G108.75 5 & 5.2(13)$^{+1.0(14)}_{-4.5(13)}$ & 3.9(15)$^{+2.4(14)}_{-2.6(15)}$ & $<$3.1(14) & $<$1.3(13) & $<$1.9(14)\\ 
G108.75 6 & 1.1(14)$^{+7.4(13)}_{-6.8(13)}$ & 8.5(14)$^{+1.0(15)}_{-2.5(14)}$ & $<$1.0(17) & 4.3(13)$^{+1.2(14)}_{-1.5(13)}$ & $<$1.5(14)\\ 
\hline 
G138.2957 1 & 3.6(13)$^{+5.6(13)}_{-1.4(13)}$ & 4.8(14)$^{+3.2(14)}_{-2.9(14)}$ & $<$1.2(15) & $<$7.4(13) & $<$1.0(14)\\ 
G138.2957 2 & 7.8(13)$^{+2.2(14)}_{-3.4(13)}$ & 5.0(14)$^{+3.7(14)}_{-2.5(14)}$ & $<$5.1(14) & $<$3.1(13) & $<$1.5(14)\\ 
G138.2957 3 & 1.1(14)$^{+1.2(14)}_{-1.4(13)}$ & 1.6(15)$^{+1.5(15)}_{-4.0(14)}$ & $<$2.9(14) & 6.6(13)$^{+1.5(14)}_{-3.1(13)}$ & $<$2.0(14)\\ 
G138.2957 4 & 9.7(13)$^{+9.6(13)}_{-8.5(13)}$ & 4.3(14)$^{+1.4(14)}_{-1.8(14)}$ & 2.9(14)$^{+6.1(14)}_{-5.2(13)}$ & $<$1.9(13) & $<$5.4(16)\\ 
G138.2957 5 & 1.8(13)$^{+7.2(13)}_{-1.0(13)}$ & 6.4(14)$^{+3.4(14)}_{-1.9(14)}$ & $<$2.2(15) & $<$2.1(13) & $<$1.4(14)\\ 
G138.2957 6 & 4.8(13)$^{+8.9(13)}_{-3.1(13)}$ & 1.1(15)$^{+1.8(15)}_{-5.2(14)}$ & $<$4.3(14) & $<$1.4(13) & $<$2.4(16)\\ 
G138.2957 7 & 1.2(14)$^{+7.1(13)}_{-5.2(13)}$ & 1.2(15)$^{+1.5(15)}_{-4.6(14)}$ & $<$1.0(17) & $<$5.4(13) & $<$1.0(14)\\ 
G138.2957 8 & $<$1.3(13) & 9.4(14)$^{+6.6(14)}_{-3.9(14)}$ & $<$2.2(14) & $<$1.6(15) & $<$1.0(15)\\ 
G138.2957 9 & 1.9(13)$^{+5.4(13)}_{-1.3(13)}$ & 1.5(15)$^{+2.3(15)}_{-1.9(14)}$ & $<$1.1(14) & $<$3.3(13) & $<$9.7(16)\\ 
G138.2957 10 & $<$1.4(13) & 7.2(14)$^{+7.9(14)}_{-8.8(13)}$ & $<$1.1(14) & $<$1.7(13) & $<$3.8(14)\\ 
G138.2957 11 & $<$7.2(12) & 1.7(15)$^{+1.9(15)}_{-9.0(14)}$ & $<$3.9(14) & $<$3.9(13) & $<$1.4(14)\\ 
G138.2957 12 & 5.6(13)$^{+3.5(13)}_{-4.8(13)}$ & 7.1(14)$^{+1.3(15)}_{-3.7(14)}$ & $<$1.0(14) & $<$5.9(13) & $<$1.0(14)\\ 
\hline 
G139.9091 1 & $<$8.4(12) & 1.1(15)$^{+1.3(14)}_{-9.6(14)}$ & $<$1.1(14) & $<$7.4(13) & $<$6.5(14)\\ 
G139.9091 2 & $<$2.2(13) & 1.3(15)$^{+6.5(14)}_{-1.0(15)}$ & $<$2.2(14) & $<$2.0(13) & $<$9.7(15)\\ 
G139.9091 3 & 5.8(13)$^{+2.5(14)}_{-3.0(13)}$ & 3.6(14)$^{+9.9(14)}_{-1.0(14)}$ & $<$1.9(14) & $<$7.7(13) & $<$9.7(15)\\ 
G139.9091 4 & $<$1.2(13) & 7.2(14)$^{+5.8(14)}_{-5.1(14)}$ & $<$2.9(14) & $<$1.0(13) & $<$1.8(14)\\ 
\hline 
G075.78 1 & 1.2(14)$^{+2.6(13)}_{-3.1(13)}$ & 5.8(16)$^{+2.2(16)}_{-7.5(15)}$ & 1.5(16)$^{+1.1(15)}_{-1.8(15)}$ & 6.2(15)$^{+5.3(14)}_{-5.9(14)}$ & 1.6(16)$^{+7.7(14)}_{-5.5(15)}$\\ 
G075.78 2 & 9.6(13)$^{+7.3(13)}_{-2.1(13)}$ & 1.9(15)$^{+1.4(15)}_{-5.9(14)}$ & $<$1.8(14) & $<$1.3(13) & $<$8.3(14)\\ 
G075.78 3 & $<$6.3(12) & 7.3(14)$^{+3.4(14)}_{-2.1(14)}$ & $<$3.1(14) & $<$1.1(13) & $<$7.5(15)\\ 
G075.78 4 & 1.3(14)$^{+8.1(13)}_{-3.1(13)}$ & 8.3(15)$^{+2.2(15)}_{-1.0(15)}$ & $<$1.1(14) & $<$1.2(13) & $<$3.6(14)\\ 
\hline 
IRAS\,21078 1 & 9.4(13)$^{+9.6(13)}_{-3.5(13)}$ & 4.8(17)$^{+3.2(16)}_{-3.0(16)}$ & 4.1(16)$^{+4.6(15)}_{-2.8(15)}$ & 6.5(15)$^{+1.7(15)}_{-1.1(15)}$ & 4.3(15)$^{+1.4(15)}_{-1.1(15)}$\\ 
IRAS\,21078 2 & 1.6(14)$^{+1.4(14)}_{-2.0(13)}$ & 4.4(16)$^{+8.8(15)}_{-6.1(15)}$ & 5.0(15)$^{+2.5(15)}_{-1.0(15)}$ & 3.7(14)$^{+6.5(14)}_{-2.1(13)}$ & 2.6(14)$^{+3.0(14)}_{-8.7(13)}$\\ 
IRAS\,21078 3 & 9.8(13)$^{+9.1(13)}_{-1.9(13)}$ & 6.5(16)$^{+1.2(16)}_{-1.2(16)}$ & 1.2(15)$^{+6.4(14)}_{-1.6(14)}$ & 5.0(14)$^{+2.4(14)}_{-1.6(14)}$ & $<$4.7(14)\\ 
IRAS\,21078 4 & 4.1(13)$^{+2.4(13)}_{-2.4(13)}$ & 1.6(15)$^{+1.3(15)}_{-2.5(14)}$ & $<$1.0(14) & 1.2(14)$^{+1.2(14)}_{-6.8(13)}$ & $<$4.7(14)\\ 
IRAS\,21078 5 & 1.6(13)$^{+1.6(13)}_{-1.3(13)}$ & 2.1(15)$^{+1.2(15)}_{-5.6(14)}$ & $<$1.5(14) & $<$8.3(13) & $<$5.7(14)\\ 
IRAS\,21078 6 & 7.6(13)$^{+4.2(13)}_{-3.2(13)}$ & 1.2(16)$^{+1.0(16)}_{-4.1(15)}$ & $<$1.9(14) & 1.8(14)$^{+2.2(14)}_{-8.9(13)}$ & $<$9.7(15)\\ 
IRAS\,21078 7 & 1.3(14)$^{+7.0(13)}_{-3.8(13)}$ & 4.8(15)$^{+1.2(15)}_{-1.4(15)}$ & $<$5.2(14) & $<$9.1(13) & $<$9.4(14)\\ 
IRAS\,21078 8 & 1.3(14)$^{+8.5(13)}_{-3.7(13)}$ & 3.9(15)$^{+1.8(15)}_{-7.2(14)}$ & $<$2.5(14) & 1.8(14)$^{+7.5(13)}_{-1.6(14)}$ & $<$1.0(14)\\ 
IRAS\,21078 9 & 4.6(13)$^{+6.1(13)}_{-2.0(12)}$ & 5.2(15)$^{+5.9(14)}_{-1.4(15)}$ & $<$1.1(14) & 1.5(14)$^{+5.1(13)}_{-1.0(14)}$ & $<$4.5(14)\\ 
IRAS\,21078 10 & 5.0(13)$^{+5.9(13)}_{-3.4(13)}$ & 2.7(16)$^{+2.0(16)}_{-5.7(15)}$ & 7.9(14)$^{+2.4(14)}_{-6.2(14)}$ & 1.9(14)$^{+3.0(14)}_{-8.9(13)}$ & $<$8.7(14)\\ 
IRAS\,21078 11 & $<$8.5(14) & 4.0(15)$^{+3.6(15)}_{-7.7(14)}$ & $<$3.3(14) & $<$8.3(13) & $<$1.1(15)\\ 
IRAS\,21078 12 & 8.5(13)$^{+6.3(13)}_{-2.6(13)}$ & 1.1(16)$^{+4.1(15)}_{-1.7(15)}$ & $<$1.4(14) & 1.6(14)$^{+1.1(14)}_{-1.4(14)}$ & $<$1.0(14)\\ 
IRAS\,21078 13 & 1.3(13)$^{+9.9(12)}_{-2.2(12)}$ & 4.7(15)$^{+1.3(15)}_{-4.7(14)}$ & $<$8.4(14) & $<$1.1(14) & $<$1.3(14)\\ 
IRAS\,21078 14 & 2.1(13)$^{+2.1(13)}_{-1.7(13)}$ & 2.0(16)$^{+6.9(15)}_{-4.3(15)}$ & $<$1.9(14) & 9.4(13)$^{+2.1(14)}_{-5.0(13)}$ & $<$1.4(14)\\ 
\hline 
NGC7538\,IRS9 1 & 1.1(14)$^{+2.8(13)}_{-2.7(13)}$ & 3.1(17)$^{+5.5(16)}_{-3.2(16)}$ & 9.8(16)$^{+7.9(14)}_{-2.7(14)}$ & 3.8(15)$^{+4.1(15)}_{-1.0(15)}$ & 4.7(15)$^{+1.2(15)}_{-9.7(14)}$\\ 
NGC7538\,IRS9 2 & 4.6(13)$^{+4.8(13)}_{-4.1(13)}$ & 4.1(15)$^{+1.5(15)}_{-1.1(15)}$ & $<$1.9(14) & $<$2.0(13) & $<$3.6(14)\\ 
NGC7538\,IRS9 3 & 1.3(14)$^{+5.6(13)}_{-3.9(13)}$ & 4.3(15)$^{+2.2(15)}_{-6.9(14)}$ & $<$1.2(14) & 8.1(13)$^{+5.1(13)}_{-7.0(13)}$ & $<$1.8(14)\\ 
NGC7538\,IRS9 4 & 1.1(14)$^{+5.3(13)}_{-2.5(13)}$ & 1.2(15)$^{+9.5(14)}_{-6.1(14)}$ & $<$1.0(14) & $<$2.8(14) & $<$2.0(14)\\ 
NGC7538\,IRS9 5 & 5.9(13)$^{+8.1(13)}_{-3.0(13)}$ & 2.1(16)$^{+7.6(15)}_{-5.6(15)}$ & $<$1.6(14) & 1.0(14)$^{+1.9(14)}_{-5.1(13)}$ & $<$1.0(14)\\ 
NGC7538\,IRS9 6 & 5.7(13)$^{+9.3(13)}_{-4.4(13)}$ & 1.2(16)$^{+1.5(15)}_{-1.6(15)}$ & $<$6.6(14) & 2.5(14)$^{+4.4(14)}_{-1.4(14)}$ & $<$2.4(14)\\ 
NGC7538\,IRS9 7 & 5.0(13)$^{+2.1(13)}_{-3.2(13)}$ & 1.3(16)$^{+5.4(15)}_{-3.8(15)}$ & $<$7.8(16) & $<$9.1(13) & $<$1.1(16)\\ 
NGC7538\,IRS9 8 & 4.3(13)$^{+4.7(13)}_{-2.0(13)}$ & 1.4(15)$^{+1.1(15)}_{-6.0(14)}$ & $<$1.6(14) & 5.8(13)$^{+1.1(14)}_{-3.3(13)}$ & $<$2.2(14)\\ 
\hline 
S87\,IRS1 1 & 2.7(13)$^{+2.9(13)}_{-2.0(13)}$ & 2.6(15)$^{+3.3(15)}_{-1.3(15)}$ & $<$3.2(16) & $<$1.5(13) & $<$2.4(14)\\ 
S87\,IRS1 2 & 7.7(13)$^{+1.3(14)}_{-3.2(13)}$ & 1.0(15)$^{+1.3(15)}_{-1.9(14)}$ & $<$3.6(14) & 6.7(13)$^{+6.6(12)}_{-5.6(13)}$ & $<$3.1(15)\\ 
S87\,IRS1 3 & $<$4.3(12) & 6.6(14)$^{+6.7(14)}_{-2.5(14)}$ & $<$1.0(14) & $<$2.0(13) & $<$2.9(14)\\ 
S87\,IRS1 4 & 2.0(13)$^{+9.1(12)}_{-1.7(13)}$ & 6.5(14)$^{+1.9(14)}_{-2.0(14)}$ & $<$2.7(14) & 1.2(14)$^{+6.5(13)}_{-9.9(13)}$ & $<$2.8(14)\\ 
S87\,IRS1 5 & $<$1.1(12) & 1.0(15)$^{+1.6(15)}_{-2.1(14)}$ & $<$3.3(16) & $<$1.7(13) & $<$3.1(14)\\ 
S87\,IRS1 6 & 1.1(14)$^{+2.1(14)}_{-8.2(13)}$ & 1.2(15)$^{+1.7(15)}_{-5.3(14)}$ & $<$1.0(14) & $<$2.2(13) & $<$1.4(14)\\ 
S87\,IRS1 7 & 9.7(13)$^{+1.9(13)}_{-4.1(13)}$ & 1.1(15)$^{+1.1(15)}_{-4.1(14)}$ & 1.5(14)$^{+1.4(14)}_{-5.2(12)}$ & 1.6(14)$^{+2.0(14)}_{-7.9(13)}$ & $<$1.4(14)\\ 
S87\,IRS1 8 & 8.5(13)$^{+4.3(14)}_{-5.9(13)}$ & 1.1(15)$^{+1.2(15)}_{-3.0(14)}$ & $<$3.1(14) & $<$1.8(13) & $<$2.3(16)\\ 
S87\,IRS1 9 & $<$1.2(13) & 2.5(15)$^{+1.2(15)}_{-1.1(15)}$ & $<$1.5(14) & $<$1.3(13) & $<$1.5(14)\\ 
\hline 
S106 1 & $<$2.2(12) & 3.0(14)$^{+2.4(14)}_{-1.3(14)}$ & $<$1.1(14) & $<$8.6(13) & $<$4.5(16)\\ 
S106 2 & 1.9(14)$^{+7.4(13)}_{-5.3(13)}$ & 1.7(16)$^{+3.8(15)}_{-2.6(15)}$ & 2.2(15)$^{+5.5(14)}_{-7.5(14)}$ & 2.0(15)$^{+1.0(15)}_{-5.7(14)}$ & $<$6.5(14)\\ 
S106 3 & 2.1(14)$^{+9.0(13)}_{-2.4(13)}$ & 2.8(15)$^{+9.7(14)}_{-7.1(14)}$ & $<$5.4(14) & 2.3(14)$^{+7.3(13)}_{-3.8(13)}$ & $<$3.3(14)\\ 
S106 4 & 1.3(14)$^{+5.2(13)}_{-3.1(13)}$ & 9.3(14)$^{+4.4(14)}_{-5.6(14)}$ & $<$1.0(14) & 8.2(13)$^{+5.1(14)}_{-8.1(12)}$ & $<$2.0(14)\\ 
\hline 
W3\,H2O 1 & 1.2(14)$^{+2.6(13)}_{-2.6(13)}$ & 1.9(16)$^{+5.2(15)}_{-3.4(15)}$ & 2.4(15)$^{+1.3(15)}_{-4.1(14)}$ & 4.0(14)$^{+6.0(14)}_{-1.1(14)}$ & $<$7.1(14)\\ 
W3\,H2O 2 & 3.7(13)$^{+2.2(13)}_{-1.6(13)}$ & 1.0(16)$^{+3.1(15)}_{-8.6(14)}$ & 4.8(14)$^{+2.5(14)}_{-3.4(14)}$ & 2.2(14)$^{+9.0(14)}_{-5.6(13)}$ & 9.9(14)$^{+5.3(15)}_{-1.8(14)}$\\ 
W3\,H2O 3 & 2.4(14)$^{+2.8(13)}_{-1.7(13)}$ & 1.3(17)$^{+2.7(16)}_{-1.5(16)}$ & 5.2(16)$^{+2.1(14)}_{-3.9(15)}$ & 8.8(15)$^{+9.4(14)}_{-1.0(15)}$ & 1.2(16)$^{+1.2(15)}_{-1.1(15)}$\\ 
W3\,H2O 4 & 4.4(14)$^{+3.2(13)}_{-2.0(13)}$ & 2.2(17)$^{+2.5(16)}_{-2.2(16)}$ & 6.7(16)$^{+1.6(15)}_{-2.9(15)}$ & 6.9(15)$^{+1.1(15)}_{-3.3(14)}$ & 4.2(16)$^{+1.0(16)}_{-1.2(16)}$\\ 
W3\,H2O 5 & 1.3(14)$^{+8.9(13)}_{-1.6(13)}$ & 5.7(16)$^{+1.6(16)}_{-1.5(16)}$ & 6.7(14)$^{+1.8(15)}_{-4.4(14)}$ & 3.1(15)$^{+5.9(14)}_{-7.1(14)}$ & $<$5.4(14)\\ 
\hline 
W3\,IRS4 1 & 3.9(13)$^{+8.1(13)}_{-3.2(13)}$ & 4.0(16)$^{+6.3(15)}_{-7.9(15)}$ & 1.3(16)$^{+1.8(15)}_{-8.9(14)}$ & 1.9(15)$^{+4.9(14)}_{-3.9(14)}$ & 7.8(14)$^{+1.7(15)}_{-3.6(14)}$\\ 
W3\,IRS4 2 & 8.4(13)$^{+1.6(14)}_{-4.5(12)}$ & 1.3(15)$^{+4.5(14)}_{-4.7(14)}$ & $<$8.7(16) & 2.5(14)$^{+2.4(14)}_{-7.5(13)}$ & $<$1.7(14)\\ 
W3\,IRS4 3 & 3.4(13)$^{+1.8(14)}_{-1.2(13)}$ & 1.2(15)$^{+5.6(14)}_{-2.7(14)}$ & $<$1.0(15) & 2.2(14)$^{+3.0(14)}_{-8.0(13)}$ & $<$1.7(14)\\ 
W3\,IRS4 4 & $<$5.9(12) & 2.1(15)$^{+1.2(15)}_{-8.9(14)}$ & $<$1.0(14) & $<$5.4(13) & $<$1.7(14)\\ 
W3\,IRS4 5 & 5.3(13)$^{+4.0(13)}_{-3.0(13)}$ & 2.4(15)$^{+5.1(14)}_{-1.0(15)}$ & $<$1.0(14) & 1.2(14)$^{+1.5(14)}_{-7.3(13)}$ & $<$6.7(14)\\ 
W3\,IRS4 6 & 3.4(13)$^{+5.5(13)}_{-2.6(13)}$ & 2.5(15)$^{+2.2(15)}_{-1.0(15)}$ & $<$6.1(14) & 7.8(13)$^{+1.3(14)}_{-3.2(13)}$ & $<$2.5(14)\\ 
\end{longtable}

\begin{longtable}{lllll}
\caption{Molecular column densities derived with \texttt{XCLASS}.}
\label{tab:XCLASSresults3}
\\
\hline\hline
Position & HC$_{3}$N;$\varv_{7}$=2 & CH$_{3}$OH & CH$_{3}$OH;$\varv_{t}$=1 & CH$_{3}$CN \\
\hline
\endfirsthead
\caption{continued.}\\
\hline\hline
Position & HC$_{3}$N;$\varv_{7}$=2 & CH$_{3}$OH & CH$_{3}$OH;$\varv_{t}$=1 & CH$_{3}$CN \\
\hline
\endhead
\hline
Notes. a(b) = a$\times$10$^{\mathrm{b}}$.
\endfoot
\hline
IRAS\,23033 1 & $<$1.0(17) & 1.5(16)$^{+9.0(15)}_{-6.5(15)}$ & $<$1.9(16) & $<$2.7(13)\\ 
IRAS\,23033 2 & $<$1.0(16) & 5.3(16)$^{+5.3(15)}_{-1.1(16)}$ & 7.5(16)$^{+7.7(16)}_{-6.6(16)}$ & 4.6(14)$^{+6.9(13)}_{-3.3(13)}$\\ 
IRAS\,23033 3 & $<$5.5(16) & 8.5(16)$^{+3.6(16)}_{-1.2(16)}$ & 1.8(17)$^{+6.0(16)}_{-6.4(16)}$ & 1.8(15)$^{+2.7(14)}_{-2.4(14)}$\\ 
IRAS\,23033 4 & $<$6.0(16) & 1.1(16)$^{+2.9(16)}_{-7.0(15)}$ & $<$1.3(16) & $<$5.0(13)\\ 
IRAS\,23033 5 & $<$2.3(16) & $<$8.6(14) & $<$5.9(15) & $<$3.1(13)\\ 
\hline 
IRAS\,23151 1 & $<$1.0(16) & 1.8(17)$^{+4.8(16)}_{-3.2(16)}$ & 7.5(17)$^{+3.7(17)}_{-2.5(17)}$ & 1.2(16)$^{+6.2(15)}_{-2.9(15)}$\\ 
IRAS\,23151 2 & $<$1.0(16) & 1.9(16)$^{+2.5(16)}_{-1.2(16)}$ & 1.6(17)$^{+4.6(16)}_{-1.6(17)}$ & 2.1(14)$^{+1.8(14)}_{-9.8(13)}$\\ 
IRAS\,23151 3 & $<$4.0(16) & 8.6(15)$^{+2.6(16)}_{-4.5(15)}$ & $<$5.4(15) & 9.8(13)$^{+1.1(14)}_{-2.1(13)}$\\ 
IRAS\,23151 4 & $<$3.2(16) & 9.1(15)$^{+1.8(16)}_{-8.8(15)}$ & $<$5.1(16) & $<$1.2(14)\\ 
IRAS\,23151 5 & $<$6.1(16) & $<$1.3(15) & $<$1.5(18) & $<$7.2(13)\\ 
\hline 
IRAS\,23385 1 & $<$3.8(16) & 2.6(16)$^{+4.2(15)}_{-3.7(15)}$ & 3.2(17)$^{+2.5(18)}_{-2.5(17)}$ & 4.1(14)$^{+7.6(13)}_{-3.6(13)}$\\ 
IRAS\,23385 2 & $<$4.1(15) & 2.5(16)$^{+9.3(15)}_{-6.2(15)}$ & $<$2.9(16) & 7.2(13)$^{+1.3(14)}_{-3.6(13)}$\\ 
IRAS\,23385 3 & $<$2.2(15) & 2.6(16)$^{+2.3(16)}_{-1.0(16)}$ & 3.2(16)$^{+1.8(17)}_{-2.5(16)}$ & 1.4(14)$^{+2.0(14)}_{-4.4(13)}$\\ 
IRAS\,23385 4 & $<$1.4(15) & 2.6(16)$^{+1.5(16)}_{-5.6(15)}$ & 5.4(16)$^{+9.0(16)}_{-5.1(16)}$ & 1.0(14)$^{+1.0(14)}_{-5.1(13)}$\\ 
IRAS\,23385 5 & $<$2.4(16) & 2.4(16)$^{+2.1(16)}_{-8.3(15)}$ & $<$5.1(16) & $<$9.4(13)\\ 
IRAS\,23385 6 & $<$1.5(16) & 2.1(16)$^{+4.1(16)}_{-7.7(15)}$ & 2.6(16)$^{+4.1(15)}_{-2.5(16)}$ & 9.8(13)$^{+3.8(14)}_{-3.2(13)}$\\ 
IRAS\,23385 7 & $<$2.6(15) & 1.8(16)$^{+1.4(16)}_{-2.4(15)}$ & $<$1.9(16) & $<$1.1(14)\\ 
IRAS\,23385 8 & $<$2.1(15) & 1.6(16)$^{+1.3(16)}_{-5.3(15)}$ & $<$4.8(16) & $<$9.2(13)\\ 
\hline 
AFGL\,2591 1 & 4.2(15)$^{+7.2(14)}_{-5.8(14)}$ & 2.0(17)$^{+5.2(16)}_{-2.4(16)}$ & 3.0(17)$^{+6.8(16)}_{-6.7(16)}$ & 1.6(16)$^{+2.6(15)}_{-1.9(15)}$\\ 
AFGL\,2591 2 & $<$6.3(16) & $<$4.5(15) & $<$4.8(16) & $<$3.4(14)\\ 
AFGL\,2591 3 & $<$7.8(16) & $<$1.3(16) & $<$4.3(16) & $<$1.0(13)\\ 
AFGL\,2591 4 & $<$8.0(16) & 2.3(16)$^{+3.0(16)}_{-2.2(16)}$ & $<$3.0(16) & $<$1.6(13)\\ 
\hline 
CepA\,HW2 1 & 2.1(16)$^{+3.5(15)}_{-5.2(15)}$ & 3.6(17)$^{+4.3(16)}_{-1.3(16)}$ & 1.3(18)$^{+4.6(17)}_{-9.4(16)}$ & 1.4(16)$^{+8.9(15)}_{-1.2(15)}$\\ 
CepA\,HW2 2 & $<$6.6(15) & 2.2(16)$^{+2.6(16)}_{-8.9(15)}$ & $<$1.4(15) & 3.9(15)$^{+1.3(15)}_{-7.0(14)}$\\ 
CepA\,HW2 3 & $<$1.1(15) & 5.4(17)$^{+8.6(16)}_{-3.3(16)}$ & 5.6(18)$^{+7.6(17)}_{-5.5(17)}$ & 1.7(16)$^{+4.0(15)}_{-2.2(15)}$\\ 
CepA\,HW2 4 & $<$6.0(16) & $<$1.2(14) & $<$2.4(15) & $<$1.0(13)\\ 
CepA\,HW2 5 & $<$1.0(16) & 6.8(16)$^{+1.5(17)}_{-2.7(16)}$ & 5.4(16)$^{+3.9(16)}_{-5.2(16)}$ & $<$1.2(13)\\ 
\hline 
G084.9505 1 & $<$1.0(16) & 4.6(16)$^{+1.6(16)}_{-1.2(16)}$ & 1.0(17)$^{+1.5(17)}_{-7.4(16)}$ & 5.8(14)$^{+2.5(14)}_{-1.2(14)}$\\ 
G084.9505 2 & $<$2.0(15) & 4.4(15)$^{+2.8(16)}_{-3.6(15)}$ & $<$1.3(15) & $<$1.1(13)\\ 
G084.9505 3 & $<$1.1(16) & 5.3(15)$^{+5.1(15)}_{-1.3(15)}$ & $<$2.3(17) & $<$1.5(13)\\ 
G084.9505 4 & $<$1.0(16) & 2.3(15)$^{+1.7(15)}_{-1.5(15)}$ & $<$1.2(17) & $<$1.7(13)\\ 
G084.9505 5 & $<$1.0(16) & 2.1(15)$^{+4.5(15)}_{-1.7(15)}$ & $<$1.1(17) & $<$2.1(13)\\ 
G084.9505 6 & $<$6.3(16) & 5.1(15)$^{+5.6(15)}_{-2.7(15)}$ & 1.4(16)$^{+1.9(16)}_{-1.2(16)}$ & $<$2.9(13)\\ 
G084.9505 7 & $<$3.4(16) & $<$5.2(15) & $<$4.1(17) & $<$2.0(13)\\ 
G084.9505 8 & $<$4.4(16) & $<$2.0(16) & $<$5.3(15) & $<$1.2(17)\\ 
\hline 
G094.6028 1 & $<$1.0(16) & 1.9(17)$^{+5.0(16)}_{-1.4(16)}$ & 1.1(18)$^{+6.5(17)}_{-2.4(17)}$ & 2.3(15)$^{+3.2(14)}_{-4.0(14)}$\\ 
G094.6028 2 & $<$5.5(15) & $<$7.8(15) & 2.8(16)$^{+7.8(16)}_{-2.3(16)}$ & $<$8.7(16)\\ 
G094.6028 3 & $<$1.3(15) & $<$1.7(15) & $<$3.1(16) & $<$2.0(17)\\ 
G094.6028 4 & $<$1.6(15) & $<$1.4(15) & $<$4.9(16) & $<$1.7(13)\\ 
G094.6028 5 & $<$4.5(16) & $<$1.8(15) & $<$1.1(17) & $<$1.9(17)\\ 
G094.6028 6 & $<$4.7(15) & $<$2.9(15) & $<$1.5(17) & $<$1.0(13)\\ 
G094.6028 7 & $<$1.0(16) & $<$3.6(15) & 3.4(16)$^{+9.6(16)}_{-3.2(16)}$ & $<$5.3(13)\\ 
G094.6028 8 & $<$1.5(15) & $<$1.4(15) & $<$2.3(16) & $<$2.3(13)\\ 
\hline 
G100.38 1 & $<$1.1(16) & 1.6(16)$^{+4.1(16)}_{-1.5(16)}$ & $<$4.7(15) & 2.1(14)$^{+1.1(14)}_{-5.2(13)}$\\ 
G100.38 2 & $<$2.4(16) & $<$1.0(15) & $<$1.1(16) & $<$7.3(13)\\ 
G100.38 3 & $<$1.2(16) & $<$1.9(15) & $<$4.5(15) & $<$1.3(13)\\ 
G100.38 4 & $<$1.0(15) & $<$2.8(15) & $<$1.8(16) & $<$5.4(16)\\ 
G100.38 5 & $<$1.0(15) & $<$9.0(14) & 2.2(16)$^{+2.8(16)}_{-1.7(16)}$ & $<$3.4(13)\\ 
\hline 
G108.75 1 & $<$1.0(15) & 1.2(16)$^{+2.5(16)}_{-1.0(16)}$ & $<$3.8(17) & 7.8(13)$^{+1.2(14)}_{-5.3(13)}$\\ 
G108.75 2 & $<$2.8(16) & $<$8.6(17) & $<$6.0(16) & $<$1.6(13)\\ 
G108.75 3 & $<$1.5(16) & 2.2(15)$^{+1.8(15)}_{-2.1(15)}$ & $<$7.9(16) & $<$3.9(13)\\ 
G108.75 4 & $<$1.2(15) & 5.7(15)$^{+1.6(16)}_{-4.6(15)}$ & $<$6.6(16) & $<$3.5(13)\\ 
G108.75 5 & $<$1.5(15) & 3.7(15)$^{+1.8(16)}_{-3.1(15)}$ & $<$6.7(15) & $<$3.5(13)\\ 
G108.75 6 & $<$1.2(15) & $<$1.5(16) & $<$2.1(16) & $<$8.4(16)\\ 
\hline 
G138.2957 1 & $<$1.6(15) & $<$1.0(19) & $<$1.2(16) & $<$1.3(13)\\ 
G138.2957 2 & $<$1.0(17) & $<$3.3(15) & $<$4.5(16) & $<$1.9(13)\\ 
G138.2957 3 & $<$8.5(16) & $<$6.1(15) & $<$2.6(16) & $<$1.9(13)\\ 
G138.2957 4 & $<$1.5(15) & $<$2.3(15) & $<$6.0(16) & $<$1.1(14)\\ 
G138.2957 5 & $<$6.1(16) & $<$2.5(15) & $<$2.3(16) & $<$8.5(17)\\ 
G138.2957 6 & $<$1.3(16) & $<$3.8(15) & $<$8.6(15) & $<$1.5(15)\\ 
G138.2957 7 & $<$2.1(15) & $<$1.3(16) & $<$1.3(17) & $<$2.3(13)\\ 
G138.2957 8 & $<$6.8(16) & 1.3(16)$^{+1.2(16)}_{-1.2(16)}$ & $<$1.0(19) & $<$1.0(13)\\ 
G138.2957 9 & $<$1.7(16) & $<$1.0(16) & $<$7.2(16) & $<$1.6(16)\\ 
G138.2957 10 & $<$1.5(16) & $<$3.4(15) & $<$8.4(17) & $<$8.1(14)\\ 
G138.2957 11 & $<$1.5(16) & $<$3.6(15) & $<$3.2(16) & $<$1.1(14)\\ 
G138.2957 12 & $<$1.0(16) & $<$1.4(15) & $<$5.9(15) & $<$1.3(13)\\ 
\hline 
G139.9091 1 & $<$2.0(15) & $<$8.7(14) & $<$3.5(16) & $<$6.0(13)\\ 
G139.9091 2 & $<$1.4(15) & $<$2.8(15) & $<$2.9(18) & $<$2.6(17)\\ 
G139.9091 3 & $<$1.5(16) & $<$7.4(14) & $<$2.4(15) & $<$1.0(15)\\ 
G139.9091 4 & $<$1.5(16) & $<$2.3(15) & $<$4.7(16) & $<$6.2(13)\\ 
\hline 
G075.78 1 & 1.1(16)$^{+3.4(15)}_{-1.8(15)}$ & 1.8(17)$^{+2.2(16)}_{-2.9(16)}$ & 3.9(17)$^{+2.4(17)}_{-5.7(16)}$ & 8.2(15)$^{+1.6(15)}_{-1.1(15)}$\\ 
G075.78 2 & $<$1.0(16) & $<$2.1(15) & $<$2.3(15) & $<$1.0(13)\\ 
G075.78 3 & $<$1.9(15) & $<$1.3(15) & $<$7.9(18) & $<$1.0(13)\\ 
G075.78 4 & $<$4.8(15) & 1.3(16)$^{+2.9(15)}_{-1.2(16)}$ & $<$6.1(16) & $<$3.3(13)\\ 
\hline 
IRAS\,21078 1 & $<$1.4(15) & 1.3(18)$^{+1.4(17)}_{-7.0(16)}$ & 2.3(18)$^{+1.7(17)}_{-2.3(17)}$ & 1.5(16)$^{+2.2(15)}_{-1.6(15)}$\\ 
IRAS\,21078 2 & $<$1.0(16) & 1.9(17)$^{+5.7(16)}_{-3.7(16)}$ & 4.0(17)$^{+5.1(16)}_{-4.5(16)}$ & 2.5(15)$^{+8.6(14)}_{-5.4(14)}$\\ 
IRAS\,21078 3 & 9.0(15)$^{+4.2(16)}_{-3.1(15)}$ & 6.6(17)$^{+2.7(17)}_{-1.3(17)}$ & 1.2(18)$^{+1.2(17)}_{-2.9(17)}$ & 4.0(15)$^{+8.3(14)}_{-1.2(15)}$\\ 
IRAS\,21078 4 & $<$1.0(16) & $<$1.2(16) & $<$9.3(17) & $<$1.5(13)\\ 
IRAS\,21078 5 & $<$1.1(15) & 7.5(15)$^{+4.0(15)}_{-1.5(15)}$ & $<$6.0(16) & $<$9.0(13)\\ 
IRAS\,21078 6 & $<$1.2(16) & 3.1(16)$^{+2.2(16)}_{-1.3(16)}$ & $<$5.5(15) & 5.1(13)$^{+4.4(13)}_{-1.7(13)}$\\ 
IRAS\,21078 7 & $<$1.0(16) & 7.1(15)$^{+8.6(15)}_{-2.5(15)}$ & $<$1.4(16) & $<$1.6(16)\\ 
IRAS\,21078 8 & $<$1.0(16) & 2.9(16)$^{+6.6(15)}_{-1.3(16)}$ & $<$8.8(15) & 1.5(14)$^{+1.0(14)}_{-7.1(13)}$\\ 
IRAS\,21078 9 & $<$1.0(16) & 2.6(16)$^{+2.5(16)}_{-9.4(15)}$ & $<$6.1(18) & $<$2.6(13)\\ 
IRAS\,21078 10 & $<$7.2(16) & 2.5(16)$^{+1.4(16)}_{-5.5(15)}$ & 2.3(16)$^{+8.3(15)}_{-2.1(16)}$ & 2.1(14)$^{+8.8(13)}_{-8.5(13)}$\\ 
IRAS\,21078 11 & $<$1.2(15) & 5.5(17)$^{+8.1(17)}_{-4.8(17)}$ & $<$4.4(16) & 1.5(14)$^{+1.0(14)}_{-6.1(13)}$\\ 
IRAS\,21078 12 & $<$2.9(16) & 1.8(16)$^{+1.2(16)}_{-4.3(15)}$ & $<$3.5(16) & 1.5(14)$^{+3.9(14)}_{-5.2(13)}$\\ 
IRAS\,21078 13 & $<$1.0(15) & 2.9(16)$^{+3.4(16)}_{-1.6(16)}$ & $<$3.7(16) & $<$2.9(14)\\ 
IRAS\,21078 14 & $<$5.4(15) & 2.6(16)$^{+3.6(16)}_{-9.7(15)}$ & $<$1.9(16) & $<$3.8(13)\\ 
\hline 
NGC7538\,IRS9 1 & $<$5.2(15) & 9.1(16)$^{+4.2(16)}_{-2.0(16)}$ & 3.3(17)$^{+2.4(17)}_{-1.2(17)}$ & 8.7(15)$^{+2.5(15)}_{-1.3(15)}$\\ 
NGC7538\,IRS9 2 & $<$7.3(16) & 5.2(15)$^{+2.5(16)}_{-2.8(15)}$ & $<$3.4(15) & $<$4.2(17)\\ 
NGC7538\,IRS9 3 & $<$1.3(16) & 7.6(15)$^{+4.6(16)}_{-6.0(15)}$ & $<$2.6(16) & $<$6.2(16)\\ 
NGC7538\,IRS9 4 & $<$3.0(16) & 3.4(15)$^{+7.7(15)}_{-2.9(15)}$ & $<$9.9(18) & $<$1.4(14)\\ 
NGC7538\,IRS9 5 & $<$6.6(16) & 9.1(15)$^{+9.2(15)}_{-4.7(15)}$ & $<$9.9(15) & $<$5.1(13)\\ 
NGC7538\,IRS9 6 & $<$1.0(16) & 4.6(16)$^{+3.9(16)}_{-8.5(15)}$ & $<$2.1(16) & 1.3(14)$^{+1.4(13)}_{-3.2(13)}$\\ 
NGC7538\,IRS9 7 & $<$3.2(16) & 3.5(15)$^{+2.9(15)}_{-3.2(15)}$ & $<$2.2(15) & $<$7.0(13)\\ 
NGC7538\,IRS9 8 & $<$4.0(16) & 4.4(15)$^{+1.4(16)}_{-3.9(15)}$ & $<$6.3(15) & $<$2.5(13)\\ 
\hline 
S87\,IRS1 1 & $<$1.1(15) & $<$1.8(15) & $<$1.5(16) & $<$1.7(17)\\ 
S87\,IRS1 2 & $<$1.0(16) & $<$1.7(16) & $<$2.5(17) & $<$3.1(13)\\ 
S87\,IRS1 3 & $<$1.5(16) & $<$5.1(18) & $<$1.3(16) & $<$3.8(13)\\ 
S87\,IRS1 4 & $<$1.4(15) & 3.4(15)$^{+8.4(15)}_{-2.9(15)}$ & $<$5.1(16) & $<$2.8(13)\\ 
S87\,IRS1 5 & $<$1.5(16) & $<$1.4(14) & $<$2.2(18) & $<$9.0(13)\\ 
S87\,IRS1 6 & $<$1.3(15) & 2.7(15)$^{+5.0(15)}_{-2.4(15)}$ & $<$2.0(16) & $<$9.3(13)\\ 
S87\,IRS1 7 & $<$3.4(16) & $<$4.4(15) & $<$1.5(18) & $<$2.5(13)\\ 
S87\,IRS1 8 & $<$1.2(15) & $<$1.1(16) & 7.5(17)$^{+5.5(17)}_{-7.4(17)}$ & $<$1.0(13)\\ 
S87\,IRS1 9 & $<$2.9(16) & $<$1.4(16) & $<$1.8(16) & $<$1.0(18)\\ 
\hline 
S106 1 & $<$2.8(16) & $<$1.6(15) & $<$1.5(16) & $<$3.7(14)\\ 
S106 2 & $<$3.8(15) & $<$3.4(16) & $<$2.2(16) & 4.1(14)$^{+8.0(13)}_{-6.7(13)}$\\ 
S106 3 & $<$2.1(16) & $<$1.4(16) & $<$1.1(17) & $<$2.6(14)\\ 
S106 4 & $<$1.0(16) & $<$6.4(14) & $<$8.3(15) & $<$1.0(14)\\ 
\hline 
W3\,H2O 1 & $<$1.8(16) & 2.0(18)$^{+3.2(17)}_{-1.2(17)}$ & 1.4(18)$^{+4.7(17)}_{-1.7(17)}$ & 2.2(15)$^{+7.1(14)}_{-5.3(14)}$\\ 
W3\,H2O 2 & $<$1.9(15) & 1.1(17)$^{+2.0(16)}_{-1.2(16)}$ & 4.4(17)$^{+2.5(17)}_{-2.1(17)}$ & 5.4(14)$^{+6.4(13)}_{-4.7(13)}$\\ 
W3\,H2O 3 & 1.2(16)$^{+3.3(15)}_{-2.1(15)}$ & 1.4(18)$^{+8.5(16)}_{-4.4(16)}$ & 3.1(18)$^{+2.6(17)}_{-2.0(17)}$ & 6.0(16)$^{+1.0(16)}_{-1.0(16)}$\\ 
W3\,H2O 4 & 1.6(16)$^{+3.0(15)}_{-1.1(15)}$ & 2.9(18)$^{+2.9(17)}_{-1.8(17)}$ & 2.5(18)$^{+7.5(16)}_{-1.3(17)}$ & 1.5(17)$^{+1.7(16)}_{-1.2(16)}$\\ 
W3\,H2O 5 & $<$4.5(16) & 4.1(16)$^{+1.3(16)}_{-9.8(15)}$ & $<$1.5(16) & 5.0(14)$^{+7.9(13)}_{-5.7(13)}$\\ 
\hline 
W3\,IRS4 1 & $<$3.3(16) & 1.8(17)$^{+1.7(16)}_{-3.5(16)}$ & 3.0(17)$^{+1.4(17)}_{-4.6(16)}$ & 6.7(15)$^{+2.5(15)}_{-1.3(15)}$\\ 
W3\,IRS4 2 & $<$8.7(15) & 2.1(16)$^{+1.1(17)}_{-1.9(16)}$ & $<$2.0(18) & $<$2.2(13)\\ 
W3\,IRS4 3 & $<$2.5(15) & 1.7(16)$^{+3.8(16)}_{-1.0(16)}$ & $<$1.2(17) & $<$3.2(14)\\ 
W3\,IRS4 4 & $<$3.8(15) & $<$5.4(15) & $<$2.6(16) & $<$2.6(17)\\ 
W3\,IRS4 5 & $<$3.1(16) & $<$1.6(15) & $<$2.1(18) & $<$1.3(14)\\ 
W3\,IRS4 6 & $<$2.3(15) & 5.0(15)$^{+3.6(15)}_{-1.8(15)}$ & $<$1.2(18) & 1.7(14)$^{+1.2(14)}_{-8.3(13)}$\\ 
\end{longtable}

\section{Observed spectra}

	Figure \ref{fig:spectrum} shows the observed spectrum and corresponding \texttt{XCLASS} fit for all 120 positions analyzed in Sect. \ref{sec:XCLASSfitting}.
	
\begin{figure*}
\centering
\includegraphics[]{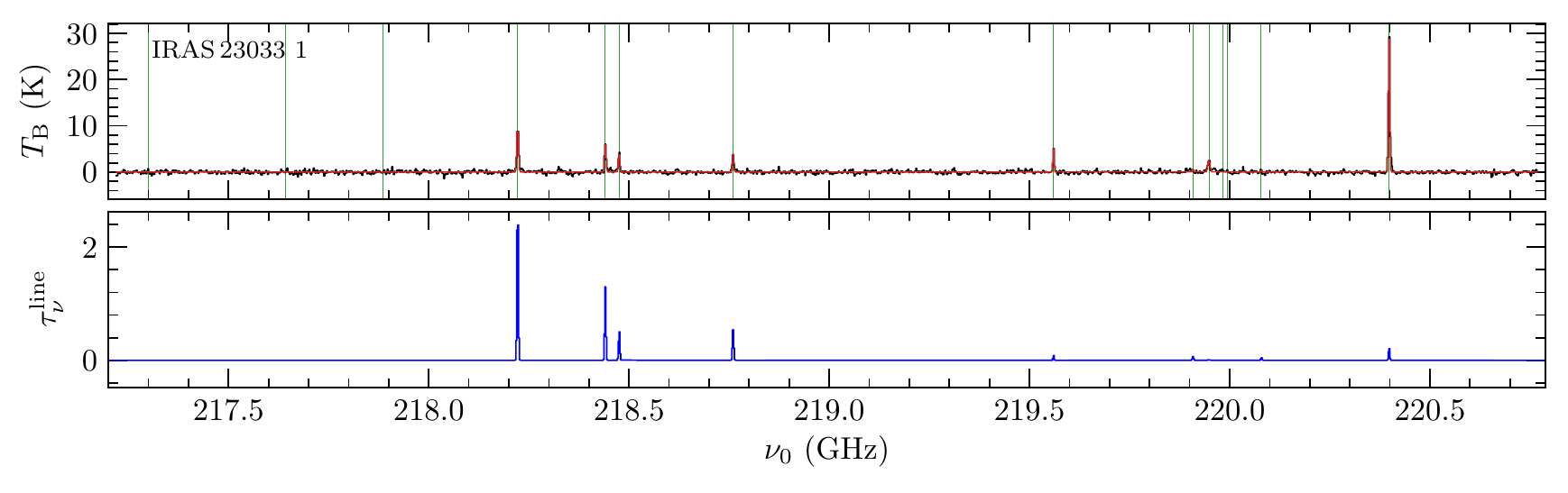}
\includegraphics[]{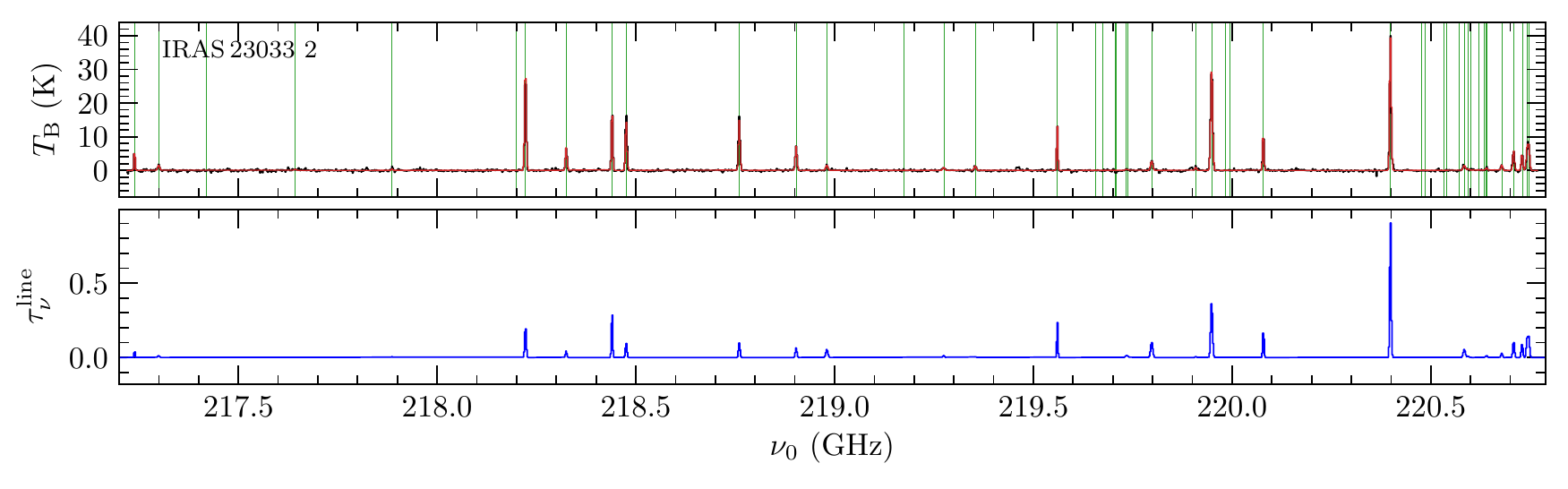}
\includegraphics[]{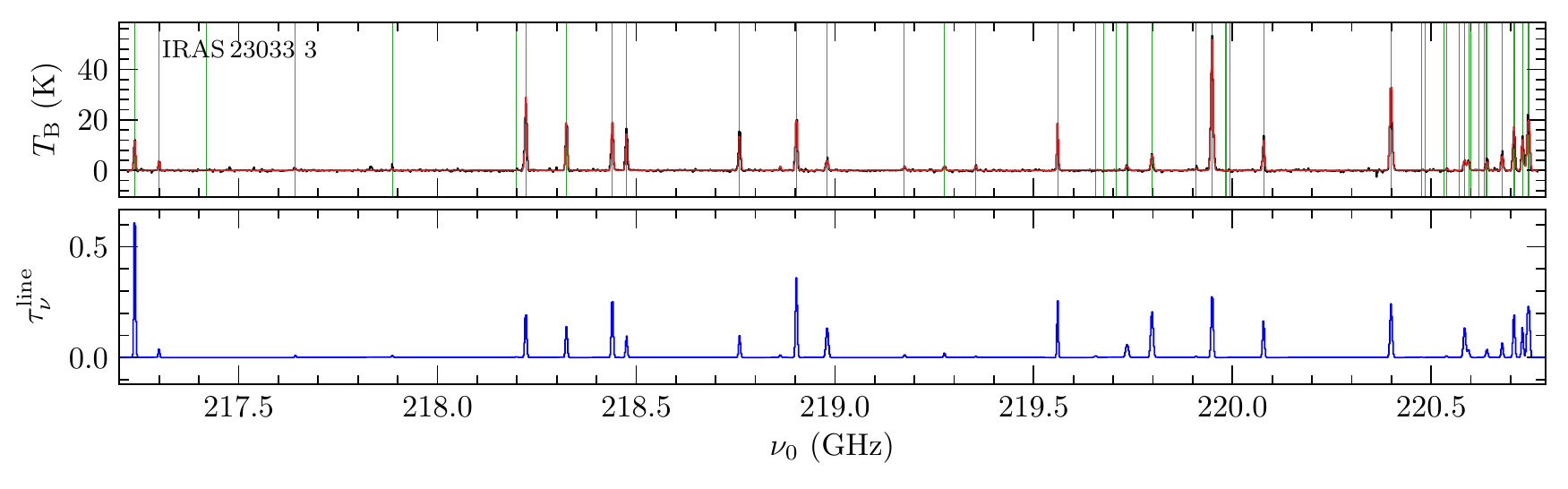}
\includegraphics[]{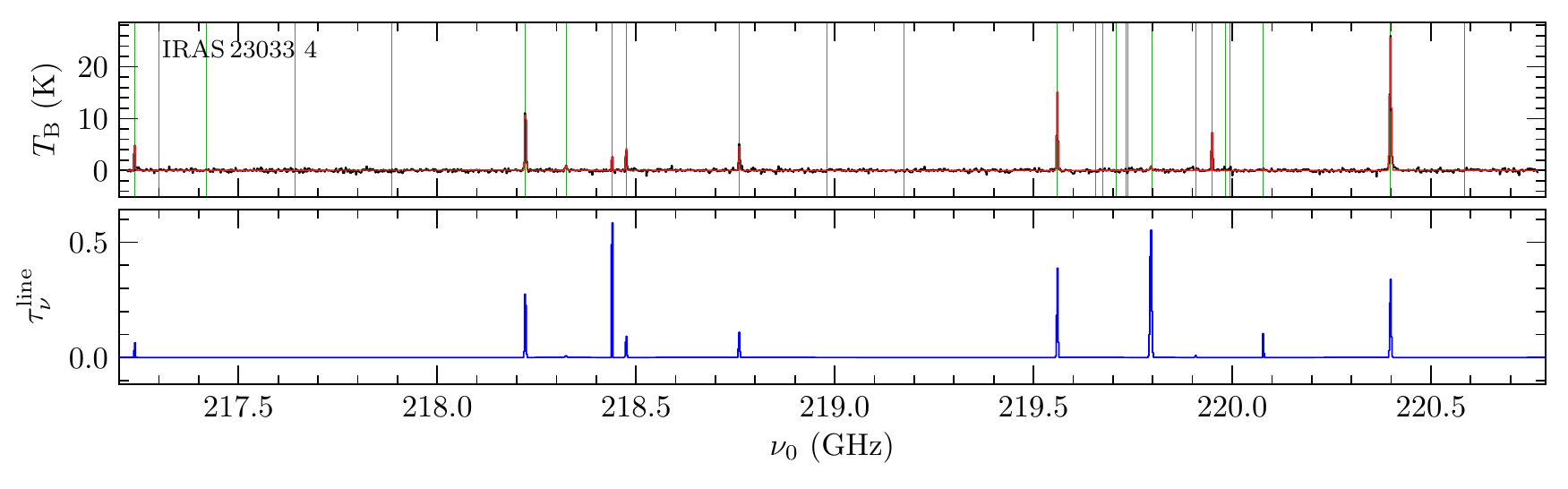}
\caption{\textit{Top panel:} Observed (black line) spectrum and \texttt{XCLASS} fit (red line) for all 120 analyzed positions. Fitted molecular transitions are indicated by green vertical lines. \textit{Bottom panel:} Optical depth profile (blue line) of all fitted transitions for all 120 analyzed positions.}
\label{fig:spectrum}
\end{figure*}
 
\begin{figure*}
\ContinuedFloat
\captionsetup{list=off,format=cont}
\centering
\includegraphics[]{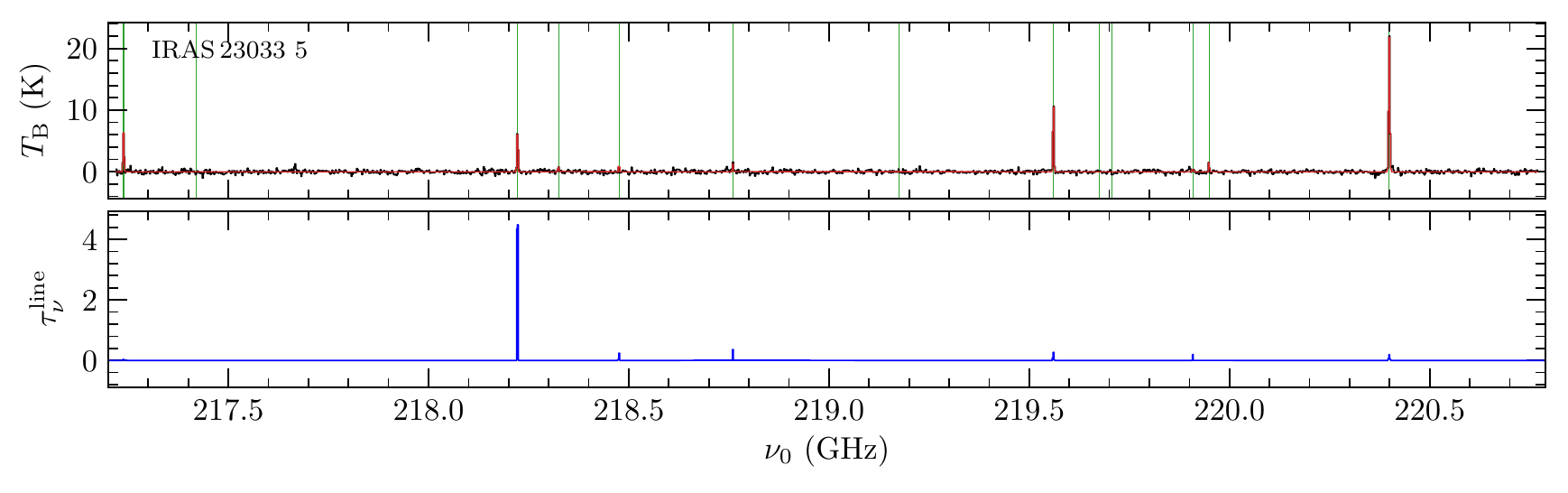}
\includegraphics[]{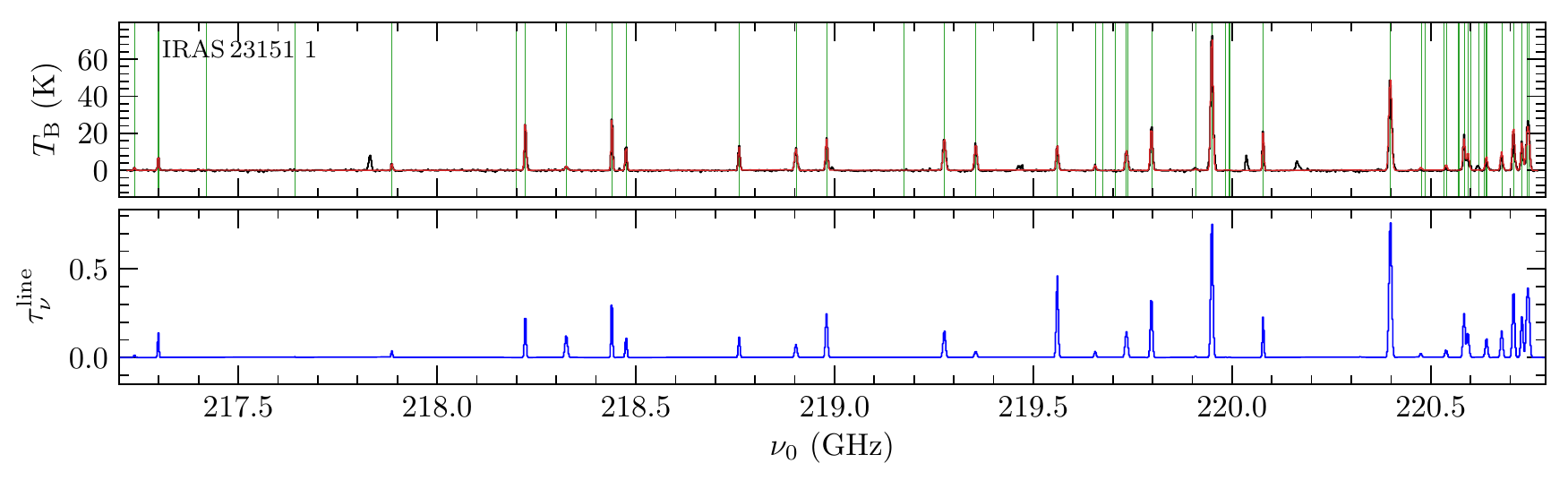}
\includegraphics[]{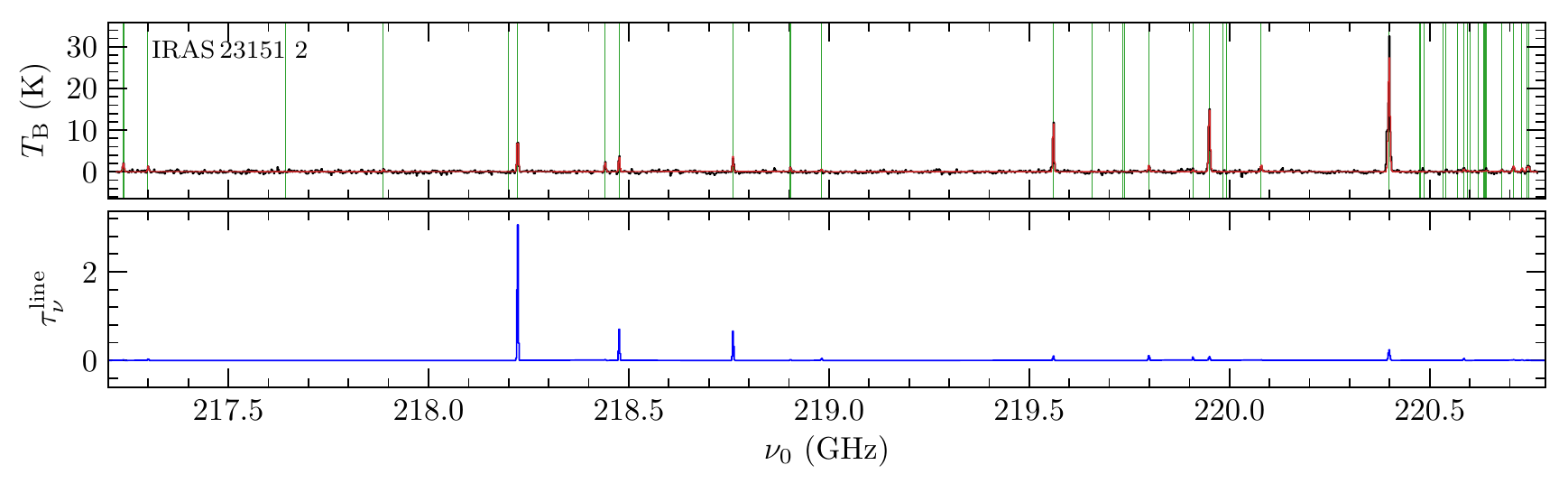}
\includegraphics[]{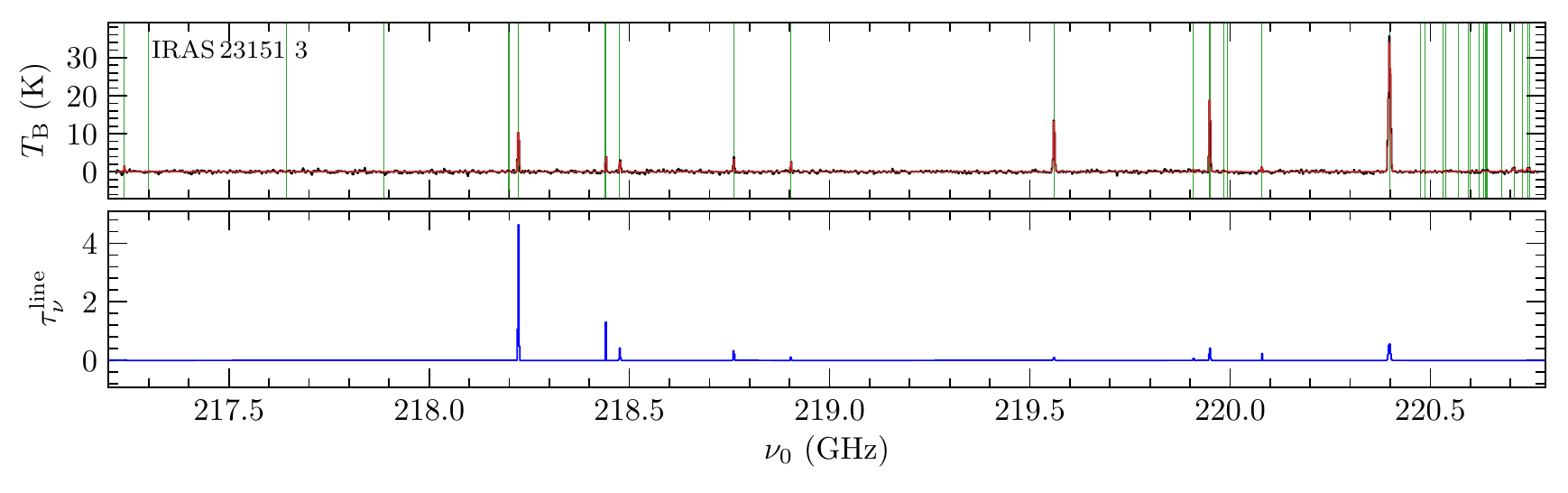}
\caption{\textit{Top panel:} Observed (black line) spectrum and \texttt{XCLASS} fit (red line) for all 120 analyzed positions. Fitted molecular transitions are indicated by green vertical lines. \textit{Bottom panel:} Optical depth profile (blue line) of all fitted transitions for all 120 analyzed positions.}
\end{figure*}
 
\begin{figure*}
\ContinuedFloat
\captionsetup{list=off,format=cont}
\centering
\includegraphics[]{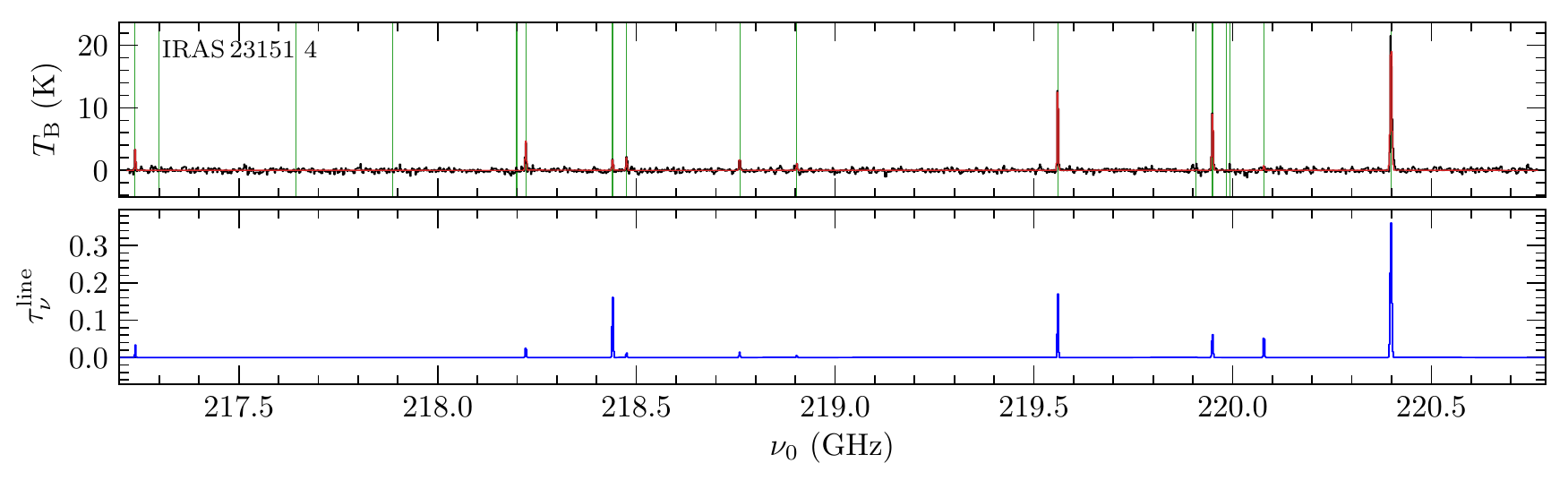}
\includegraphics[]{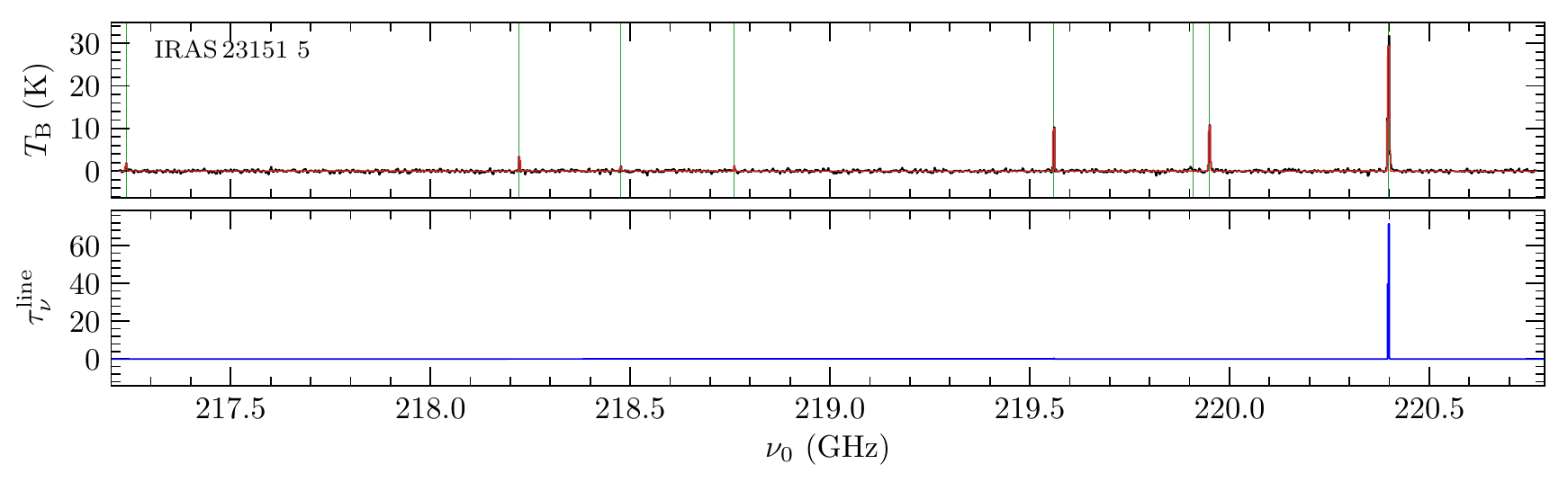}
\includegraphics[]{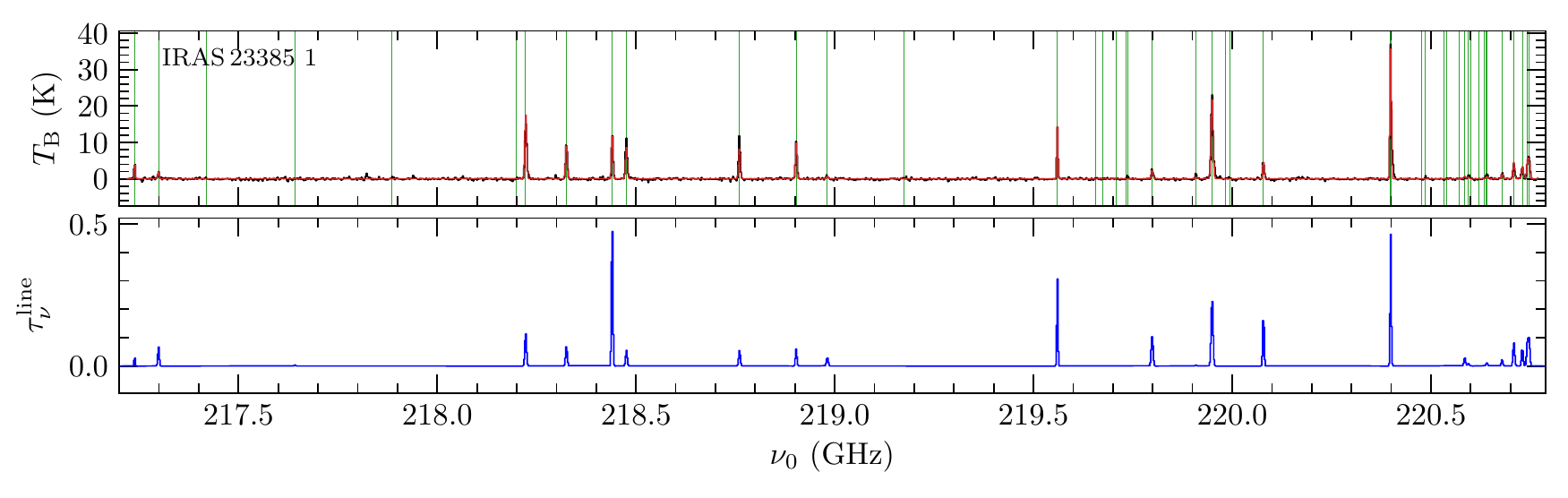}
\includegraphics[]{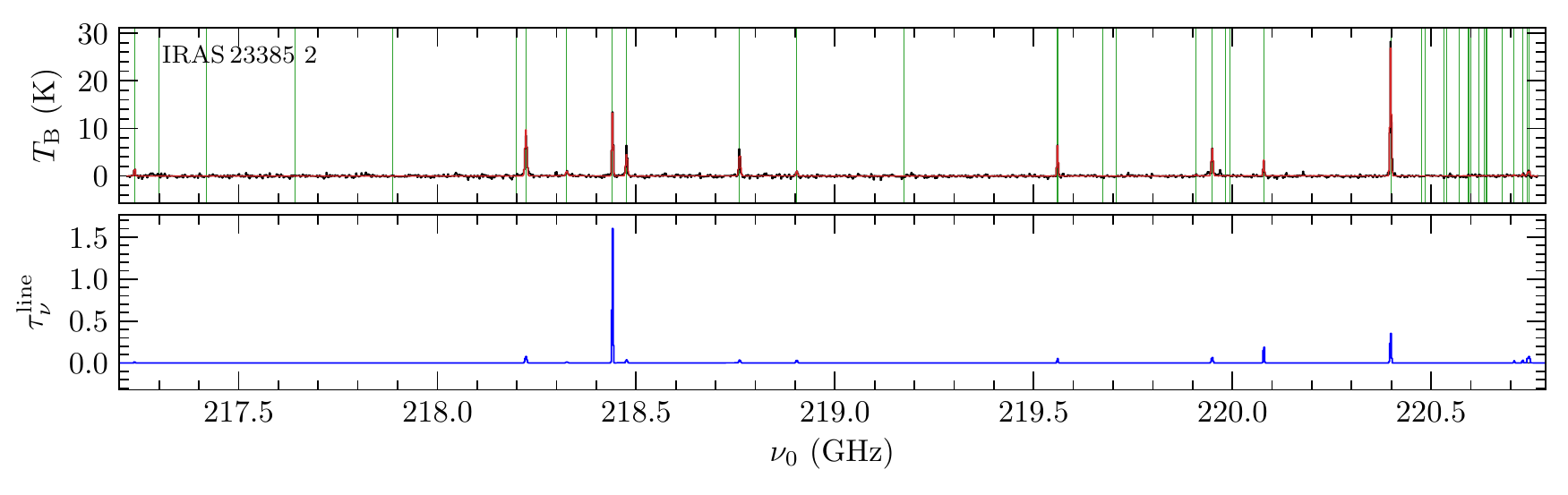}
\caption{\textit{Top panel:} Observed (black line) spectrum and \texttt{XCLASS} fit (red line) for all 120 analyzed positions. Fitted molecular transitions are indicated by green vertical lines. \textit{Bottom panel:} Optical depth profile (blue line) of all fitted transitions for all 120 analyzed positions.}
\end{figure*}
 
\begin{figure*}
\ContinuedFloat
\captionsetup{list=off,format=cont}
\centering
\includegraphics[]{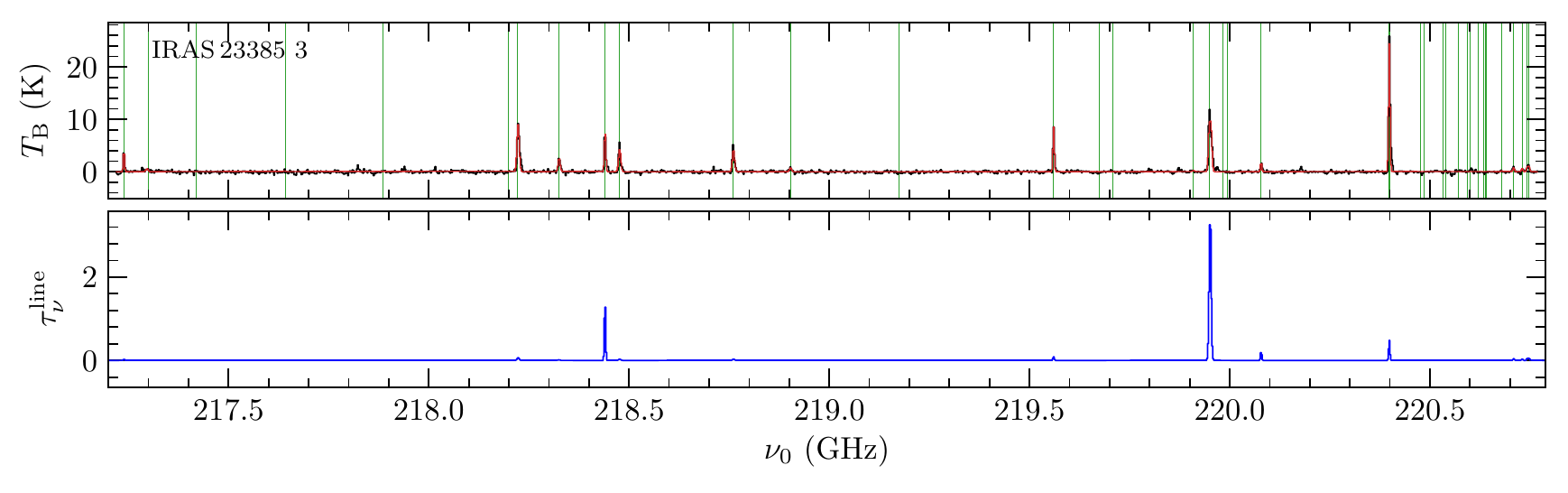}
\includegraphics[]{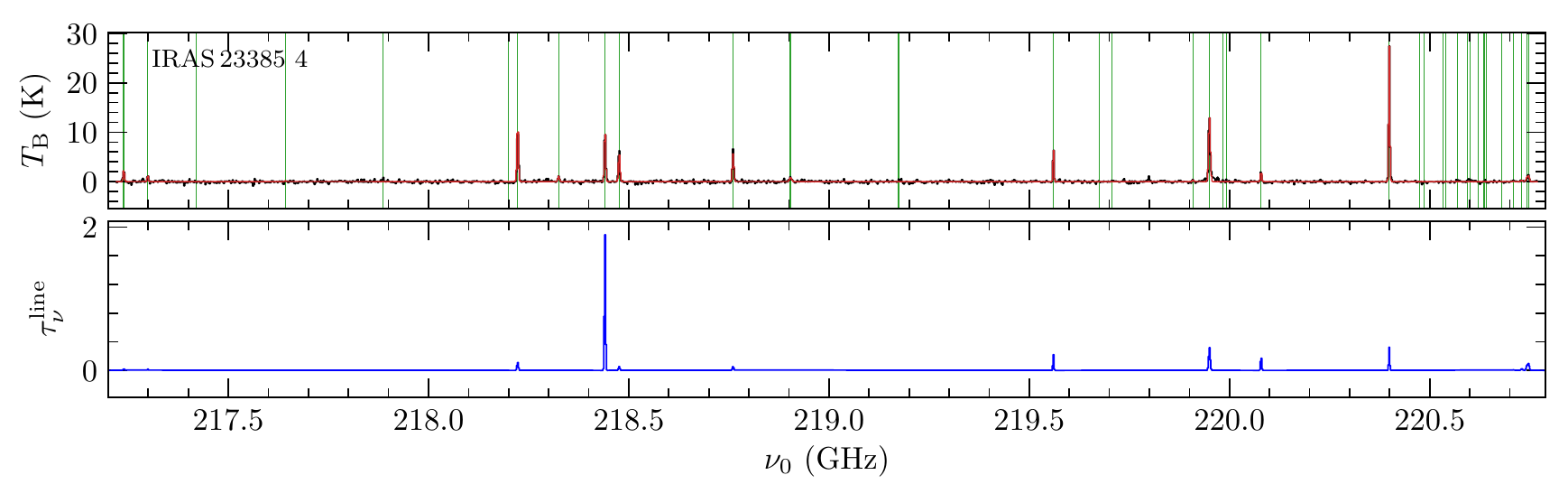}
\includegraphics[]{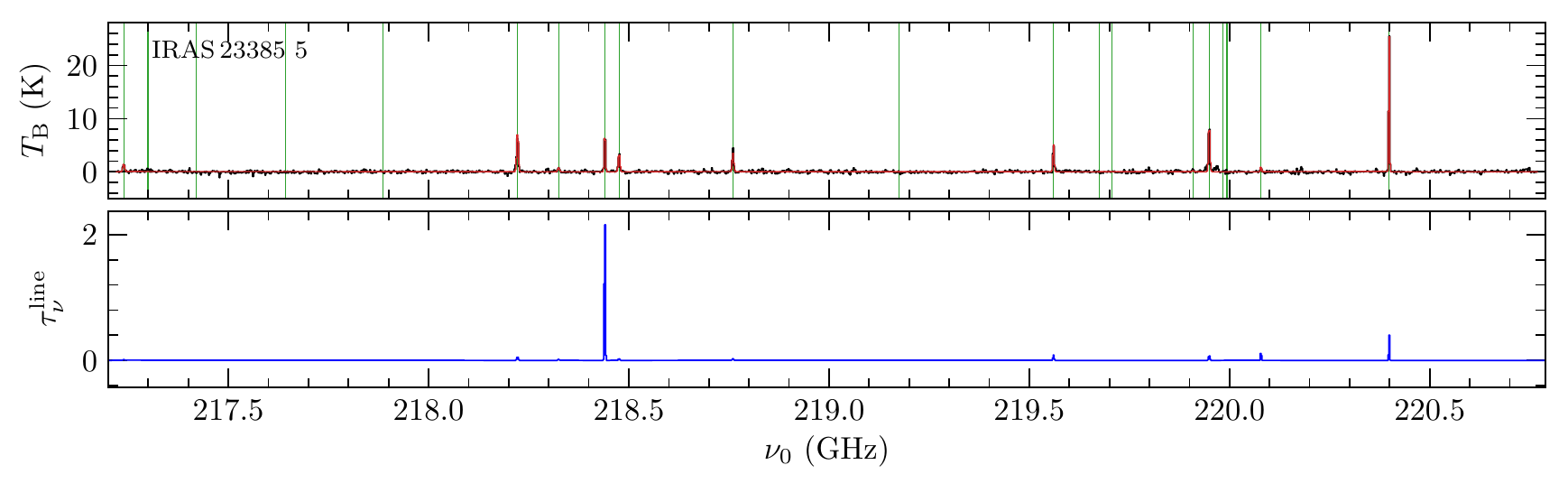}
\includegraphics[]{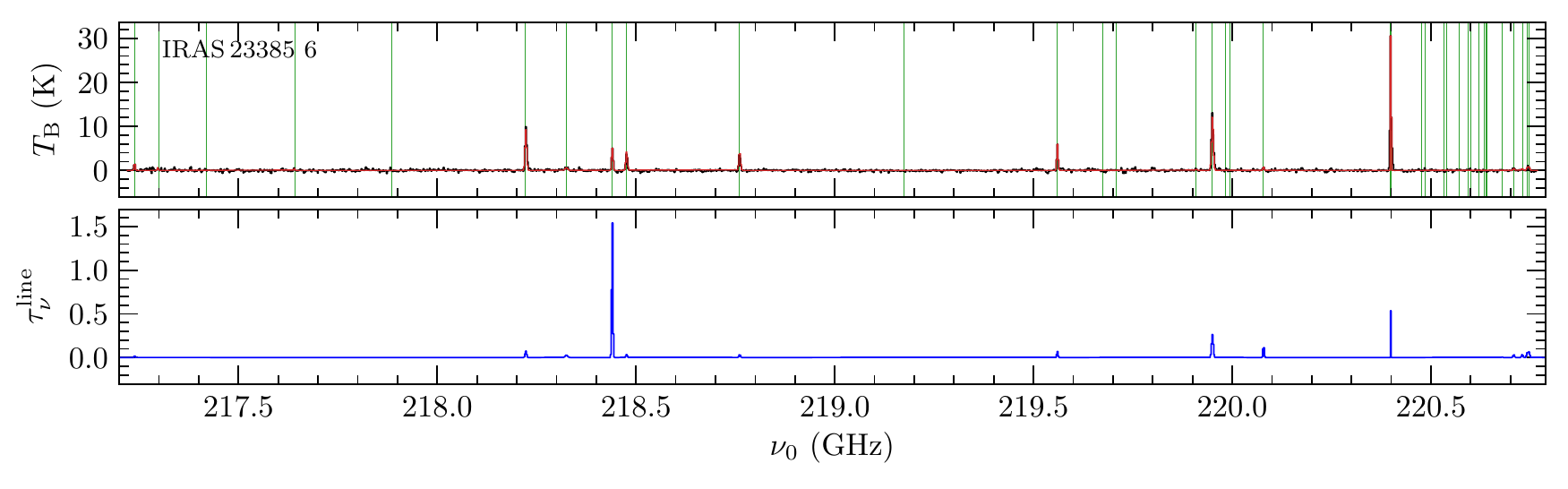}
\caption{\textit{Top panel:} Observed (black line) spectrum and \texttt{XCLASS} fit (red line) for all 120 analyzed positions. Fitted molecular transitions are indicated by green vertical lines. \textit{Bottom panel:} Optical depth profile (blue line) of all fitted transitions for all 120 analyzed positions.}
\end{figure*}
 
\begin{figure*}
\ContinuedFloat
\captionsetup{list=off,format=cont}
\centering
\includegraphics[]{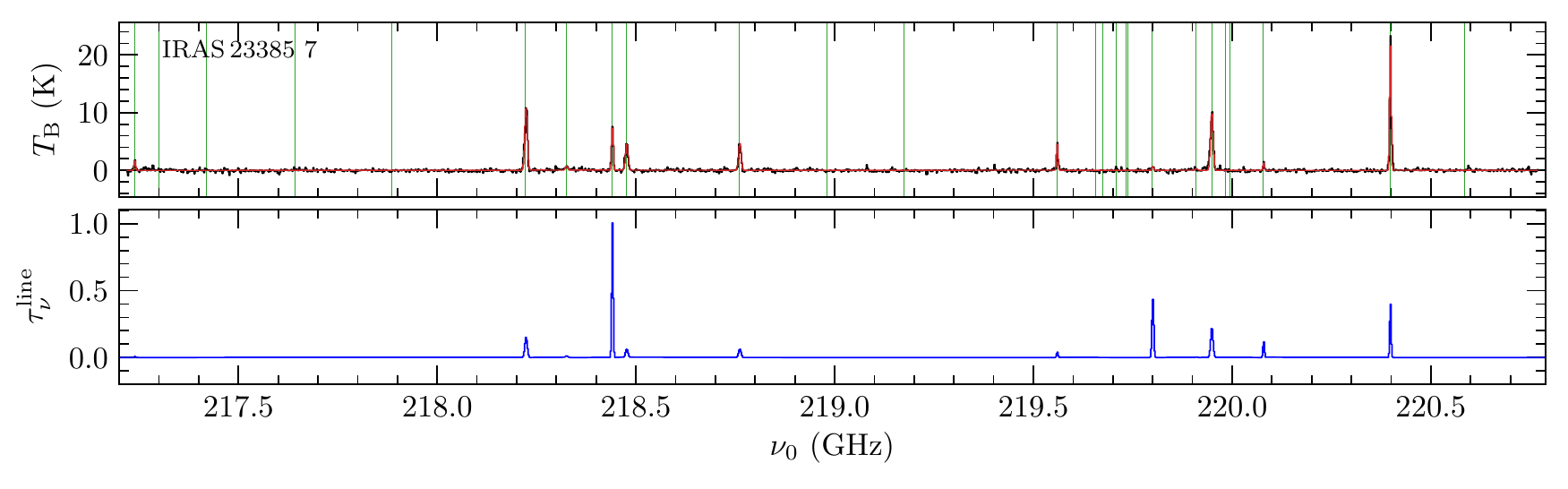}
\includegraphics[]{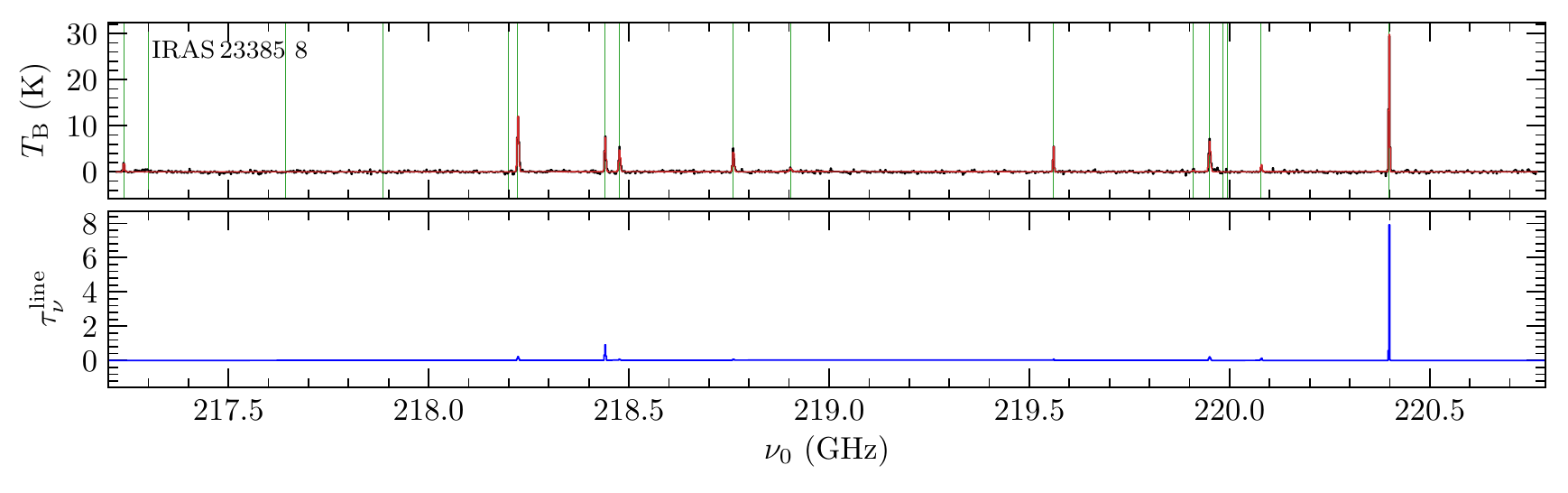}
\includegraphics[]{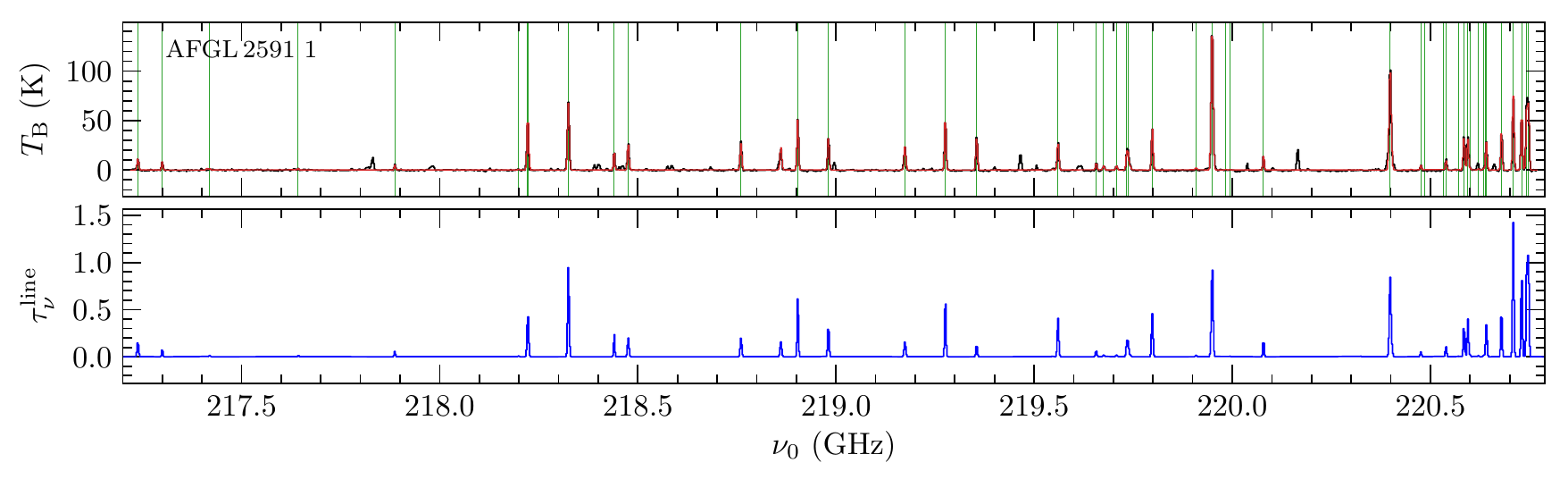}
\includegraphics[]{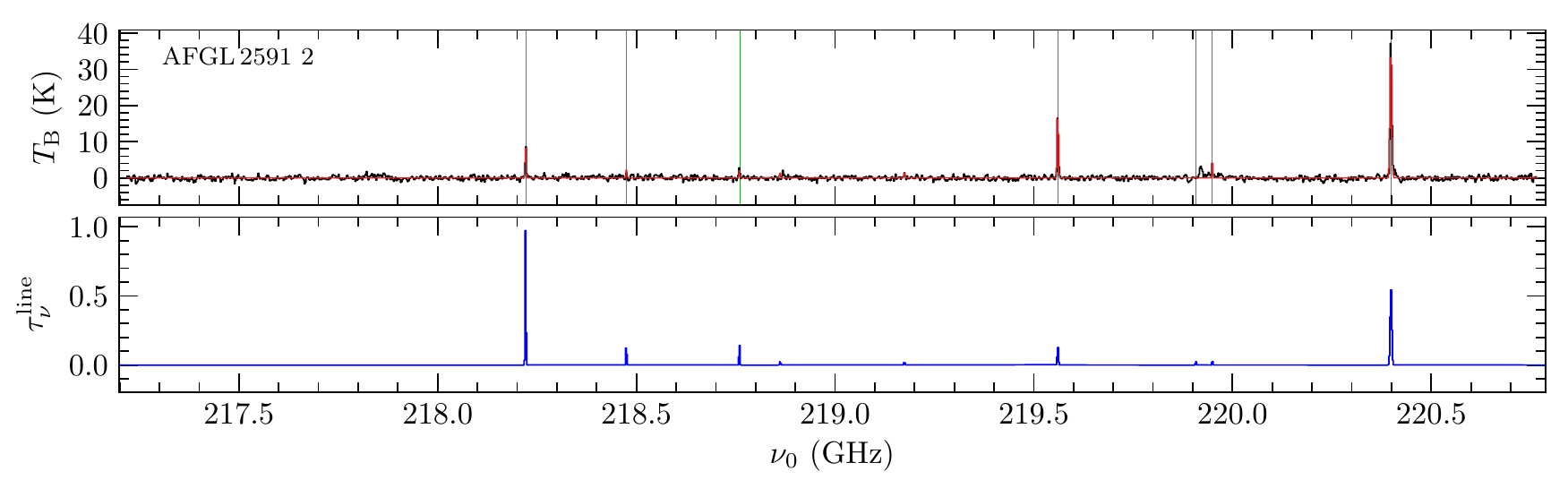}
\caption{\textit{Top panel:} Observed (black line) spectrum and \texttt{XCLASS} fit (red line) for all 120 analyzed positions. Fitted molecular transitions are indicated by green vertical lines. \textit{Bottom panel:} Optical depth profile (blue line) of all fitted transitions for all 120 analyzed positions.}
\end{figure*}
 
\begin{figure*}
\ContinuedFloat
\captionsetup{list=off,format=cont}
\centering
\includegraphics[]{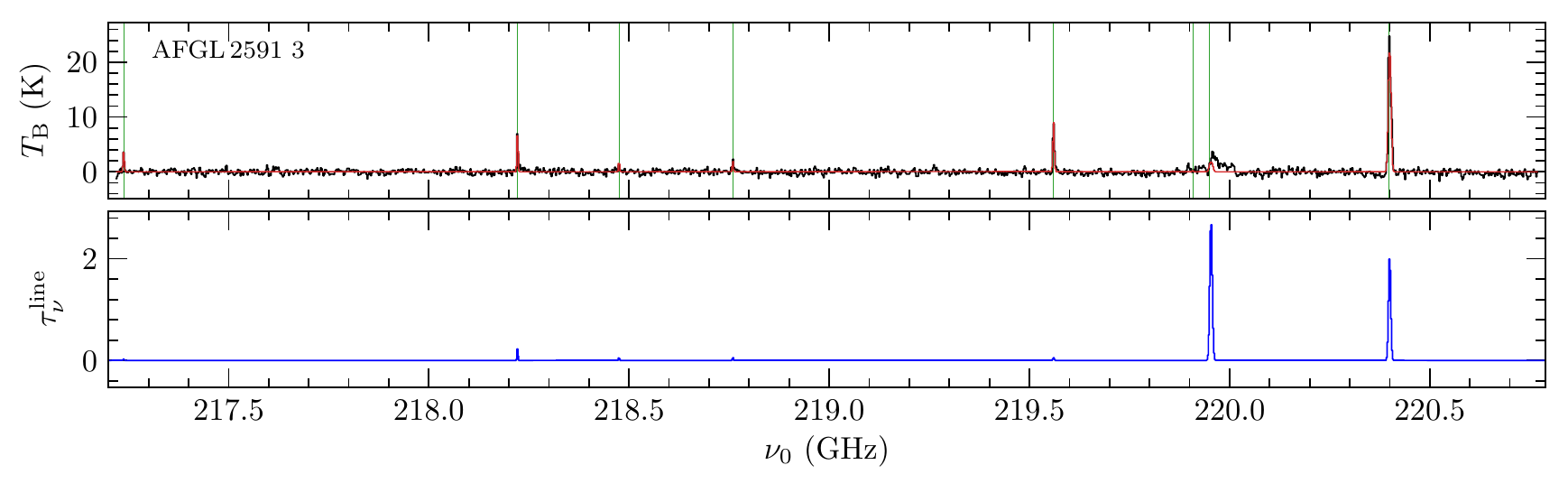}
\includegraphics[]{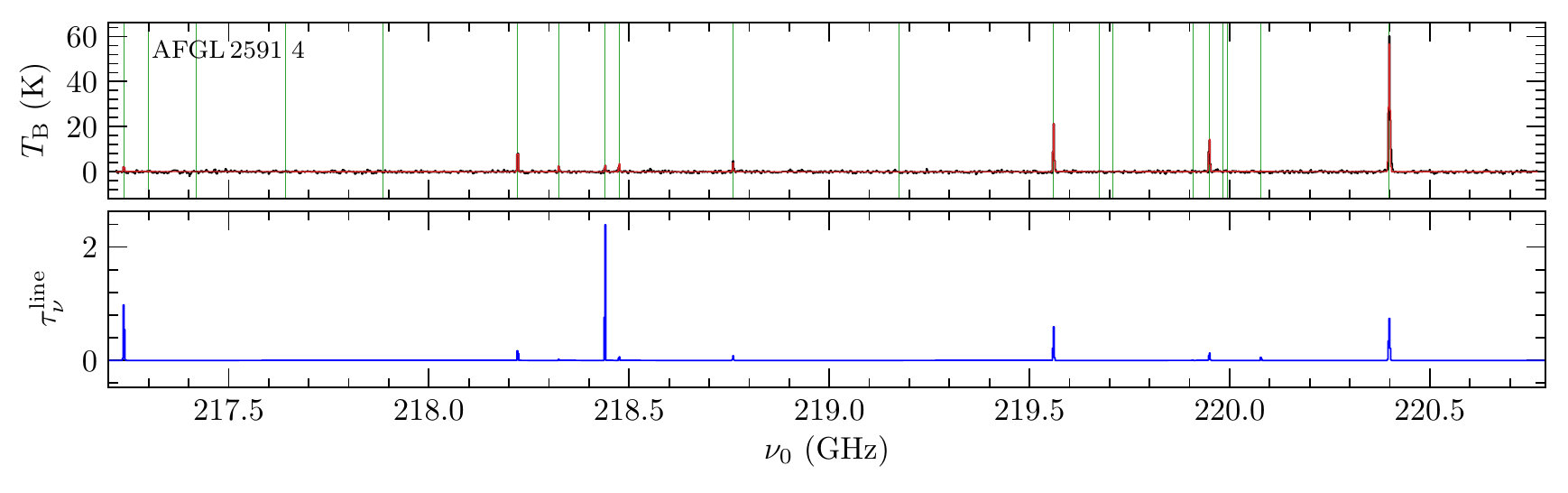}
\includegraphics[]{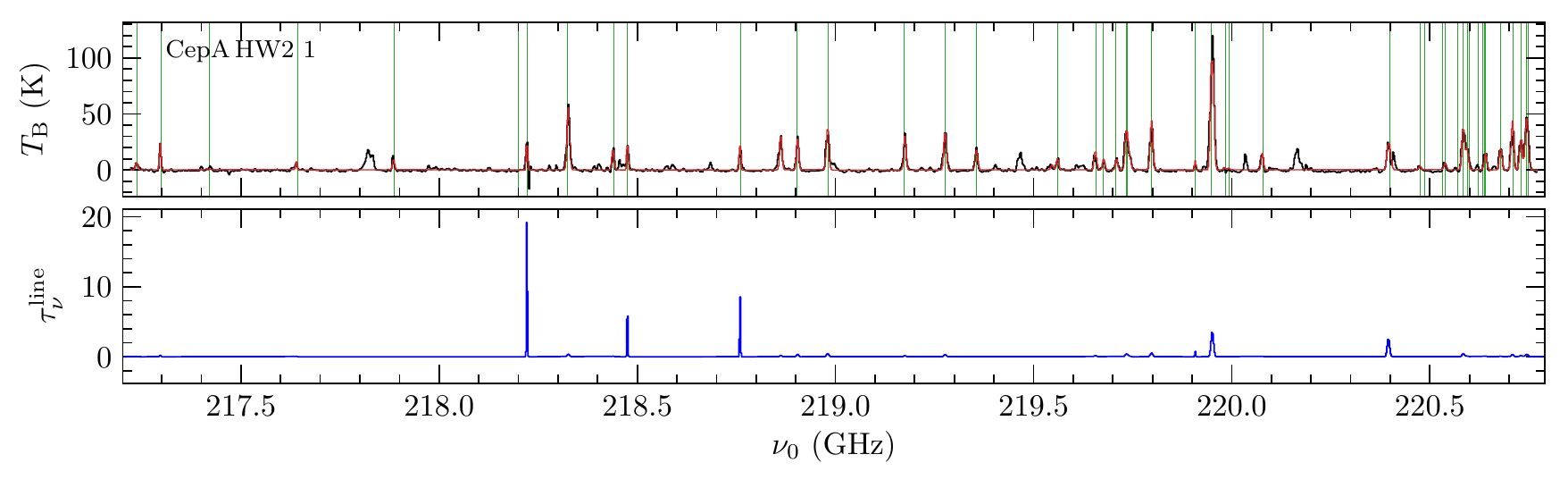}
\includegraphics[]{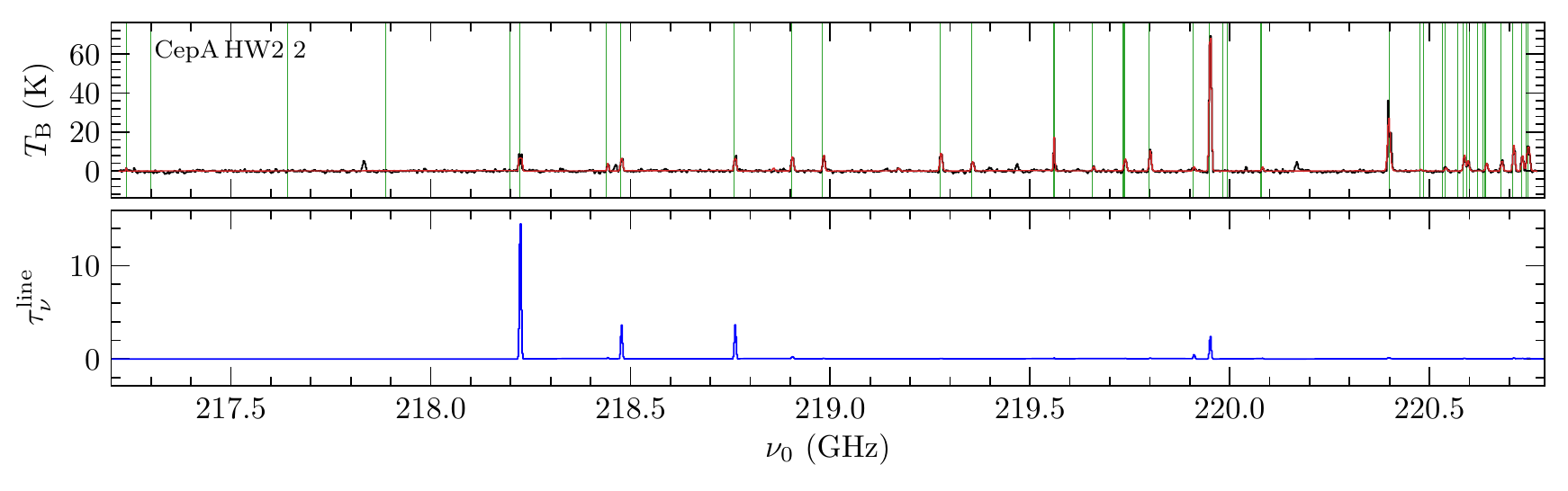}
\caption{\textit{Top panel:} Observed (black line) spectrum and \texttt{XCLASS} fit (red line) for all 120 analyzed positions. Fitted molecular transitions are indicated by green vertical lines. \textit{Bottom panel:} Optical depth profile (blue line) of all fitted transitions for all 120 analyzed positions.}
\end{figure*}
 
\begin{figure*}
\ContinuedFloat
\captionsetup{list=off,format=cont}
\centering
\includegraphics[]{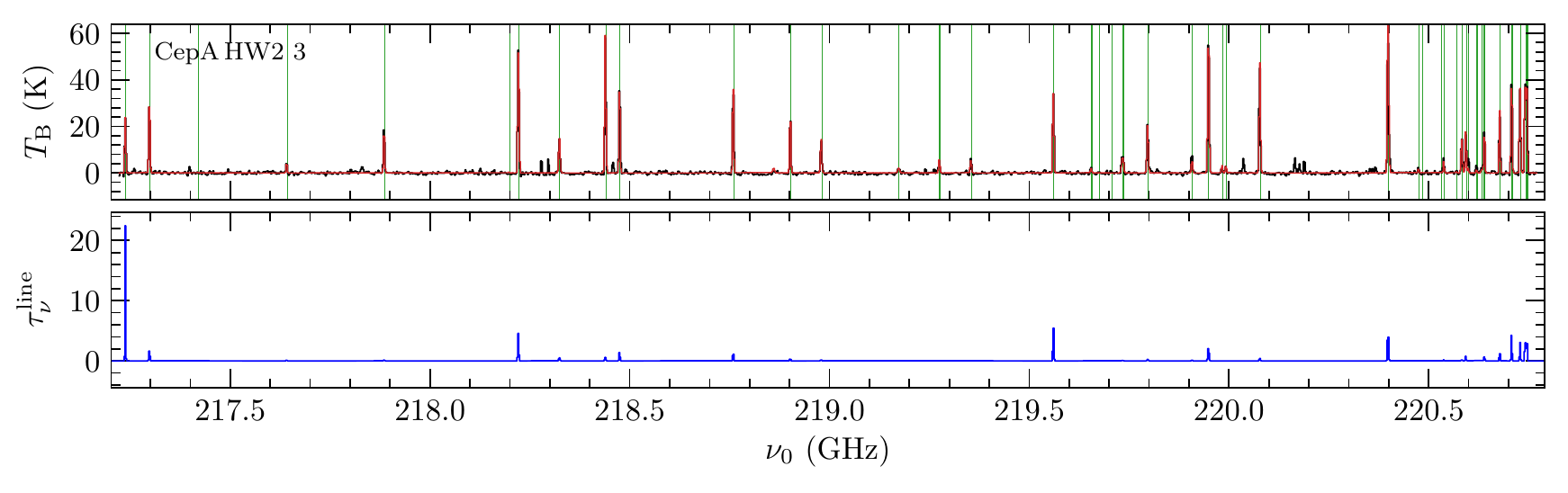}
\includegraphics[]{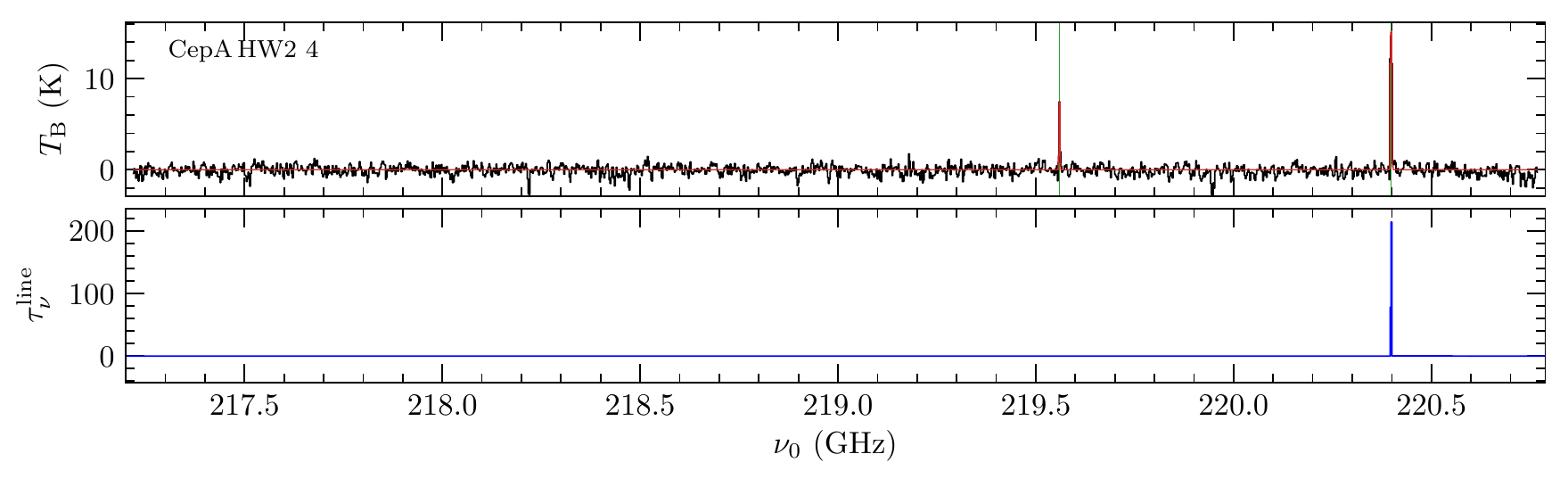}
\includegraphics[]{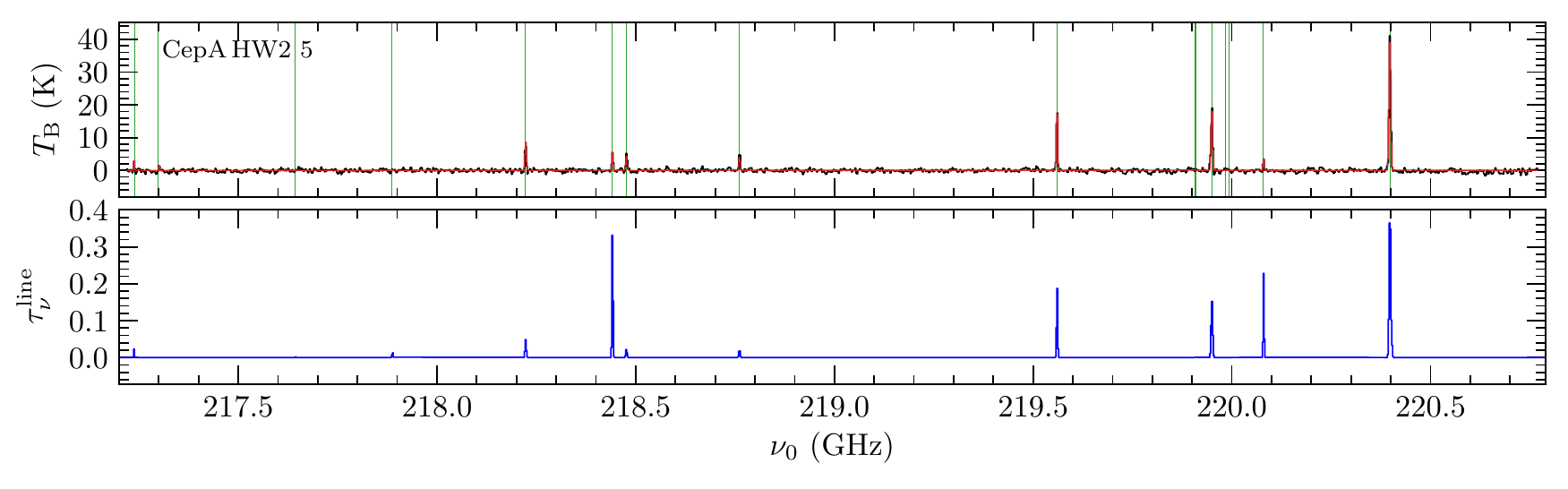}
\includegraphics[]{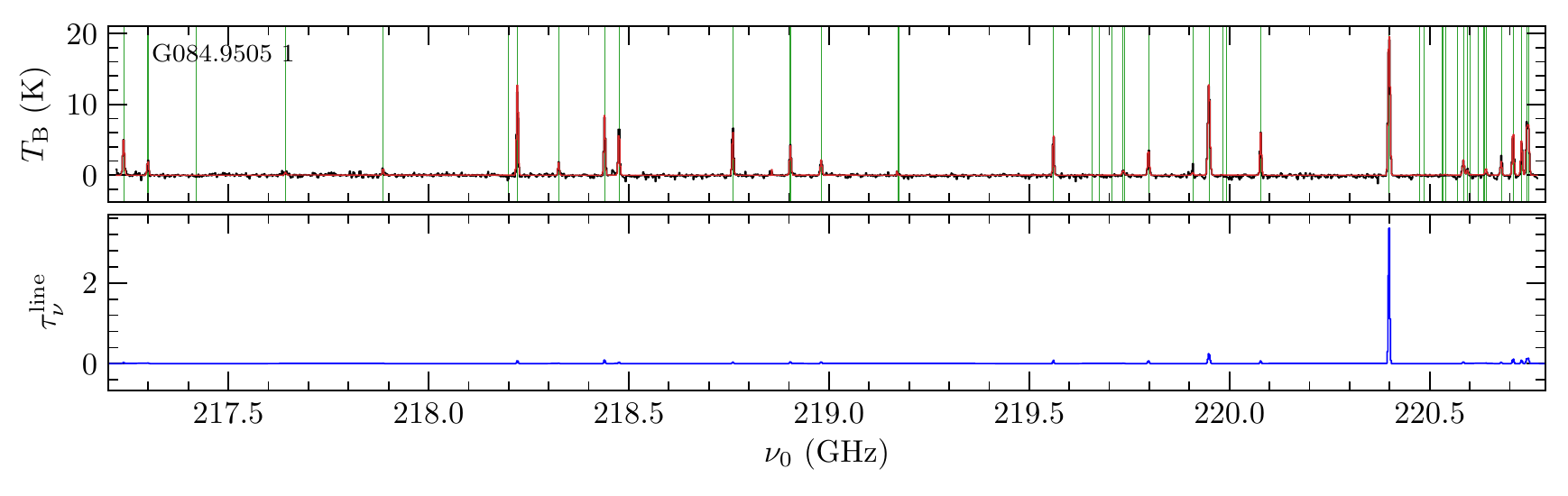}
\caption{\textit{Top panel:} Observed (black line) spectrum and \texttt{XCLASS} fit (red line) for all 120 analyzed positions. Fitted molecular transitions are indicated by green vertical lines. \textit{Bottom panel:} Optical depth profile (blue line) of all fitted transitions for all 120 analyzed positions.}
\end{figure*}
 
\begin{figure*}
\ContinuedFloat
\captionsetup{list=off,format=cont}
\centering
\includegraphics[]{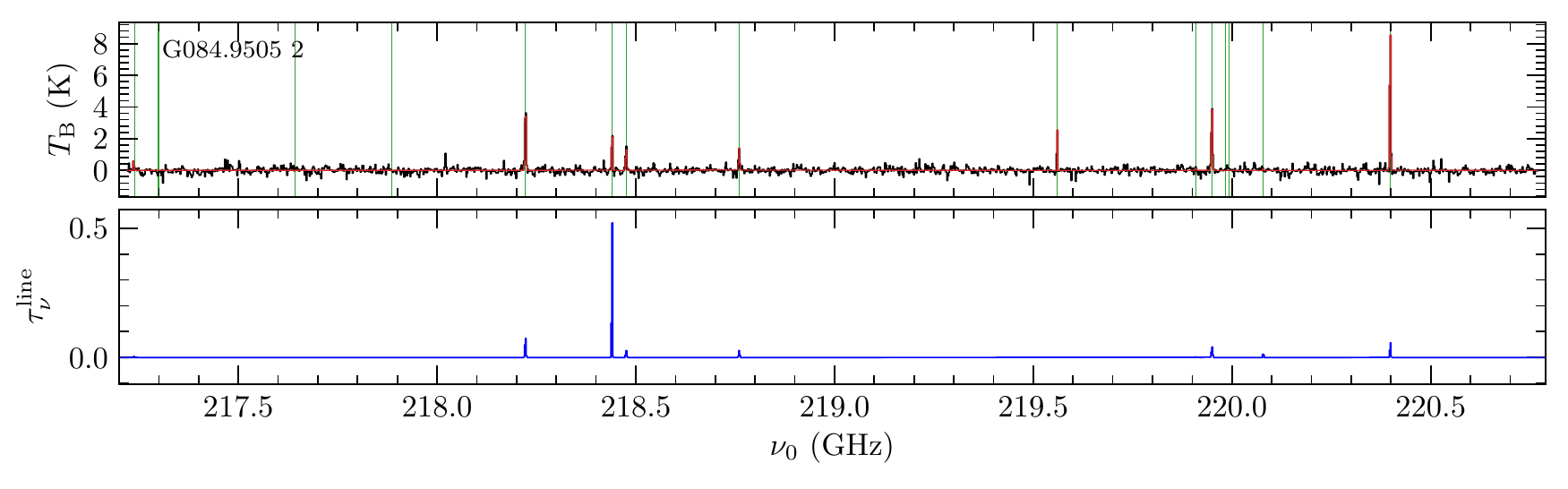}
\includegraphics[]{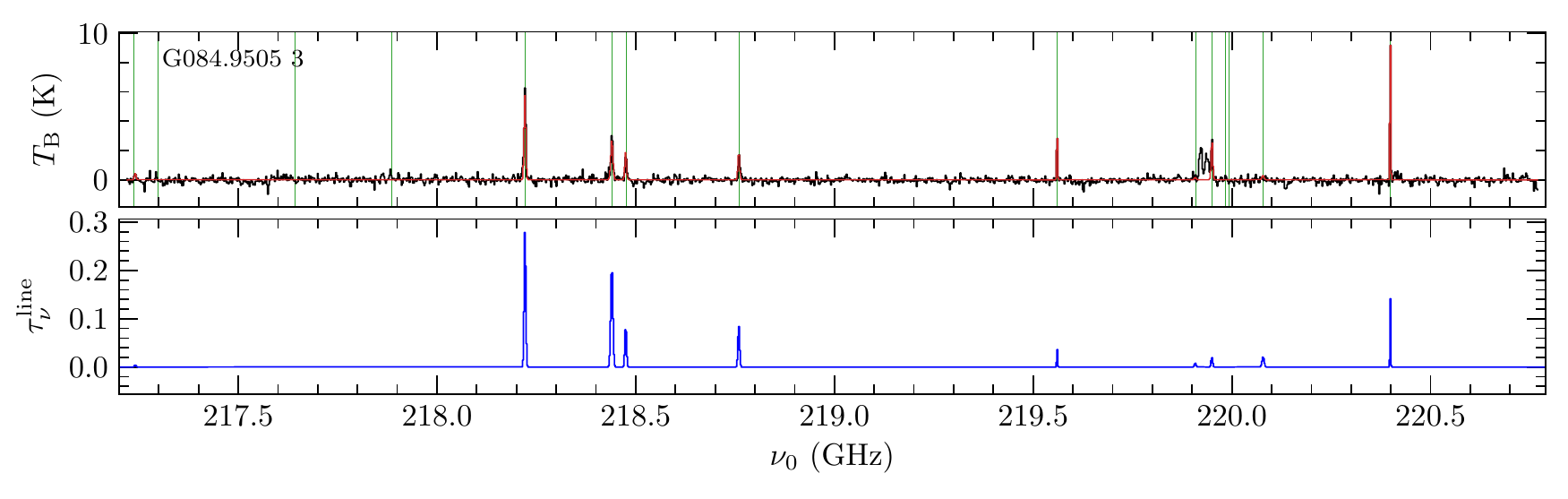}
\includegraphics[]{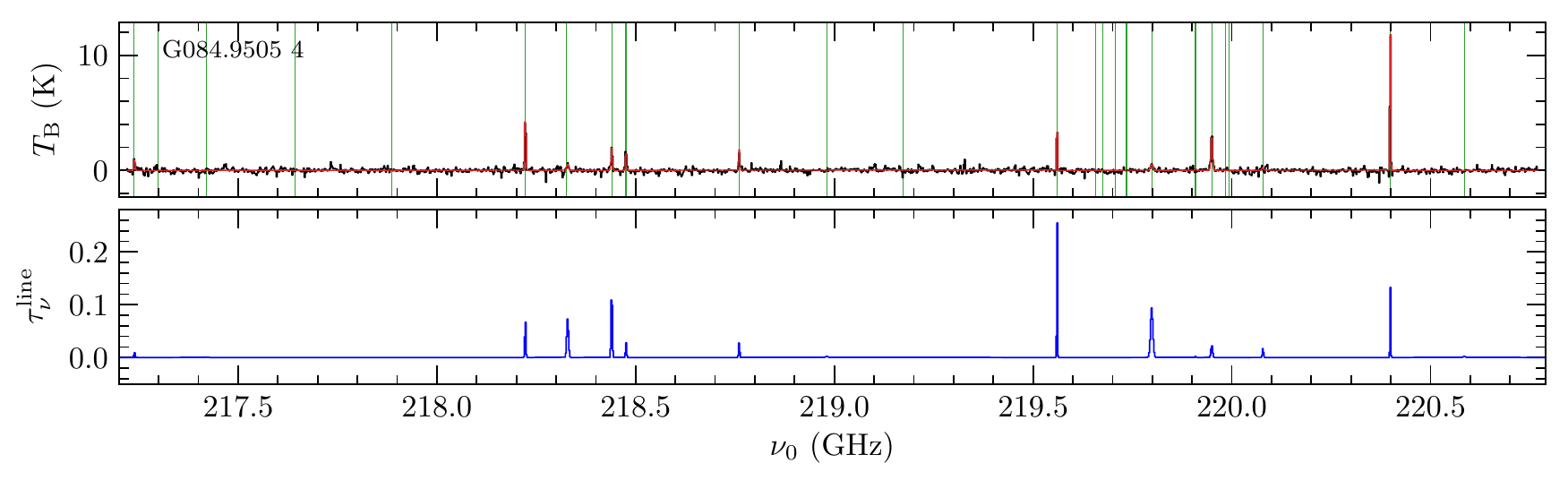}
\includegraphics[]{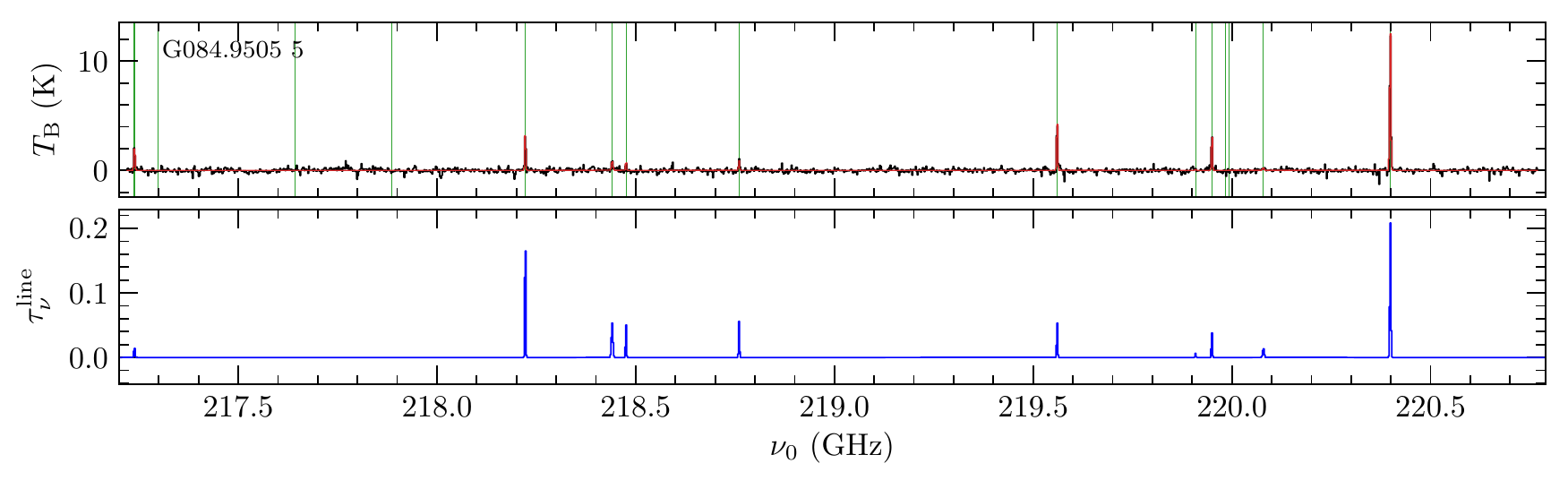}
\caption{\textit{Top panel:} Observed (black line) spectrum and \texttt{XCLASS} fit (red line) for all 120 analyzed positions. Fitted molecular transitions are indicated by green vertical lines. \textit{Bottom panel:} Optical depth profile (blue line) of all fitted transitions for all 120 analyzed positions.}
\end{figure*}
 
\begin{figure*}
\ContinuedFloat
\captionsetup{list=off,format=cont}
\centering
\includegraphics[]{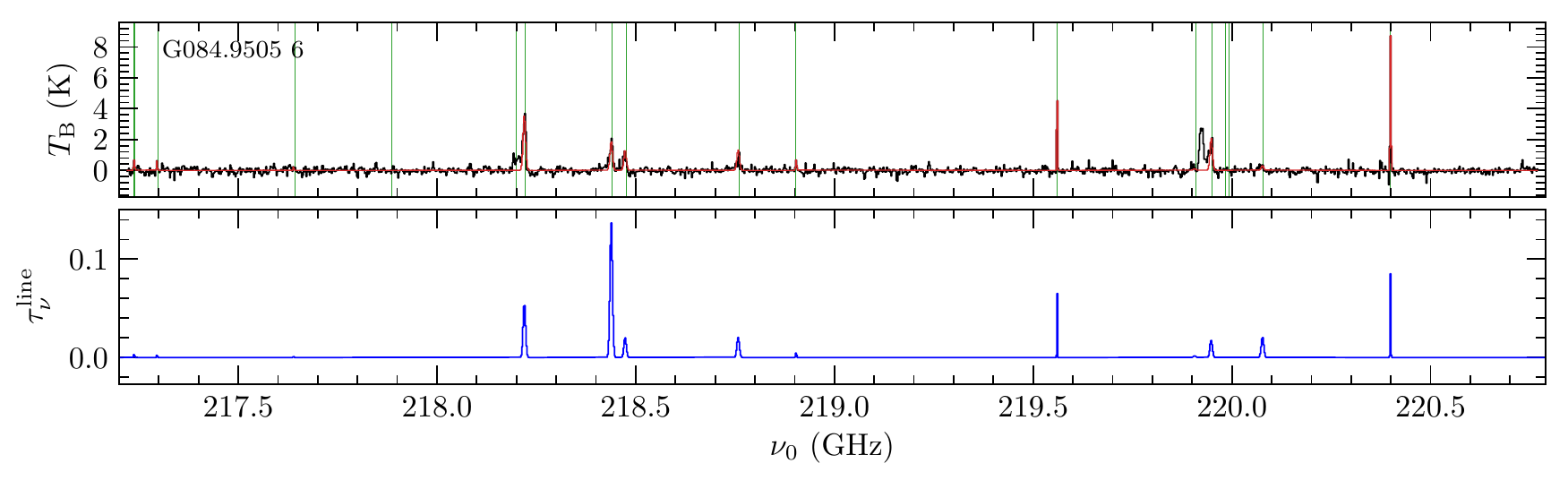}
\includegraphics[]{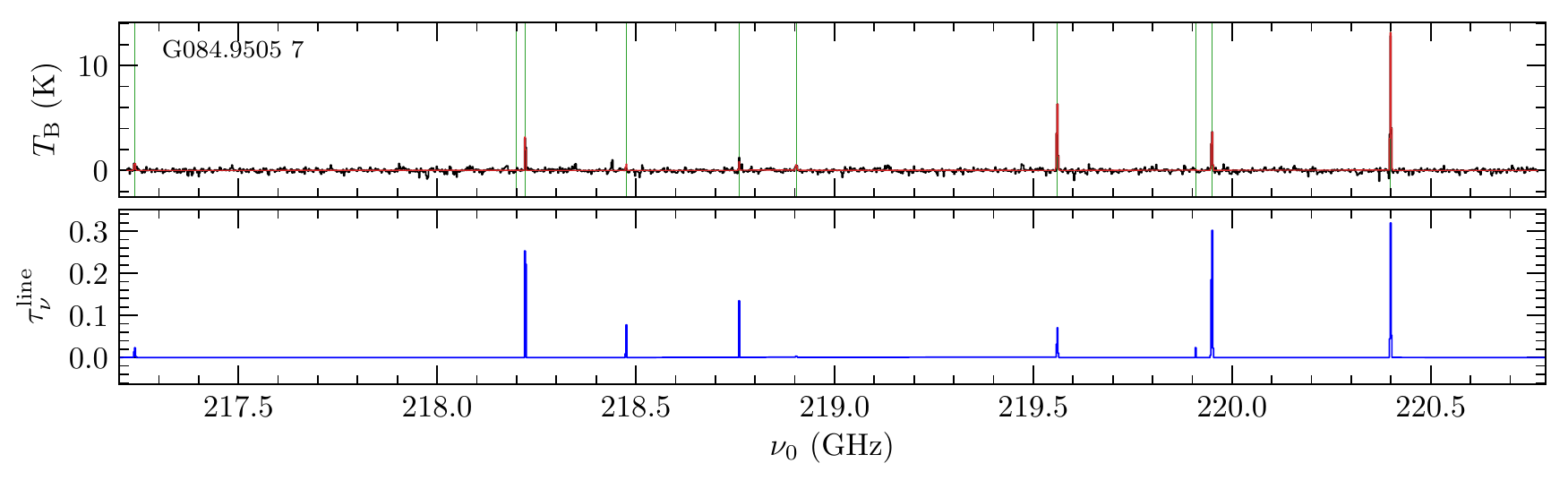}
\includegraphics[]{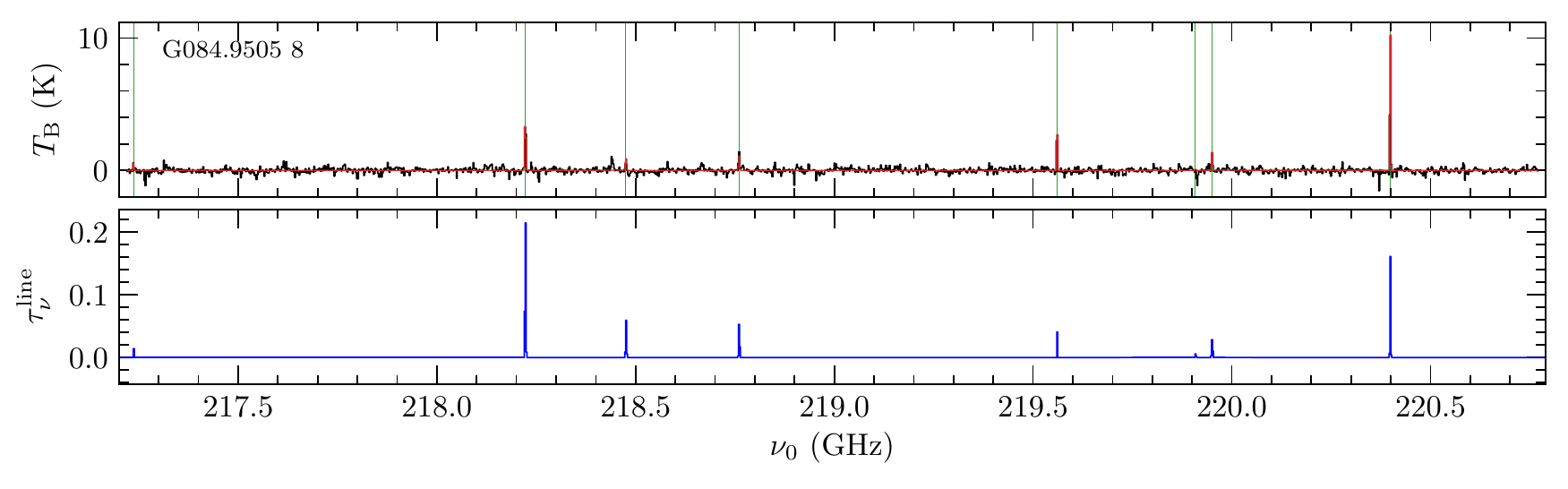}
\includegraphics[]{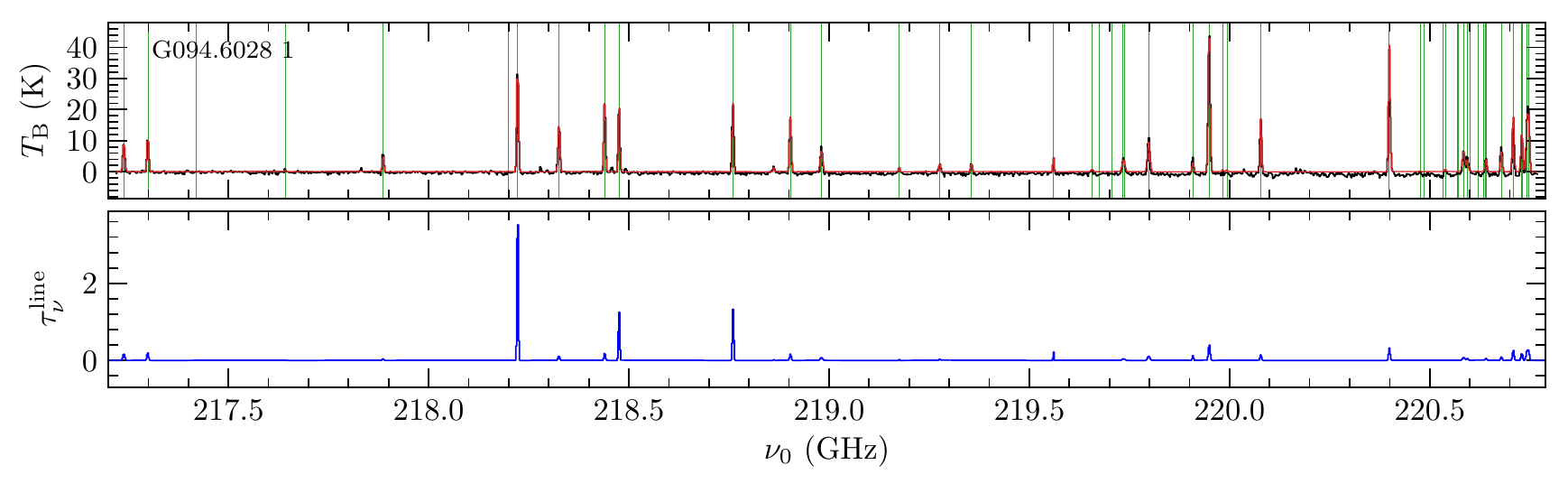}
\caption{\textit{Top panel:} Observed (black line) spectrum and \texttt{XCLASS} fit (red line) for all 120 analyzed positions. Fitted molecular transitions are indicated by green vertical lines. \textit{Bottom panel:} Optical depth profile (blue line) of all fitted transitions for all 120 analyzed positions.}
\end{figure*}
 
\begin{figure*}
\ContinuedFloat
\captionsetup{list=off,format=cont}
\centering
\includegraphics[]{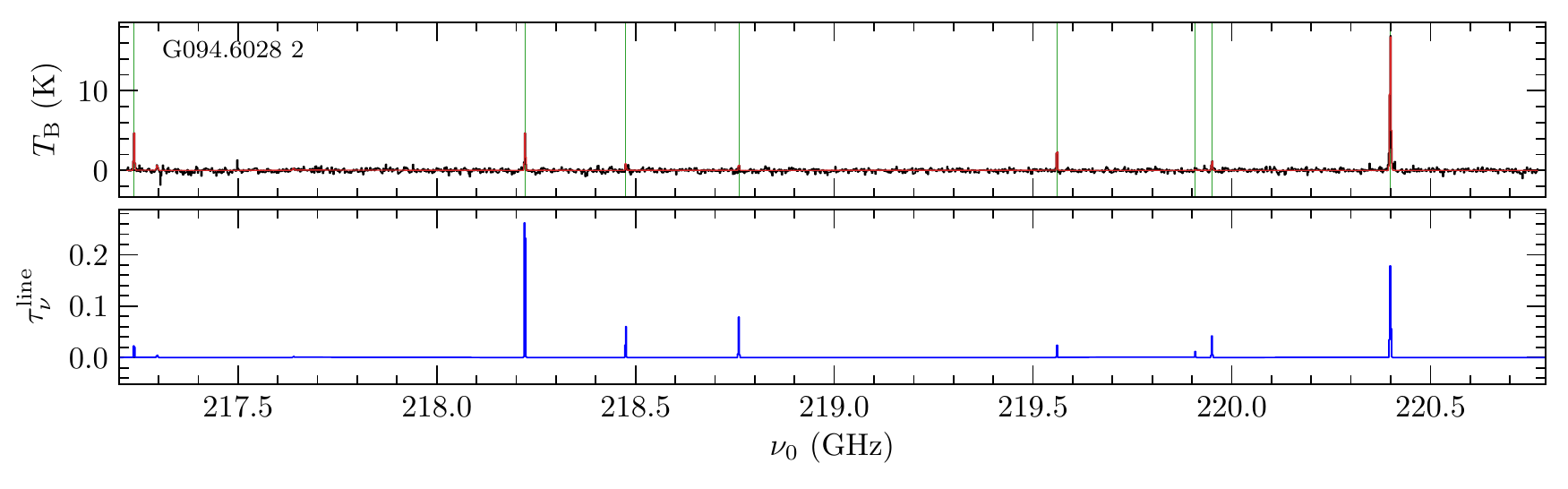}
\includegraphics[]{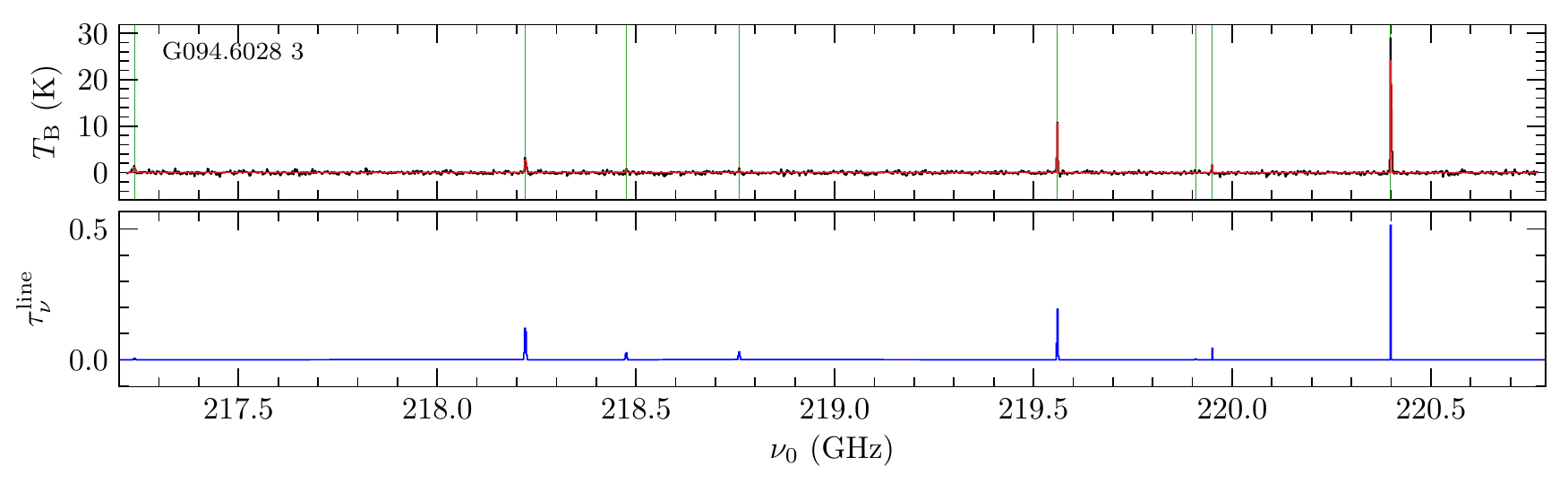}
\includegraphics[]{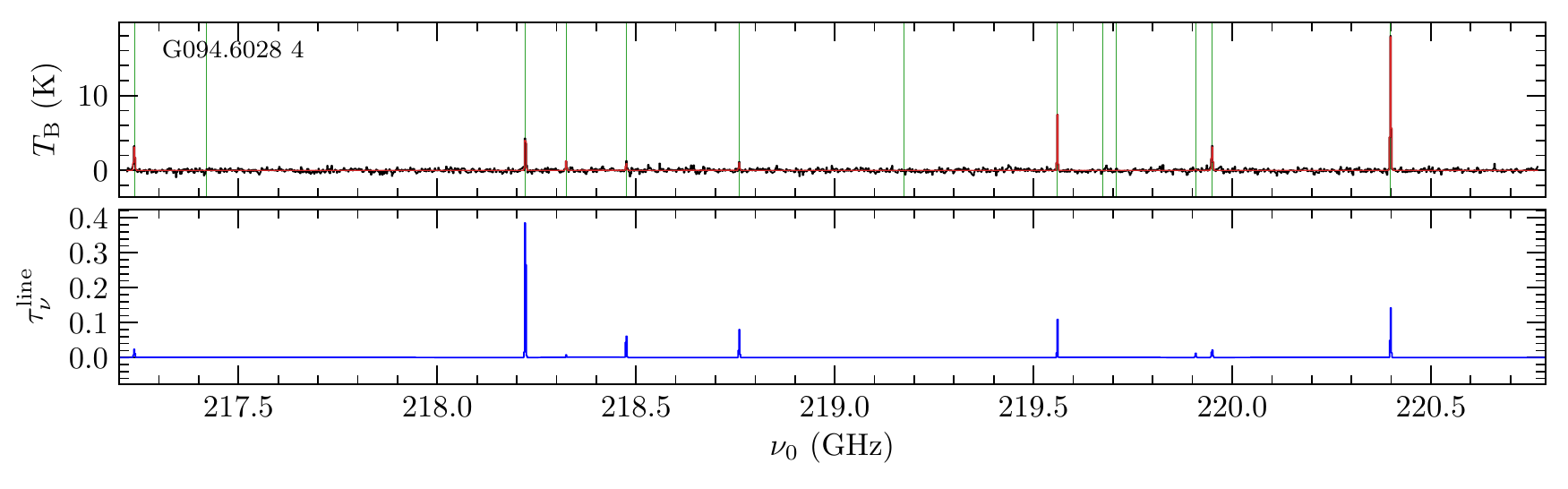}
\includegraphics[]{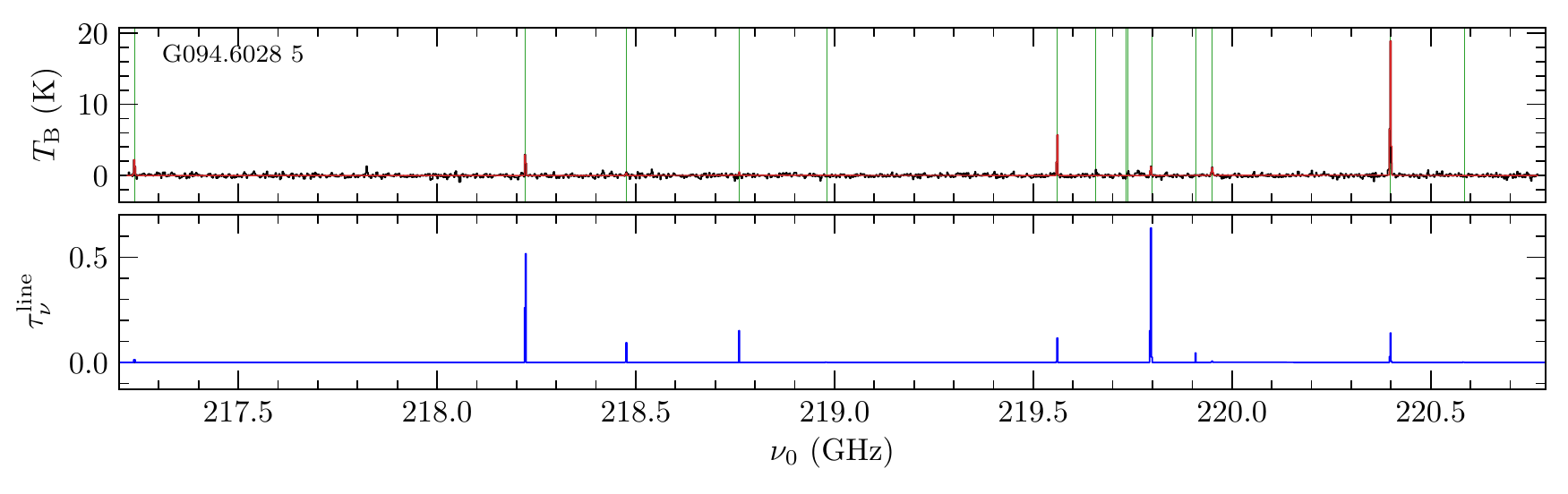}
\caption{\textit{Top panel:} Observed (black line) spectrum and \texttt{XCLASS} fit (red line) for all 120 analyzed positions. Fitted molecular transitions are indicated by green vertical lines. \textit{Bottom panel:} Optical depth profile (blue line) of all fitted transitions for all 120 analyzed positions.}
\end{figure*}
 
\begin{figure*}
\ContinuedFloat
\captionsetup{list=off,format=cont}
\centering
\includegraphics[]{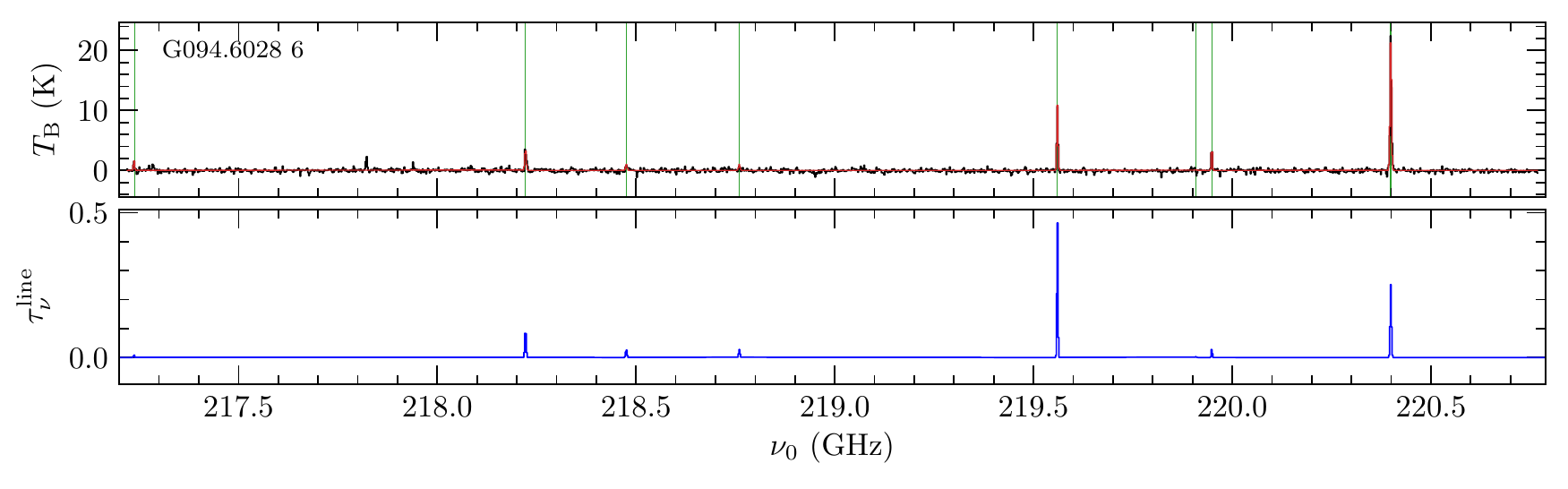}
\includegraphics[]{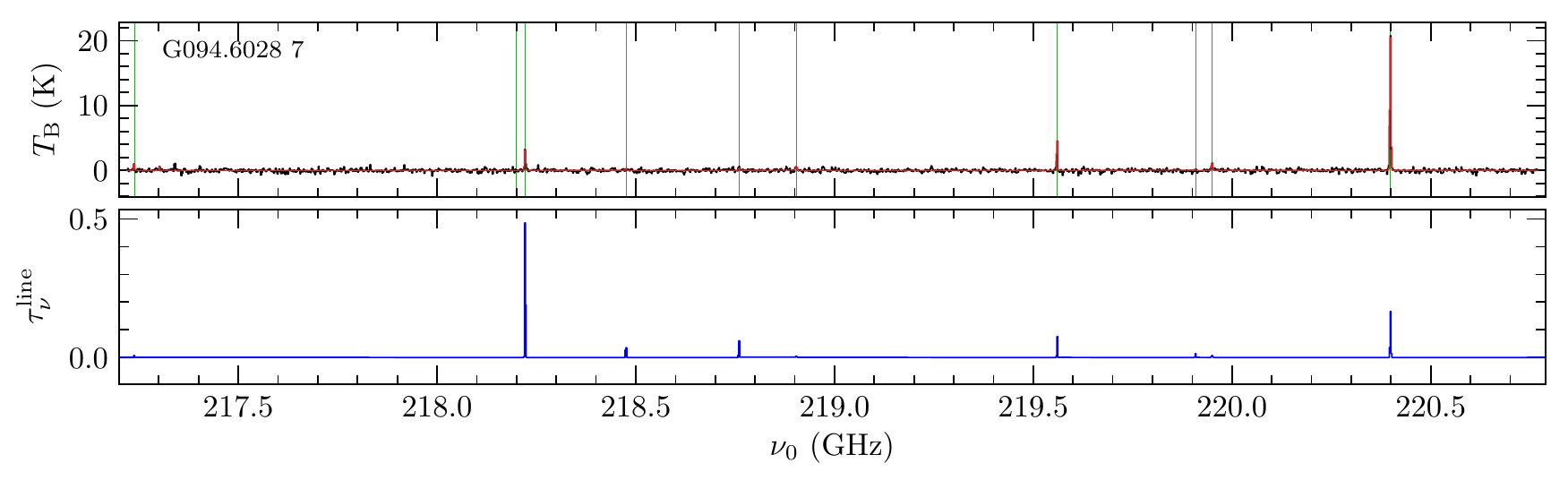}
\includegraphics[]{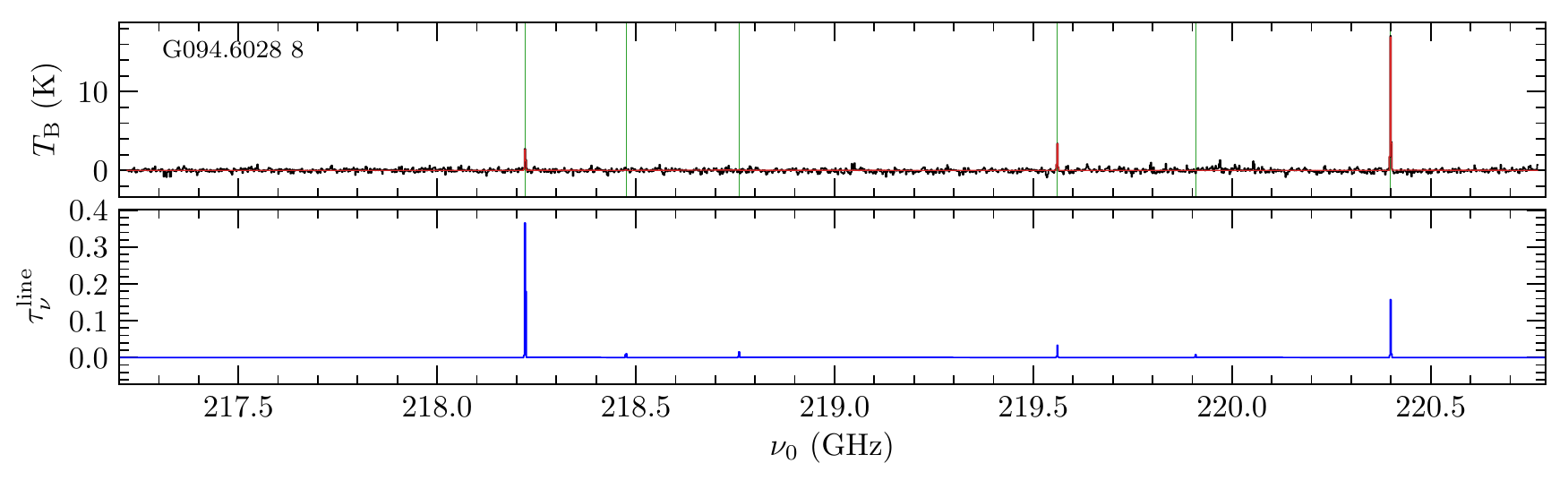}
\includegraphics[]{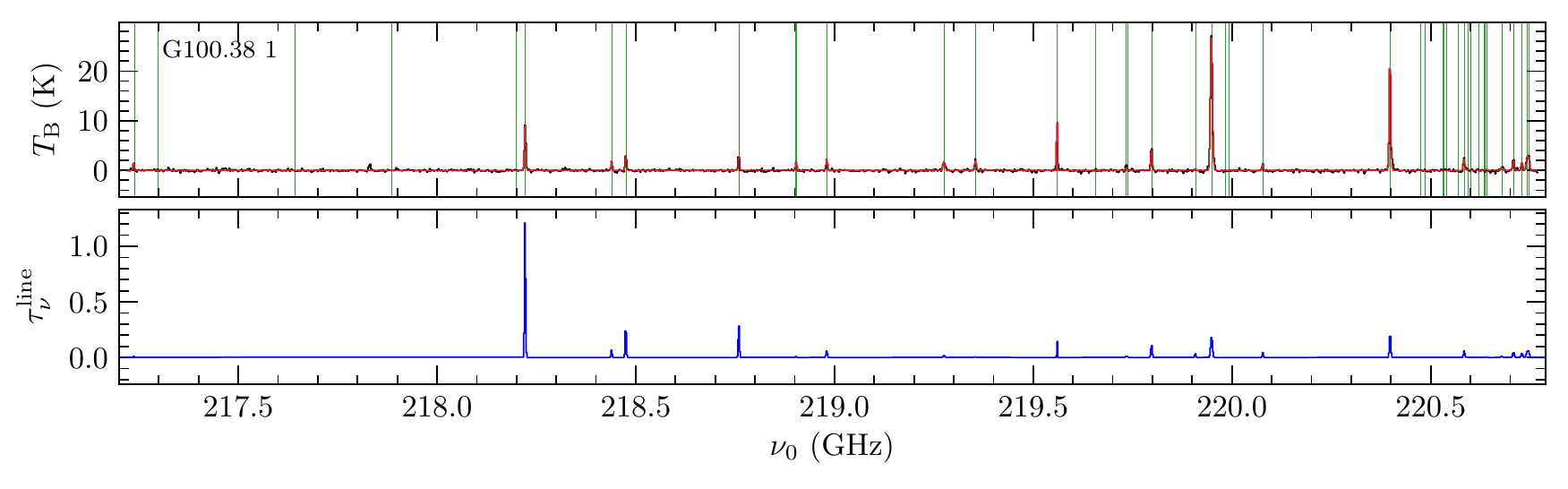}
\caption{\textit{Top panel:} Observed (black line) spectrum and \texttt{XCLASS} fit (red line) for all 120 analyzed positions. Fitted molecular transitions are indicated by green vertical lines. \textit{Bottom panel:} Optical depth profile (blue line) of all fitted transitions for all 120 analyzed positions.}
\end{figure*}
 
\begin{figure*}
\ContinuedFloat
\captionsetup{list=off,format=cont}
\centering
\includegraphics[]{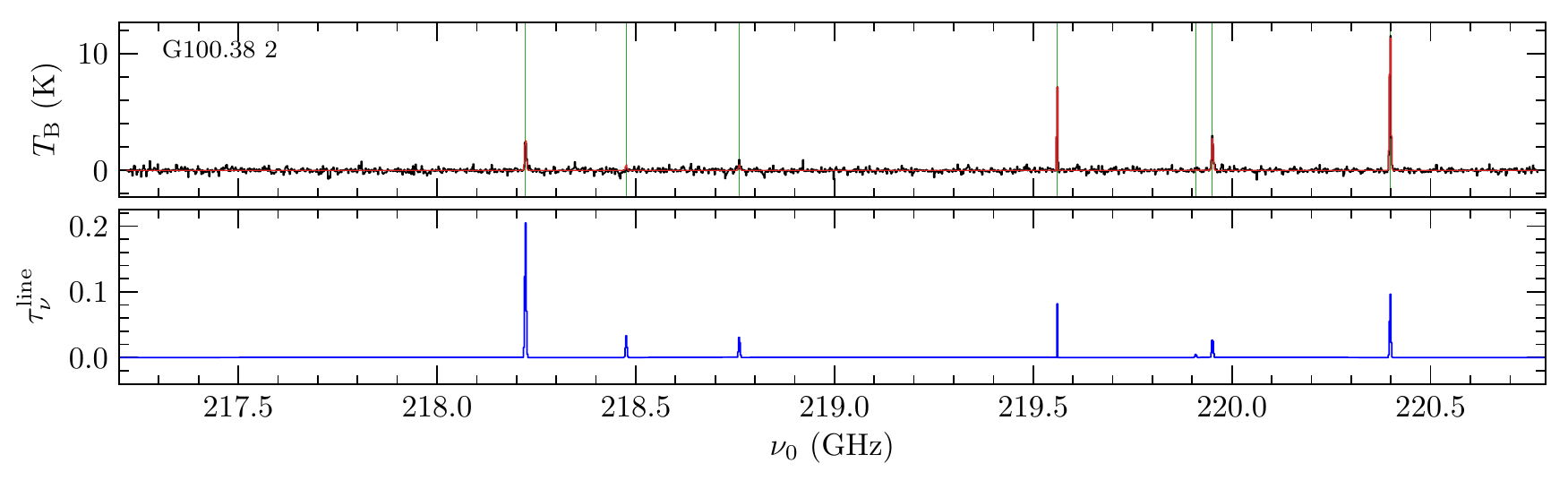}
\includegraphics[]{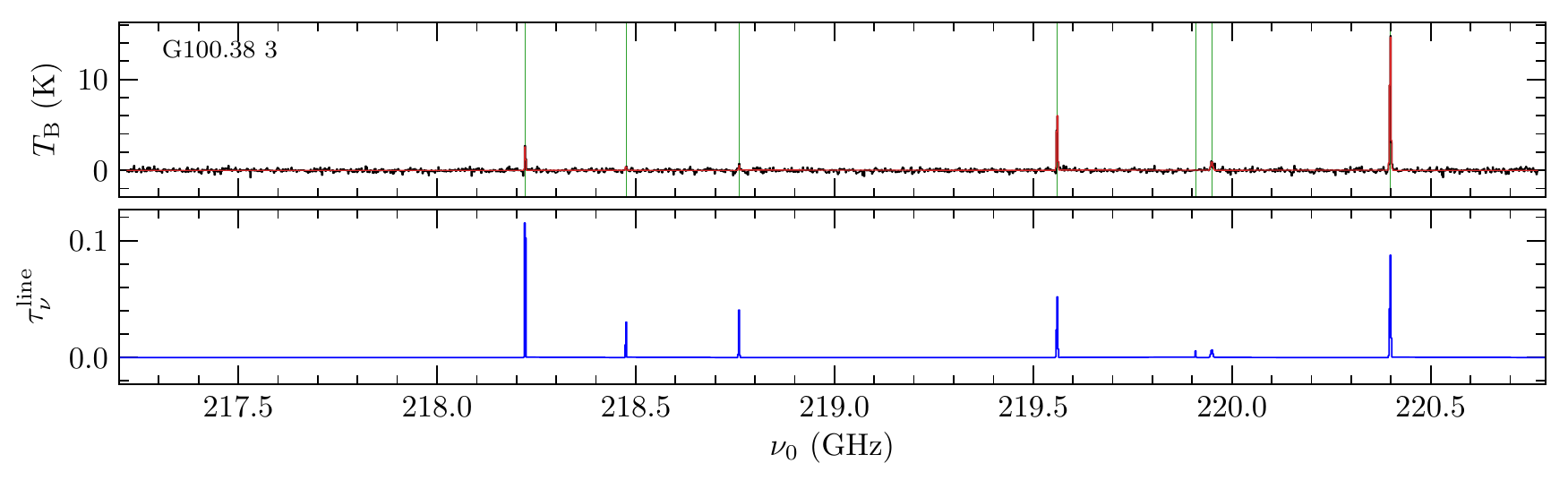}
\includegraphics[]{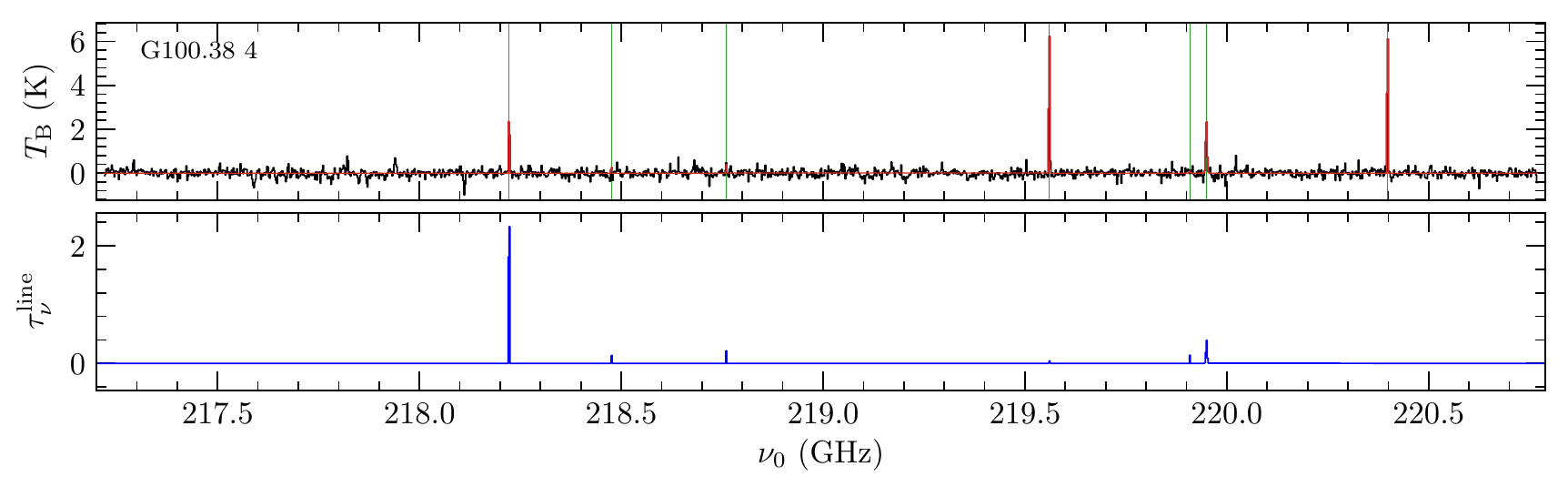}
\includegraphics[]{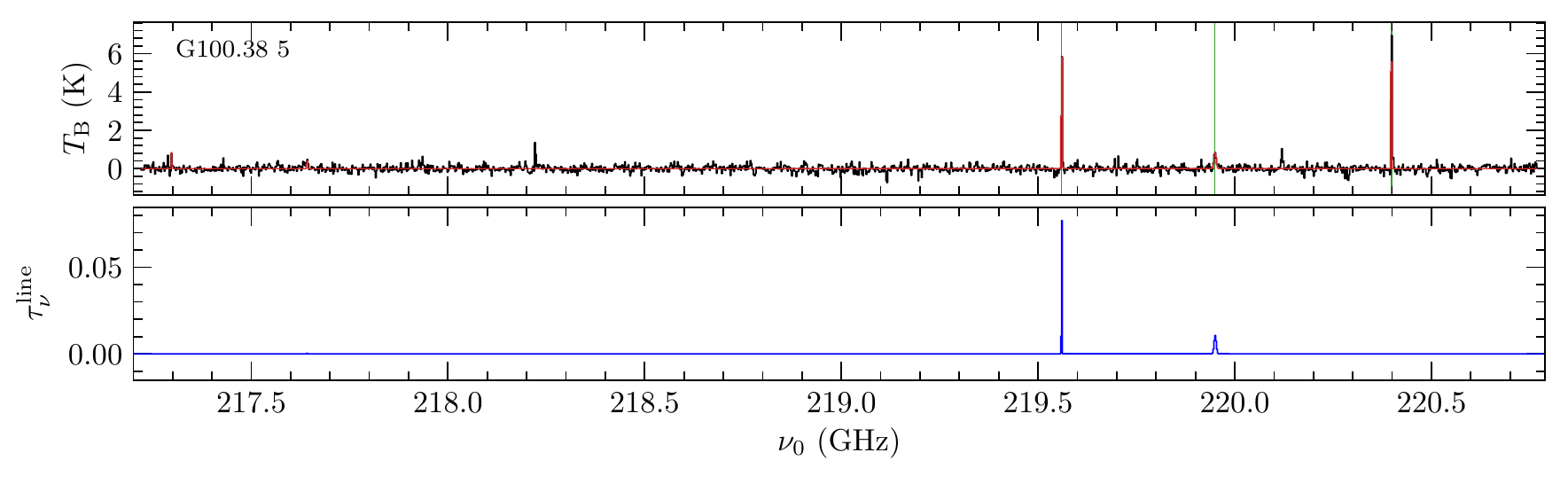}
\caption{\textit{Top panel:} Observed (black line) spectrum and \texttt{XCLASS} fit (red line) for all 120 analyzed positions. Fitted molecular transitions are indicated by green vertical lines. \textit{Bottom panel:} Optical depth profile (blue line) of all fitted transitions for all 120 analyzed positions.}
\end{figure*}
 
\begin{figure*}
\ContinuedFloat
\captionsetup{list=off,format=cont}
\centering
\includegraphics[]{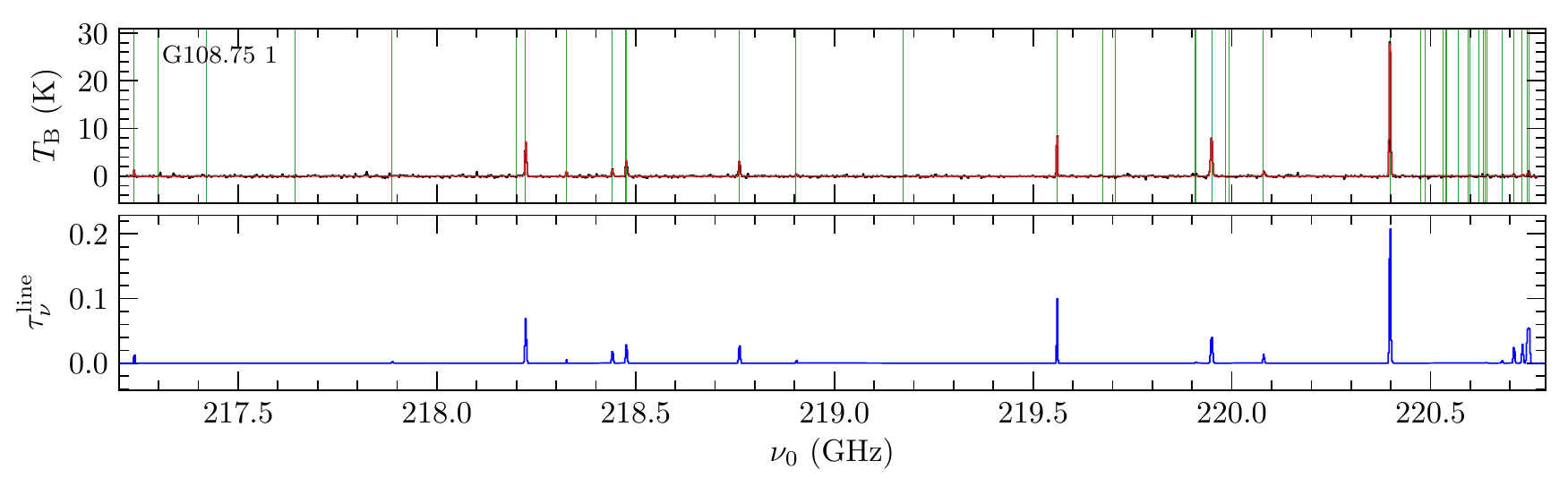}
\includegraphics[]{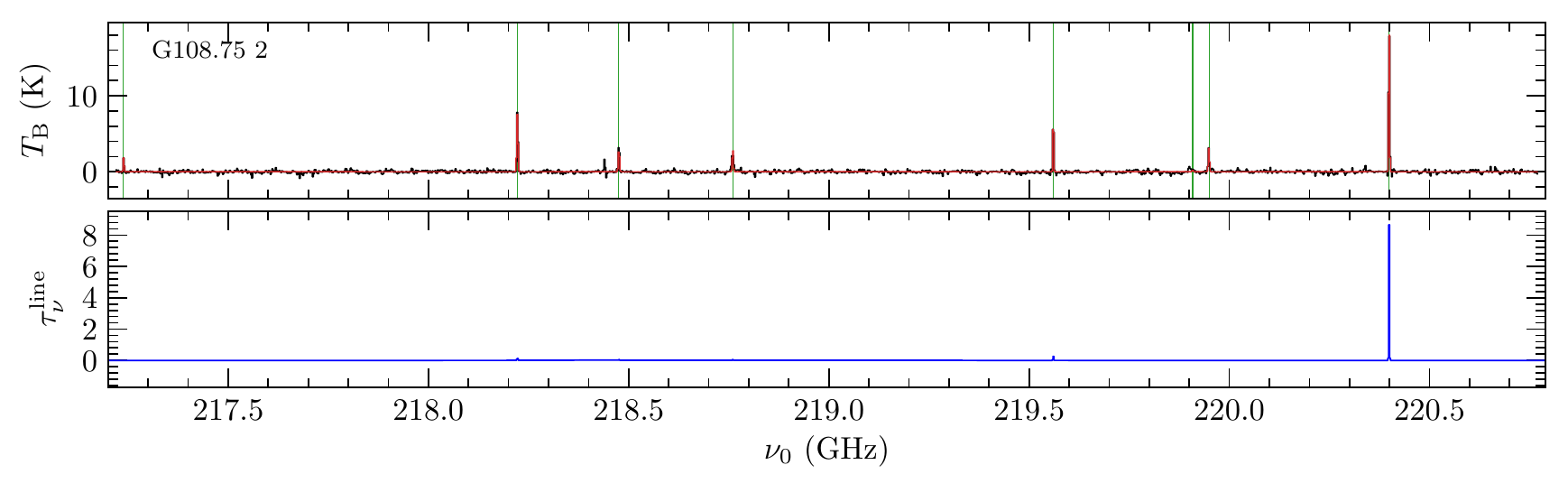}
\includegraphics[]{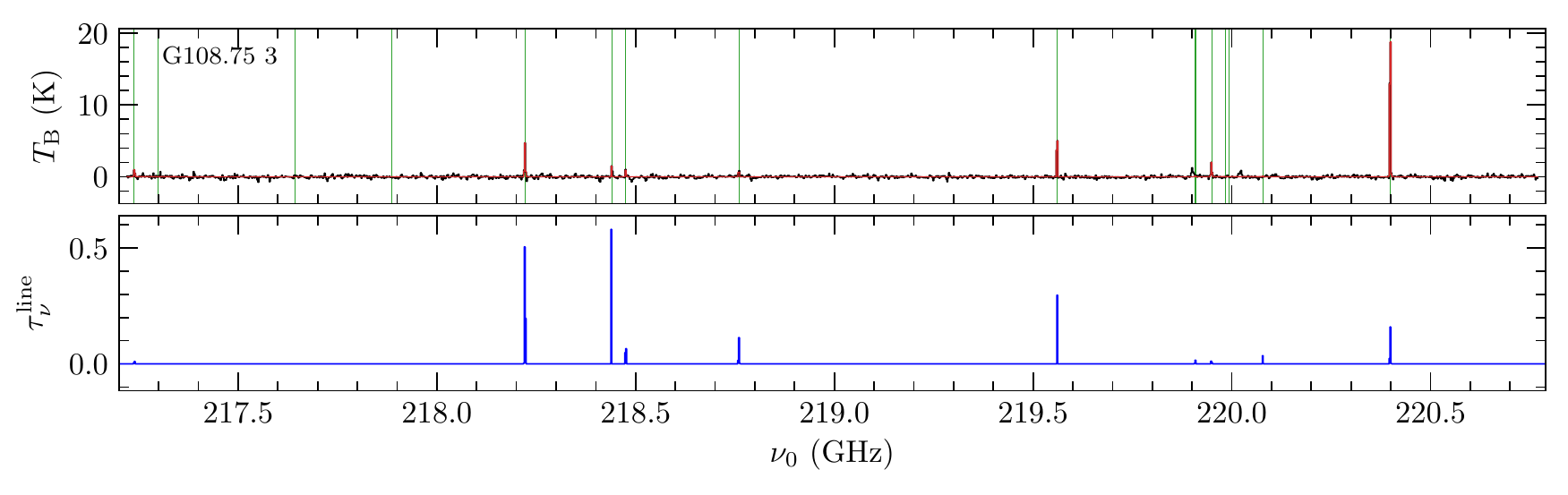}
\includegraphics[]{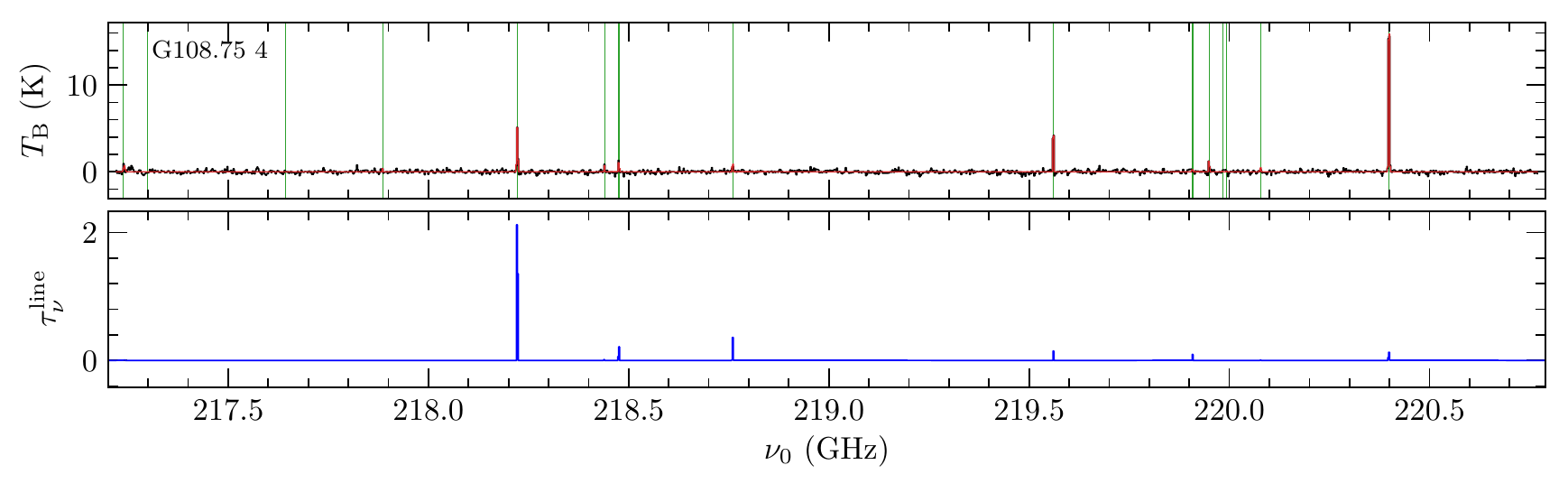}
\caption{\textit{Top panel:} Observed (black line) spectrum and \texttt{XCLASS} fit (red line) for all 120 analyzed positions. Fitted molecular transitions are indicated by green vertical lines. \textit{Bottom panel:} Optical depth profile (blue line) of all fitted transitions for all 120 analyzed positions.}
\end{figure*}
 
\begin{figure*}
\ContinuedFloat
\captionsetup{list=off,format=cont}
\centering
\includegraphics[]{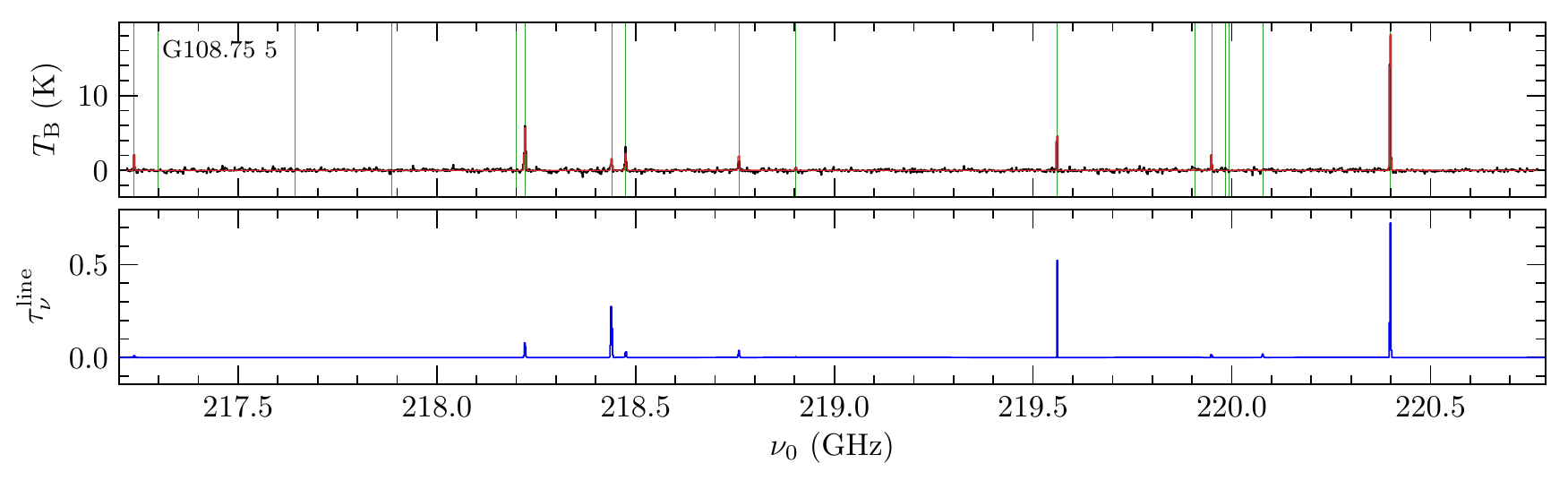}
\includegraphics[]{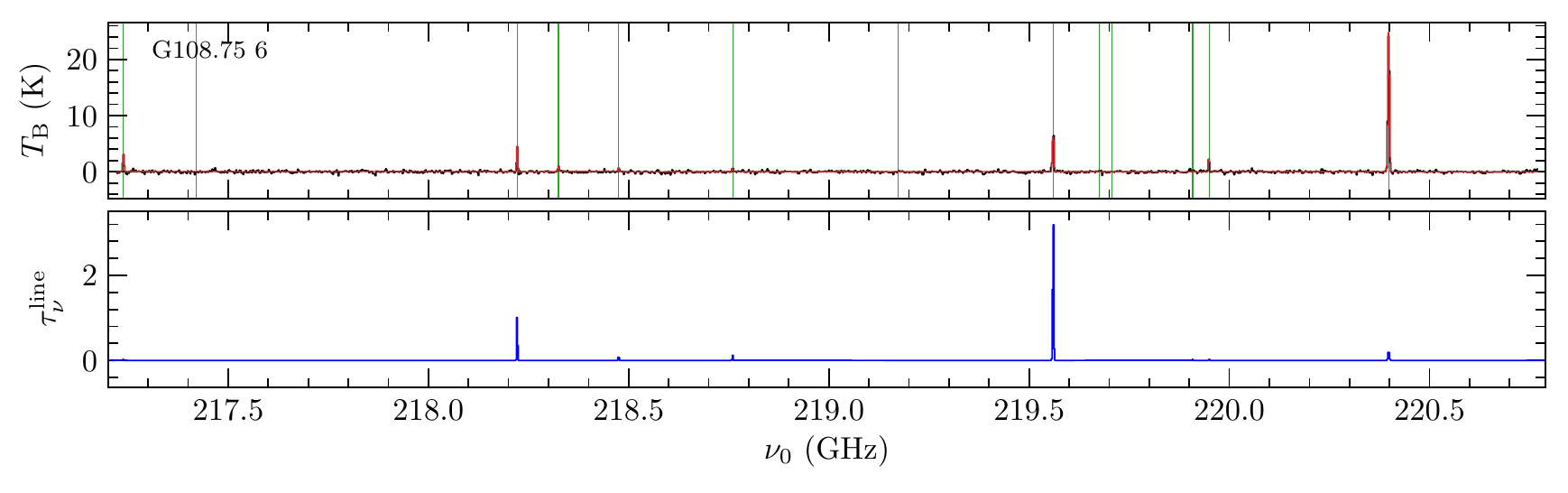}
\includegraphics[]{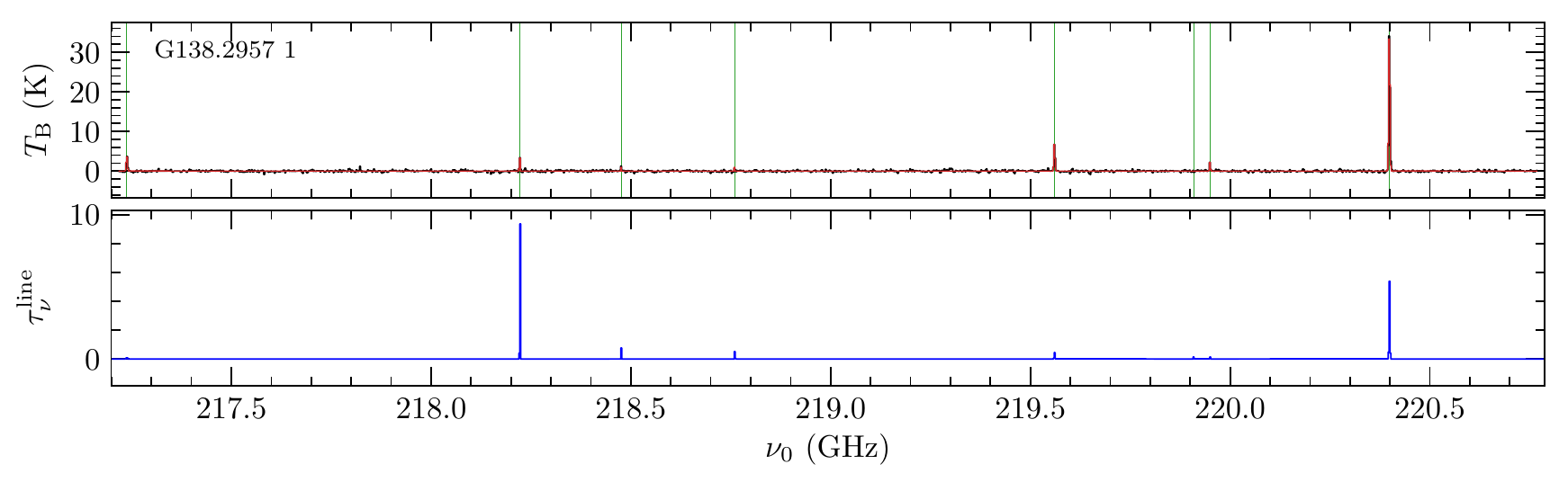}
\includegraphics[]{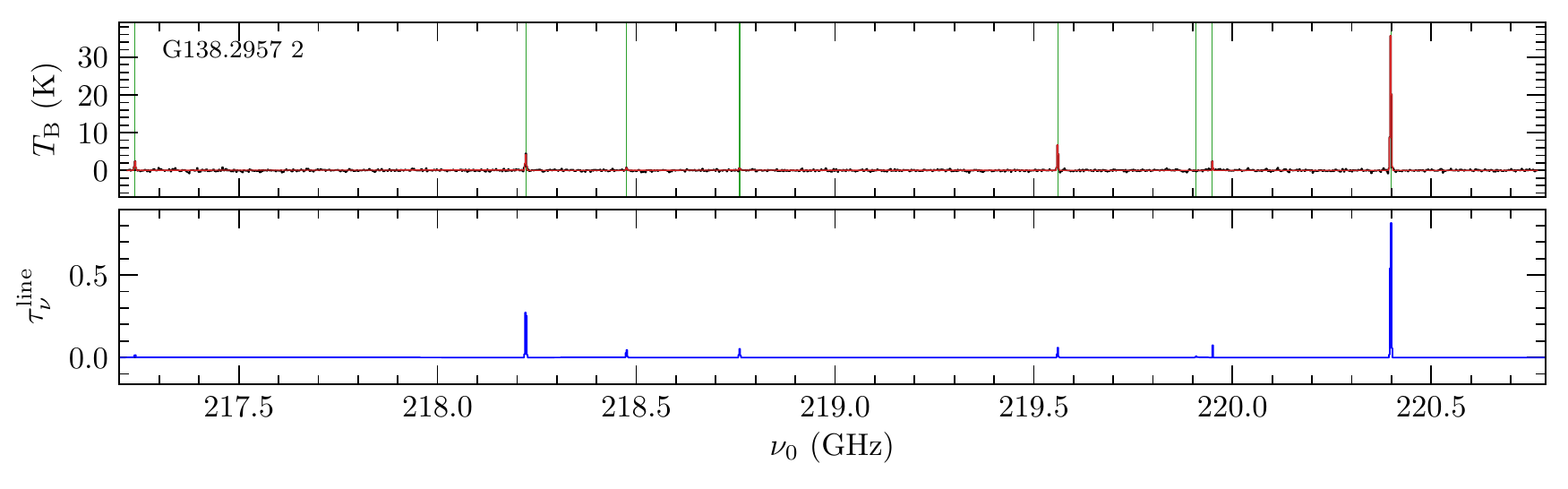}
\caption{\textit{Top panel:} Observed (black line) spectrum and \texttt{XCLASS} fit (red line) for all 120 analyzed positions. Fitted molecular transitions are indicated by green vertical lines. \textit{Bottom panel:} Optical depth profile (blue line) of all fitted transitions for all 120 analyzed positions.}
\end{figure*}
 
\begin{figure*}
\ContinuedFloat
\captionsetup{list=off,format=cont}
\centering
\includegraphics[]{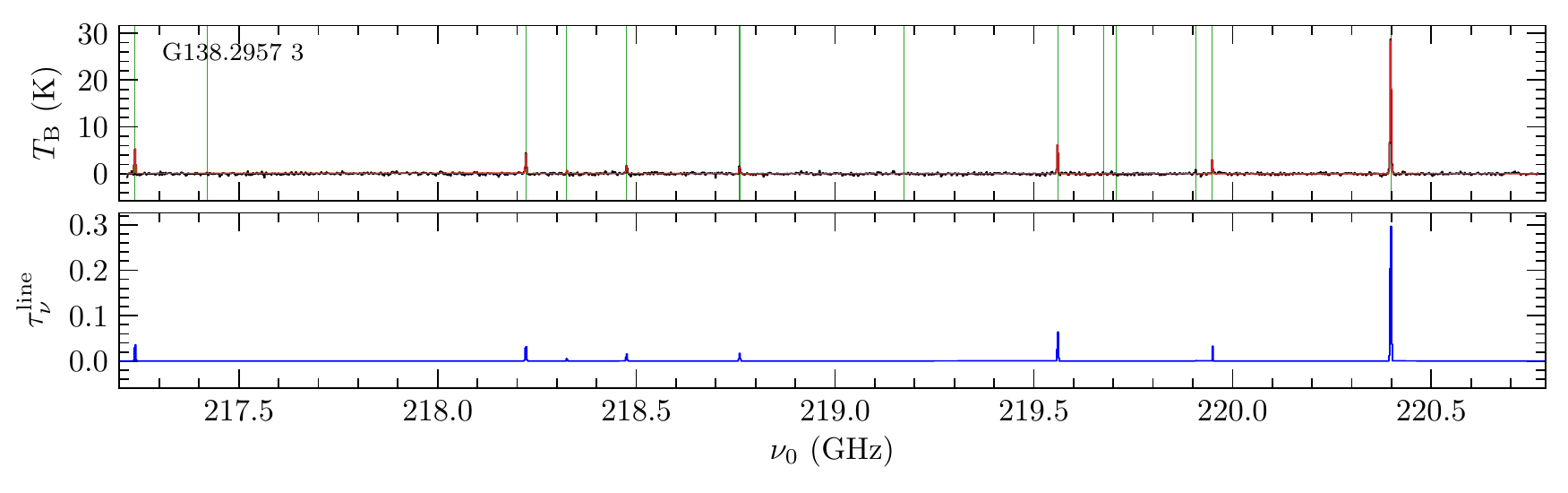}
\includegraphics[]{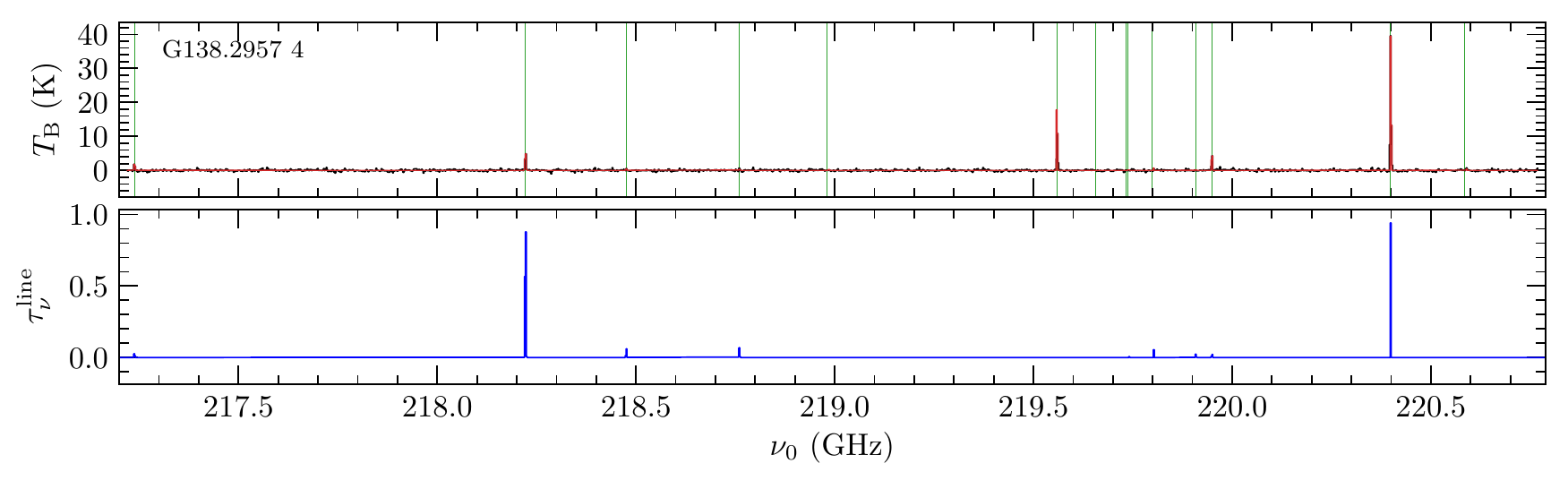}
\includegraphics[]{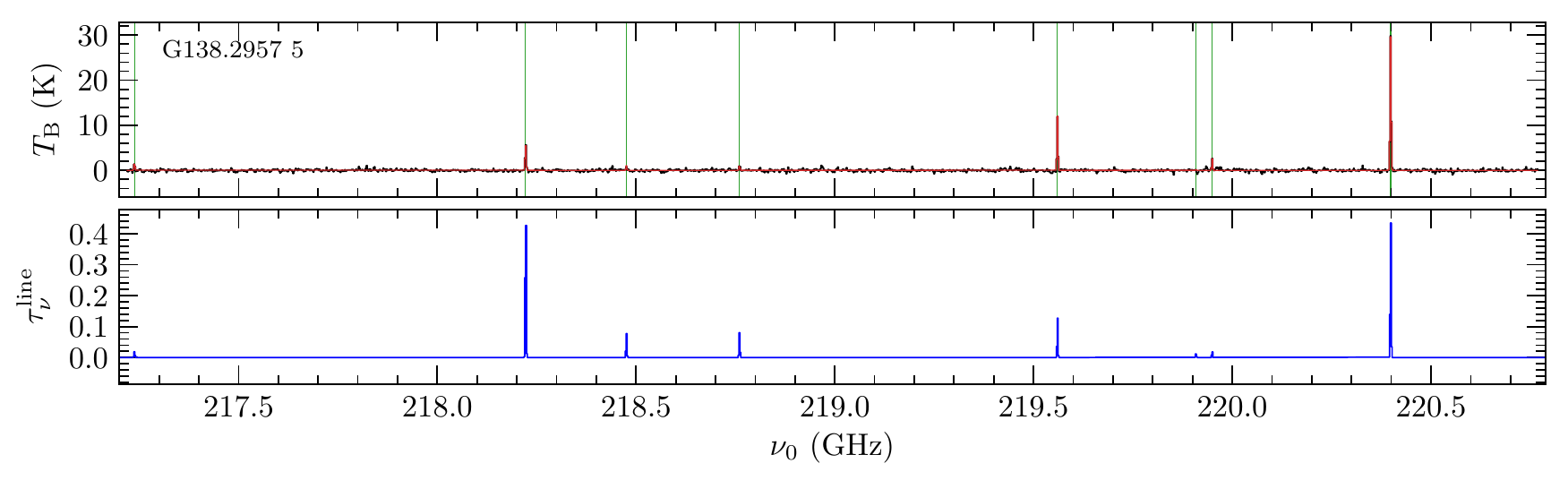}
\includegraphics[]{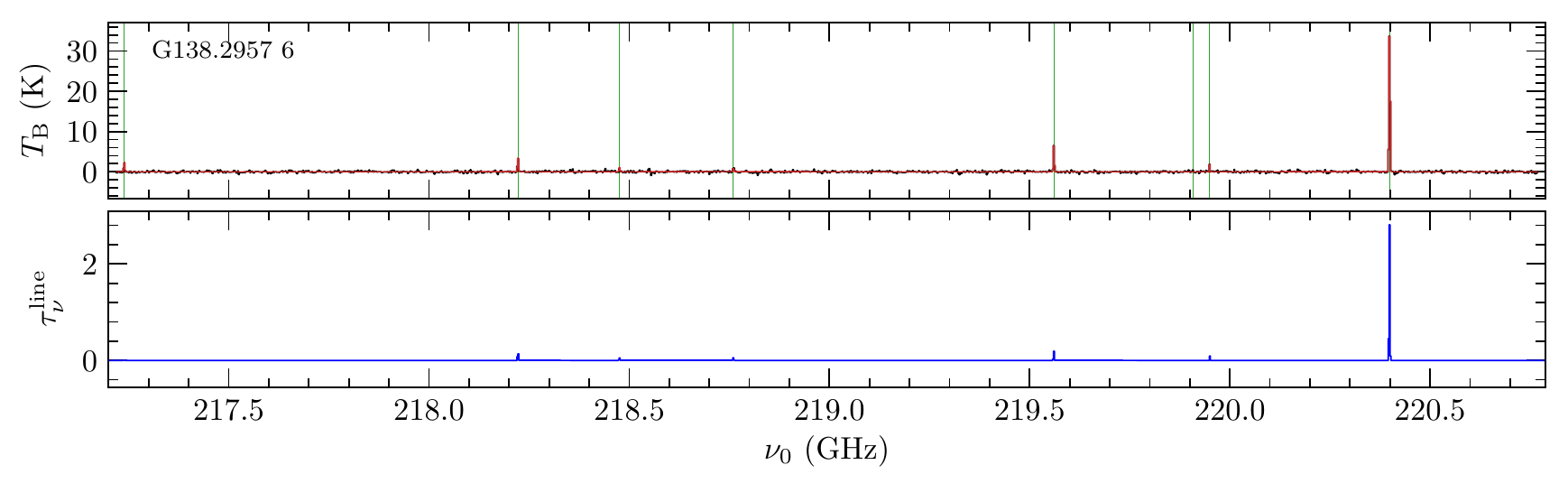}
\caption{\textit{Top panel:} Observed (black line) spectrum and \texttt{XCLASS} fit (red line) for all 120 analyzed positions. Fitted molecular transitions are indicated by green vertical lines. \textit{Bottom panel:} Optical depth profile (blue line) of all fitted transitions for all 120 analyzed positions.}
\end{figure*}
 
\begin{figure*}
\ContinuedFloat
\captionsetup{list=off,format=cont}
\centering
\includegraphics[]{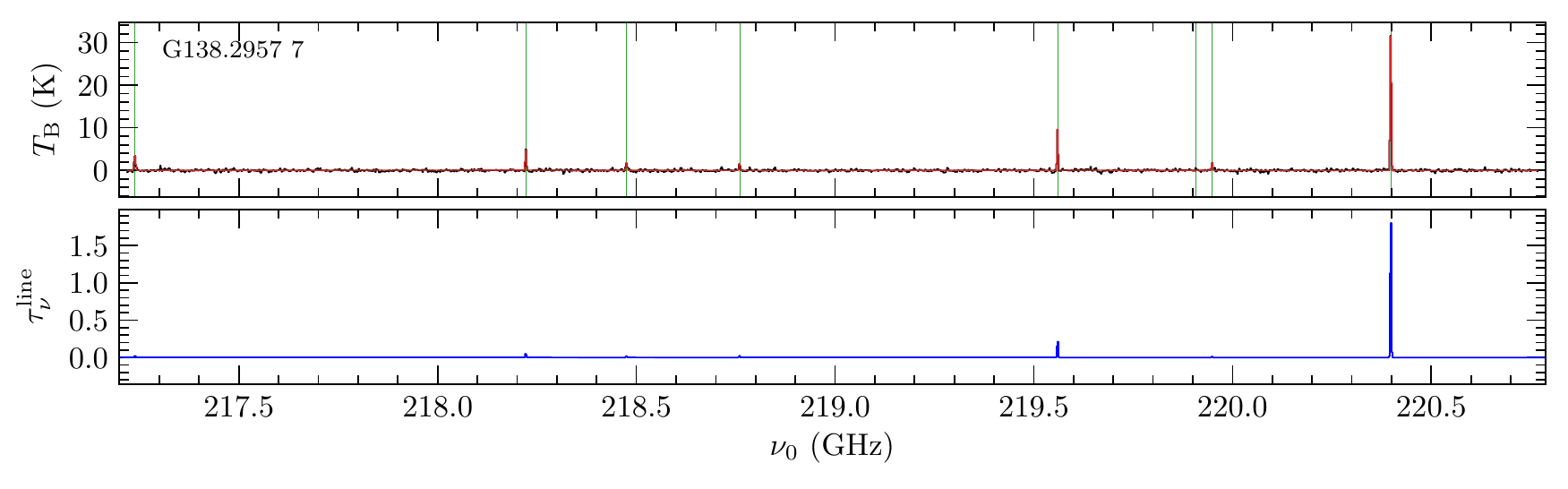}
\includegraphics[]{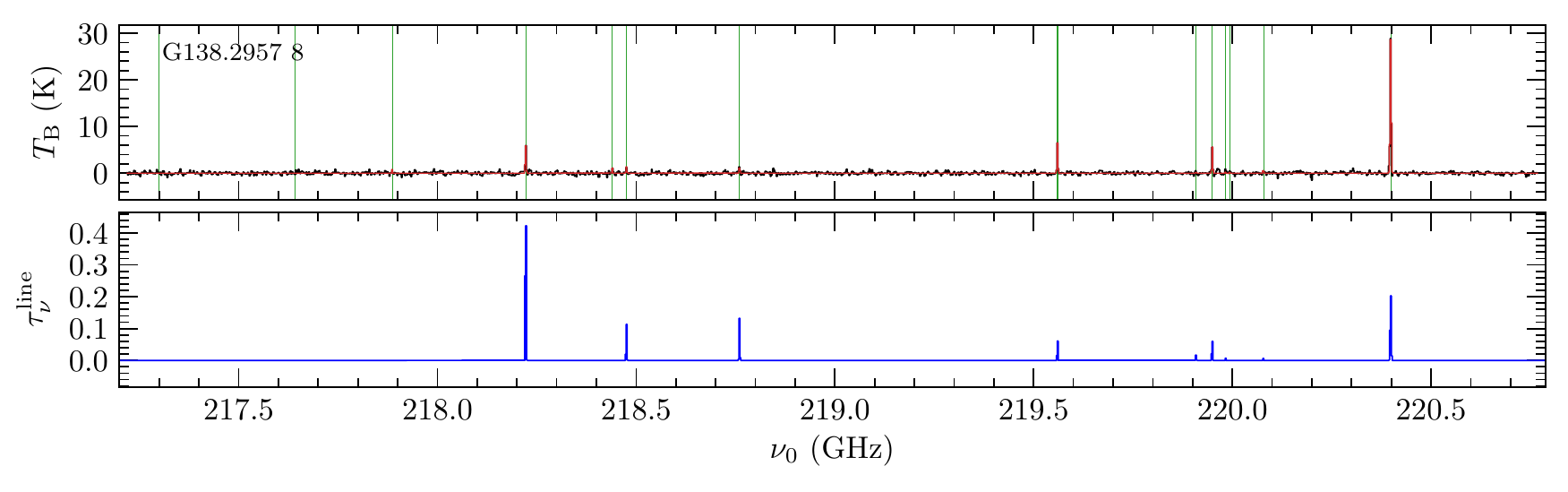}
\includegraphics[]{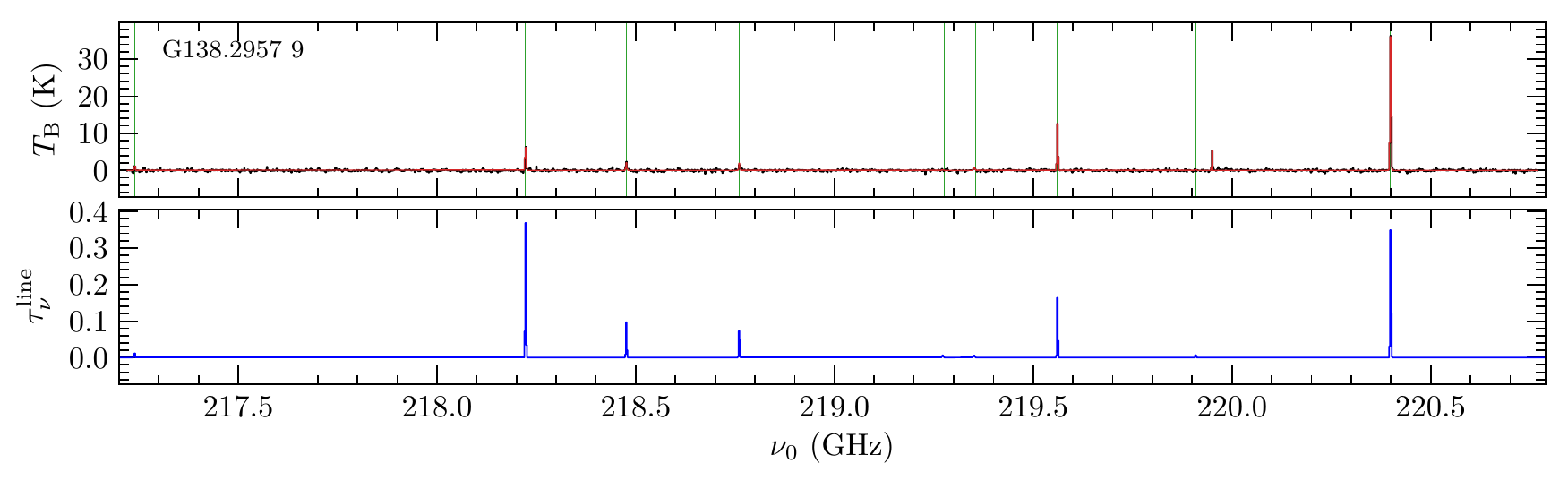}
\includegraphics[]{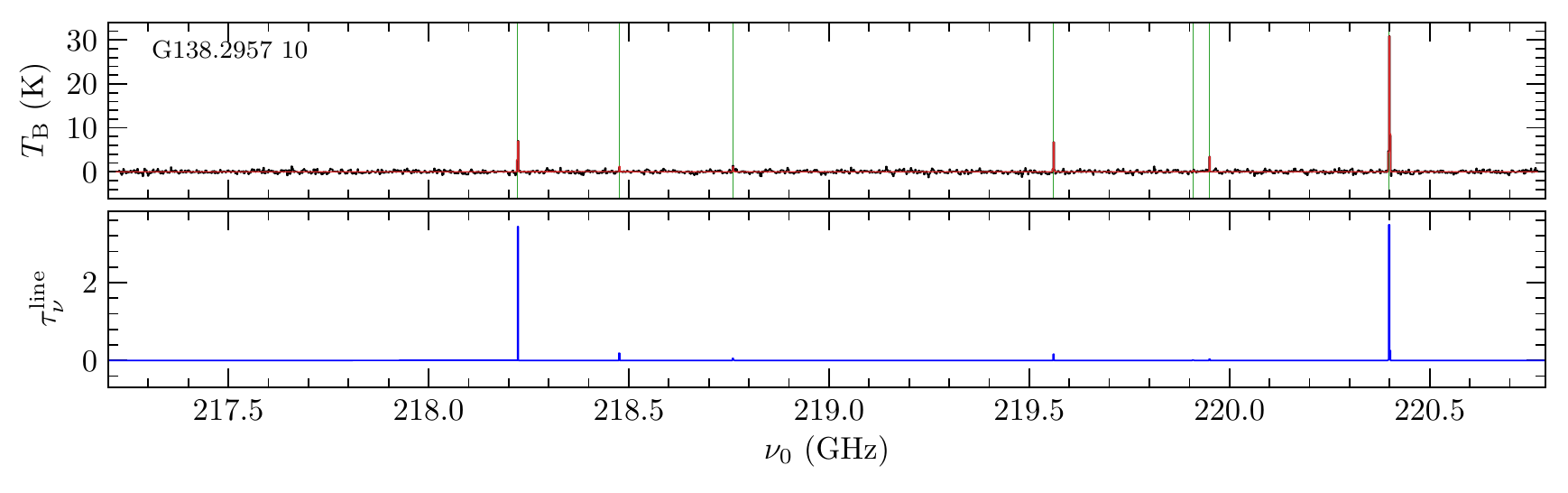}
\caption{\textit{Top panel:} Observed (black line) spectrum and \texttt{XCLASS} fit (red line) for all 120 analyzed positions. Fitted molecular transitions are indicated by green vertical lines. \textit{Bottom panel:} Optical depth profile (blue line) of all fitted transitions for all 120 analyzed positions.}
\end{figure*}
 
\begin{figure*}
\ContinuedFloat
\captionsetup{list=off,format=cont}
\centering
\includegraphics[]{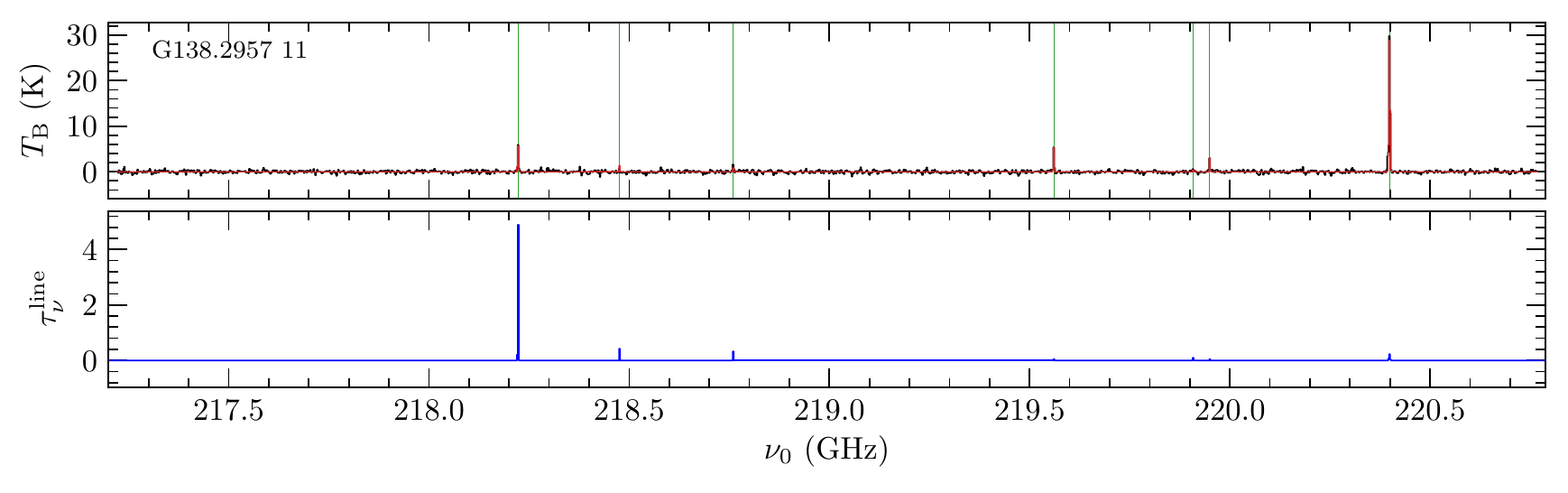}
\includegraphics[]{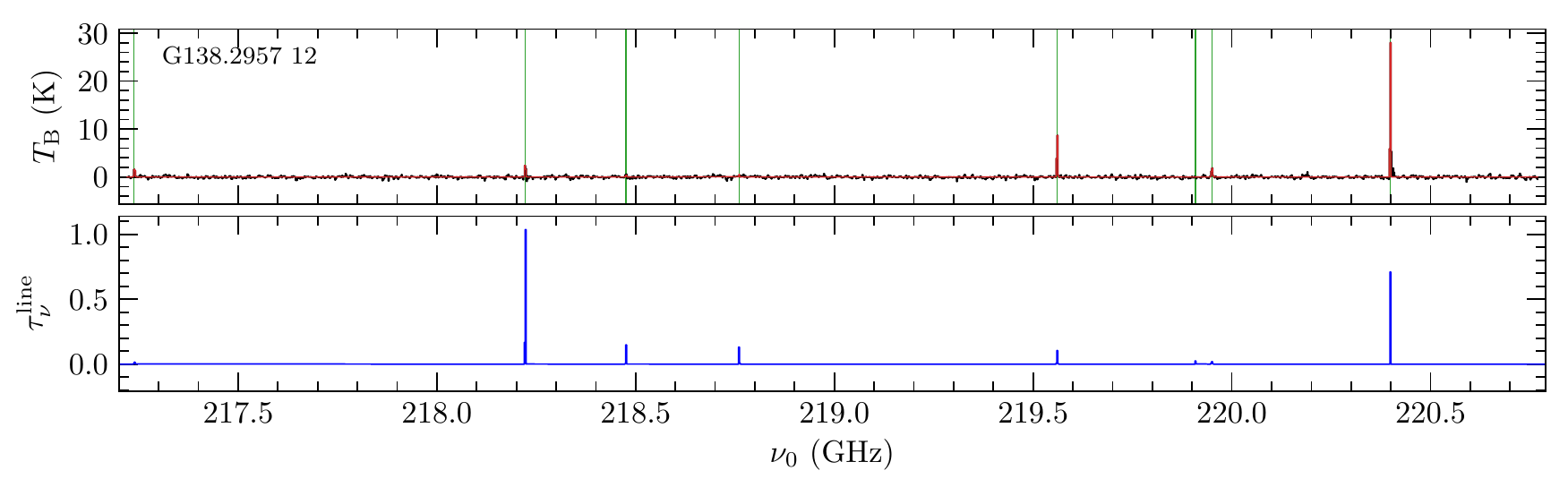}
\includegraphics[]{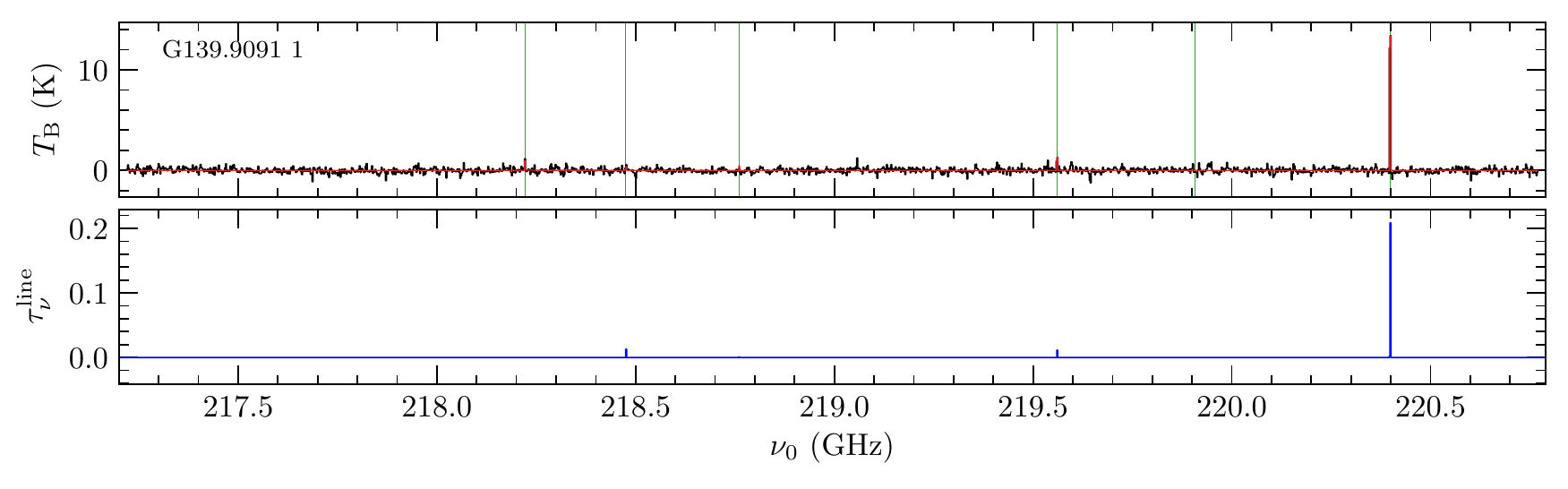}
\includegraphics[]{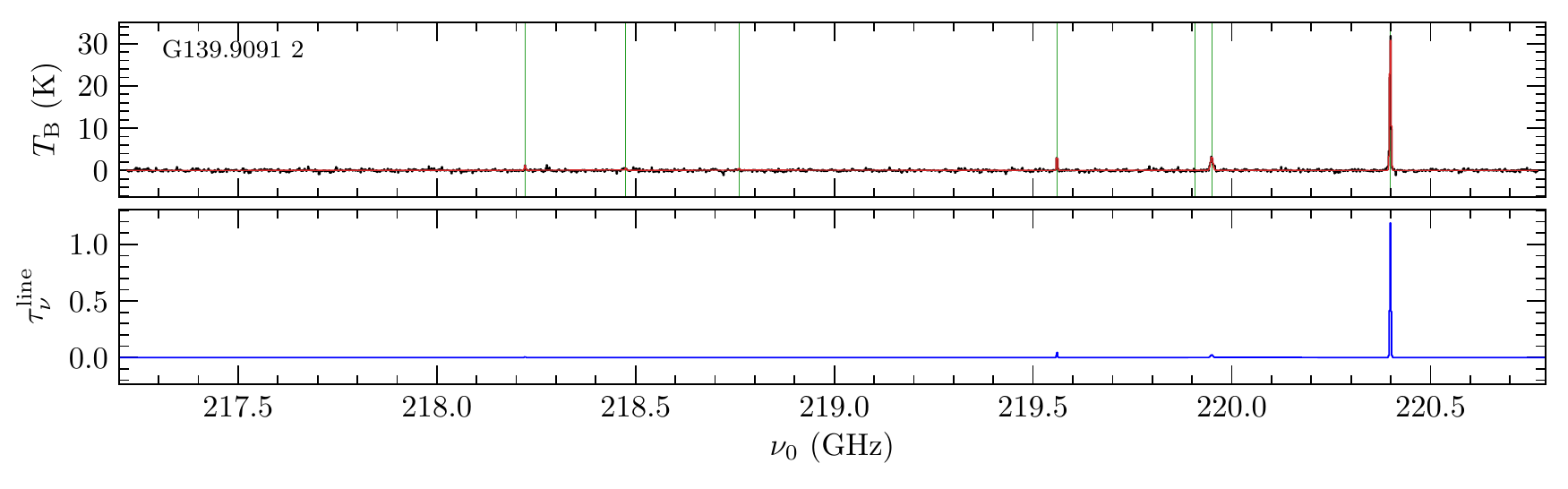}
\caption{\textit{Top panel:} Observed (black line) spectrum and \texttt{XCLASS} fit (red line) for all 120 analyzed positions. Fitted molecular transitions are indicated by green vertical lines. \textit{Bottom panel:} Optical depth profile (blue line) of all fitted transitions for all 120 analyzed positions.}
\end{figure*}
 
\begin{figure*}
\ContinuedFloat
\captionsetup{list=off,format=cont}
\centering
\includegraphics[]{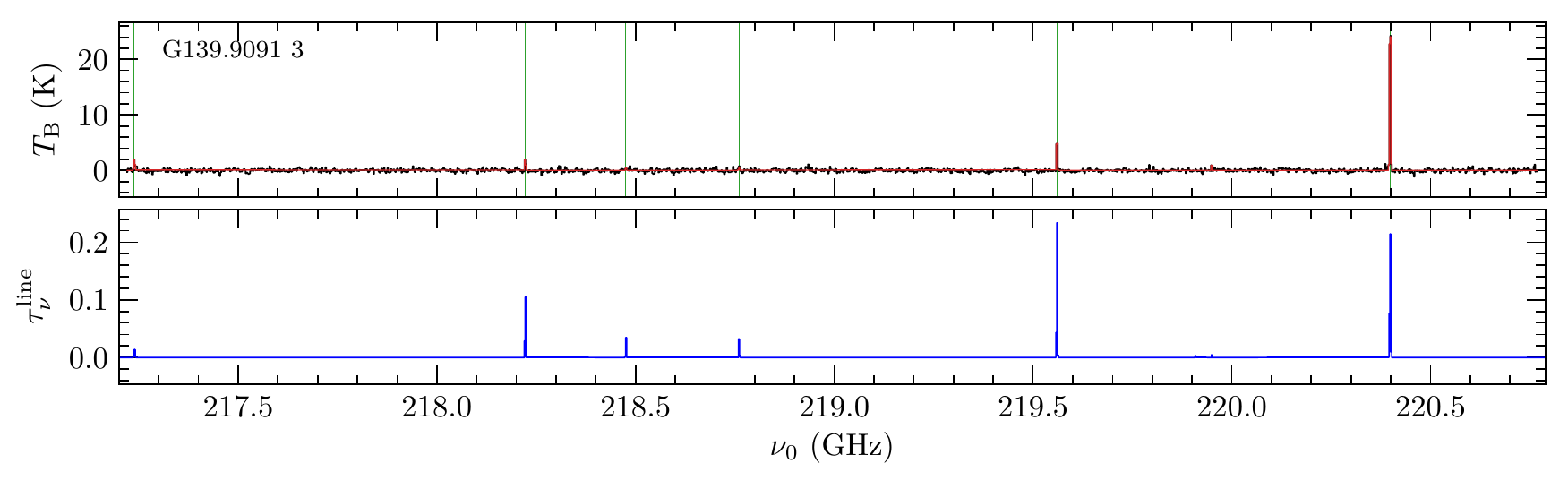}
\includegraphics[]{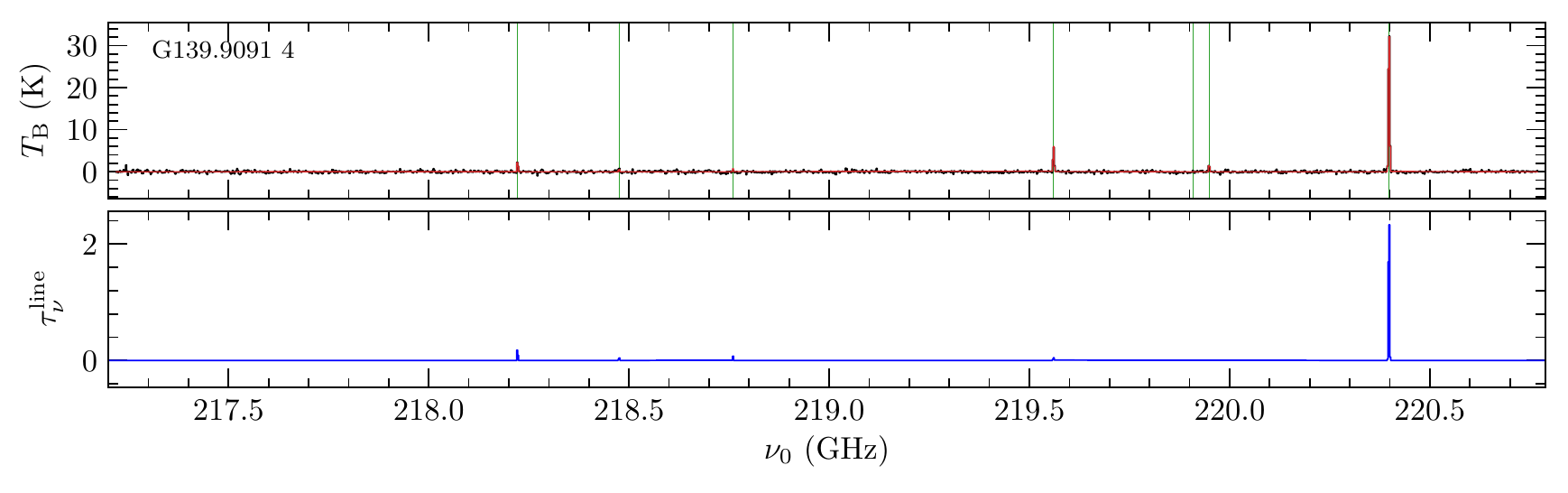}
\includegraphics[]{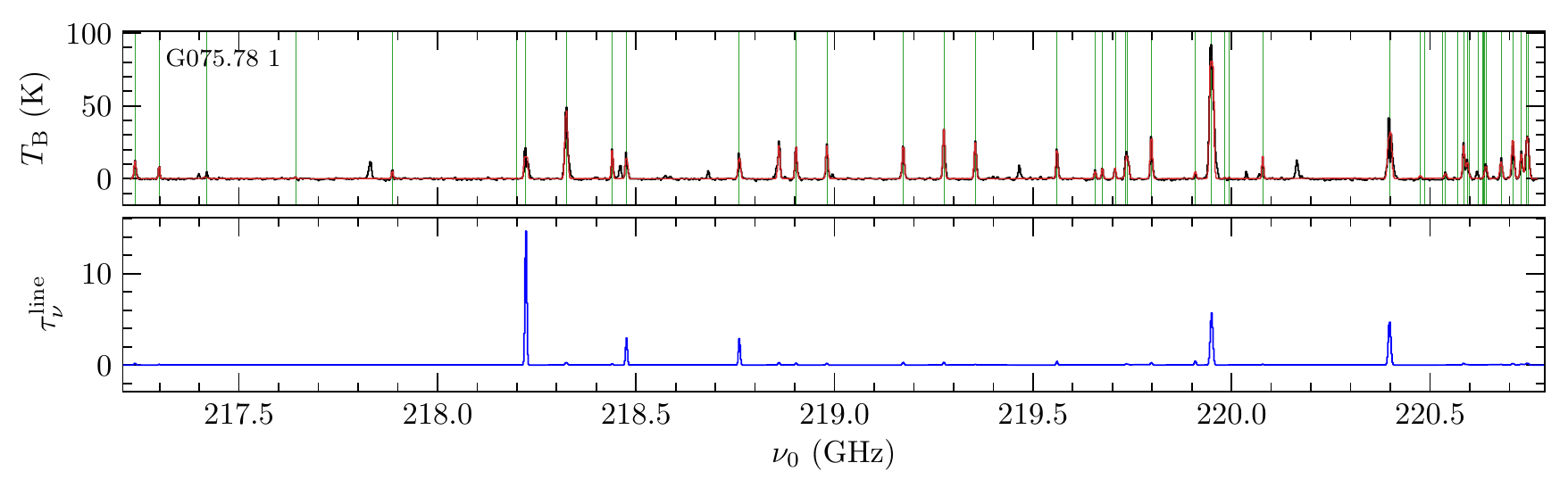}
\includegraphics[]{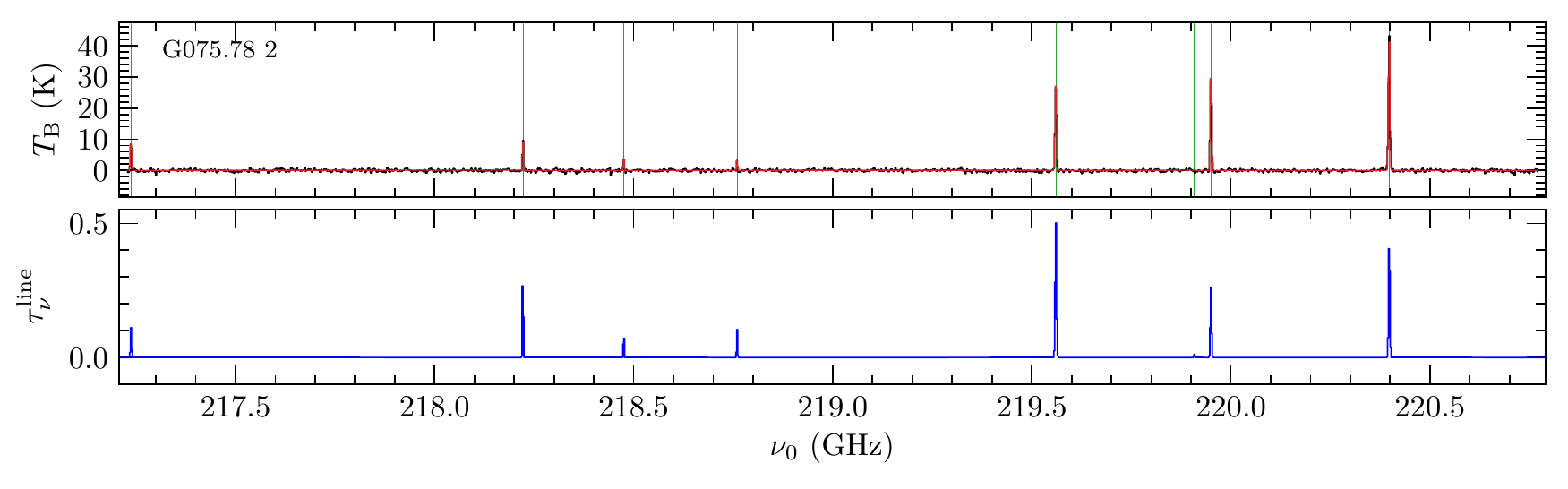}
\caption{\textit{Top panel:} Observed (black line) spectrum and \texttt{XCLASS} fit (red line) for all 120 analyzed positions. Fitted molecular transitions are indicated by green vertical lines. \textit{Bottom panel:} Optical depth profile (blue line) of all fitted transitions for all 120 analyzed positions.}
\end{figure*}
 
\begin{figure*}
\ContinuedFloat
\captionsetup{list=off,format=cont}
\centering
\includegraphics[]{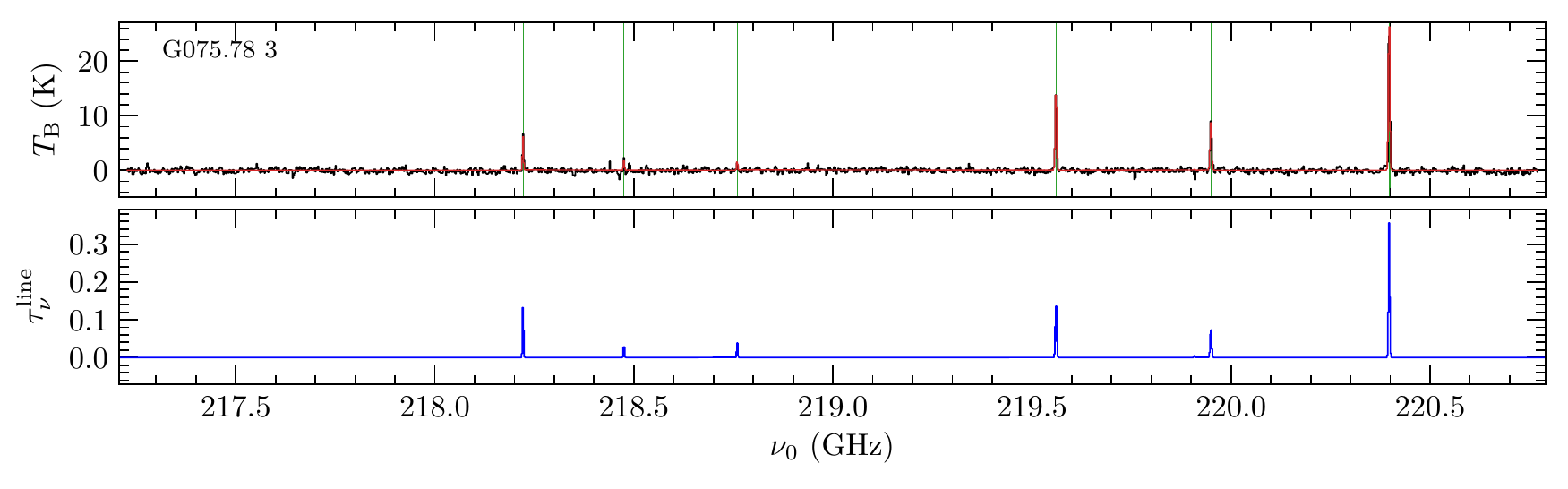}
\includegraphics[]{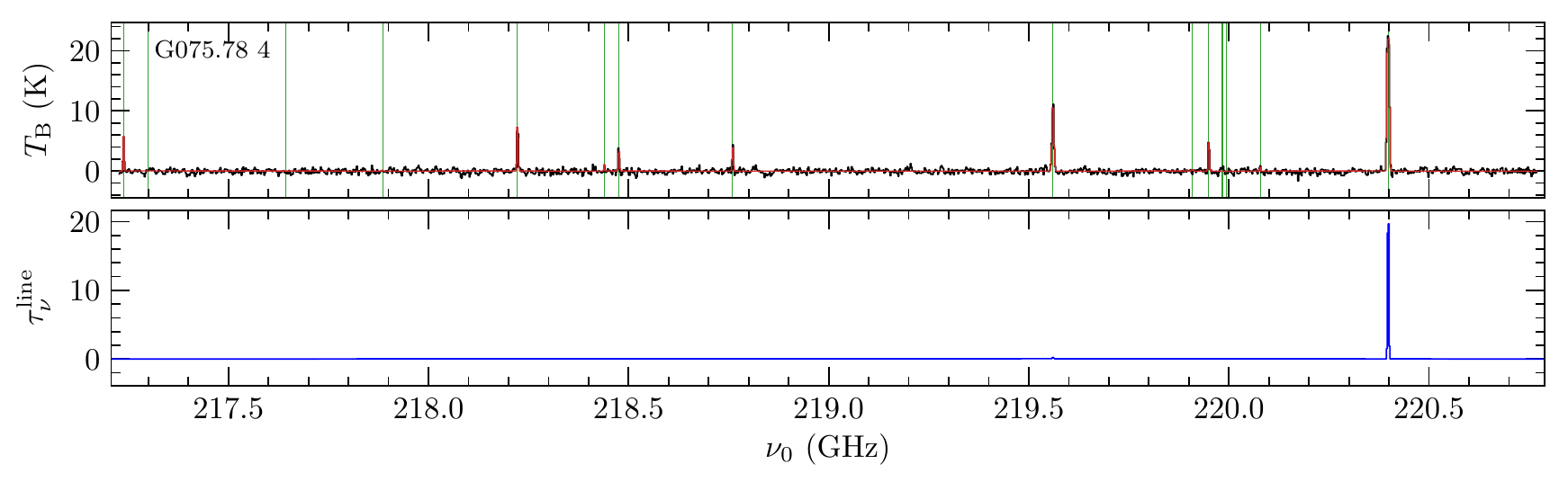}
\includegraphics[]{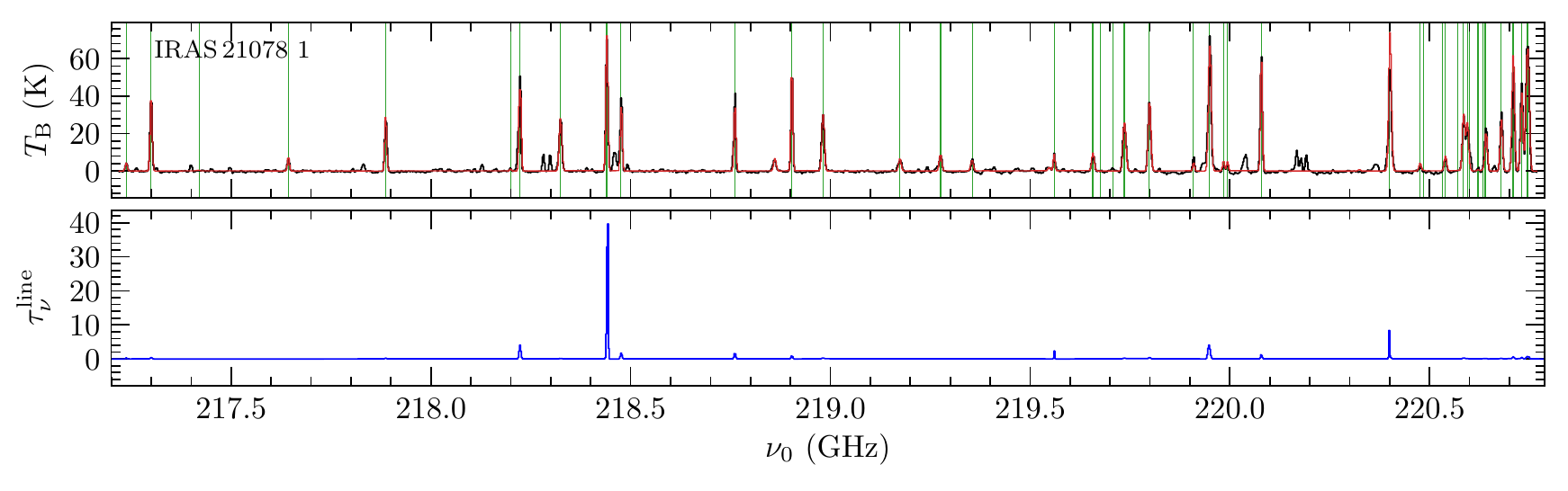}
\includegraphics[]{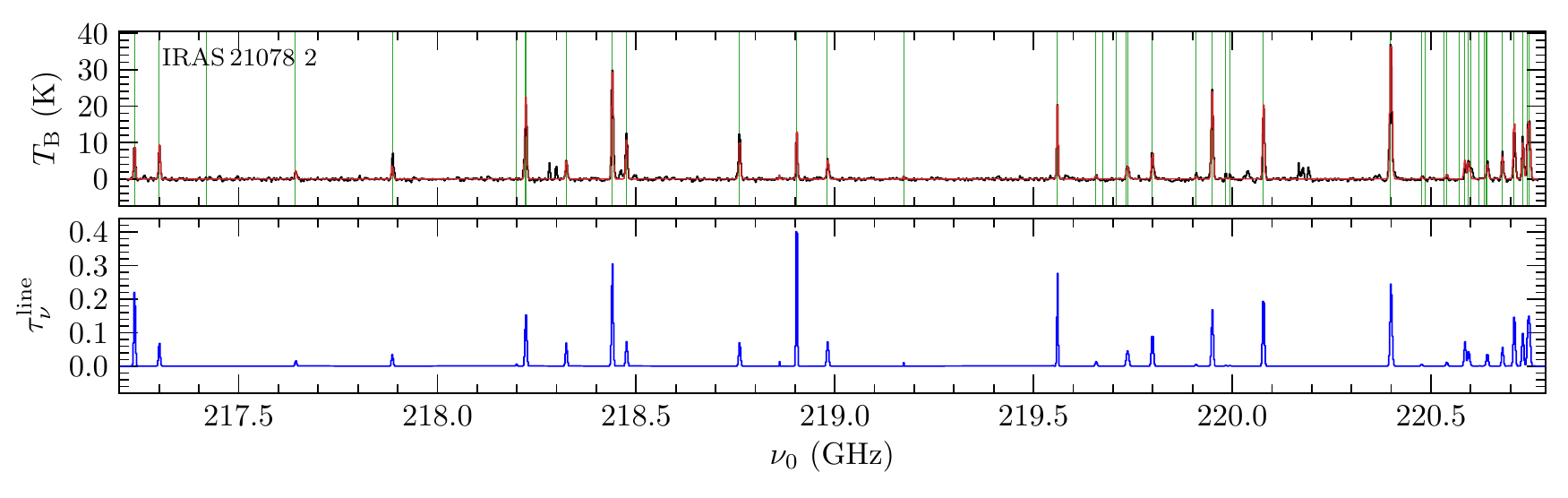}
\caption{\textit{Top panel:} Observed (black line) spectrum and \texttt{XCLASS} fit (red line) for all 120 analyzed positions. Fitted molecular transitions are indicated by green vertical lines. \textit{Bottom panel:} Optical depth profile (blue line) of all fitted transitions for all 120 analyzed positions.}
\end{figure*}
 
\begin{figure*}
\ContinuedFloat
\captionsetup{list=off,format=cont}
\centering
\includegraphics[]{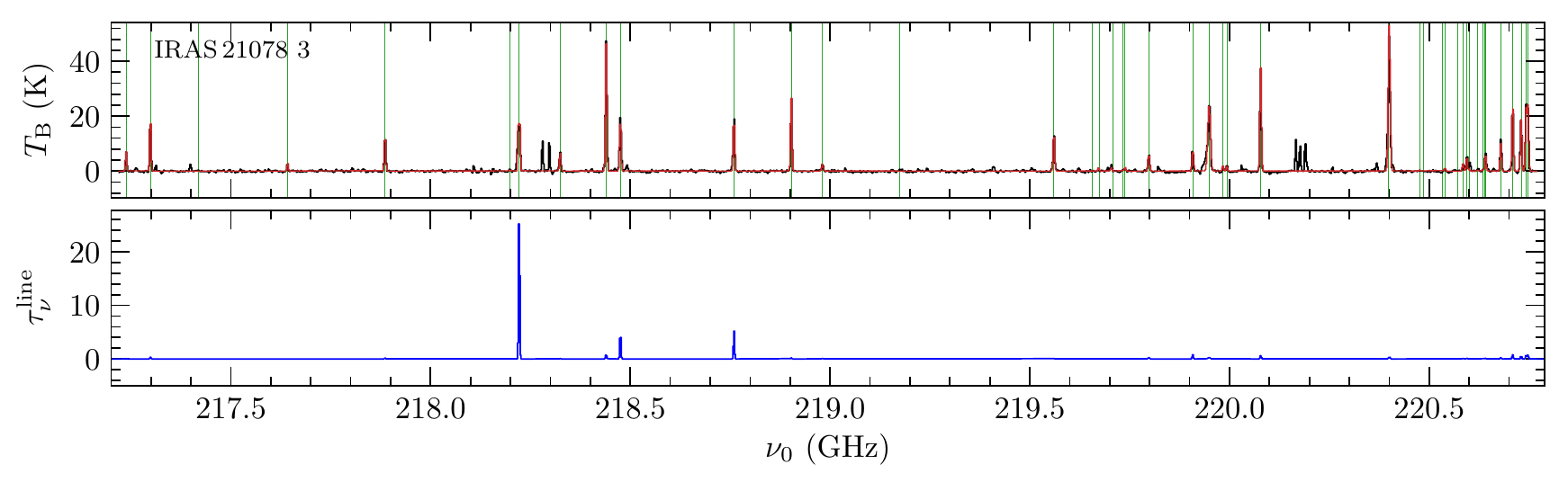}
\includegraphics[]{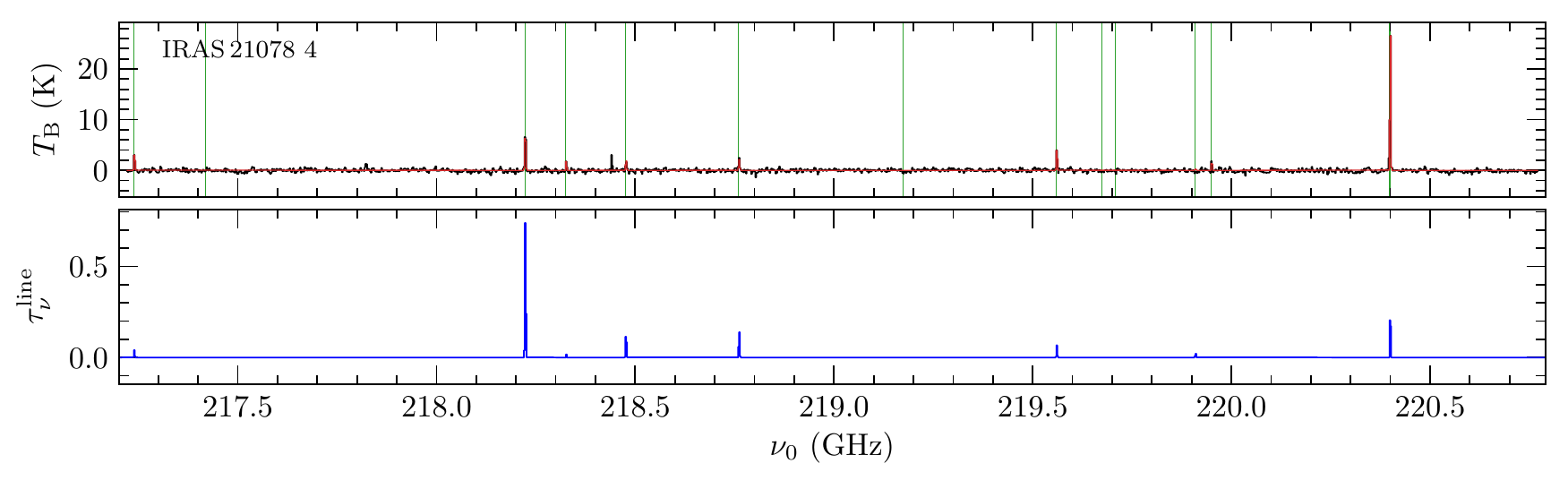}
\includegraphics[]{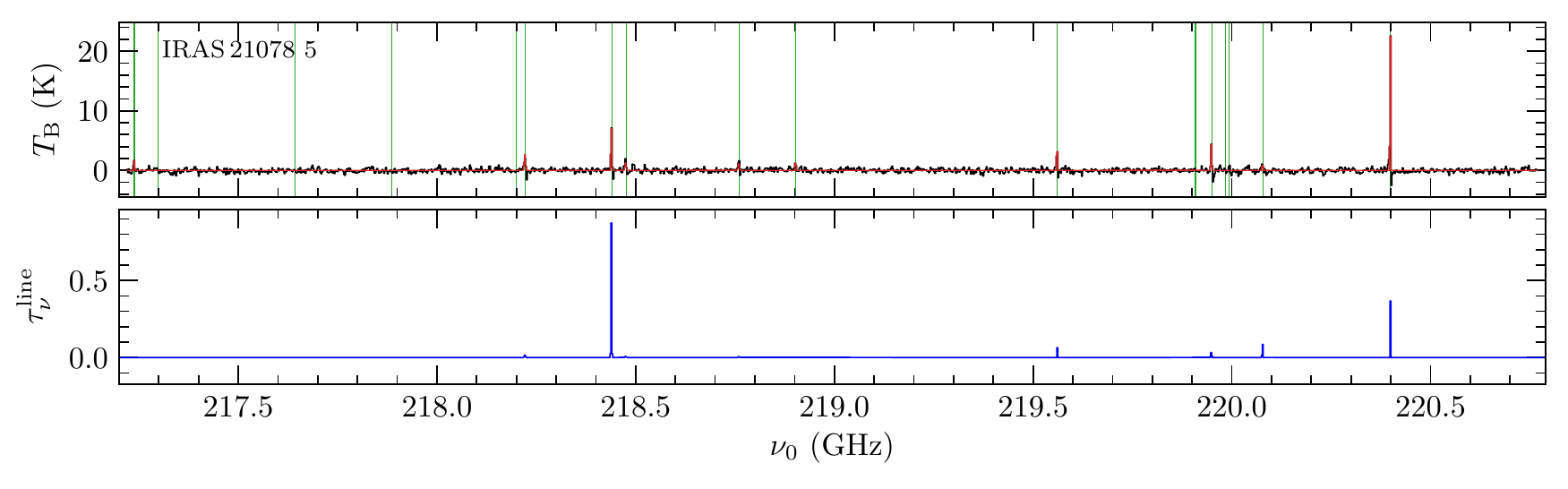}
\includegraphics[]{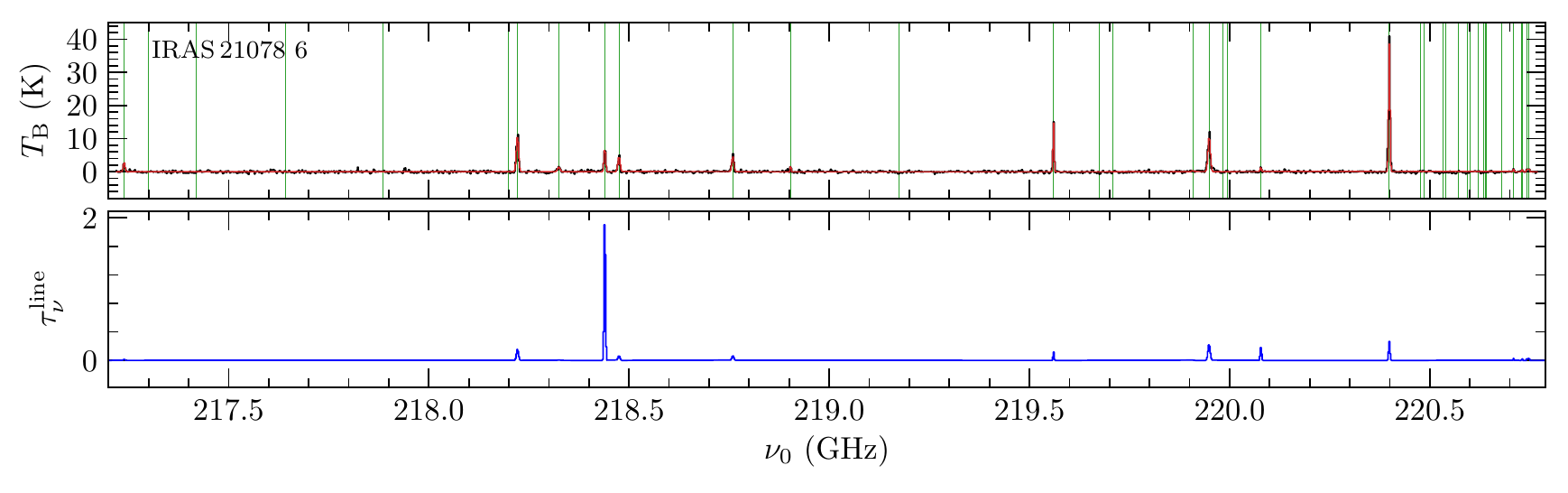}
\caption{\textit{Top panel:} Observed (black line) spectrum and \texttt{XCLASS} fit (red line) for all 120 analyzed positions. Fitted molecular transitions are indicated by green vertical lines. \textit{Bottom panel:} Optical depth profile (blue line) of all fitted transitions for all 120 analyzed positions.}
\end{figure*}
 
\begin{figure*}
\ContinuedFloat
\captionsetup{list=off,format=cont}
\centering
\includegraphics[]{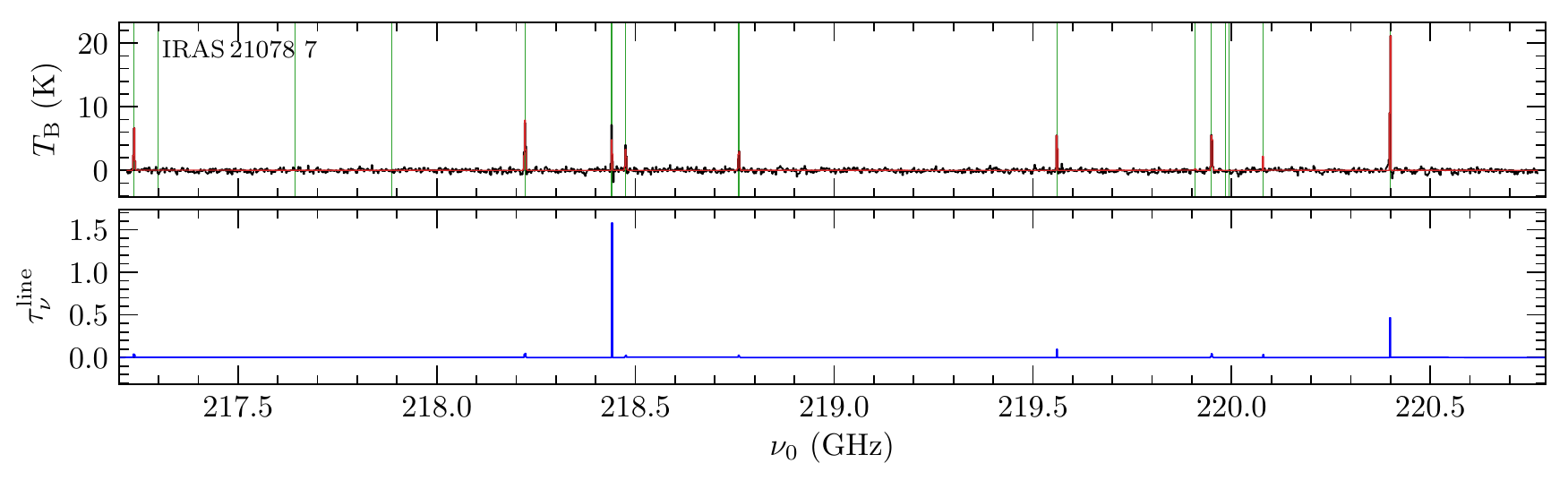}
\includegraphics[]{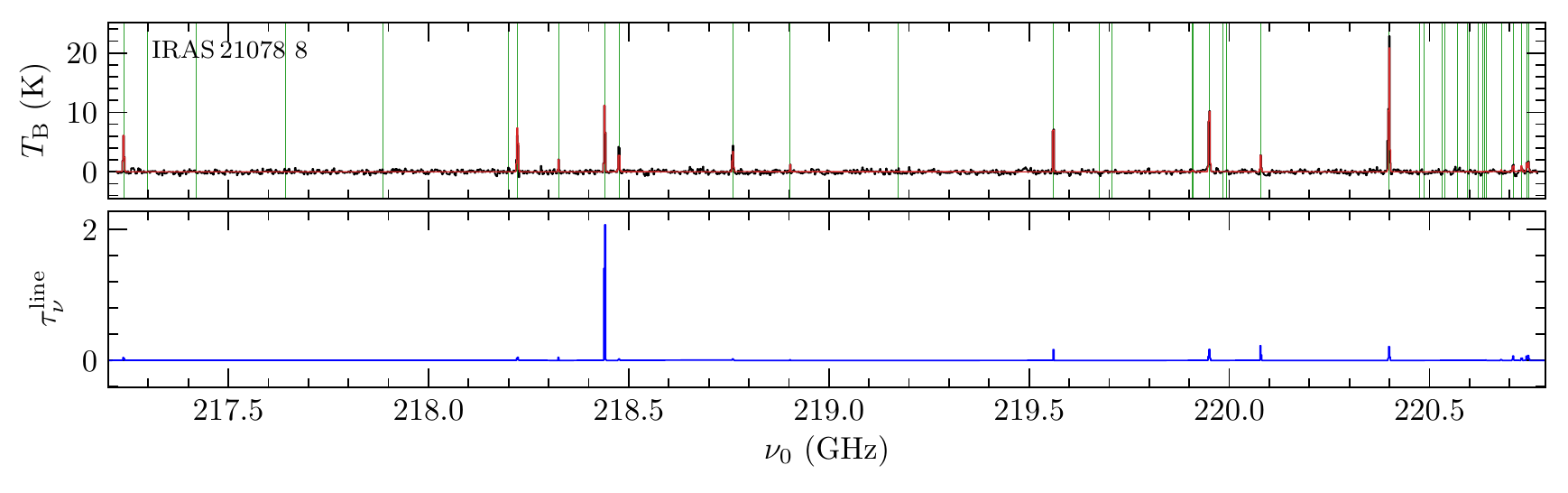}
\includegraphics[]{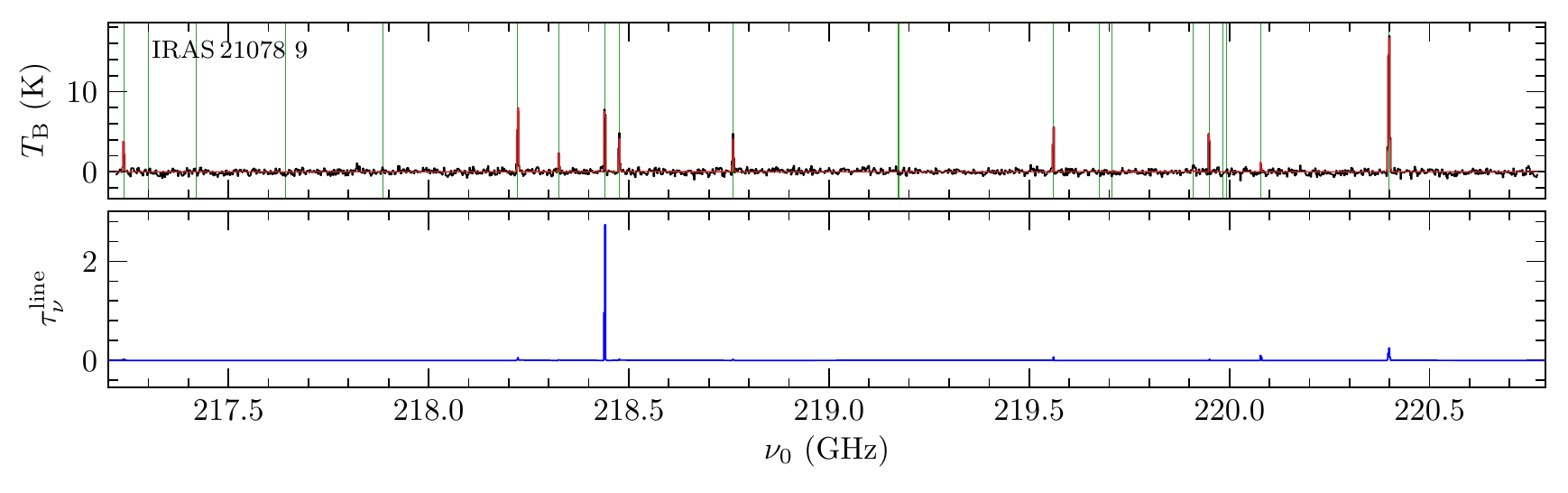}
\includegraphics[]{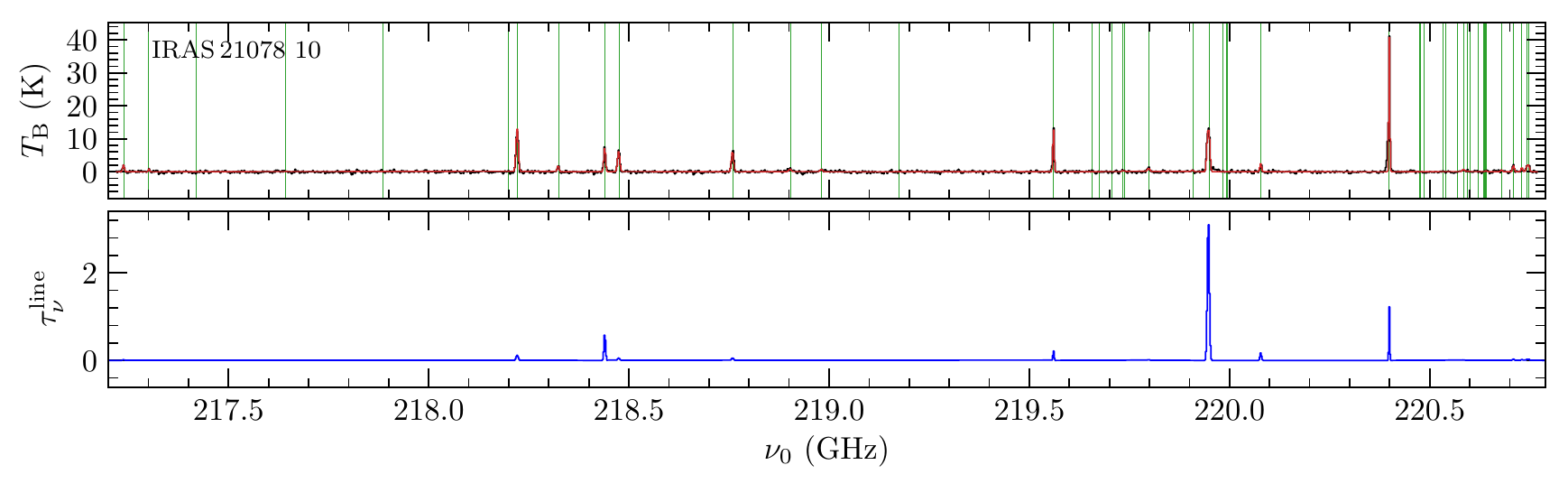}
\caption{\textit{Top panel:} Observed (black line) spectrum and \texttt{XCLASS} fit (red line) for all 120 analyzed positions. Fitted molecular transitions are indicated by green vertical lines. \textit{Bottom panel:} Optical depth profile (blue line) of all fitted transitions for all 120 analyzed positions.}
\end{figure*}
 
\begin{figure*}
\ContinuedFloat
\captionsetup{list=off,format=cont}
\centering
\includegraphics[]{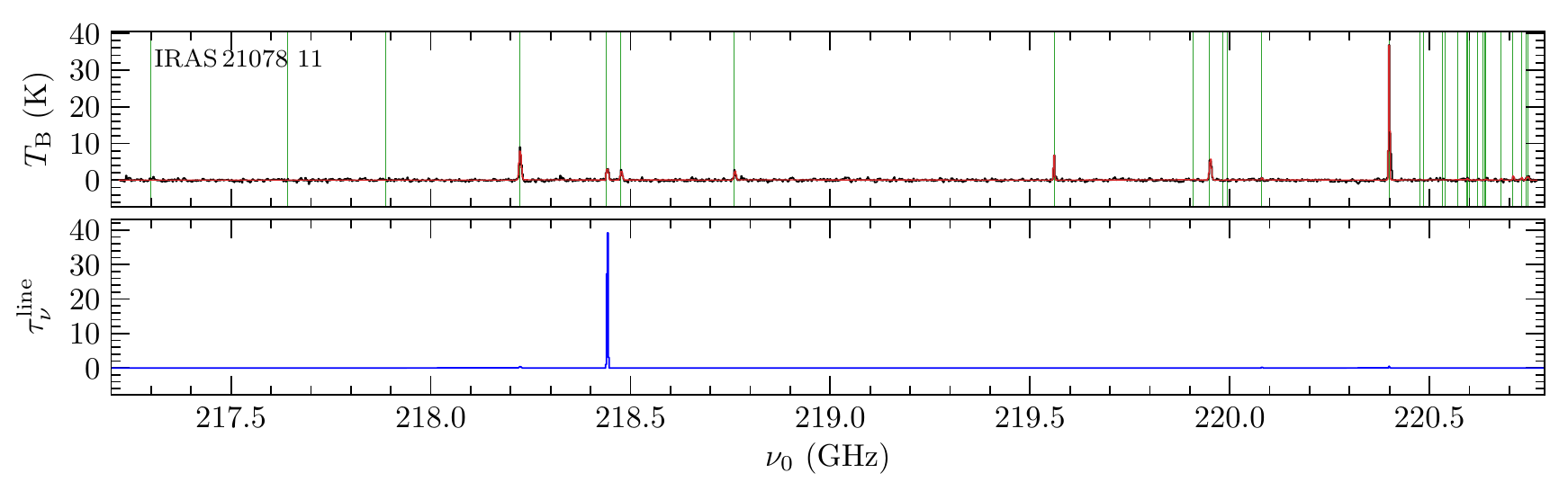}
\includegraphics[]{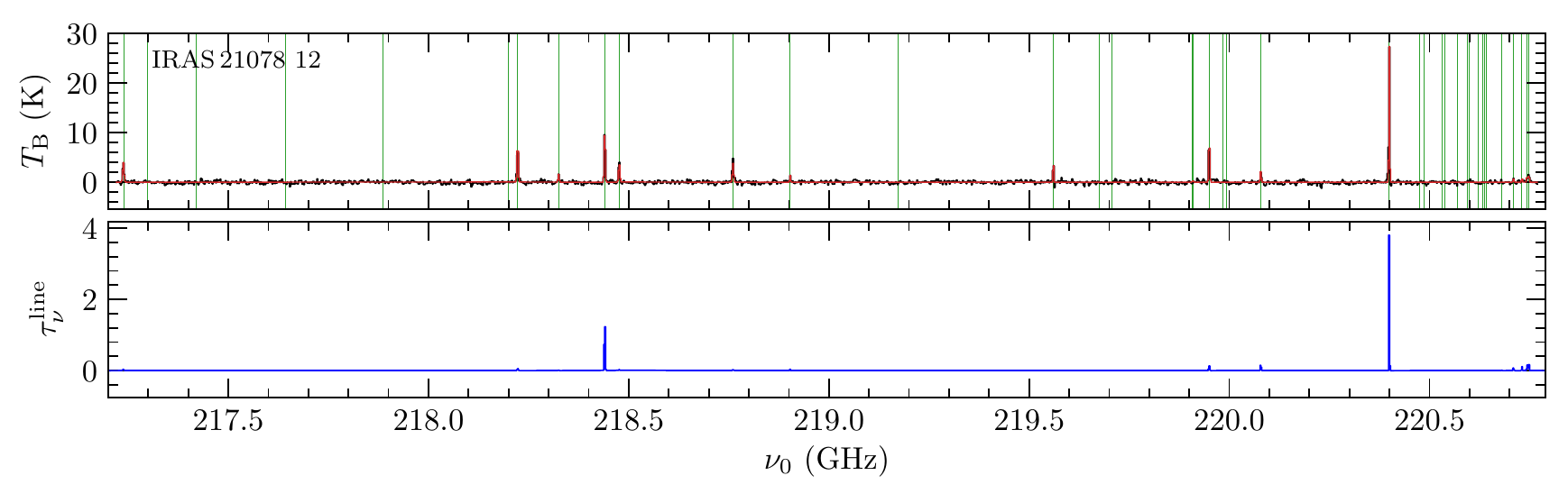}
\includegraphics[]{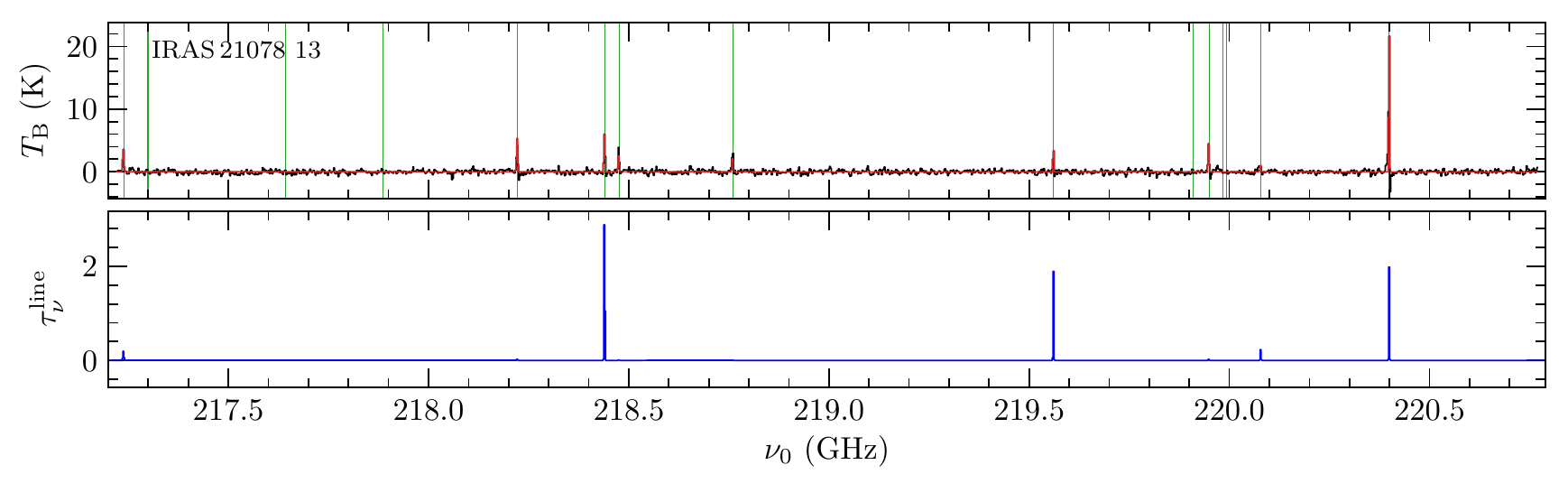}
\includegraphics[]{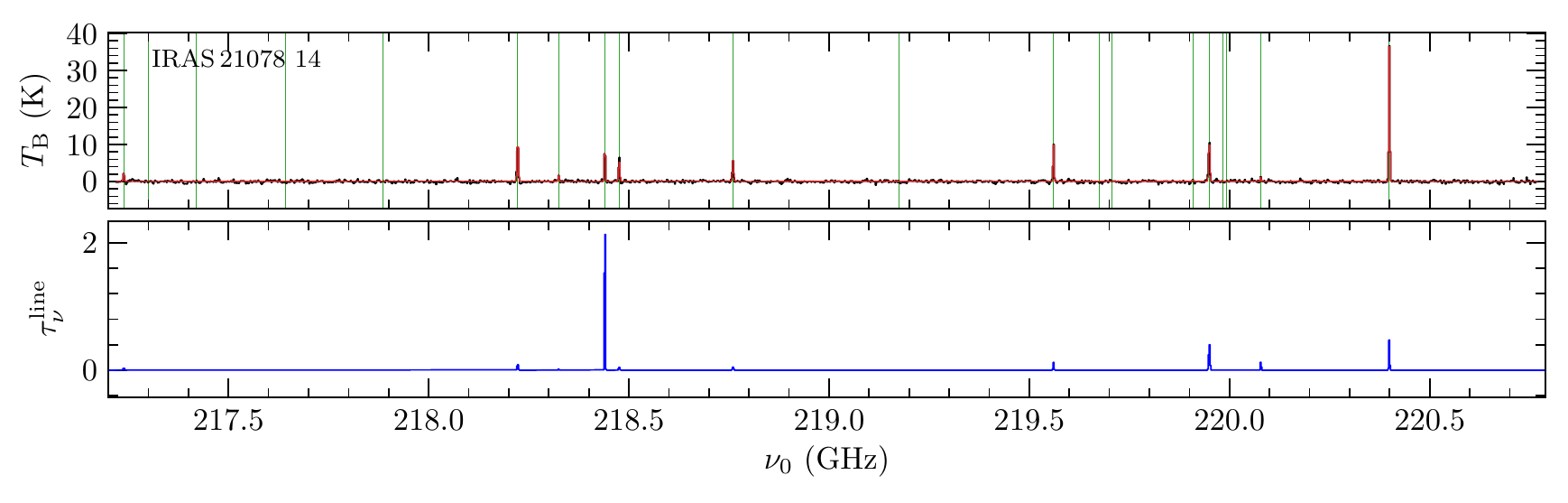}
\caption{\textit{Top panel:} Observed (black line) spectrum and \texttt{XCLASS} fit (red line) for all 120 analyzed positions. Fitted molecular transitions are indicated by green vertical lines. \textit{Bottom panel:} Optical depth profile (blue line) of all fitted transitions for all 120 analyzed positions.}
\end{figure*}
 
\begin{figure*}
\ContinuedFloat
\captionsetup{list=off,format=cont}
\centering
\includegraphics[]{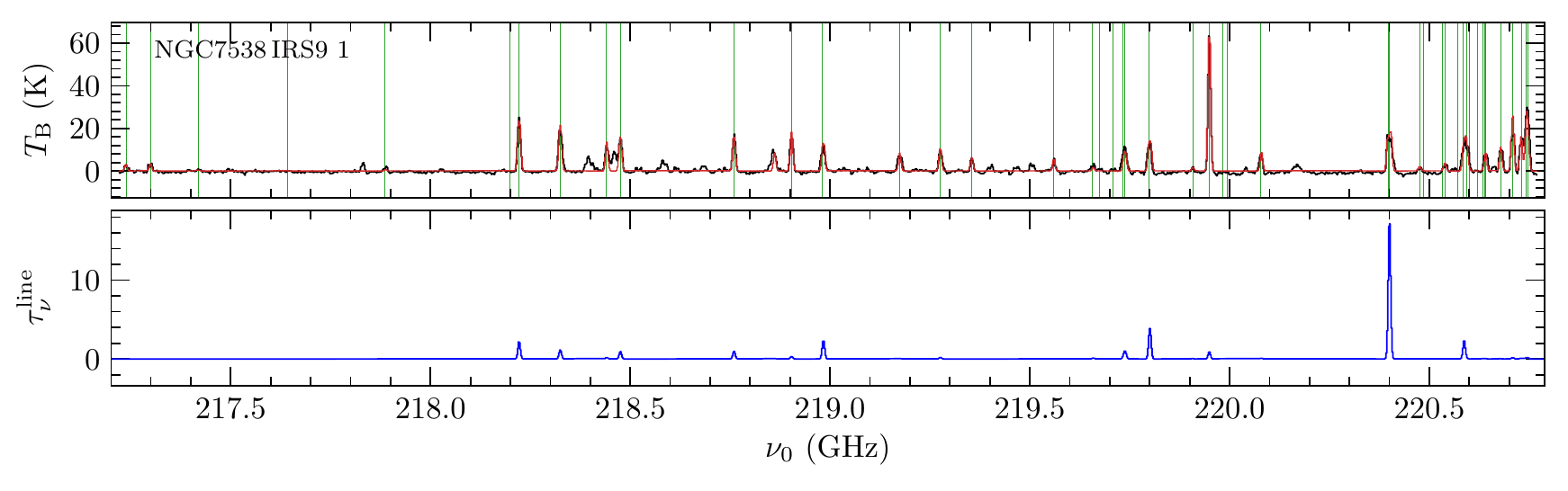}
\includegraphics[]{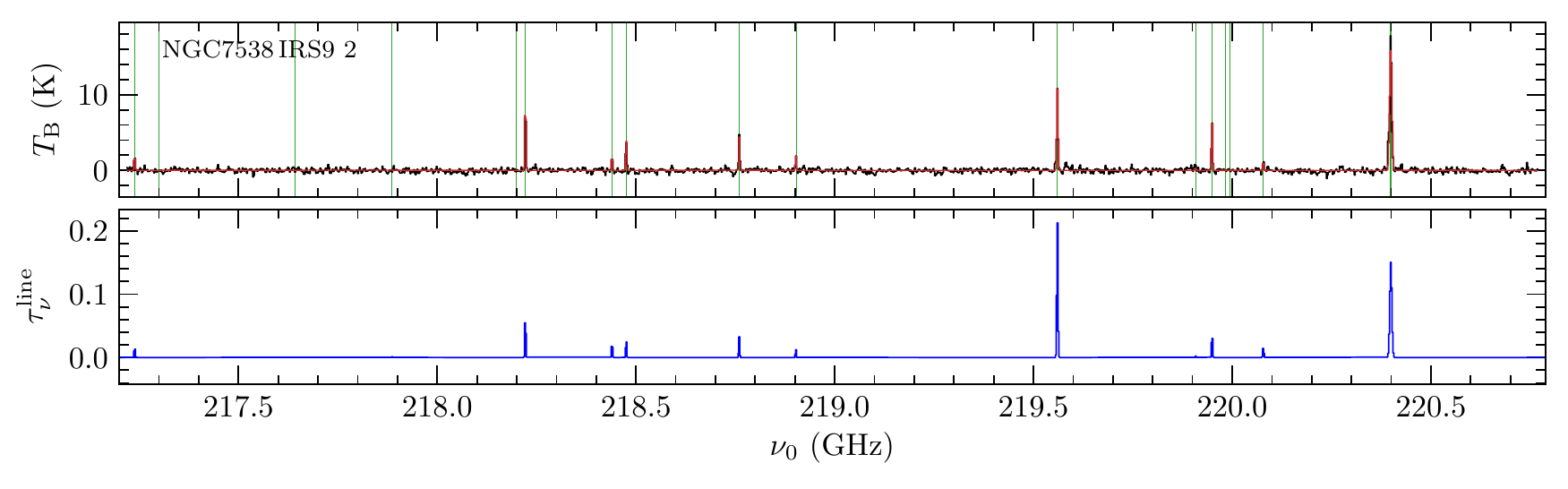}
\includegraphics[]{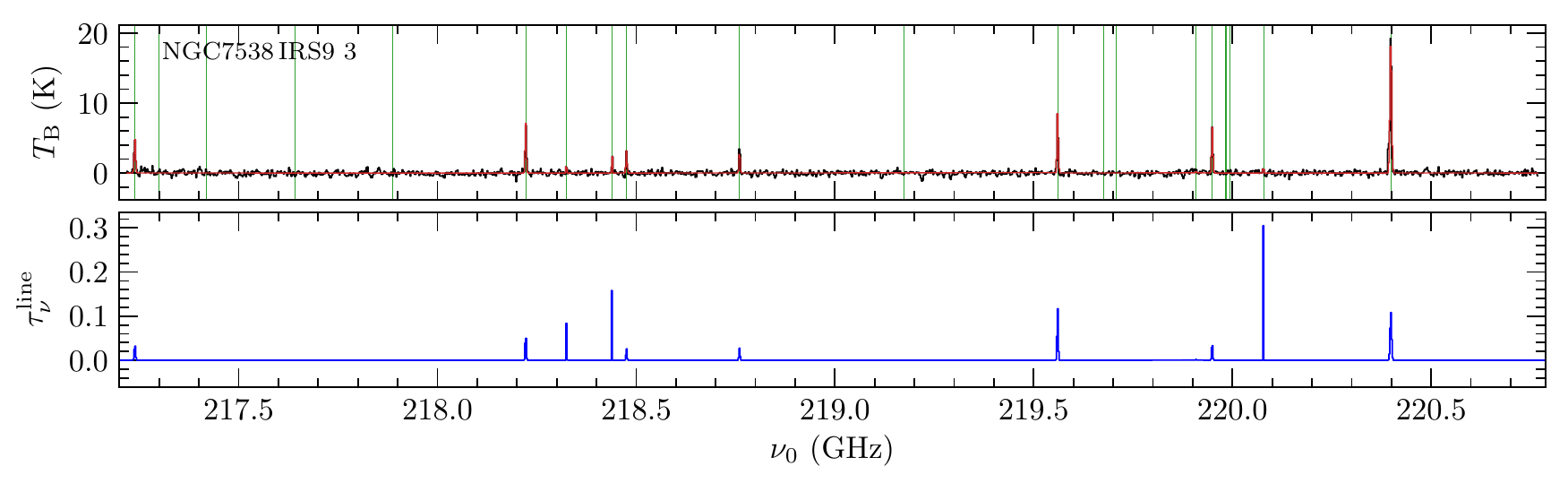}
\includegraphics[]{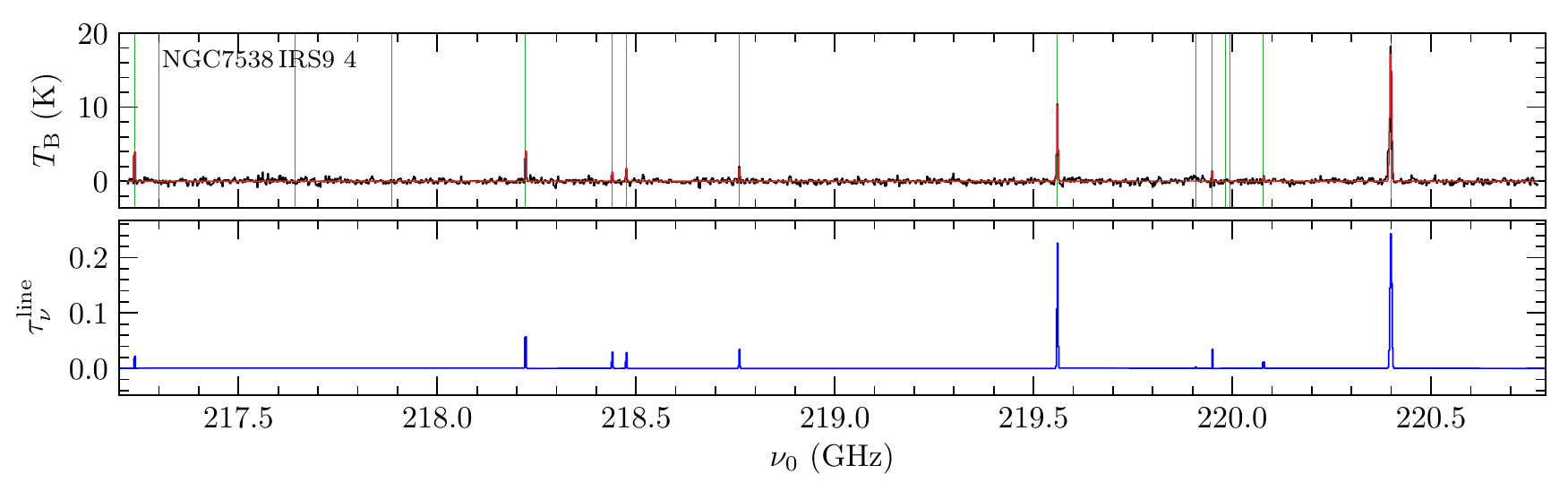}
\caption{\textit{Top panel:} Observed (black line) spectrum and \texttt{XCLASS} fit (red line) for all 120 analyzed positions. Fitted molecular transitions are indicated by green vertical lines. \textit{Bottom panel:} Optical depth profile (blue line) of all fitted transitions for all 120 analyzed positions.}
\end{figure*}
 
\begin{figure*}
\ContinuedFloat
\captionsetup{list=off,format=cont}
\centering
\includegraphics[]{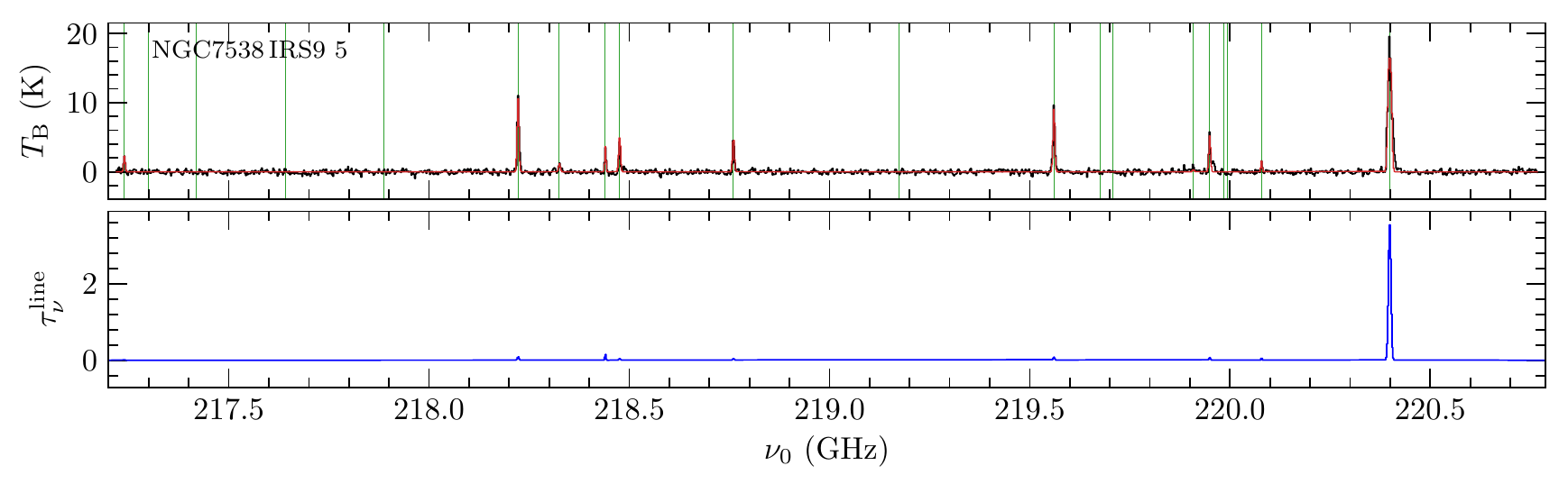}
\includegraphics[]{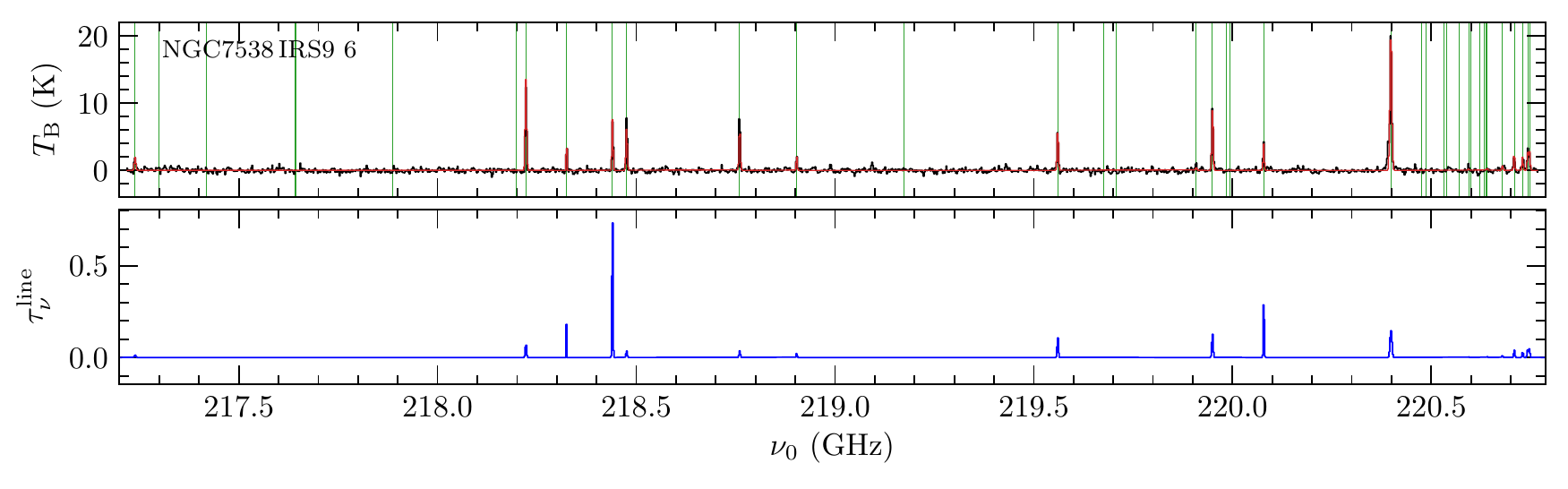}
\includegraphics[]{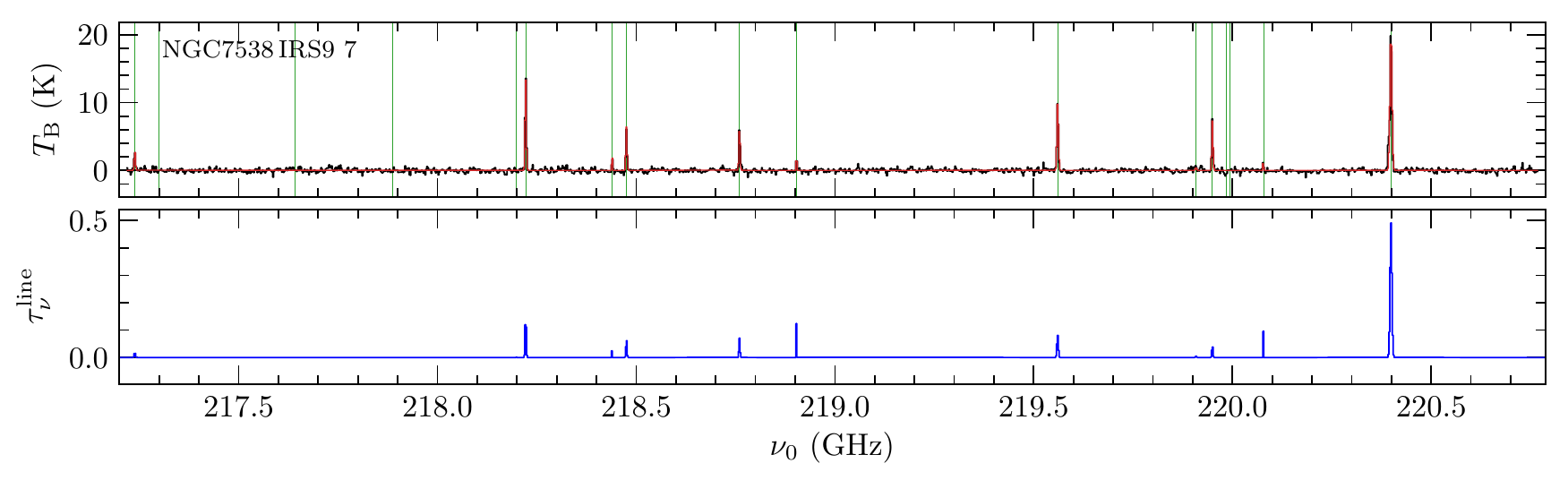}
\includegraphics[]{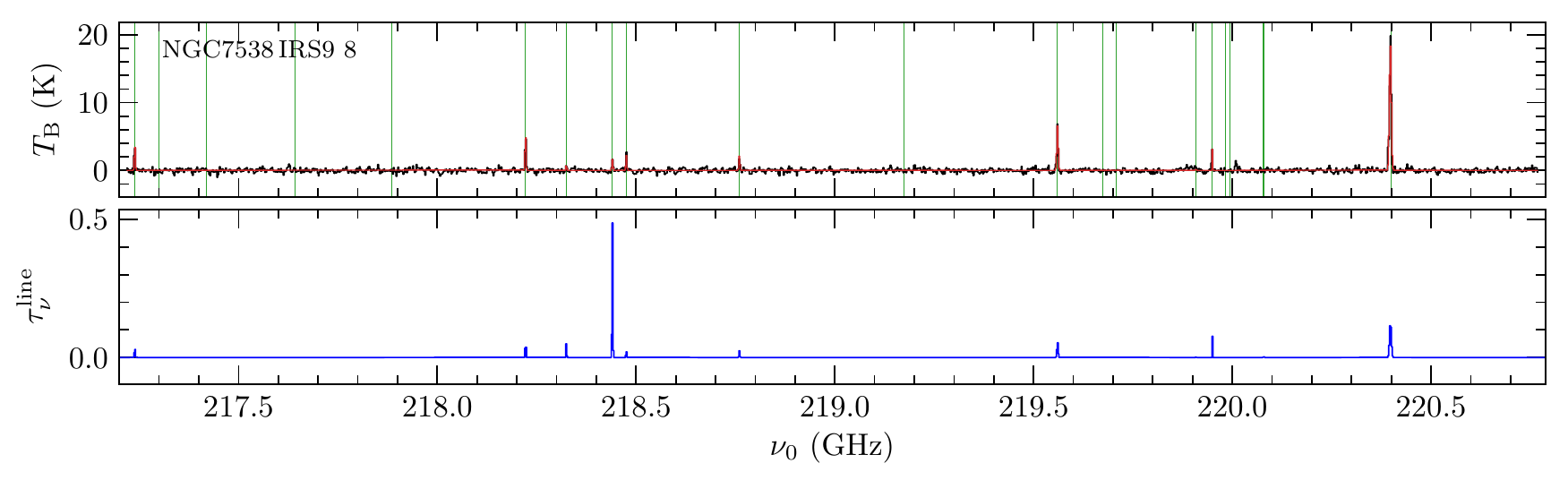}
\caption{\textit{Top panel:} Observed (black line) spectrum and \texttt{XCLASS} fit (red line) for all 120 analyzed positions. Fitted molecular transitions are indicated by green vertical lines. \textit{Bottom panel:} Optical depth profile (blue line) of all fitted transitions for all 120 analyzed positions.}
\end{figure*}
 
\begin{figure*}
\ContinuedFloat
\captionsetup{list=off,format=cont}
\centering
\includegraphics[]{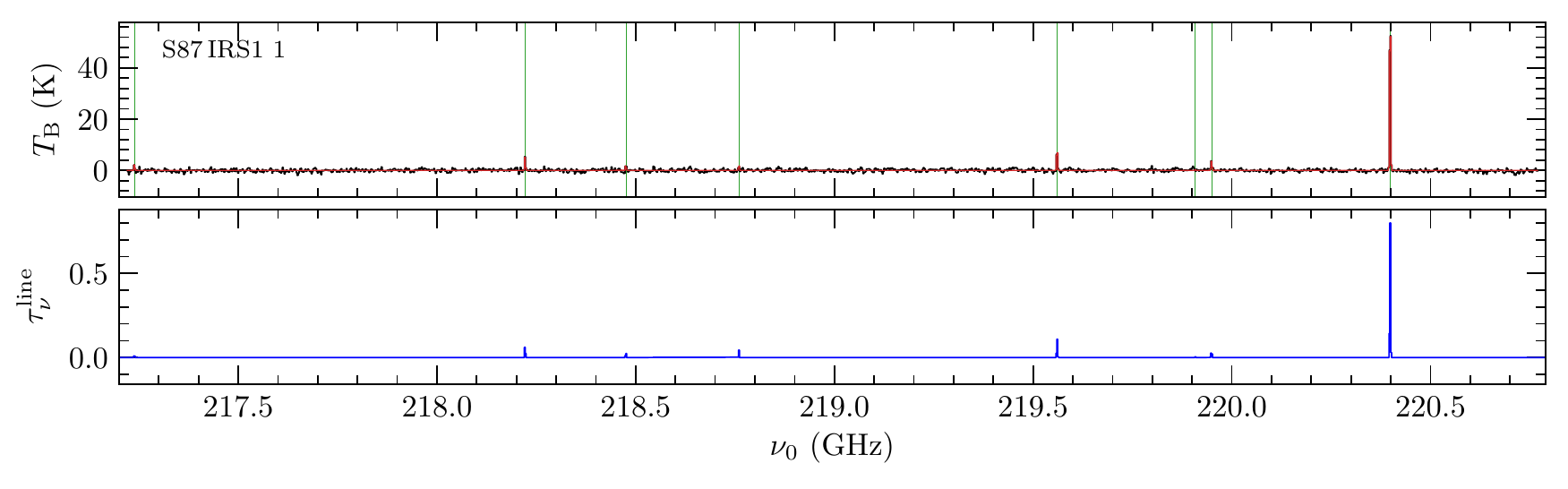}
\includegraphics[]{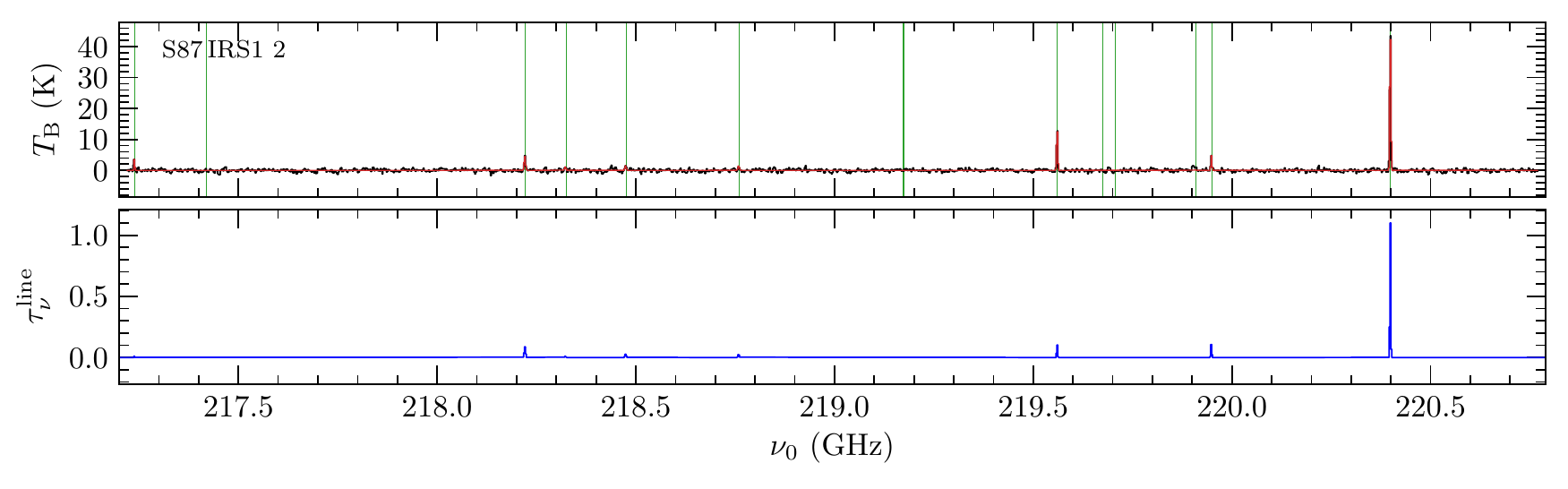}
\includegraphics[]{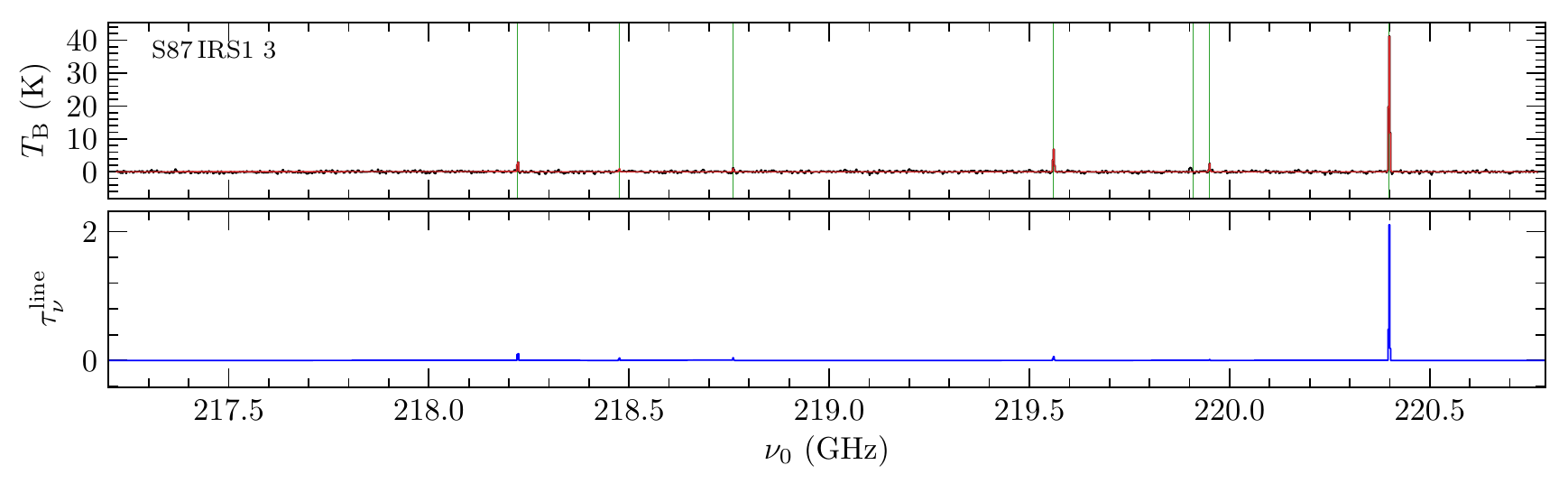}
\includegraphics[]{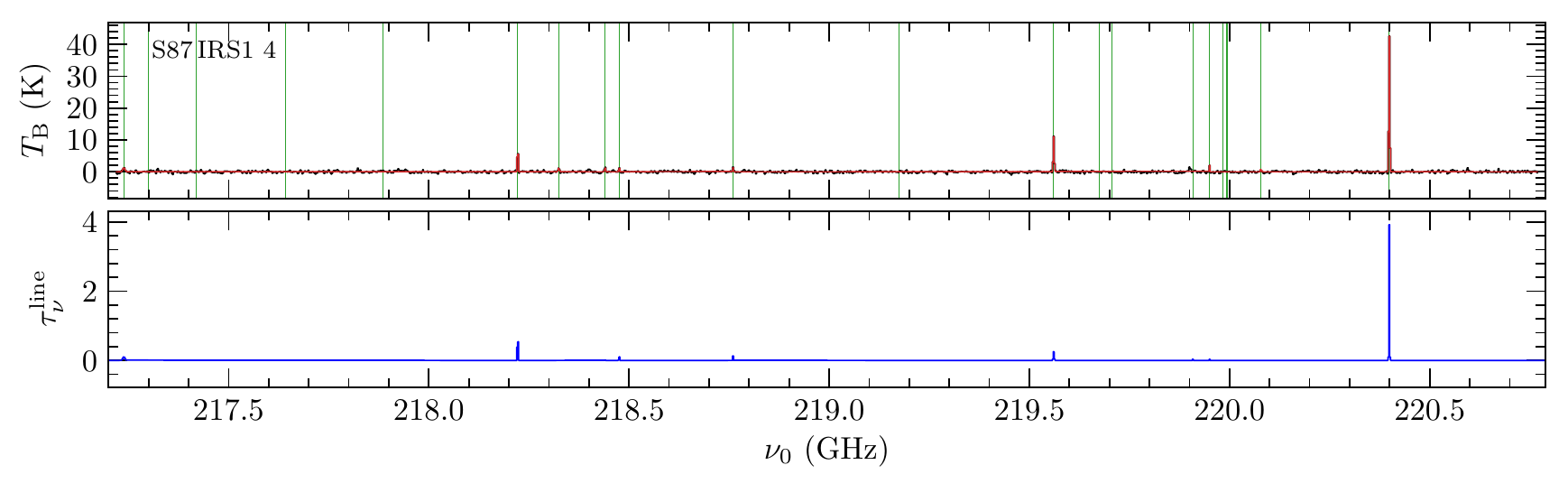}
\caption{\textit{Top panel:} Observed (black line) spectrum and \texttt{XCLASS} fit (red line) for all 120 analyzed positions. Fitted molecular transitions are indicated by green vertical lines. \textit{Bottom panel:} Optical depth profile (blue line) of all fitted transitions for all 120 analyzed positions.}
\end{figure*}
 
\begin{figure*}
\ContinuedFloat
\captionsetup{list=off,format=cont}
\centering
\includegraphics[]{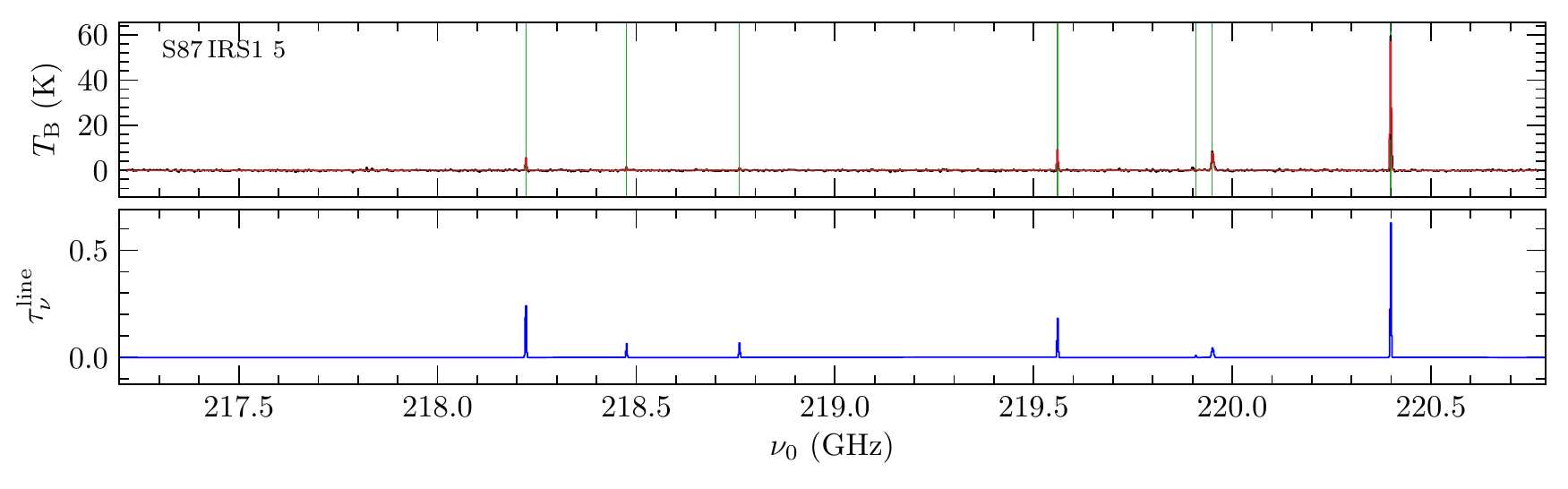}
\includegraphics[]{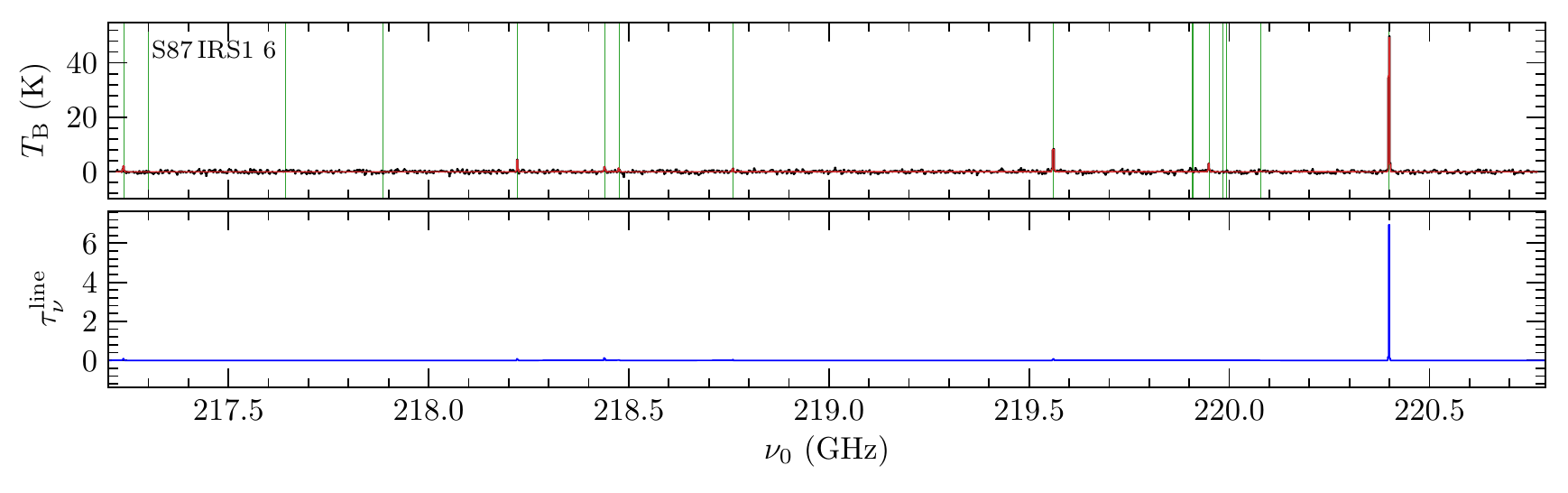}
\includegraphics[]{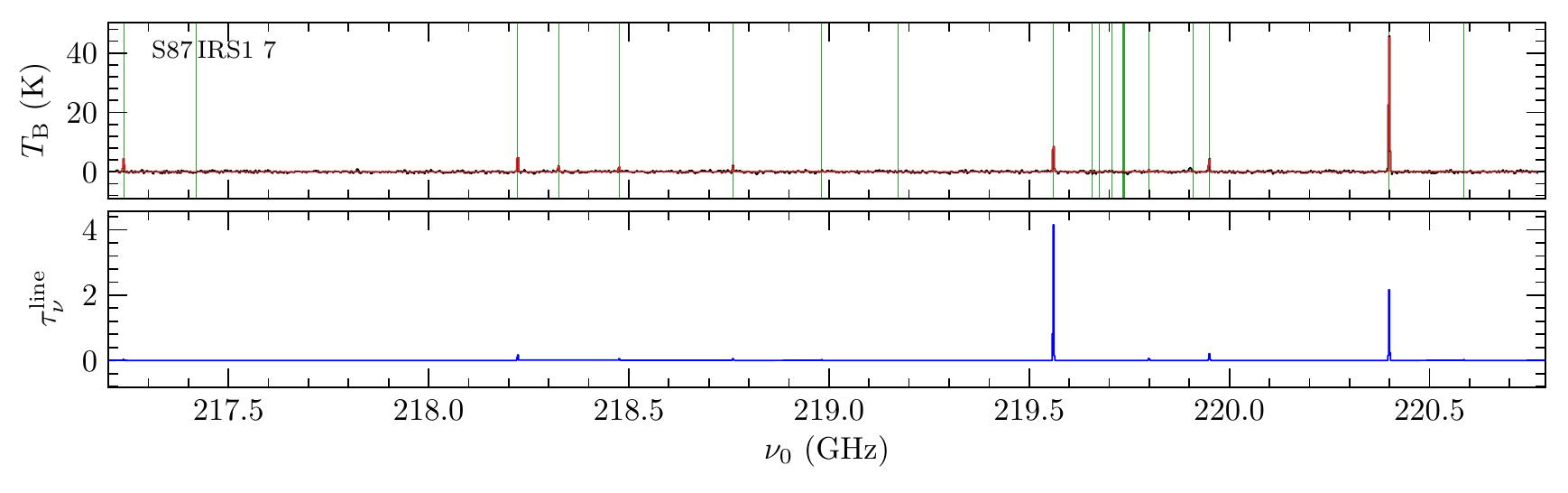}
\includegraphics[]{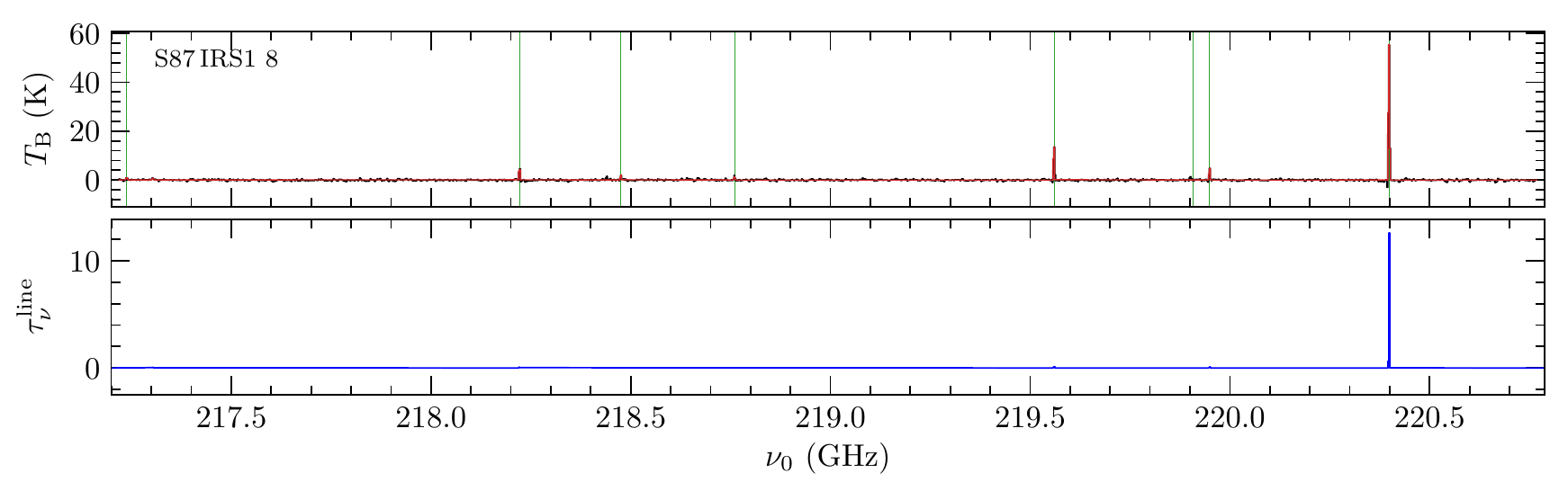}
\caption{\textit{Top panel:} Observed (black line) spectrum and \texttt{XCLASS} fit (red line) for all 120 analyzed positions. Fitted molecular transitions are indicated by green vertical lines. \textit{Bottom panel:} Optical depth profile (blue line) of all fitted transitions for all 120 analyzed positions.}
\end{figure*}
 
\begin{figure*}
\ContinuedFloat
\captionsetup{list=off,format=cont}
\centering
\includegraphics[]{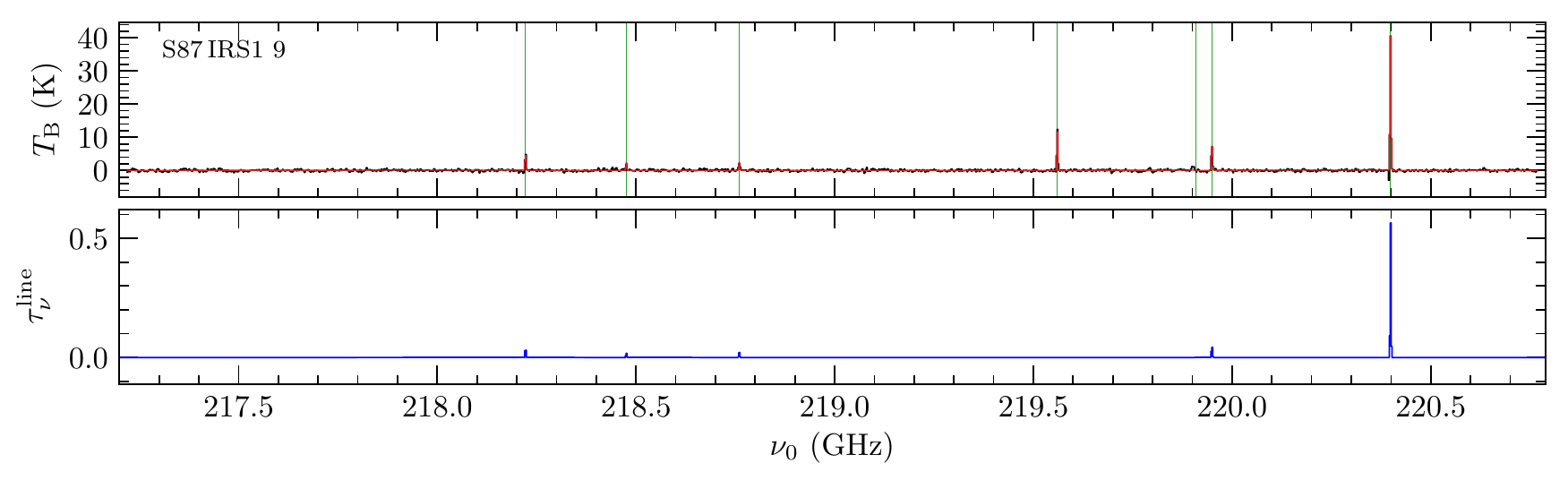}
\includegraphics[]{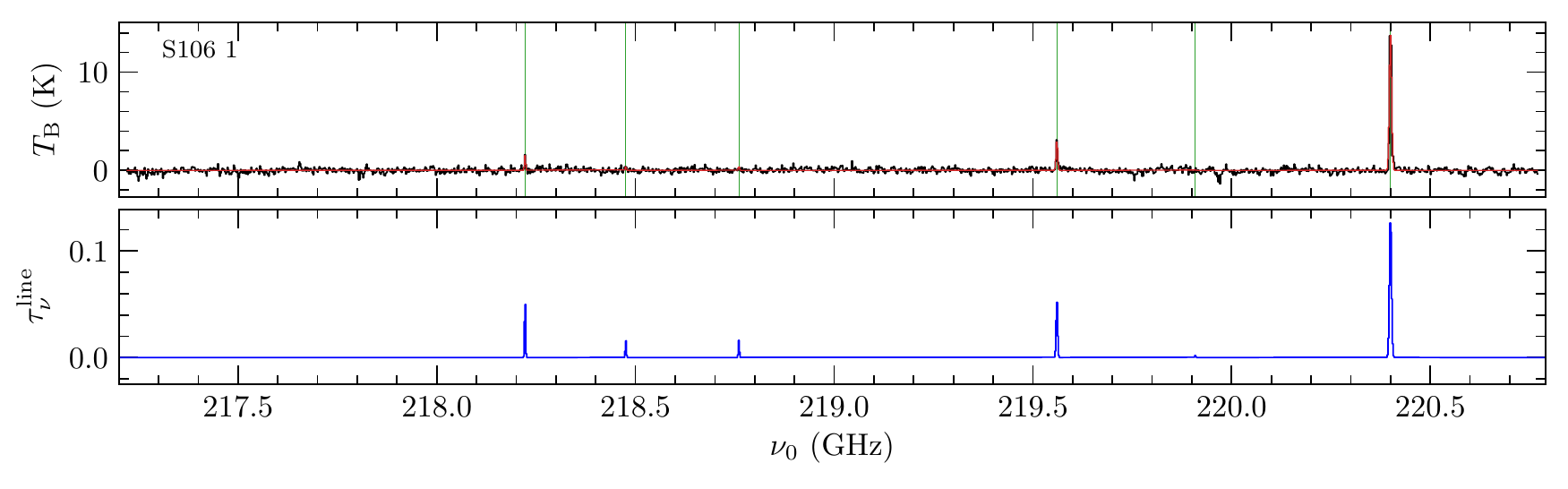}
\includegraphics[]{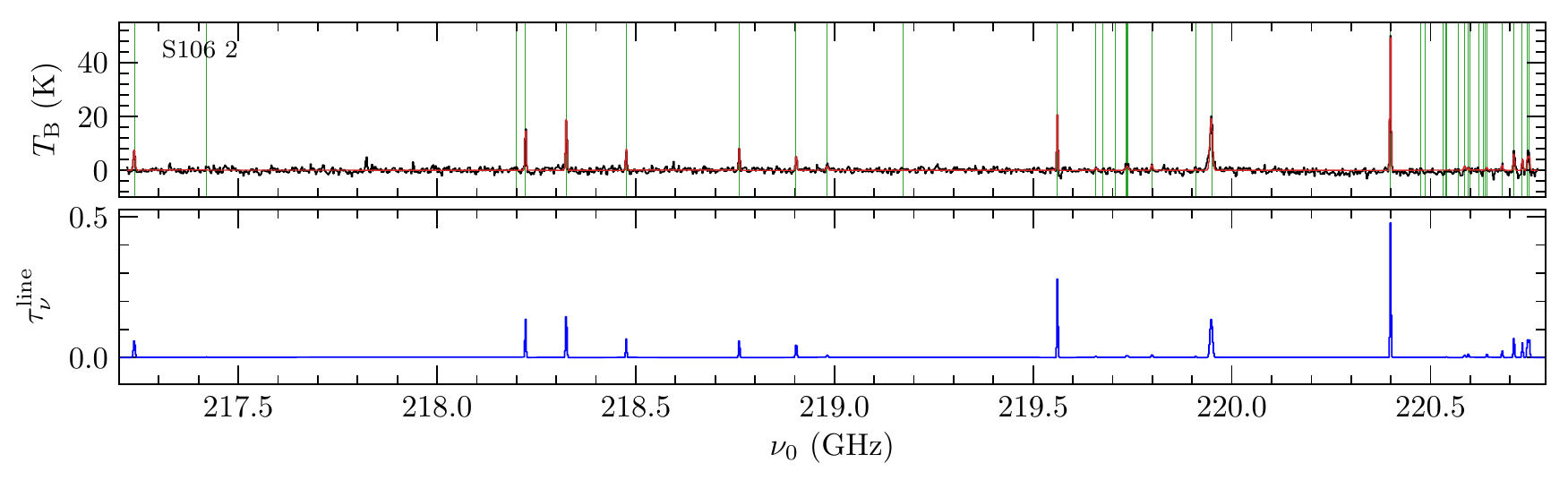}
\includegraphics[]{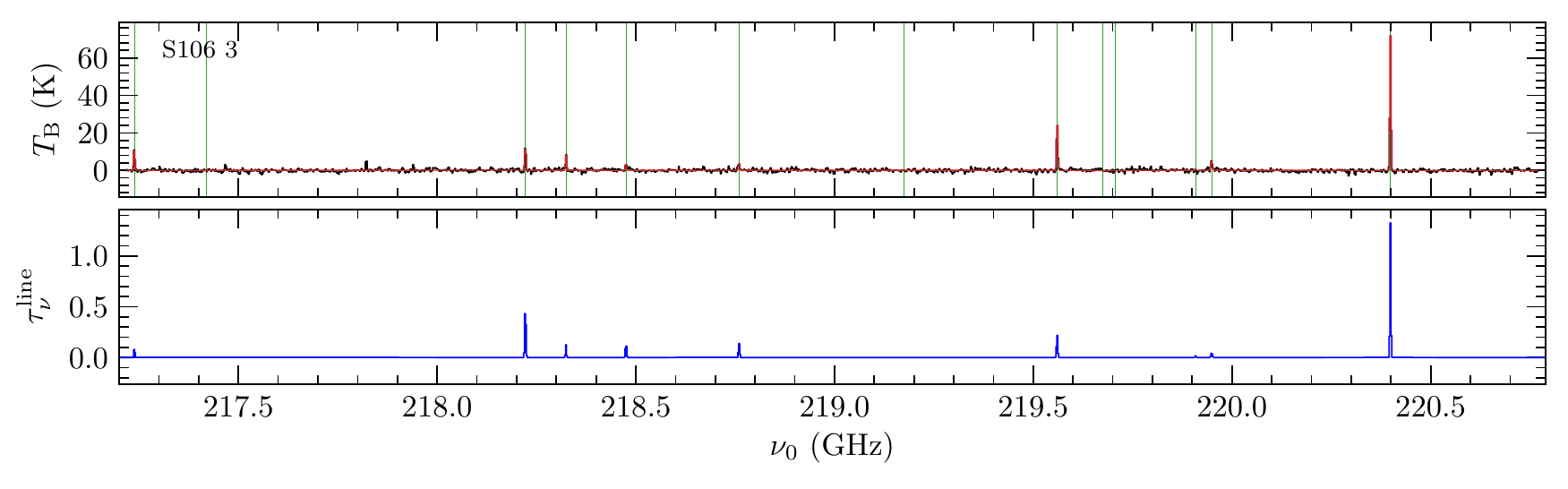}
\caption{\textit{Top panel:} Observed (black line) spectrum and \texttt{XCLASS} fit (red line) for all 120 analyzed positions. Fitted molecular transitions are indicated by green vertical lines. \textit{Bottom panel:} Optical depth profile (blue line) of all fitted transitions for all 120 analyzed positions.}
\end{figure*}
 
\begin{figure*}
\ContinuedFloat
\captionsetup{list=off,format=cont}
\centering
\includegraphics[]{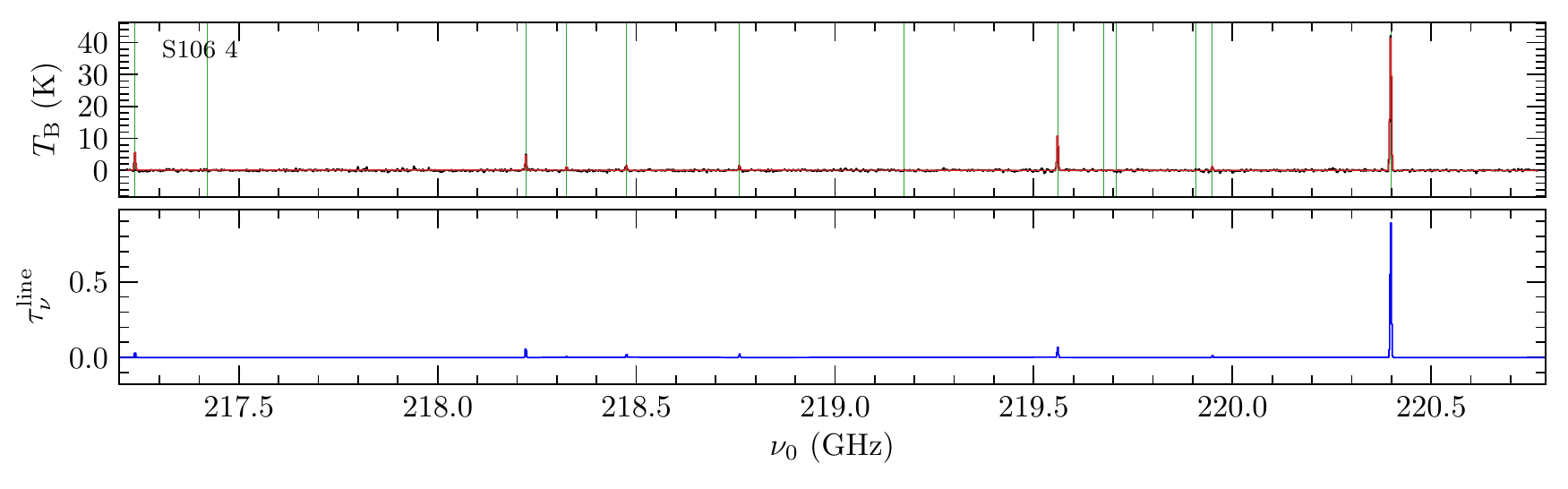}
\includegraphics[]{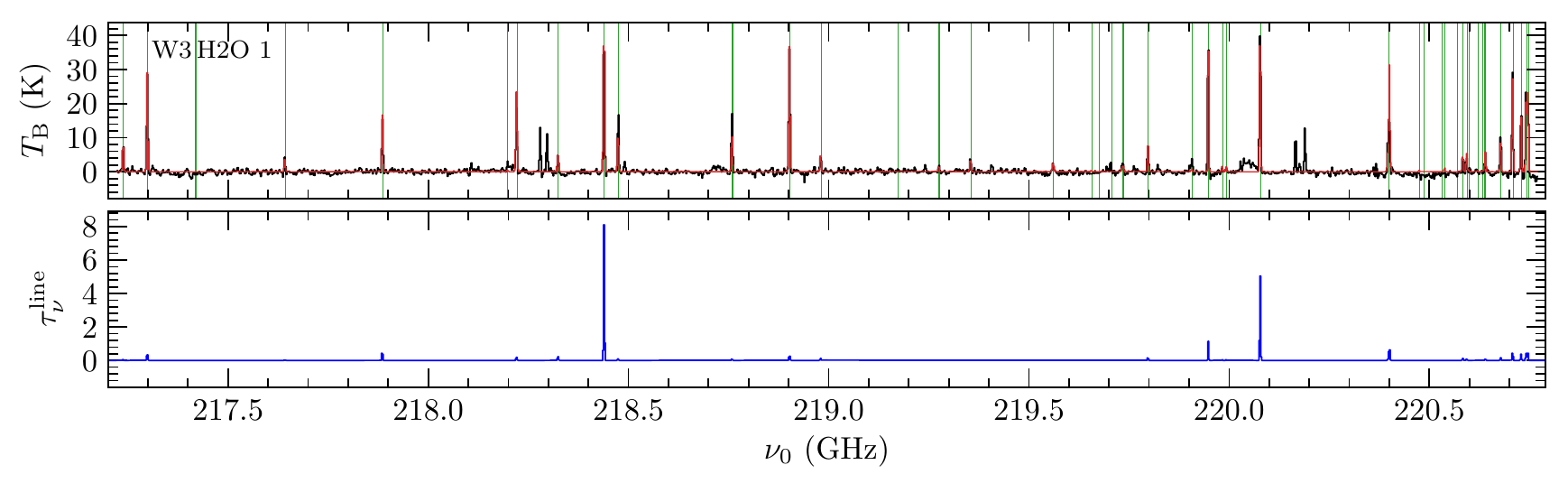}
\includegraphics[]{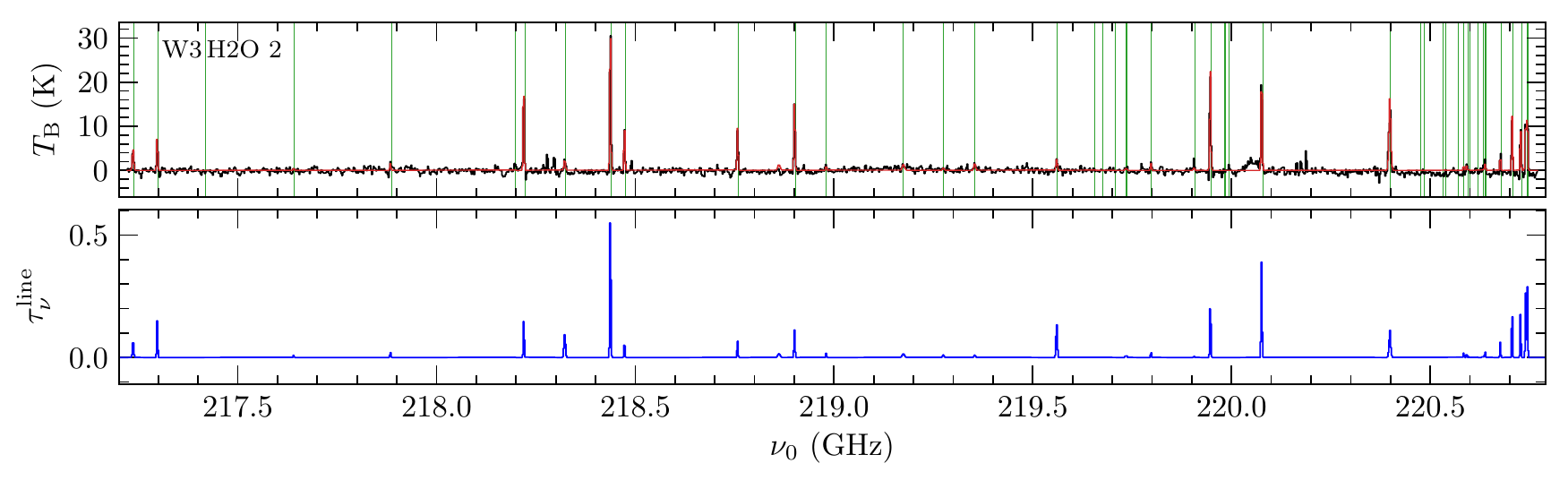}
\includegraphics[]{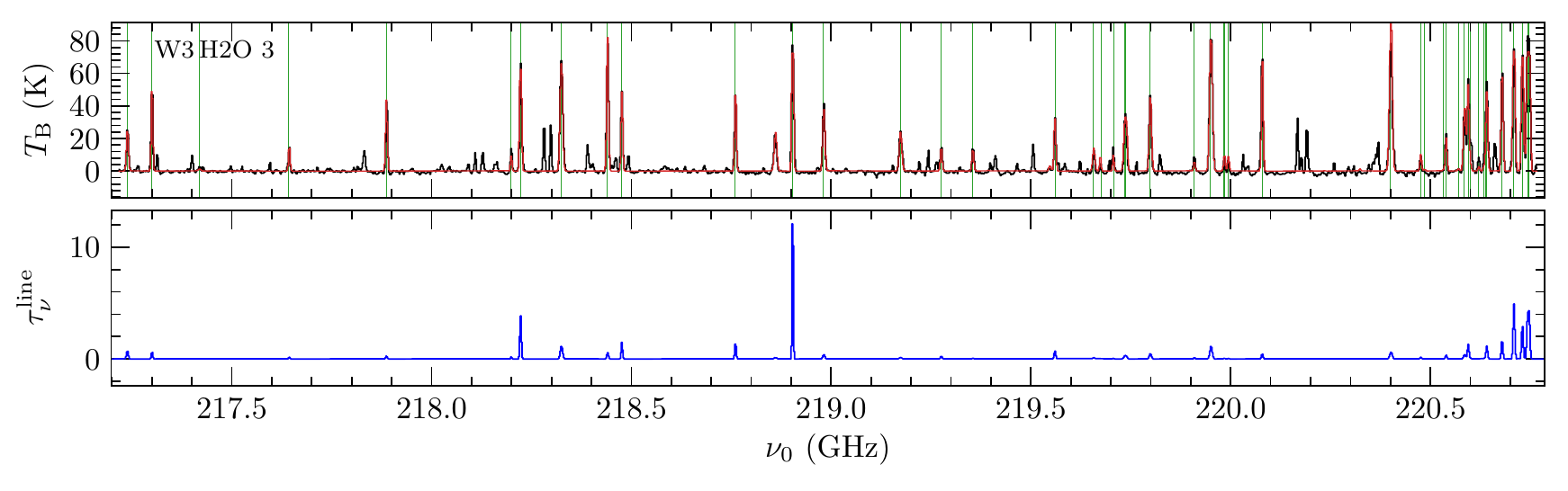}
\caption{\textit{Top panel:} Observed (black line) spectrum and \texttt{XCLASS} fit (red line) for all 120 analyzed positions. Fitted molecular transitions are indicated by green vertical lines. \textit{Bottom panel:} Optical depth profile (blue line) of all fitted transitions for all 120 analyzed positions.}
\end{figure*}
 
\begin{figure*}
\ContinuedFloat
\captionsetup{list=off,format=cont}
\centering
\includegraphics[]{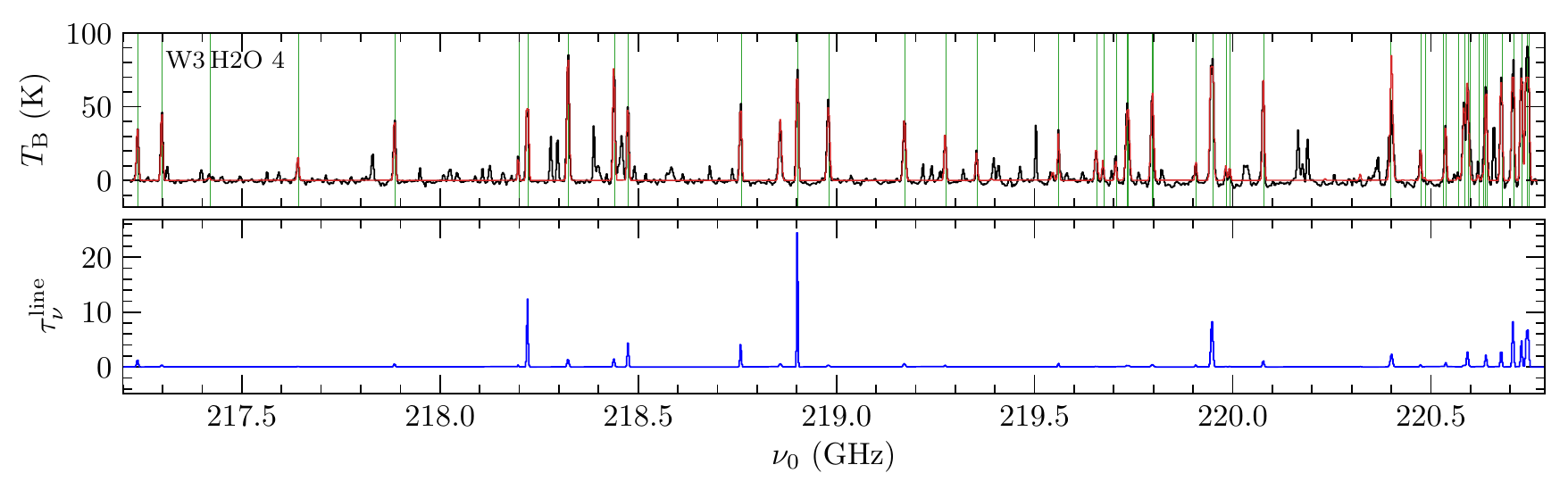}
\includegraphics[]{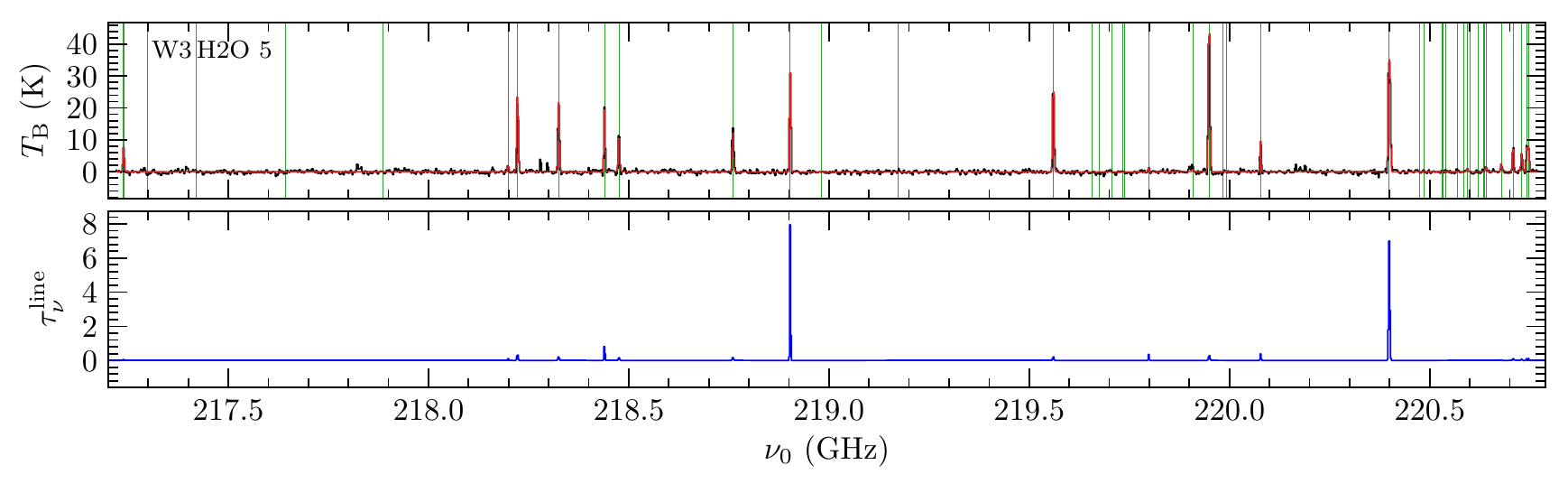}
\includegraphics[]{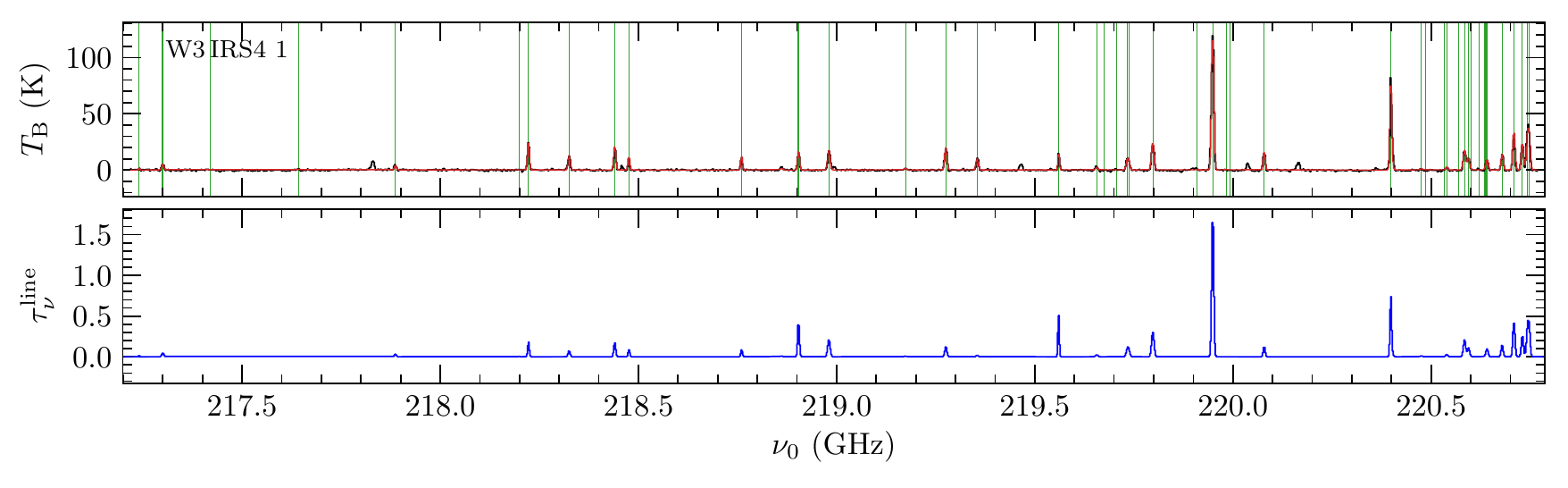}
\includegraphics[]{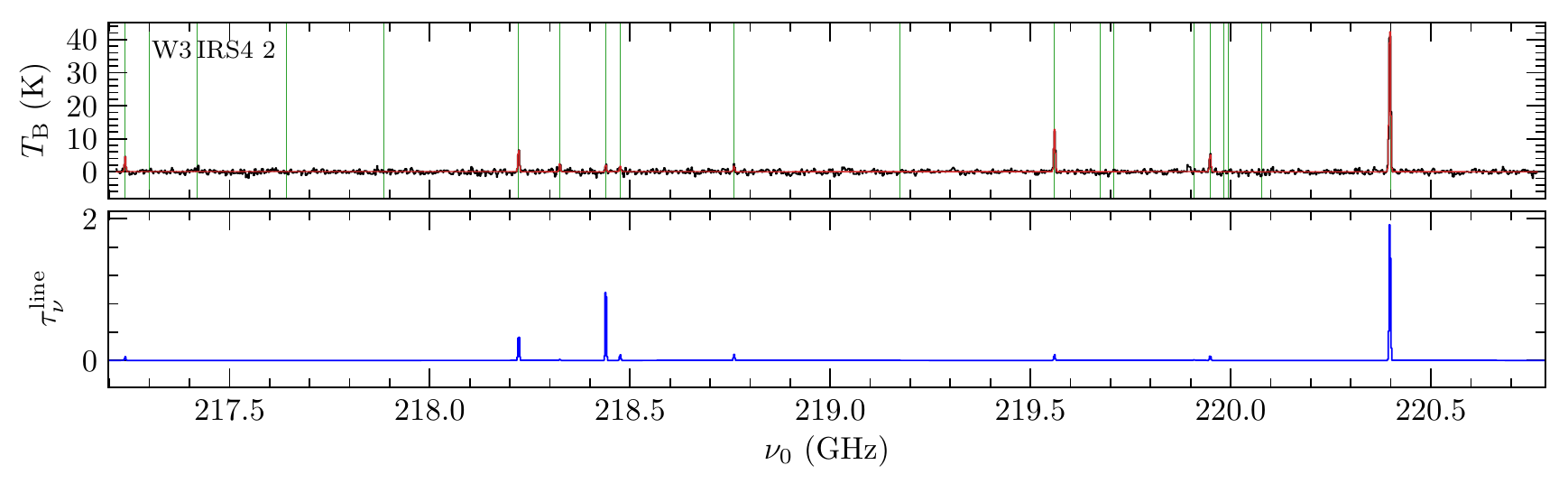}
\caption{\textit{Top panel:} Observed (black line) spectrum and \texttt{XCLASS} fit (red line) for all 120 analyzed positions. Fitted molecular transitions are indicated by green vertical lines. \textit{Bottom panel:} Optical depth profile (blue line) of all fitted transitions for all 120 analyzed positions.}
\end{figure*}
 
\begin{figure*}
\ContinuedFloat
\captionsetup{list=off,format=cont}
\centering
\includegraphics[]{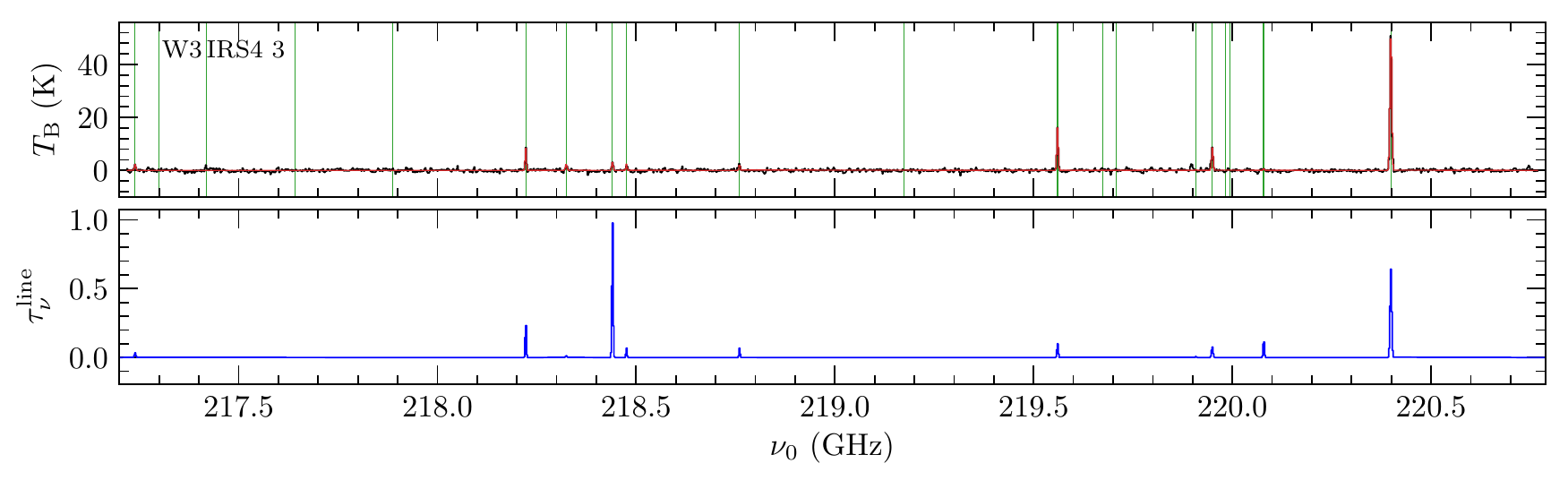}
\includegraphics[]{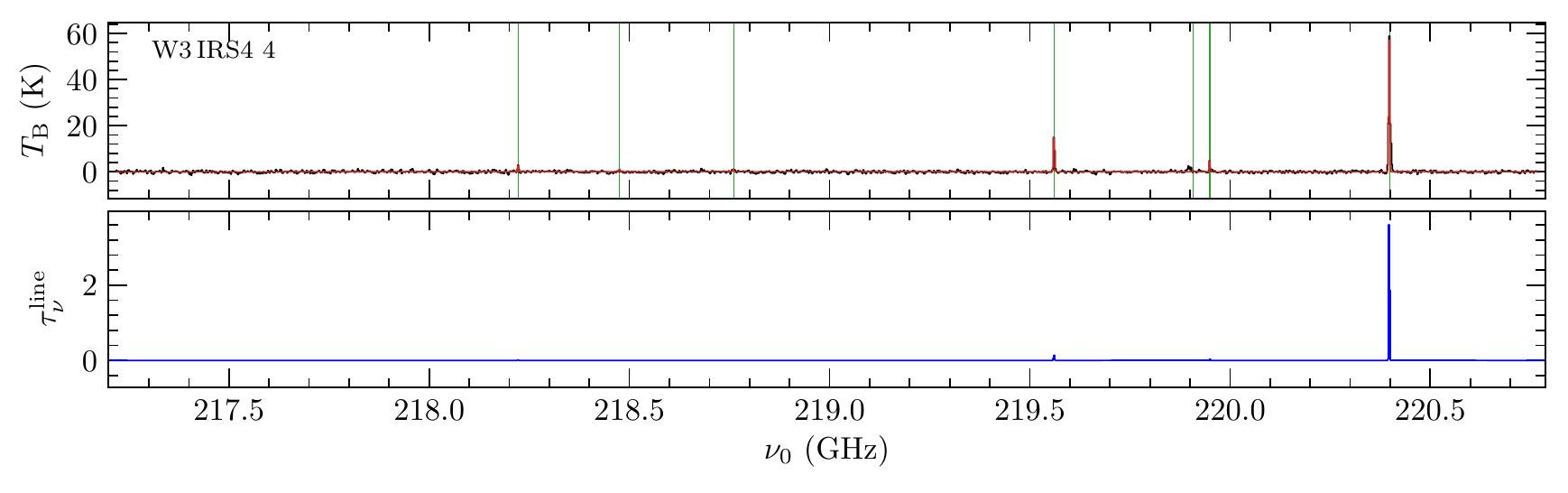}
\includegraphics[]{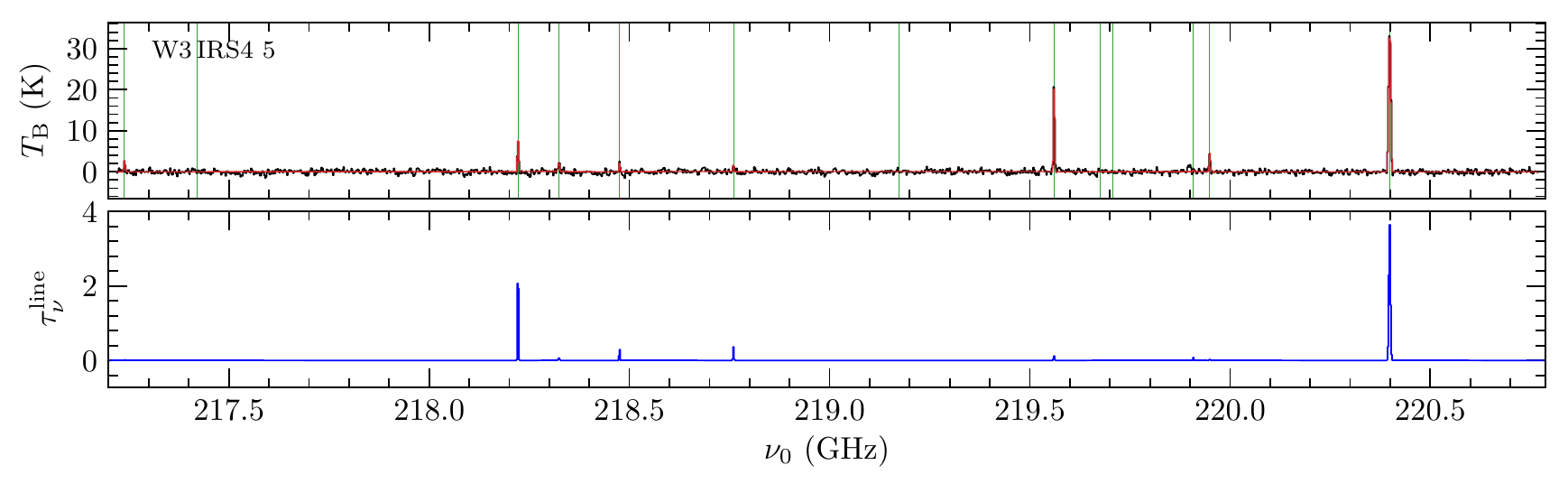}
\includegraphics[]{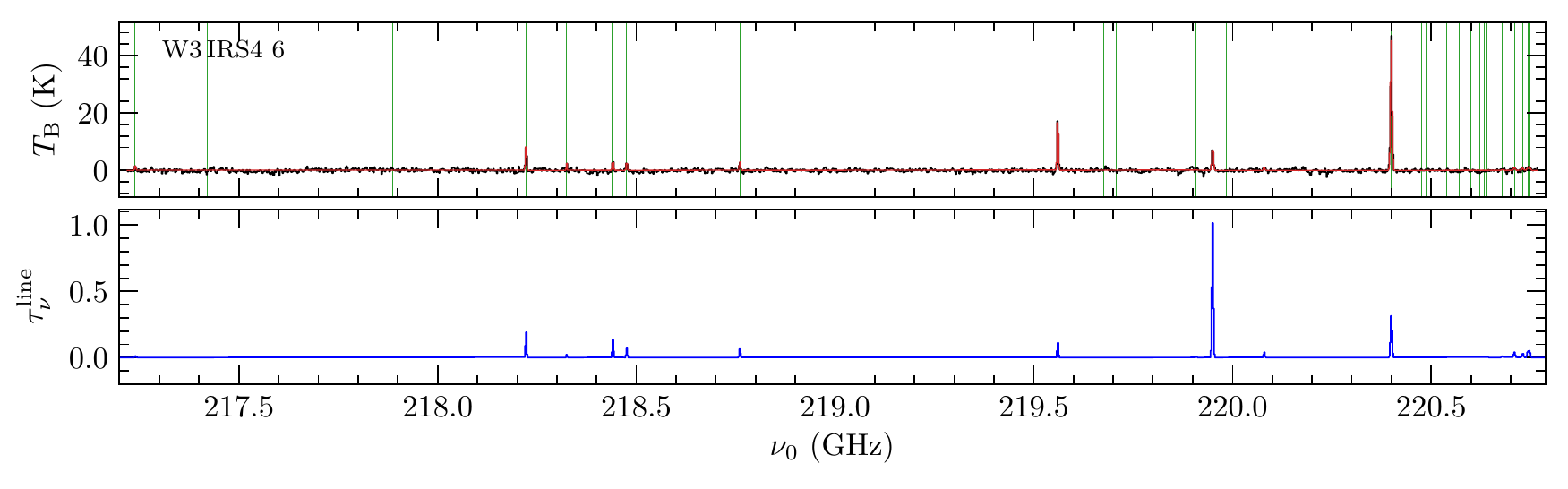}
\caption{\textit{Top panel:} Observed (black line) spectrum and \texttt{XCLASS} fit (red line) for all 120 analyzed positions. Fitted molecular transitions are indicated by green vertical lines. \textit{Bottom panel:} Optical depth profile (blue line) of all fitted transitions for all 120 analyzed positions.}
\end{figure*}

\section{Model results}

	Table \ref{tab:MUSCLE_results} shows the best-fit chemical age $\tau_\mathrm{chem}$, $\chi^2$ value, and percentage of well modeled molecules for each initial condition model (HMPO, HMC, and UCH{\sc ii} model) for each core.

\begin{table*}[!htb]
\caption{\texttt{MUSCLE} results for all three initial condition models. The best-fit chemical age $\tau_\mathrm{chem}$, $\chi^2$ value, and percentage of well modeled molecules $\Upsilon$ are shown for each initial condition model.}
\label{tab:MUSCLE_results}
\centering
\begin{tabular}{l | l l l | l l l | l l l}
\hline\hline
 & \multicolumn{3}{|c}{HMPO model} & \multicolumn{3}{|c}{HMC model} & \multicolumn{3}{|c}{UCH{\sc ii} model}\\
Core & $\tau_\mathrm{chem}$ & $\chi^2$ & $\Upsilon$ & $\tau_\mathrm{chem}$ & $\chi^2$ & $\Upsilon$ & $\tau_\mathrm{chem}$ & $\chi^2$ & $\Upsilon$\\
 & (yrs) & & (\%) & (yrs) & & (\%) & (yrs) & & (\%) \\
\hline
IRAS\,23033 1 & \,\,\,33\,574 & 0.339 & 64 & \,\,\,73\,271 & 0.346 & 64 & \,\,\,97\,675 & 0.336 & 64\\ 
IRAS\,23033 2 & \,\,\,18\,019 & 0.525 & 64 & \,\,\,49\,046 & 0.589 & 64 & 104\,065 & 0.589 & 55\\ 
IRAS\,23033 3 & \,\,\,18\,705 & 0.517 & 64 & \,\,\,50\,019 & 0.559 & 64 & \,\,\,96\,415 & 0.567 & 64\\ 
IRAS\,23151 1 & \,\,\,19\,415 & 0.458 & 64 & \,\,\,49\,222 & 0.557 & 64 & \,\,\,83\,510 & 0.444 & 82\\ 
IRAS\,23385 1 & \,\,\,17\,547 & 0.581 & 55 & \,\,\,48\,510 & 0.498 & 55 & \,\,\,87\,011 & 0.644 & 45\\ 
IRAS\,23385 2 & \,\,\,20\,729 & 0.545 & 55 & \,\,\,48\,510 & 0.602 & 45 & \,\,\,95\,268 & 0.465 & 55\\ 
AFGL\,2591 1 & \,\,\,17\,884 & 0.508 & 64 & \,\,\,54\,635 & 0.562 & 64 & \,\,\,83\,510 & 0.441 & 82\\ 
CepA\,HW2 1 & \,\,\,18\,705 & 0.602 & 55 & \,\,\,48\,530 & 0.605 & 55 & \,\,\,83\,510 & 0.478 & 64\\ 
CepA\,HW2 2 & \,\,\,19\,415 & 0.459 & 73 & \,\,\,48\,812 & 0.616 & 45 & \,\,\,88\,141 & 0.454 & 73\\ 
G084.9505 1 & \,\,\,16\,510 & 0.475 & 55 & \,\,\,48\,510 & 0.533 & 55 & \,\,\,83\,510 & 0.537 & 55\\ 
G094.6028 1 & \,\,\,17\,884 & 0.593 & 55 & \,\,\,56\,611 & 0.521 & 64 & \,\,\,83\,784 & 0.536 & 55\\ 
G100.38 1 & \,\,\,18\,705 & 0.483 & 64 & \,\,\,54\,090 & 0.459 & 64 & \,\,\,83\,602 & 0.432 & 73\\ 
G108.75 1 & \,\,\,20\,011 & 0.485 & 73 & \,\,\,48\,602 & 0.608 & 55 & \,\,\,87\,011 & 0.610 & 55\\ 
G108.75 2 & \,\,\,20\,729 & 0.319 & 73 & \,\,\,60\,268 & 0.325 & 73 & 106\,070 & 0.296 & 73\\ 
G075.78 1 & \,\,\,18\,330 & 0.471 & 64 & \,\,\,49\,547 & 0.532 & 64 & \,\,\,84\,222 & 0.410 & 82\\ 
IRAS\,21078 1 & \,\,\,17\,157 & 0.602 & 73 & \,\,\,55\,234 & 0.509 & 64 & \,\,\,92\,402 & 0.516 & 73\\ 
IRAS\,21078 2 & \,\,\,18\,168 & 0.601 & 55 & \,\,\,48\,876 & 0.542 & 64 & \,\,\,84\,884 & 0.575 & 55\\ 
NGC7538\,IRS9 1 & \,\,\,17\,884 & 0.502 & 73 & \,\,\,71\,070 & 0.521 & 73 & \,\,\,84\,292 & 0.447 & 82\\ 
S87\,IRS1 1 & \,\,\,16\,784 & 0.383 & 64 & \,\,\,59\,223 & 0.337 & 64 & \,\,\,83\,510 & 0.413 & 64\\ 
W3\,H2O 3 & \,\,\,17\,547 & 0.590 & 55 & \,\,\,48\,876 & 0.579 & 45 & \,\,\,83\,510 & 0.532 & 64\\ 
W3\,H2O 4 & \,\,\,17\,292 & 0.514 & 55 & \,\,\,54\,635 & 0.563 & 64 & \,\,\,85\,920 & 0.485 & 73\\ 
W3\,IRS4 1 & \,\,\,20\,011 & 0.492 & 64 & \,\,\,67\,238 & 0.570 & 64 & \,\,\,86\,415 & 0.469 & 82\\ 
\hline
\end{tabular}
\end{table*}

\end{appendix}

\end{document}